\documentclass[11pt]{article}

\usepackage{graphicx}

\usepackage{amssymb,amsfonts,amsmath}
\newcommand{\ecoli}{\emph{E.\ coli}}
\newcommand{\abs}[1]{\left\vert #1 \right\vert}
\newcommand{\norm}[1]{\left\Vert #1 \right\Vert}

\newcommand{\beq}[1]{\begin{equation}\label{#1}}
\newcommand{\eeq}{\end{equation}}
\newcommand{\beqn}{\begin{eqnarray*}}
\newcommand{\eeqn}{\end{eqnarray*}}
\newcommand{\bes}[1]{\begin{subequations}\label{#1}\begin{eqnarray}}
\newcommand{\ees}{\end{eqnarray}\end{subequations}}

\newcommand{\wfcd}{uniform linearizations with fast output}
\newcommand{\WFCD}{ULFO}

\newcommand{\xA}{x_A}
\newcommand{\xB}{x_B}
\newcommand{\xC}{x_C}

\newcommand{\xFB}{x_{F_B}}

\newcommand{\xEC}{x_{E_C}}
\newcommand{\xFC}{x_{F_C}}

\newcommand{\xCbar}{\overline{x}_C}
\newcommand{\txA}{\widetilde {x}_A}
\newcommand{\txB}{\widetilde {x}_B}
\newcommand{\txC}{\widetilde {x}_C}

\newcommand{\dxA}{\dot {x}_A}
\newcommand{\dxB}{\dot {x}_B}
\newcommand{\dxC}{\dot {x}_C}

\newcommand{\barG}{\overline{G}}

\newcommand{\inp}{u}

\newcommand{\x}{v}
\newcommand{\y}{w}

\newcommand{\X}{V}
\newcommand{\Y}{W}

\newcommand{\rref}[1]{Eq.\ref{#1}}

\newcommand{\ulow}{\underline{u}}
\newcommand{\uhigh}{\overline{u}}

\newcommand{\ubar}{\bar{u^*}}
\newcommand{\xbar}{\bar{x^*}}

\newcounter{figcount}
\newcommand{\myfig}[1]{\refstepcounter{figcount} \label{#1}}

\newcommand{\M}[1]{$#1$}

\newcommand{\Ro}{R_1}
\newcommand{\Rt}{R_2}
\newcommand{\ro}{r_1}
\newcommand{\rt}{r_2}
\newcommand{\GEF}{GEF}
\newcommand{\GAP}{GAP}
\newcommand{\kGEF}{k_{GEF}}
\newcommand{\kmGEF}{k_{-GEF}}
\newcommand{\kGAP}{k_{GAP}}
\newcommand{\kmGAP}{k_{GAP}}
\newcommand{\RASGTP}{Ras^{GTP}}
\newcommand{\RBDCYT}{RBD^{\mbox{cyt}}}
\newcommand{\kRAS}{k_{RAS}}
\newcommand{\kmRAS}{k_{-RAS}}
\newcommand{\kRBDoff}{k_{RBD}^{\mbox{off}}}
\newcommand{\kRBDon}{k_{RBD}^{\mbox{on}}}
\newcommand{\RAS}{RAS}
\newcommand{\RBD}{RBD}
\newcommand{\camp}{v}

\usepackage{epsf,latexsym,amssymb}
\usepackage{graphicx}
\usepackage{subfig}
\usepackage{amsmath}  %will give errors with \be \ee !!!
\newcommand{\comment}[1]{}
\newcommand{\edo}{\end{document}}

\newcommand{\be}[1]{\begin{equation}\label{#1}}
\newcommand{\ee}{\end{equation}}

\newcommand{\bi}{\begin{itemize}}
\newcommand{\ei}{\end{itemize}}
\newcommand{\ben}{\begin{enumerate}}
\newcommand{\een}{\end{enumerate}}

\begin{document}

\title{A Characterization of Scale Invariant Responses\\
 in Enzymatic Networks}

\author{Maja Skataric\\
{Department of Electrical Engineering, Rutgers University, Piscataway, NJ, USA} \and
Eduardo Sontag\footnote{Corresponding author}\\
{Department of Mathematics, Rutgers University, Piscataway, NJ, USA}}

\maketitle

\medskip

\noindent{\bf Running head:} Scale Invariant Responses in Enzymatic Networks

%\begin{article}

\begin{abstract}
%%%%%%%%%%%%%%%%%%%%
An ubiquitous property of biological sensory systems is adaptation: a step
increase in stimulus triggers an initial change in a biochemical or
physiological response, followed by a more gradual relaxation toward a basal,
pre-stimulus level.  Adaptation helps maintain essential variables within
acceptable bounds and allows organisms to readjust themselves to an
optimum and non-saturating sensitivity range when faced with a prolonged
change in their environment.
%%%%%%%%%%%%%%%%%%%%%%%%%%%%%%%%%%%%%%%%%%%%%%%%%%%%%%%%%%%%%%%%%
Recently, it was shown theoretically and experimentally that many adapting
systems, both at the organism and single-cell level, enjoy a remarkable
additional feature: scale invariance, meaning that the initial, transient
behavior remains (approximately) the same even when the background signal
level is scaled.
%%%%%%%%%%%%%%%%%%%%%%%%%%%%%%%%%%%%%%%%%%%%%%%%%%%%%%%%%%%%%%%%%
In this work, we set out to investigate under what conditions a broadly used
model of biochemical enzymatic networks will exhibit scale-invariant behavior.
An exhaustive computational study led us to discover a new property of
surprising simplicity and generality, {\wfcd} ({\WFCD}), whose validity we
%EDS18feb12: changed tense here and below to present
show is both 
%EDS18feb12, emph
\emph{necessary and sufficient}
for scale invariance of enzymatic networks.
Based on this study, we 
go
%went 
on to develop a mathematical explanation of how
{\WFCD} results in scale invariance.
Our work provides a surprisingly consistent, simple, and general framework for
understanding this phenomenon,
%EDS18feb12, added
and results in concrete experimental predictions.
%  Our analysis also revealed

\end{abstract}

%{\bf Keywords: enzymatic networks, adaptation, scale-invariance, fold-change detection}

\noindent{{\bf Author summary:}
An ubiquitous property of biological sensory systems is adaptation: a step
increase in stimulus triggers an initial change in a biochemical or
physiological response, followed by a more gradual relaxation toward a basal,
pre-stimulus level.  Adaptation helps maintain essential variables within
acceptable bounds and allows organisms to readjust themselves to an
optimum and non-saturating sensitivity range when faced with a prolonged
change in their environment.
%%%%%%%%%%%%%%%%%%%%%%%%%%%%%%%%%%%%%%%%%%%%%%%%%%%%%%%%%%%%%%%%%
Recently, it was shown theoretically and experimentally that many adapting
systems, both at the organism and single-cell level, enjoy a remarkable
additional feature: scale invariance, meaning that the initial, transient
behavior remains (approximately) the same even when the background signal
level is scaled.
%%%%%%%%%%%%%%%%%%%%%%%%%%%%%%%%%%%%%%%%%%%%%%%%%%%%%%%%%%%%%%%%%
In this work, we develop a mathematical characterization of biochemical
enzymatic networks that exhibit scale-invariant behavior
and make concrete experimental predictions.

\newpage

\section{Introduction}

The survival of organisms depends critically upon their capacity to
formulate appropriate responses to sensed chemical and physical environmental
cues. 
These responses manifest themselves at multiple levels, from human sight,
hearing, taste, touch, and smell, to individual cells in which signal
transduction and gene regulatory networks mediate the processing of measured
external chemical concentrations and physical conditions, such as ligand
concentrations or stresses, eventually leading to regulatory changes in
metabolism and gene expression.

An ubiquitous property of biological sensory systems at all levels
%%%%%%%%%%%%%%%%%%%%%%%%%%%%%%%%%%%%%%%%%%%%%%%%%%%%%%%%%%%%%%%%%%%%%%%%%%%%%%%%
% leave word scale for ``scale-invariance''
%scales 
%%%%%%%%%%%%%%%%%%%%%%%%%%%%%%%%%%%%%%%%%%%%%%%%%%%%%%%%%%%%%%%%%%%%%%%%%%%%%%%%
is that of \emph{adaptation}: a step increase in stimulus triggers an initial,
and often rapid, change in a biochemical or physiological response, followed
by a more gradual relaxation toward a basal, pre-stimulus level
\cite{alonbook}.
%%%%%%%%%%%%%%%%%%%%%%%%%%%%%%%%%%%%%%%%%%%%%%%%%%%%%%%%%%%%%%%%%%%%%%%%%%%%%%%%
% could break paragraph here
%%%%%%%%%%%%%%%%%%%%%%%%%%%%%%%%%%%%%%%%%%%%%%%%%%%%%%%%%%%%%%%%%%%%%%%%%%%%%%%%
Adaptation plays a role in ensuring that essential variables stay within
acceptable bounds, and it also allows organisms to readjust themselves to an
optimum and non-saturating sensitivity range even when faced with a prolonged
change in their operating environment, thus making them capable of detecting
changes in signals while ignoring background information.

Physiological examples of adaptation in higher organisms include phenomena
such as the control of the amount of light entering eyes through the
contraction and relaxation of the pupil by the nervous system, which brings
intensities of illumination within the retinal working range, or the
regulation of key metabolites in the face of environmental variations
\cite{Keener}.  
%%%%%%%%%%%%%%%%%%%%%%%%%%%%%%%%%%%%%%%%%%%%%%%%%%%%%%%%%%%%%%%%%%%%%%%%%%%%%%%%
% could break paragraph here
%%%%%%%%%%%%%%%%%%%%%%%%%%%%%%%%%%%%%%%%%%%%%%%%%%%%%%%%%%%%%%%%%%%%%%%%%%%%%%%%
At the single-cell level, one of the best understood examples of adaptation is
exhibited by the {\ecoli} chemotaxis sensory system, which responds to
gradients of nutrient and ignores constant (and thus uninformative)
concentrations \cite{BlockEtlAl,ShimizuTuBerg2010}.
%%%%%%%%%%%%%%%%%%%%%%%%%%%%%%%%%%%%%%%%%%%%%%%%%%%%%%%%%%%%%%%%%%%%%%%%%%%%%%%%
% could break paragraph here
%%%%%%%%%%%%%%%%%%%%%%%%%%%%%%%%%%%%%%%%%%%%%%%%%%%%%%%%%%%%%%%%%%%%%%%%%%%%%%%%
The term ``exact'' or ``perfect'' adaptation is employed to describe processes
which, after a transient, return with very high accuracy to the same
input-independent level.  In practice, however, an approximate adaptation
property is usually adequate for proper physiological response
\cite{MelloTu2003}. %context reference from aerotaxis mogilner paper

By definition, neither the concepts of perfect nor approximate adaptation
address the characteristics of the transient signaling which occurs prior to
a return to steady state.  The amplitude and other characteristics of
transient behaviors, however, are physiologically relevant.
In this more general context, a remarkable phenomenon exhibited by several
human and animal sensory systems is 
\emph{scale invariance} or logarithmic sensing
\cite{Keener} %pnas shoval bib
\cite{Laming} %pnas shoval bib
\cite{Thompson}.
% this ref taken from kalinin et al:
%Thompson R.F. Harper and Row; New York: 1967. Foundations of physiological psychology.
This means that responses are functions of upon ratios (in contrast to
actual magnitudes), of a stimulus relative to the background.
There is evidence for this phenomenon at an intracellular level as well.
It appears in bacterial chemotaxis
\cite{kalinin2009} %pnas shoval bib
\cite{Mesibov1973}, %pnas shoval bib
in the sensitivity of \emph{S.\ cerevisiae} to fractional rather
than absolute pheromone gradients
\cite{Paliwal},
%Paliwal, S. and Iglesias, P. and Campbell, K. and Hilioti, Z. and Groisman, A. and Levchenko, A.
%MAPK-mediated bimodal gene expression and adaptive gradient sensing in yeast
%Nature
%446
%46--51
%2007
and in two mammalian signaling systems: 
transcriptional as well as embryonic phenotype responses to
$\beta$-catenin levels in Wnt signaling pathways
\cite{Goentoro2009},
and nuclear ERK localization in response to EGF signaling
\cite{Cellina2009}.
%%%%%%%%%%%%%%%%%%%%%%%%%%%%%%%%%%%%%%%%%%%%%%%%%%%%%%%%%%%%%%%%%%%%%%%%%%%%%%%%
% could break paragraph here
%%%%%%%%%%%%%%%%%%%%%%%%%%%%%%%%%%%%%%%%%%%%%%%%%%%%%%%%%%%%%%%%%%%%%%%%%%%%%%%%
Scale invariance allows systems to react to inputs ranging over several orders
of magnitude, and is speculated to help make behaviors robust to external
noise as well as to stochastic variations in total expressed concentrations of
signaling proteins
\cite{shoval10}.

Mathematically, scale invariance is defined by the following property of
transient behaviors
\cite{shoval10}:
if a stimulus changes from a background level $u_0$ to a new level $u$, then
the entire time response of the system is the same as if the stimulus had
changed, instead, from a background level $pu_0$ to $pu$.
%, no matter what is the positive number $p$.
In other words, only the ratio (or ``fold-change'')
$pu/pu_0 = u/u_0$ is relevant to the response; the ``scale'' $p$ is
irrelevant.  
%EDS18feb12:
For this reason, the term ``fold change detection'' is interchangeably used
instead of scale-invariance.
%%%%%%%%%%%%%%%%%%%%%%%%%%%%%%%%%%%%%%%%%%%%%%%%%%%%%%%%%%%%%%%%%%%%%%%%%%%%%%%%
% could break paragraph here
%%%%%%%%%%%%%%%%%%%%%%%%%%%%%%%%%%%%%%%%%%%%%%%%%%%%%%%%%%%%%%%%%%%%%%%%%%%%%%%%
Scale invariance implies adaptation, but not every adaptive system is scale
invariant
\cite{shoval10}.
A 
%general 
mathematical analysis of scale-invariance was initiated in
\cite{shoval10},
\cite{shoval_alon_sontag_2011}.
Predictions regarding scale-invariance of
  \emph{E.\ coli} chemotaxis were subsequently experimentally verified
\cite{ShimizuStocker2011}.
%A necessary and sufficient condition for a network to have scale invariant
%behavior has been found
%\cite{shoval_alon_sontag_2011}.
While adaptation can be often understood in terms of control-theoretic tools
based on linearizations
\cite{mct}
\cite{doyleIMP}
\cite{imp03}
\cite{iglesias_ejc03}
\cite{Ma},
scale invariance is a genuinely {\bf nonlinear} property; as a matter of fact, a
linear system can never display scale-invariance, since the response to an
input scaled by $p$ will also be scaled by this same factor $p$.

In this work, 
we focus on enzymatic signal transduction systems,
which involve the activation/deactivation cycles that typically mediate
transmission of external signals to transcription factors and other effectors.
%%%%%%%%%%%%%%%%%%%%%%%%%%%%%%%%%%%%%%%%%%%%%%%%%%%%%%%%%%%%%%%%
Networks involving such enzymatic cycles are involved in signal
transduction networks from 
bacterial two-component systems and phosphorelays 
\cite{groisman, grossman}
to actin treadmilling
\cite{chen},
guanosine triphosphatase cycles
\cite{Donovan}, 
glucose mobilization
\cite{karp},
metabolic control
\cite{stryer},
cell division and apoptosis
\cite{sulis},
cell-cycle checkpoint control
\cite{lew},
and 
the eukaryotic Mitogen-Activated Protein Kinase (MAPK) cascades which mediate
growth factor inputs and determine proliferation, differentiation, and apoptosis
\cite{lauffenburger,%
Chang,%
ferrell,%
widman,%
pnasangeliferrellsontag04}.

Given the biological importance of these processes, and the already observed
scale-invariance in some of these pathways
\cite{Goentoro2009}
\cite{Cellina2009},
%it is of interest to characterize the classes of enzymatic signal
%transduction networks that display scale-invariance.
we pose here the following question:
%to characterize what enzymatic networks with scale-invariant response.
%Here we further constrain the problem and ask: 
which enzymatic networks do not merely adapt, but also display scale invariance?
In order to answer this question, 
we performed an exhaustive computational
study of all 3-node networks, finely sampled in parameter space.
Only about 0.01\% of these networks are capable of (approximate) adaptation.
Testing which of these adapting networks also display scale-invariant
behavior, we found that only about 0.15\% of them did.
Once that this small subclass was identified, we turned to the problem of
determining what network characteristics would 
%completely 
explain the results of these numerical experiments.
We discovered a surprisingly simple and general property, which we call
{\wfcd} ({\WFCD}), that is displayed by all the networks in this subclass,
and here we provide a theoretical framework that explains
%EDS18feb12: emphasis; also some tense changes above
conceptually why this property is both \emph{necessary and sufficient} for
scale invariance of enzymatic networks.
%EDS18feb12: added
As an application, we consider a recently published model
\cite{takeda_et_al2012} 
of an eukaryotic enzymatic system, specifically the pathway involved in 
the social amoeba \emph{Dictyostelium discoideum}'s chemotactic response to
cAMP. and show that our conditions are satisfied in appropriate ranges of
cAMP input.

%%%%%%%%%%%%%%%%%%%%%%%%%%%%%%%%%%%%%%%%%%%%%%%%%%%%%%%%%%%%%%%%%%%%%%%%%%%%%%%%
% could break paragraph here
%%%%%%%%%%%%%%%%%%%%%%%%%%%%%%%%%%%%%%%%%%%%%%%%%%%%%%%%%%%%%%%%%%%%%%%%%%%%%%%%
Characterizations of this sort allow one to understand which networks are
robust to scale uncertainty, and constitute a powerful tool in allowing one to
discard putative mechanisms that are not consistent with experimentally
observed scale-invariant behaviors
\cite{shoval_alon_sontag_2011},
\cite{ShimizuStocker2011}.

\section{Results}

\subsection{Three-node enzymatic networks}

We consider networks consisting of three types of enzymes, denoted respectively
as $A$, $B$, and $C$.
Each of these enzymes can be in one of two states, active or inactive.
The fractional concentration of active enzyme $A$ is represented by
a variable $\xA=\xA(t)$, so $\txA = 1-\xA$ is the fraction of inactive
enzyme $A$.  Similar notations are used for $B$ and $C$.
Only enzyme $A$ is directly activated by an external input signal, and the
response of the network is reported by the fraction of active $C$.
Enzyme $B$ acts as an auxiliary element.
Each enzyme may potentially act upon each other through activation
(positive regulation), deactivation (negative regulation), or not at all.
If a given enzyme is not deactivated by any of the remaining two, we assume
that it is constitutively deactivated by a specific enzyme; similarly,
if a given enzyme is not activated by any other, there is a 
constitutively activating enzyme for it.
One represents networks by 3-node directed graphs, with
nodes labeled $A$, $B$, $C$, and with edges between two nodes labeled $+$ and
$-$ (or ``$\rightarrow $'' and ``$\dashv$'') to denote positive or negative regulation
respectively; no edge is drawn if there is no action.  There are $3^2=9$
potential directed edges among the three nodes ($A$ to $A$, $A$ to $B$, etc.),
each of whose labels may be $+$, $-$, or ``none'' if there is no edge.  
This gives a total of $3^9 = 19,683$ possible graphs.  One calls each of these
possible graphs a \emph{topology}.  Discarding the 3,645 topologies that have
no direct or indirect links from the input to the output, there remain 16,038
topologies.

\subsection{Specification of a dynamic model}

We quantify the effects of each existing regulatory interaction by a
Michaelis-Menten term and write a three-variable ordinary differential
equation (ODE) that describes the time evolution of $\xA(t)$, $\xB(t)$, and
$\xC(t)$:
\bes{eq:network}
\dxA  &=& \sum_i\frac{k_{\X_iA}\x_i \cdot  \txA}{\txA+K_{\X_iA}}
  - \sum_i\frac{k_{\Y_iA}\y_i \cdot  \xA}{\xA+K_{\Y_iA}}\\
\dxB &=& \sum_i\frac{k_{\X_iB}\x_i \cdot  \txB}{\txB+K_{\X_iB}}
  - \sum_i \frac{k_{\Y_iB}\y_i \cdot  \xB}{\xB+K_{\Y_iB}}\\
\dxC &=& \sum_i\frac{k_{\X_iC}\x_i \cdot  \txC}{\txC+K_{\X_iC}}
  - \sum_i \frac{k_{\Y_iC}\y_i \cdot  \xC}{\xC+K_{\Y_iC}}
\ees
The $K$'s denote Michaelis-Menten, and the $k$'s catalytic, rate
constants associated to each regulatory interaction.
%, and $\txA = 1-\xA$, $\txB = 1-\xB$, $\txC = 1-\xC$.
All the summations range over $i=1,\ldots ,6$.
Each ``$\X_i$'' represents one of $A$, $B$, $C$, $E_A$, $E_B$, $E_C$, the
activating enzymes in the respective equations, and each ``$\Y_i$'' one of
$A$, $B$, $C$, $F_A$, $F_B$, $F_C$, the deactivating enzymes; $E$
and $F$ are the constitutively activating and deactivation enzymes, buffered
at constant concentrations.
(Lower-case variables $v_i,w_i = \xA, \ldots  , \xFC$ denote active fractions)
%$1/2$.
As an exception, the equation for node $A$ does not include an $E_A$ term, but
instead includes a term $k_{UA}\inp\frac{\txA}{\txA+K_{UA}}$ that
models activation of $A$ by an external input whose strength at time $t$ is
given by $u=u(t)$ and whose values $u(t)$ stay within a range $[\ulow,\uhigh]$.
No enzyme appears both an activator and as a deactivator of any given
component, that is, $k_{X_iA}k_{Y_iA}=0$, $k_{X_iB}k_{Y_iB}=0$, and
$k_{X_iC}k_{Y_iC}=0$, and constitutive enzymes are included only if the
reaction would be otherwise irreversible.
For example, the topology shown in Fig.\ref{fig:topology2293}
\begin{figure}[ht]
\centering
\includegraphics[width=50mm]{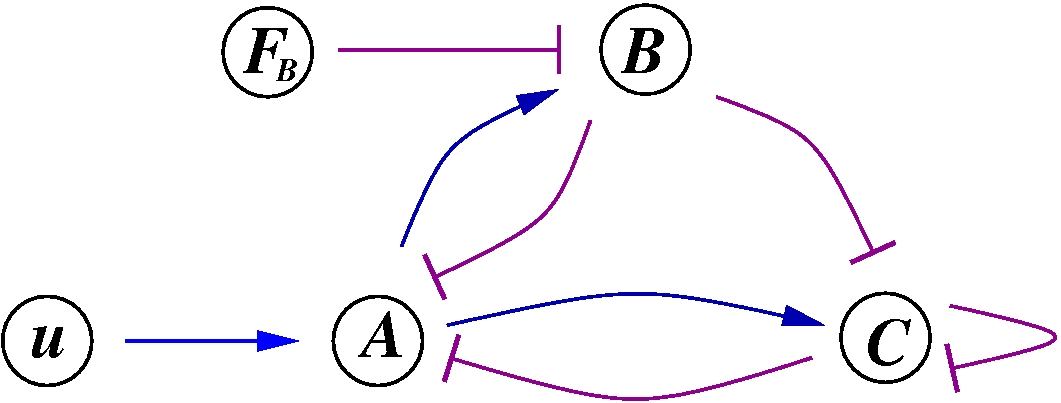}
\caption{Topology 2293}
%\label{fig:topology2293}
\end{figure}%
\myfig{fig:topology2293}%
is described by the following following set of ODE's:
\bes{eq:example1}
\dxA &=&
\frac{k_{UA} u  \cdot  \txA}{\txA+K_{UA}}-\frac{k_{BA} \xB  \cdot  \xA}{\xA+K_{BA}}
  -\frac{k_{CA} \xC \cdot  \xA}{\xA+K_{CA}}\\
\dxB &=&
\frac{k_{AB}\xA \cdot  \txB}{\txB+K_{AB}}-\frac{k_{F_BB} \xFB  \cdot  \xB}{\xB+K_{F_BB}}\\
\dxC  &=&
\frac{k_{AC}\xA \cdot  \txC}{\txC+K_{AC}}- \frac{k_{BC} \cdot  \xB\xC}{\xC+K_{BC}}
   -\frac{k_{CC}\xC \cdot  \xC}{\xC+K_{CC}}
\ees
The term \emph{circuit} is used to refer to a given topology together with a
particular choice of the $K$ and $k$ parameters.
The three-node model
in \rref{eq:network}
% and computational exploration tools
was employed by Ma et al.\ 
\cite{Ma},
%%%%%%%%%%%%%%%%%%%%%%%%%%%%%%%%%%%%%%%%%%%%%%%%%%%%%%%%%%%%%%%%%%%%%%%%
in order to classify the minimal enzymatic circuits that adapt.
%EDS18feb12:
(With the model in \cite{Ma} that we adopted, there is no direct connection
from the input to the output node, and two-node networks are not sufficient
for adaptation, while larger adapting 
networks contain these three-node networks \cite{Ma}.
If one allows direct connections from input to outputs, then two-node networks
are able to display adaptation.)
The same paradigm has since been used to investigate other network
characteristics as well
\cite{shah-sarkar2011},
\cite{yao-tan-et-al2011}.
%% refs from PUBMED to Ma et al
%{shah-sarkar2011}
%Shah NA, Sarkar CA
%Robust network topologies for generating switch-like cellular responses.
%PLoS Comput Biol.
%2011
%7(6):e1002085.
%PubMed PMID: 21731481; PubMed Central PMCID: PMC3121696.
%%%%%%%%%%%%%%%%%%%%%%%%%%%%%%%%%%%%%%%%%%%%%%%%%%%%%%%%%%%%%%%%%%%%%%%%%%%%%%
%{yao-tan-et-al2011}
%Yao G, Tan C, West M, Nevins JR, You L.
%Origin of bistability underlying mammalian cell cycle entry.
%Mol Syst Biol.
%2011
%7:485. 
%PubMed PMID: 21525871; PubMed Central PMCID: PMC3101952.

\subsection{Adaptation}

%Physical Biology 
%Paul François and Eric D Siggia
%2008
%Phys. Biol. 5 026009
Following
\cite{FrancoisSiggia2008},
we define adaptation behavior in terms of two functional metrics.
%(Fig.~\ref{fig:01ch3_3}).
%%%%%%%%%%%%%%%%%%%%%%%%%%%%%%%%%%%%%%%%%%%%%%%%%%%%%%%%%%%%%%%%%
The first metric quantifies the following effect:
if we start at steady state, and then step the input at time $t=0$ 
from a value $u_0$ to a different constant value $u_1$, then the system's
output, as reported by a response variable $y(t)$ (where $y(t)=\xC(t)$
in \rref{eq:network}), should return asymptotically to a value that is close
to the original value $y(0)$.
The relative difference in initial and final response
$\Delta ^{\infty }_y= \abs{y(+\infty )-y(0)}$ provides a measure of adaptation precision.
We say that a system is (approximately) adaptive provided that, for all inputs
in the valid range, $\Delta ^{\infty }_y/\Delta _u<0.1$, where $\Delta _u=\abs{u_1-u_0}/\abs{u_0}$
is the relative change in input. 
In particular, exact or perfect adaptation means that $\Delta ^{\infty }_y=0$.
The 10\% error tolerance is natural in applications, and the qualitative
conclusions are not changed by picking a smaller cutoff
\cite{Ma}.
A second metric relies upon the maximal transient difference in output,
normalized by the steady-state output,
$\Delta ^{\max}_y = \max \abs{y(t)-y(0)}/\abs{y(0)}$. 
%Adaptation requires that $\sigma _y$ exceed a minimal threshold.
A \emph{signal-detection} property for adaptation
\cite{imp03},
\cite{iglesias_ifac08},
should be imposed in order to rule out the trivial situation
$\Delta ^{\max}_y\approx0$ in which a system's output is independent of the input.
%($y(t)$ is almost constant, irrespective of the external input).
To avoid having to pick an arbitrary threshold, in this study we follow
the convention in
\cite{Ma}
of requiring the \emph{sensitivity} $\Delta ^{\max}_y/\Delta _u$ to be greater than one.

\subsection{Scale invariance}

Scale invariance is the property that if a system starts from a steady state
that was pre-adapted ($t<0$) to a certain background level $u_0$, and the
input is subsequently set to a new level $u$
%=u_0+\delta u$ 
at $t=0$, then the entire
time response of the system $y_{u_0,u}(t)$
%$y_{u_0,\delta }(t)$
is the same as the response
%$y_{pu_0,p\delta }(t)$
$y_{pu_0,pu}(t)$
that would result if the stimulus had changed, instead, from
$pu_0$ to $pu$.
This property should hold for scale changes $p>0$ that respect the
bounds $\ulow\leq u\leq \uhigh$ on inputs.
%%%%%%%%%%%%%%%%%%%%%%%%%%%%%%%%%%%%%%%%%%%%%%%%%%%%%%%%%%%%%%%%%%%%%%%%
For example, recent microfluidics and FRET experimental work
\cite{ShimizuStocker2011}
verified scale-invariance predictions that had
been made in
\cite{shoval10}
for bacterial chemotaxis under the nonmetabolizable attractant
$\alpha $-methylaspartate (MeAsp) as an input.
In these experiments, {\ecoli} bacteria were pre-adapted to input
concentrations and then tested in new nutrient gradients, and it was found
experimentally that there were two different ranges of inputs 
$[\ulow_1, \uhigh_1]$ and $[\ulow_2, \uhigh_2]$
in which scale-invariance holds, the ``FCD1'' and ``FCD2'' regimes, repectively.
(The term fold-change detection, or FCD, is used to reflect the fact that only
the ratio or fold-change $pu/pu_0 = u/u_0$ can be detected by the response
$y(t)$.)
More generally, the mathematical definition of (perfect) scale invariance
\cite{shoval_alon_sontag_2011}
imposes the ideal requirement that the same response invariance property is
exhibited if $u=u(t)$, $t\geq 0$ is any time-varying input.
The experiments in
\cite{ShimizuStocker2011}
included excitation by certain oscillatory inputs, for example.
In practice, however, this property will always break down for high-frequency
inputs, since there are limits to the speed of response of biological systems.

\subsection{%Perfectly a
Adaptive systems need not be scale-invariant} 
As an illustration of a (perfectly) adaptive yet not scale-invariant system,
consider the following equations:
\bes{eq:example2}
\dxA &=& k_1 u -k_2 \xB  \\
\dxB &=& k_3 \xA -k_4 \xB \\
\dxC &=& k_5\xA - k_6\xB\xC
\ees
which is a limiting case of the system described by
\rref{eq:example1}
when
%$k_{CA}$, $k_{CC}=0$,
%$K_{UA}$, $K_{BA}$, $K_{AB}$, and $K_{AC}$ are all $\approx0$,
%$k_{CA}=k_{CC}=K_{UA}=K_{BA}=K_{AB}=K_{AC}\approx0$,
$k_{CA},k_{CC},K_{UA},K_{BA},K_{AB},K_{AC}\approx0$,
$k_{B C} = k_6 K_{B C}$, $K_{B C}\gg1$ (so
$-k_{B C} \xB \xC / (\xC+K_{B C}) \approx -k_6\xB\xC$),
and
$k_{F_BB} \xFB = k_2 K_{F_BB}$ and $K_{F_BB}\gg1$.
This network perfectly adapts, since at steady state the output
is $\xC=\xCbar = k_4k_5/(k_3k_6)$, no matter what is the magnitude of the
constant input $u$, and in fact the system returns to steady state after a
step change in input $u$, with $\xC(t)\rightarrow \xCbar$ as $t\rightarrow \infty $
(general stability properties of feedforward circuits shown in
\cite{feedforward_circuits_adaptation}).
% see proposition 1; the z system is affine
On the other hand, the example in
\rref{eq:example2}
does not display scale invariance.
Indeed, consider the solution from an initial state pre-adapted to an input
level $u_0$, that is $\xA(0)=k_1k_4u_0/(k_2k_3)$, $\xB(0)=k_1u_0/k_2$, and
$\xC(0)=k_4k_5/(k_3k_6)$, and the input $u(t)\equiv u_1$ for $t\geq 0$.  Then,
$\xC(t) =  k_4k_5/(k_3k_6) + k_1k_5 (u_1-u_0) t^2/2 + O(t^3)$ for small $t\geq 0$.
% note: relative degree is geq 2 in this system$
Since the $t^2$ coefficient in this Taylor expansion gets multiplied by $p$
when $u_0$ is replaced by $pu_0$ and $u_1$ is replaced by $pu_1$, it follows
that the transient behavior of the output $\xC(t)$ depends on $p$.
% ADDED:
Interestingly, if the equation for the third node is replaced by
$\dxC = k_5\xA/\xB - k_6\xC$, that is to say the activation of $C$ is
repressed by $A$, instead of its de-activation being enhanced by $A$,
then scale invariance does hold true, because $\xA(t)$ and $\xB(t)$ both scale
by $p$ when $u_0\mapsto pu_0$, $u_1\mapsto u_0$, and $C(t)$ depends on the ratio of these
two functions (in particular, the $t^2/2$ term is $k_2k_5(u_1-u_0)/u_0$).
Such a repression is typical of genetic interaction networks, but is not
natural in enzymatic reactions.

It turns out that the example described by
\rref{eq:example2}
is typical: no enzymatic network described by
\rref{eq:network} can display perfect scale-invariant behavior.  This fact
is a consequence of the equivariance theorem proved in
\cite{shoval_alon_sontag_2011}
(see \emph{Materials and Methods}).
Thus, a meaningful study of enzymatic networks, even for perfectly adaptive
ones, must rely upon a test of approximate scale invariance.  Instead of
asking that
%$y_{u_0,\delta }(t)=y_{pu_0,p\delta }(t)$,
$y_{u_0,u}(t)=y_{pu_0,pu}(t)$,
as was the case in the theory developed in
\cite{shoval10}
\cite{shoval_alon_sontag_2011},
one should require only that the difference be small.
To investigate this issue, we computationally screened all
% 16,038 possible
3-node topologies through a high-throughput random parameter scan,
testing for small differences in responses to scaled steps.
We found that approximately 0.01\% of the samples showed adaptation, but of
them, only about 0.15\% passed the additional criterion of approximate scale 
invariance
(see \emph{Materials and Methods}).
These samples belonged to 21 (out of 16,038 possible) topologies.
As an example of the behavior of one of these,
Fig.\ref{fig:plotFCD2293} shows a response resulting from a 20\% step, from
$3$ to $3.6$, compared to the response obtained when stepping from $5$ to $6$;
the graphs are almost indistinguishable.
(See \emph{SI Text} for an enumeration of circuits and corresponding plots). 
In the following discussion, we will refer to these surviving circuits, and
their topologies, as being ``approximately scale invariant'' (ASI).

We found that all ASI networks possess a feedforward motif,
meaning that there are connections $A\rightarrow B\rightarrow C$ and as well as $A\rightarrow C$.
Such feedforward motifs have been the subject of extensive analysis in the
systems biology literature
\cite{alonbook}.
%%%%%%%%%%%%%%%%%%%%%%%%%%%%%%%%%%%%%%%%%%%%%%%%%%%%%%%%%%%%%%%%%%%%%%%%%
% refs from feedforward-adapt paper:
and are often involved in detecting changes in signals
\cite{mangan06}.
They appear in pathways as varied as
{\ecoli} carbohydrate uptake via the carbohydrate phosphotransferase system
\cite{Kremling},
control mechanisms in mammalian cells
\cite{maayan_science05},
nitric oxide to NF-$\kappa$B activation
\cite{kim33,kim32},
EGF to ERK activation
\cite{kim16,kim18},
glucose to insulin release
\cite{kim28,kim30},
ATP to intracellular calcium release
\cite{kim34},
and
microRNA regulation
\cite{kim12}.
%and are over-represented
%in genetic/metabolic interfaces
%~\cite{semsey}
%%%%%%%%%%%%%%%%%%%%%%%%%%%%%%%%%%%%%%%%%%%%%%%%%%%%%%%%%%%%%%%%%%%%%%%%%
The feedforward motifs in all ASI networks are incoherent, meaning
such that the direct effect $A\rightarrow C$ has an opposite sign to the
net indirect effect through $B$.
An example of an incoherent feedforward connection is provided by the simple
system described by
\rref{eq:example2},
where the direct effect of $A$ on $C$ is positive, but the indirect effect is
negative: $A$ activates $B$ which in turn deactivates $C$.
(Not every incoherent feedforward network provides scale invariance; a 
classification of those that provide exact scale invariance is known
\cite{shoval_alon_sontag_2011}.)
It is noteworthy that all ASI circuits have a positive regulation from
A to B and a negative regulation from B to A.
%Thus they all include a ``negative feedback loop with a buffer node''
%(NFBLB) motif 
%(Fig.~7 in
%\cite{Ma}).
%An NFBLB does not represent the classical adaptation motif of integral feedback
%\cite{doyleIMP},
%\cite{imp03}.
%However, when $C$ depends only on $A$ and has fast dynamics, as is the case 
%with the receptor complex including histidine kinase cheA  (``$A$'') and the
%response regulator protein cheY (``$C$'') in the {\ecoli} chemotaxis system
%\cite{Tu-shimizu-berg08},

We then discovered a surprising common feature among all ASI circuits.
This feature can best be explained by a further examination of the example in
\rref{eq:example2}.

\begin{figure}[ht]
\centering
\includegraphics[width=0.5\textwidth]{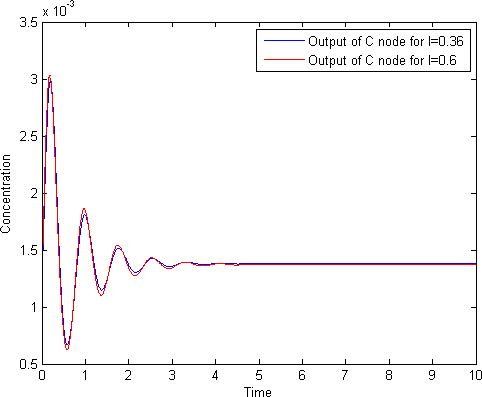}
\caption{Scale-invariance: plots overlap, for responses to steps $3$$\rightarrow $$1.2*3$
  and $5$$\rightarrow $$1.2*5$. Network is the one described by
  \protect{\rref{eq:example1}}.  Random parameter set: % 
$\scriptstyle K_{UA}= 0.093918$
$\scriptstyle k_{UA}= 11.447219$, 
$\scriptstyle K_{BA}= 0.001688$
$\scriptstyle k_{BA}= 44.802268$,
$\scriptstyle K_{CA}= 90.209027$
$\scriptstyle k_{CA}= 96.671843$,
$\scriptstyle K_{AB}=0.001191$
$\scriptstyle k_{AB}=1.466561$,
$\scriptstyle K_{F_B}=9.424319$
$\scriptstyle k_{F_B}=22.745736$,
$\scriptstyle K_{AC}= 0.113697$
$\scriptstyle k_{AC}=1.211993$,
$\scriptstyle K_{BC}=0.009891$
$\scriptstyle k_{BC}=7.239357$,
$\scriptstyle K_{CC}=0.189125$
$\scriptstyle k_{CC}= 17.910182$}
%old caption: Output from C Nonlinear Model}
%\label{fig:plotFCD2293}
\end{figure}%
\myfig{fig:plotFCD2293}

\subsection{Approximate scale invariance}
%In many cases, however, a network may exhibit an approximate property of
%scale-invariance.  
Continuing with example in
\rref{eq:example2},
let us suppose that
$k_1,k_2,k_3,k_4\ll k_5,k_6$, so that
the output variable $y=\xC$ reaches its steady state much faster than $\xA$
and $\xB$ do.
Then, we may approximate the original system by the planar linear system
represented by the differential equations for $\xA$ and $\xB$ together with
the new output variable
$\widetilde y(t) = h(\xA(t),\xB(t)) = k \xA(t)/\xB(t)$, where $k=k_5/k_6$.
This reduced planar system, obtained by a quasi-steady state approximation,
has a perfect scale-invariance property:
replacing the input $u$ by $pu$ results in the solution $(p\xA(t),p\xB(t))$,
and thus the output is the same: $h(\xA(t),\xB(t)) = h(p\xA(t),p\xB(t))$.
The exact invariance of the reduced system translates into an approximate
scale invariance property for the original three-dimensional system because,
except for a short boundary-layer behavior (the relatively short time
for $\xC$ to reach equilibrium), the outputs of both systems are essentially
the same, $y(t)\approx \widetilde y(t)$.

\subsection{Generality of the planar reduction}

We found that, just as in the example in
\rref{eq:example2}
when $k_1,k_2,k_3,k_4\ll k_5,k_6$,
in every ASI circuits the time scale of node $C$ is much shorter
than that of $A$ and $B$.  Therefore, the same two-dimensional reduction is
always valid.  It follows that one can drop the last equation, approximating
these circuits by planar systems that are described by only the two state
variables $\xA$ and $\xB$, where every occurence of $\xC$ in the first two
equations of the right-hand side of \rref{eq:network} is replaced by
$h(\xA,\xB)$, the function obtained by setting the right-hand side of the third
equation in \rref{eq:network} to zero and solving for the unique root in the
interval $[0,1]$ of the quadratic equation.
This reduced system, with $\widetilde y(t) = h(\xA(t),\xB(t))$ as an output, provides
an excellent approximation of the original dynamics.
Fig.\ref{fig:qss_2293_C_6} compares the true response with the response
obtained by the quasi-steady state approximation, for one ASI circuit
(see \emph{SI Text} for all comparisons).

\begin{figure}[ht] \centering
\includegraphics[width=0.51\textwidth]{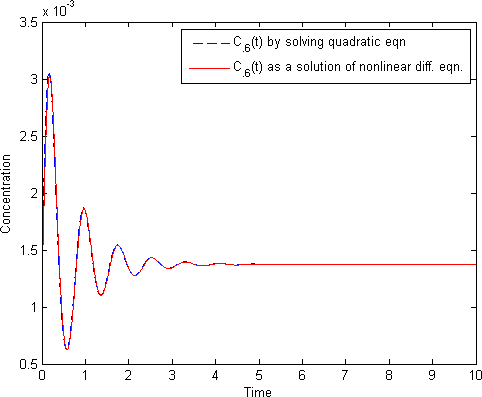}
\caption{QSS quadratic approximation.
Network is the one described by \protect{\rref{eq:example1}}.  Random
parameter set is as in Fig.\ref{fig:plotFCD2293}}.
%\label{fig:qss_2293_C_6}
\end{figure}%

\myfig{fig:qss_2293_C_6}

\subsection{Generality of dependence on $\scriptstyle \xA/\xB$}
In the example given by
\rref{eq:example2},
there were two additional key mathematical properties that made the planar
reduction scale-invariant (and hence the original system approximately so).
The first property was that, at equilibrium, the variable $\xC$ must be a
function of the ratio $\xA/\xB$, and the second one was that each of $\xA$ and
$\xB$ must scale by the same factor when the input scales by $p$.
Neither of these two properties need to hold, even approximately, for general
networks.
Surprisingly, however, we discovered that both are valid with very high
accuracy for every ASI circuit.
The equilibrium value of $\xC$ is obtained from setting the last right-hand
side of
\rref{eq:network}
to zero and solving for $\xC$.
A solution $\xC = h(\xA,\xB)$ in the interval $[0,1]$ always exists, because 
at $\xC=0$ one has $\txC=1$ and thus the term is positive, and
at $\xC=1$ one has $\txC=0$ and so the term is negative.
This right-hand side has the general form
$\xA \phi (\xC) + \xB \gamma (\xC) + \kappa (\xC,\xEC,\xFC)$, 
where $\phi $ and $\gamma $ are increasing functions, each a constant multiple of
a function of the form $\txC/(\txC+K)$ or $-\xC/(\xC+K)$.
If the term $\kappa $ is negligible, then $\xA \phi (\xC) + \xB \gamma (\xC)=0$ means that
also $(\xA/\xB) \phi (\xC) + \gamma (\xC)=0$, and therefore $\xC$ at equilibrium is a
(generally nonlinear) function of the ratio $\xA/\xB$.
There is no a priori reason for the term $\kappa $ to be negligible.
However, we discovered that in every ASI circuit, $\kappa \approx0$.
More precisely, there is no dependence on the constitutive enzymes, and
this ``self-loop'' link, when it exists, contributes to the
derivative $\dxC$ much less than the $\xA$ and $\xB$ terms, see
Fig.\ref{fig:ch4f7}.

\begin{figure}[ht]
\centering
\includegraphics[width=150mm]{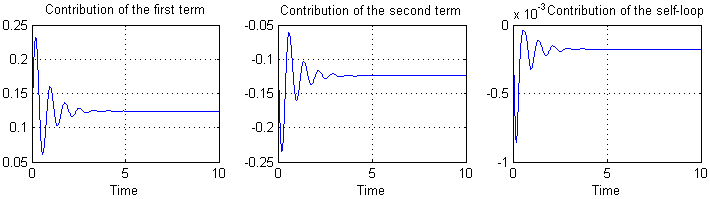}
\caption{Relative contribution of terms in the equation for node C.
  The first two terms range in $[-0.25,0.25]$ but self-loop magnitude is
  always less than $10^{-3}$.
i.e. contribution or self-loop to $\dxC$ is less than 1\%.  Similar results
hold for all ASI circuits.
  Network is the one described by
  \protect{\rref{eq:example1}}.
  Random parameter set is as in
  Fig.\protect{\ref{fig:plotFCD2293}}.
  Similar results are available for all ASI circuits. 
}
%\label{fig:ch4f7}
\end{figure}%
\myfig{fig:ch4f7}

\subsection{Generality of homogeneity of $\scriptstyle \xA,\, \xB$}

The last ingredient of the example given by
\rref{eq:example2}
that plays a role in approximate scale invariance is that each of $\xA$ and
$\xB$ must scale proportionately when the input is scaled.
In that example, the property holds simply because the equations for these two
variables are linear.
In general, however, the dynamics of $(\xA,\xB)$ are described by nonlinear
equations.
Thus it is remarkable that, in all ASI circuits, the property holds.
We tested the property by plotting $\xA(t)/\xB(t)$ in a set of experiments in
which a system was pre-adapted to an input value $u_0$ and the input was
subsequently set to a new level 
%$u=u_0+\delta u$
$u$ at $t=0$.  When going from $pu_0$ to 
%$pu=pu_0+p\delta u$,
$pu$, we found that the new value $\xA(t)/\xB(t)$ was
almost the same, meaning that $\xA$ and $\xB$ scaled in the same fashion.
A representative plot is shown in Fig.\ref{fig:ratio}.

\begin{figure}[ht]
  \centering
\includegraphics[width=0.5\textwidth]{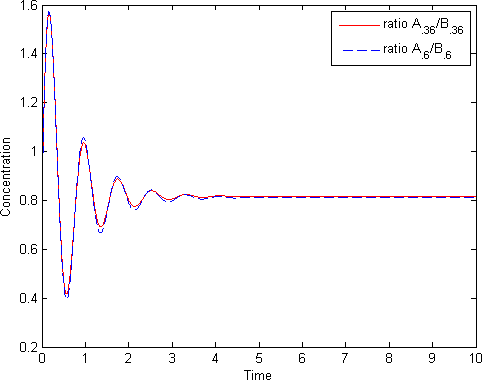}
\caption{Constant A/B ratio in responses to $3$$\rightarrow $$1.2*3$
  and $5$$\rightarrow $$1.2*5$.
  Network is the one described by
  \protect{\rref{eq:example1}}.
  Random parameter set is as in
  Fig.\protect{\ref{fig:plotFCD2293}}.
  Similar results are available for all ASI circuits (see \emph{SI Text}).}
%\label{fig:ratio}
\end{figure}%
\myfig{fig:ratio}

\subsection{A new property: \wfcd}

The (approximate) independence of $\xA(t)/\xB(t)$ on input scalings is not due
to linearity of the differential equations for $\xA$ and $\xB(t)$.
Instead, the analysis of this question led us to postulate a new property,
which we call \emph{{\wfcd} (\WFCD)}.
To define this property, we again drop the last equation, and approximate
circuits by the planar system that has only the state variables $\xA$ and $\xB$,
where every occurence of $\xC$ in their differential equations shown in
\rref{eq:network}
is replaced by $h(\xA,\xB)$. 
We denote by $f(\xA,\xB,u) = (f_1(\xA,\xB,u),f_2(\xA,\xB,u))$ the result of
these substitutions, so that the reduced system is described in vector form by
$\dot x = f(x,u)$, $x = (\xA,\xB)$.
We denote by $\sigma (u)$ the unique steady state corresponding to a constant input
$u$, that is, the solution of the algebraic equation
$f(\sigma (u),u)=0$.
We denote by ${\cal A}(u)=({\partial f}/{\partial x})(\sigma (u))$
the Jacobian matrix of $f$ with respect to $x$,
and by ${\cal B}(u)=({\partial f}/{\partial u})(\sigma (u))$
the Jacobian vector of $f$ with respect to $u$.
The property {\WFCD} is then defined by requiring time-scale separation for
$\xC$, that $h(\xA,\xB)$ depends only on the ratio $\xA/\xB$, and:
%\begin{itemize}
%\item
\beq{eq:linearizations}
\sigma (pu)=p\sigma (u),\;\;
{\cal A}(u) = {\cal A}(v),\;\;
{\cal B}(u) = {\cal B}(v)
\eeq
for every $u$ , $v$, and $p$ such that $u$, $v$, and $pu$ are in the range
$[\ulow,\uhigh]$.
Notice that we are not imposing the far stronger property that the Jacobian
matrices should be constant.  We are only requiring the same matrix at every
steady state.
%%%%%%%%%%%%%%%%%%%%%%%%%%%%%%%%%%%%%%%%%%%%%%%%%%%%%%%%%%%%%%
The first condition in
\rref{eq:linearizations}
means that the vector $\sigma (u)/u$ should be constant.
We verified that this requirement holds with very high accuracy in every one
of the ASI circuits.
With $\ulow=0.3$ and $\uhigh=0.6$, we have the following $\sigma (u)/u$ values,
rounded to 3 decimal digits:
$(0.195 ,  0.239)$,
$(0.193 , 0.237)$,
$(0.192  ,  0.236)$,
$(0.191  , 0.235)$
when $u=0.3$, $0.4$, $0.5$, and $0.6$ respectively, for the network described by
\protect{\rref{eq:example1}} and the random parameter set in
Fig.\ref{fig:plotFCD2293}.
Similar results are available for all ASI circuits (see \emph{SI Text}).
The Jacobian requirements are also verified with high accuracy for all the
ASI circuits.
We illustrate this with the same network and parameter set.
Let us we compute the linearizations
${\cal A}_{0.3}={\cal A}(0.3)$, ${\cal A}_{0.4}={\cal A}(0.4)$, \ldots , ${\cal B}_{0.6}={\cal B}(0.6)$
and the average relative differences
\[
{\cal A}^{\mbox{err}}_{ij} = \sum_{u=0.3,0.4,0.5,0.6} 
\abs{\frac{({\cal A}_u)_{ij}-({\cal A}_{0.45})_{ij}}
          {({\cal A}_{0.45})_{ij}}}
\]
and we define similarly ${\cal B}^{\mbox{err}}$.
These relative differences are very small (shown to 3 decimal digits):
\[
{\cal A}^{\mbox{err}}=
\begin{pmatrix}
0.069 & 0.004\\
0 & 0.005
\end{pmatrix}
,\;\;\;
{\cal B}^{\mbox{err}}=
\begin{pmatrix}
0.002\\
0
\end{pmatrix},
\]
thus justifying the claim that the Jacobians are practically constant.
Similar results are available for all ASI circuits (see \emph{SI Text}).

The key theoretical fact is that the property {\WFCD} implies approximate
scale-invariance, see \emph{Materials and Methods}.

\subsection{A concrete example}

In a recent paper \cite{takeda_et_al2012} Takeda and collaborators studied the
adaptation kinetics of a eukaryotic chemotaxis signaling pathway, employing a
microfluidic device to expose \emph{Dictyostelium discoideum} to changes in
chemoeffector cyclic adenosine monophosphate (cAMP).  Specifically, they
focused on the dynamics of activated Ras (Ras-GTP), which was in turn reported
by RBD-GFP (the Ras binding domain of fluorescently tagged human Raf1), and
showed almost perfect adaptation of previously unstimulated cells to cAMP
concentrations ranging from 10$^{-2}$ nM to 1 $\mu M$.
Furthermore, inspired by \cite{Ma}, the authors
proposed alternative models for adaptation, and concluded that the best fit
was obtained by using an incoherent feedforward structure.  The model that
they identified is given by the following system of 6 differential equations:
\beqn
\frac{d\Ro}{dt} &=& k_{\Ro} (\camp+\ro) (\Ro^{\mbox{tot}}-\Ro)-k_{-\Ro} \Ro\\
\frac{d\Rt}{dt} &=& k_{\Rt} (\camp+\rt) (\Rt^{\mbox{tot}}-\Rt)-k_{-\Rt} \Rt\\
u&=&\Ro+\Rt\\
\frac{d\GEF}{dt} &=& \kGEF \,u-\kmGEF \GEF\\
\frac{d\GAP}{dt} &=& \kGAP \,u-\kmGAP \GAP\\
\frac{d\RASGTP}{dt} &=& \kRAS \, \GEF \, (\RAS^{\mbox{tot}}-\RASGTP)-\kmRAS \, \GAP \, \RASGTP\\
\frac{d\RBDCYT}{dt} &=& \kRBDoff \, (\RBD^{\mbox{tot}}-\RBDCYT)-\kRBDon \,
\RASGTP \, \RBDCYT
\,.
\eeqn
The symbol $\camp$ stands for the chemoeffector cAMP, and the authors assumed the
existence of two different receptor populations ($\Ro$ and $\Rt$, with 
very different $K_d$'s) which when bound pool their signals to downstream
components (through $u$).  The constants $\ro$ and $\rt$ represent levels of
constitutive activation.  The variables $\GEF$ and $\GAP$ represent 
activation and deactivation of RasGEF and RasGAP, $\RASGTP$ represents the
activated Ras, and $\RBDCYT$ describes the cytosolic reporter molecule RBD-GFP.
Fig.~\ref{fig:takeda_diagram} shows a schematic of the main players.
\begin{figure}[ht]
\centering
\includegraphics[width=40mm]{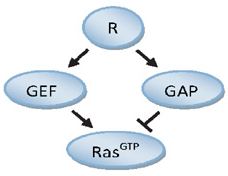}
\caption{The system studied in \protect{\cite{takeda_et_al2012}}}
%\label{fig:takeda_diagram}
\end{figure}%
\myfig{fig:takeda_diagram}

The best-fit parameters obtained in \cite{takeda_et_al2012} are as follows:
$\Ro^{\mbox{tot}}=0.1$,
$\Rt^{\mbox{tot}}=0.9$,
$\ro=0.012 \mbox{nM}$,
$\rt=0.115 \mbox{nM}$,
$k_{\Ro}=0.00267 \mbox{nM}^{-1}\mbox{sec}^{-1}$,
$k_{-\Ro}=0.16 \mbox{sec}^{-1}$,
$k_{\Rt}=0.00244 \mbox{nM}^{-1}\mbox{sec}^{-1}$,
$k_{-\Rt}=1.1 \mbox{sec}^{-1}$,
$\kGEF=0.04 \mbox{sec}^{-1}$,
$\kmGEF=0.4 \mbox{sec}^{-1}$,
$\kGAP=0.01 \mbox{sec}^{-1}$,
$\kmGAP=0.1 \mbox{sec}^{-1}$,
$\RAS^{\mbox{tot}}=1$,
$\kRAS=390 \mbox{sec}^{-1}$,
$\kmRAS=3126 \mbox{sec}^{-1}$,
$\RBD^{\mbox{tot}}=1$,
$\kRBDoff=0.53 \mbox{sec}^{-1}$,
$\kRBDon=1.0 \mbox{sec}^{-1}$.
With these parameters, and cAMP concentrations which are small yet also satisfy
$\ro\ll \camp(t)$ and $\rt\ll \camp(t)$, 
it follows that $\dot \Ro\approx k_{\Ro} \Ro^{\mbox{tot}}\camp-k_{-\Ro} \Ro$
and
$\dot \Rt\approx k_{\Rt} \Rt^{\mbox{tot}}\camp-k_{-\Rt} \Rt$,
so we may view $u(t)$ as an input (linearly dependent on the external
$\camp(t)$) to 
the three-variable system described by
$\xA = \GEF$, $\xB = \GAP$, $\xC = \RASGTP$.
Since 
$\RBDCYT$ depends only on $\xC$, we may view $\xC$ as the output.
This three-variable system (interpreted as having limiting values of
Michaelis-Menten constants) has the {\WFCD} property provided that
the dynamics of $\xC$ are fast compared to $\xA$ and $\xB$, which the
identified parameters insure.
So, we expect scale-invariant behavior.
Indeed, Fig.\ref{fig:dicty_fcd} shows a simulation of the entire
six-dimensional system (not merely of our 3-dimensional reduction)
when using a step from 1 to 2 nM of cAMP, and shows that essentially the same
response is obtained when stepping from 2 to 4 nM.
\begin{figure}[ht]
\centering
\includegraphics[width=60mm]{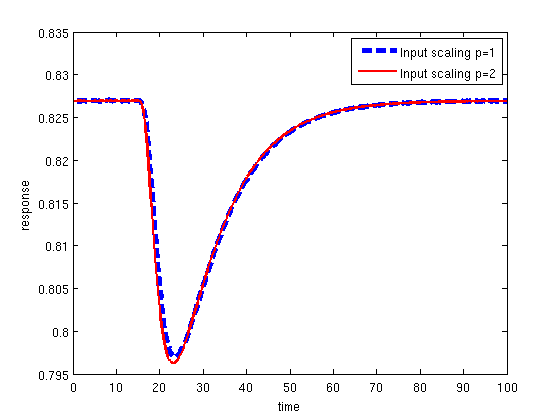}
\caption{Scale-invariance for model from \protect{\cite{takeda_et_al2012}}: responses to steps $1$$\rightarrow $$2$ and $2$$\rightarrow $$4$ coincide}
%\label{fig:dicty_fcd}
\end{figure}%
\myfig{fig:dicty_fcd}
This prediction of scale-invariant behavior is yet to be tested experimentally.

\section{Discussion}

Work in molecular systems biology seeks to unravel the basic dynamic
processes, feedback control loops, and signal processing mechanisms
in single cells and entire organisms, both for basic
scientific understanding and for guiding drug design.
One of the key questions is: how can one relate phenotype (function) to
interaction maps (gene networks, protein graphs, and so forth) derived from
experimentation, especially those obtained from high-throughput tools?
Answers to this question provide powerful tools for guiding the
reverse-engineering
of networks, by focusing on mechanisms that are consistent with experimentally
observed behaviors, and, conversely, from a synthesis viewpoint, allow one to
design artificial biological systems that are capable of adaptation
\cite{bleris10}
%\cite{othersynthbio}
and other objectives.
In particular, scale-invariance, a property that has been observed in various
systems
\cite{Goentoro2009},
\cite{Cellina2009},
can play a key role in this context, helping to discard putative mechanisms
that are not consistent with experimentally observed scale-invariant behaviors
\cite{ShimizuStocker2011}.
Through a computational study, we identified a set of simple mathematical
conditions that are used to characterize scale invariant enzymatic networks.

\section{Materials and Methods}

\subsection{Computational screen}

We generalized and extended the computational protocol developed for
adaptation in
\cite{Ma}
to an investigation of approximate scale invariance.
MATLAB$^{\mbox{\tiny \textregistered}}$ scripts were used, in conjunction with the
software developed in
\cite{Ma}.
In order to test inputs in ranges of the form \M{a\leq u(t)\leq 2a}, redefining the
constant \M{k_{UA}} if needed, we take simply \M{\ulow=0.3} and \M{\uhigh=0.6}.
We considered 160,380,000 circuits, obtained from the 16,038 nontrivial 3-node
topologies, each one with 10,000 parameters sampled in logarithmic scale using
the Latin hypercube method
\cite{LatinIman}.
(We picked the ranges \M{k_{cat}}=0.1-10 and \M{K_m}=0.001-100.  A finer sampling
does not affect conclusions in any significant way \cite{Ma}.)
Of these, 0.01\% (16,304) circuits showed adaptation, meaning that, as in
\cite{Ma},
when making a 20\% step from \M{u_0=0.5} to \M{u_1=0.6} the precision is
10\% or better, and the sensitivity is at least unity.
Approximate scale invariance (ASI) was then tested by also performing a 20\%
step experiment from \M{u_0=0.3} to \M{u_1=0.36} and requiring that the relative
difference between the responses be at most 10\%:
\M{\max_t \left\{\abs{y_{0.6}(t)-y_{3.6}(t)}/\max(y_{0.6}(t)-y_{3.6}(t))\right\}
< 0.1}
Of the adapting circuits, about 0.15\% (25 circuits, classified into 21
different topologies) were determined to be ASI.

\subsection{{\WFCD} implies approximate scale invariance}

Consider a system of \M{n} differential equations with input signal \M{u},
%\M{\dot x=f(x,u)}, 
\[
\dot x=f(x,u)
\]
with the variables \M{x} evolving on some closed bounded set and
\M{f} differentiable, and suppose that for each constant input \M{\ubar} there is a
unique steady state \M{\xbar=\sigma (\ubar)} with the conditions in
\rref{eq:linearizations}
and an output
\[
%\M{y(t)=h(x(t))}
y(t)=h(x(t))
\]
such that \M{h} is differentiable and homogeneous
of degree zero (\M{h(px)=h(x)} for nonzero \M{p}).
We view 3-node enzymatic networks as obtained from a set of \M{n+1} equations
\beqn
%\M{\dot x = F(x,z,u)}, \M{\varepsilon \dot z = G(x,z)}
\dot x &=& F(x,z,u)\\
\varepsilon \dot z &=& G(x,z)
\eeqn
with \M{n=2}, \M{x = (\xA,\xB)}, and \M{z=\xC}
(\M{0<\varepsilon \ll1} represents the faster time scale for \M{\xC}),
and we are studying the reduced system \M{\dot x=f(x,u) = F(x,\alpha (x),u)}
obtained by solving \M{G(x,z)=0} for \M{z=\alpha (x)} and substituting in \M{F}.
Consider a time interval \M{[0,T]}, a constant input \M{\ubar}, and a possibly
time-varying input \M{u(t)}, \M{t\geq 0}, as well as a scaling \M{p>0}, such that all
values \M{\ubar}, \M{p\ubar}, \M{u(t)}, \M{pu(t)} are in the input range
\M{[\ulow,\uhigh]}.
The solutions of
\M{\dot x=f(x,u)} with initial condition \M{x(0)=\sigma (\ubar)}
and of
\M{\dot z=f(z,pu)} with initial condition \M{z(0)=\sigma (p\ubar)}
are denoted respectively by \M{x(t)} and \M{z(t)}, and the respective outputs
are \M{y(t)=h(x(t))} and \M{y_p(t)=h(z(t))}.
We wish to show that these two responses are approximately equal on \M{0\leq t\leq T}.
Write \M{\delta (t) = u(t)-\ubar}.
From Theorem 1 in
\cite{mct}
we know that
\[
%\M{
x(t) = x(0) + \xi (t) + o(\norm{\delta })
%}
\]
where \M{\norm{\delta } = \sup_{0\leq t\leq T} \abs{\delta (t)}} and
\M{\xi } is the solution of the variational system
%\M{
\[
\dot \xi (t) = {\cal A}\xi (t) + {\cal B}\delta (t)
\]
%}
with \M{\xi (0)=0}, and that
\[
%\M{
z(t) = z(0) + \zeta (t) + o(\norm{pu-p\ubar}) = z(0) + \zeta (t) + o(\norm{\delta }),
\]
%},
where
%\M{
\[
\dot \zeta  = {\cal A}\zeta (t) + {\cal B}p\delta (t)
\]
%}
with \M{\zeta (0)=0}.
By linearity, \M{\zeta =p\xi }.
Using \M{z(0)\equiv \sigma (p\ubar)=p\sigma (\ubar)=px(0)},
we have that
\M{
px(t) - z(t) = o(\norm{\delta }).
}
Thus, 
%\M{y(t)=h(x(t))=h(px(t))=h(z(t)+o(\norm{\delta }))}.
\[
y(t)=h(x(t))=h(px(t))=h(z(t)+o(\norm{\delta }))\,.
\]
If \M{K} is an upper bound on the gradient of \M{h}, then
%  so \M{\abs{h(a)-h(b)}\leq K\abs{a-b}}.
%\M{
\[
\abs{y_p(t)-y(t)} = \abs{h(z(t)) - h(z(t)+o(\norm{\delta }))} \leq  K o(\norm{\delta }).
%}
\]
Thus, the relative error \M{\sup_t\abs{y_p(t)-y(t)}/\sup_t\abs{u(t)-\ubar}} converges to
zero as a function of the input perturbation \M{u(t)-\ubar}.
As a numerical illustration, we consider again the the network described by
\protect{\rref{eq:example1}} and the random parameter set in
Fig.\ref{fig:plotFCD2293}.
We compare the relative error between the original nonlinear system, with
initial state \M{\xi =(\xA,\xB)} corresponding to \M{u=0.3}, and applied input
\M{u=0.36}, and the approximation is \M{\xi + z(t)}, where the \M{z}
solves the linear system with initial condition zero and constant input \M{0.06}.
The maximum approximation error is about 5\% (to 3 decimal places, \M{0.055} for
\M{\xA} and \M{0.01} for \M{\xB}).  When stepping from \M{u=0.5} to \M{u=0.6}, the error
is less than 3\% (\M{0.028} and \M{0.005} respectively).
Similar results are available for all ASI circuits (see \emph{SI Text}).

\subsection{Impossibility of perfect scale-invariance}

Consider any system with state \M{x=(\xA,\xB,\xC)}, output \M{\xC},
and equations of the general form
\M{\dxA= f(x) + G(\xA)u},
\M{\dxB = g(x)},
\M{\dxC = h(x) = \xA a(\xC) + \xB b(\xC) + c(\xC)}.
\beqn
\dxA &=& f(x) + G(\xA)u\\
\dxB &=& g(x)\\
\dxC &=& h(x) \;=\; \xA a(\xC) + \xB b(\xC) + c(\xC) \,.
\eeqn
It is assumed that \M{a(\xC)\not= 0} for all \M{\xC}, \M{G(\xA)\not= 0} for all \M{\xA},
\M{\barG:= \sup_x G(x) <\infty }, and the system is
irreducible
\cite{shoval_alon_sontag_2011}.
We now prove that such a system cannot be scale-invariant.
Suppose by way of contradiction that it would be, and pick any fixed \M{p\not= 1}.
The main theorem in
\cite{shoval_alon_sontag_2011}
insures that there are two differentiable functions \M{\alpha (x)} and \M{\beta (x)}
such that the algebraic identities:
\beqn
\alpha _x(x) [f(x) + G(\xA)u] + \alpha _y(x) g(x) + \alpha _z(x) h(x)&=&
f(\alpha (x),\beta (x),\xC) + G(\alpha (x))pu,
\\
\beta _x(x) [f(x) + u] +\beta _y(x) g(x) + \beta _z(x) h(x)&=&g(\alpha (x),\beta (x),\xC)\\
\alpha (x) a(\xC) + \beta (x) b(\xC) + c(\xC) &=& \xA a(\xC) + \xB b(\xC) + c(\xC)
\eeqn
%%%%%%%%%%%%%%%%%%%%%%%%%%%%%%%5
%\beqn
%\scriptstyle
%\alpha _x(x) [f(x) + G(\xA)u]
%+ \alpha _y(x) g(x)
%+ \alpha _z(x) h(x)
%&=& 
%\scriptstyle
%f(\alpha (x),\beta (x),\xC) + G(\alpha (x))pu\\
%\scriptstyle
%\beta _x(x) [f(x) + u]
%+\beta _y(x) g(x)
%+ \beta _z(x) h(x)
%&=&g
%\scriptstyle
%(\alpha (x),\beta (x),\xC) \\
%\scriptstyle
%\alpha (x) a(\xC) + \beta (x) b(\xC) + c(\xC) &=& 
%\scriptstyle
%\xA a(\xC) + \xB b(\xC) + c(\xC)
%\eeqn

\null\vskip-0.5cm
\noindent
hold for all constant \M{x=(\xA,\xB,\xC)} and \M{u}, and
the vector function
\M{x\mapsto (\alpha (x),\beta (x),z)} is one-to-one and onto,
which implies in particular that 
\[
\sup_xG(\alpha (x))  = \barG \,.
\]
%\M{\sup_xG(\alpha (x))  = \barG}.
Dividing by \M{u} and taking the limit as \M{u\rightarrow \infty } in the first identity, we
conclude that \M{\alpha _x(x) G(\xA) \equiv  p G(\alpha (x))}.
Doing the same in the second identity, we conclude that \M{\beta _x(x) \equiv  0}.
Finally, taking partial derivatives with respect to \M{\xA} in the third
identity:
%and using that \M{\alpha \equiv p G(\alpha (x))/G(\xA)} and \M{\beta _x\equiv 0}, 
\[
%\M{ 
a(\xC) p G(\alpha (x))/G(\xA) = \alpha _x(x) a(\xC) + \beta _x(x) b(\xC) = a(\xC) 
%}
\]
is true for all \M{x}.
Since \M{a(\xC)\not\equiv 0}, it follows that 
\[
%\M{ 
p G(\alpha (x)) = G(\xA) 
%}
\]
for all \M{x}.
We consider two cases: (a) \M{p<1} and (b) \M{p>1}.
Suppose \M{p<1}.
Pick any sequence of points \M{x^{(i)}} with \M{G(x^{(i)})\rightarrow \barG} as \M{i\rightarrow \infty }.
Then \M{G(\alpha (x^{(i)}))\rightarrow \barG/p > \barG}, contradicting
%the assumption that 
\M{G(x)\leq \barG}.
If \M{p>1}, picking a sequence such that 
\M{G(\alpha (x^{(i)}))\rightarrow \barG} as \M{i\rightarrow \infty } gives the contradiction
\M{G(x^{(i)})\rightarrow p\barG>\barG}.
This shows that the FCD property cannot hold.

.

%\end{materials}

\subsection*{Acknowledgments}
We are grateful to Wenzhe Ma for making available and explaining his software
for generating and testing
%visualizing 
networks for adaptation.
This work was supported in part by the US National Institutes of Health and
the Air Force Office of Scientific Research.
%\end{acknowledgments}

\clearpage

\bibliographystyle{plain}

\clearpage

{\bf Supplementary Material}
\medskip

{\bf A characterization of scale invariant responses in enzymatic networks}

\medskip
\section{Circuits that exhibit ASI}

We list here the results of the computational screen as described in the Main
Text. After showing graphical representations for the 25 identified ASI
circuits (21 topologies), we provide their equations and parameters.

For each circuit, four plots are shown: 
\begin{itemize}

\item[(a)] a comparison between the plots of $\xA(t)$ and $\xB(t)$ for the
  original nonlinear system and the respective plots for the linearized
  approximations,

\item[(b)] the plots showing scale-invariant behavior for step inputs,

and the comparison between the plots of $\xC(t)$ 
for the original nonlinear system and for the quasi-steady state
approximation, for 

\item[(c)] step input change from $0.3$ to $0.36$ and 

\item[(d)] step input change from $0.5$ to $0.6$.
\end{itemize}

\clearpage

\begin{figure}[h]
  \centering
\subfloat[Circuit 1.]{\label{fig:f1}\includegraphics[width=0.35\textwidth]{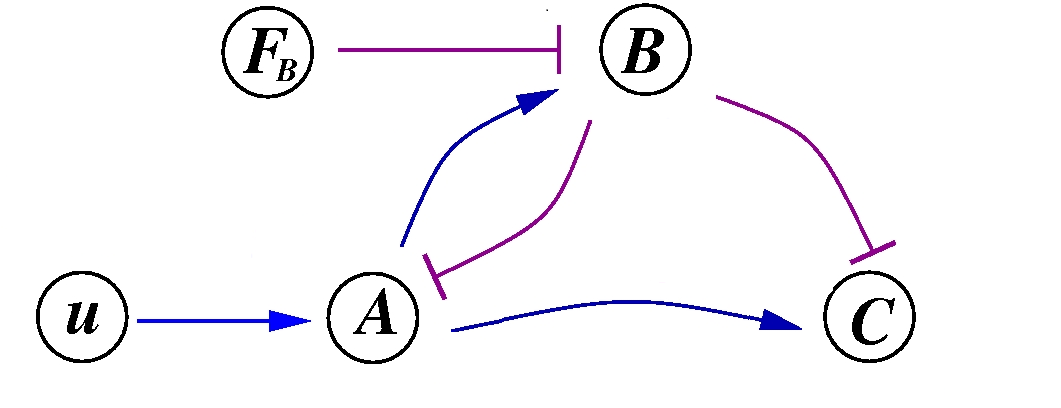}}               
  \subfloat[Circuit 2.]{\label{fig:f2}\includegraphics[width=0.35\textwidth]{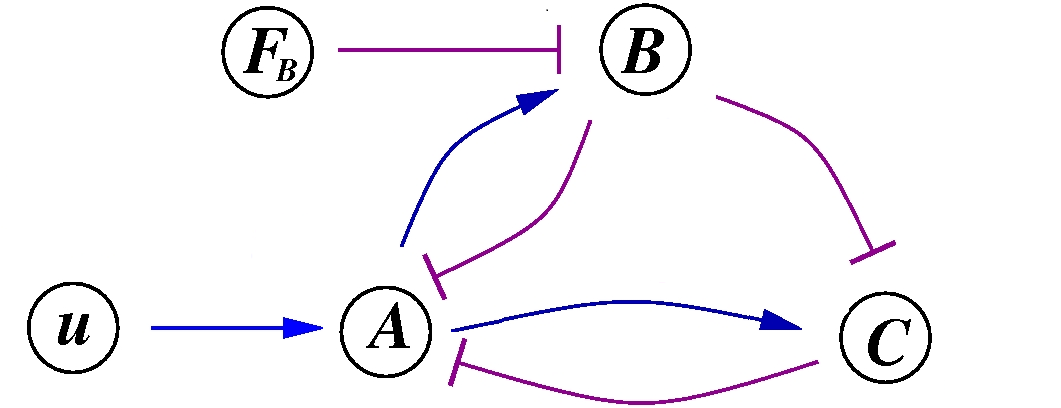}}
\subfloat[Circuit 3.]{\label{fig:f3}\includegraphics[width=0.35\textwidth]{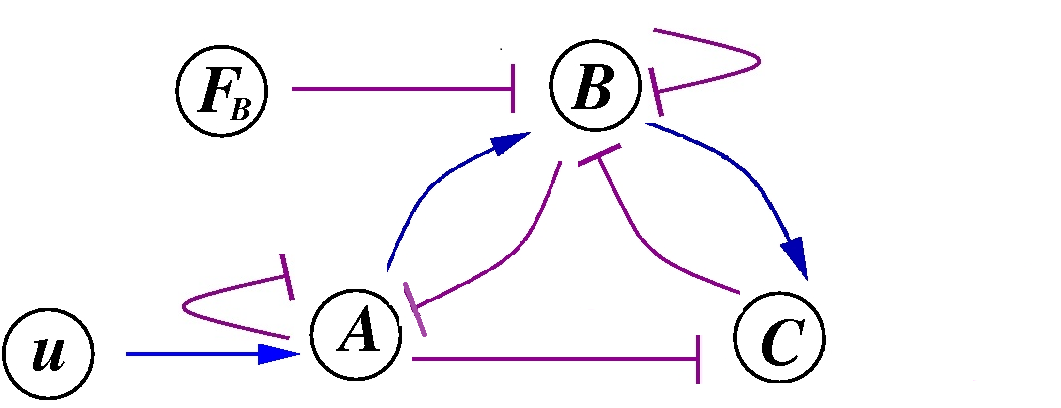}} \\          
  \subfloat[Circuit 4.]{\label{fig:f4}\includegraphics[width=0.35\textwidth]{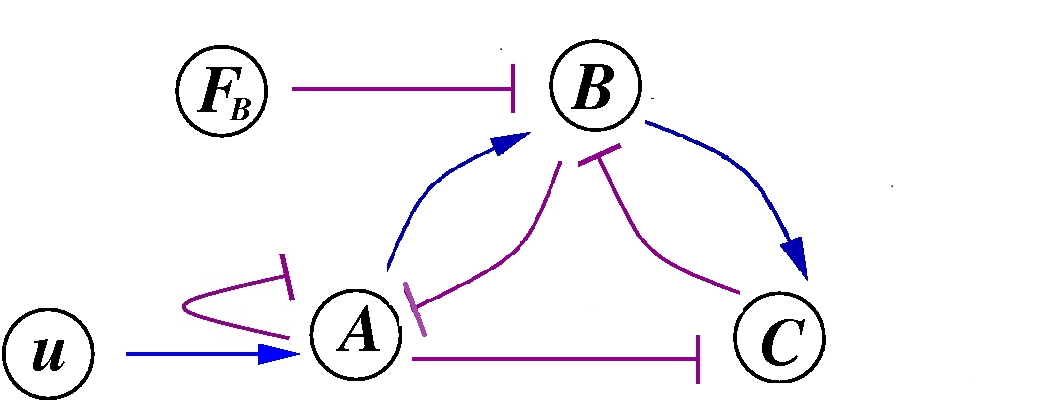}}
  \subfloat[Circuit 5.]{\label{fig:f5}\includegraphics[width=0.35\textwidth]{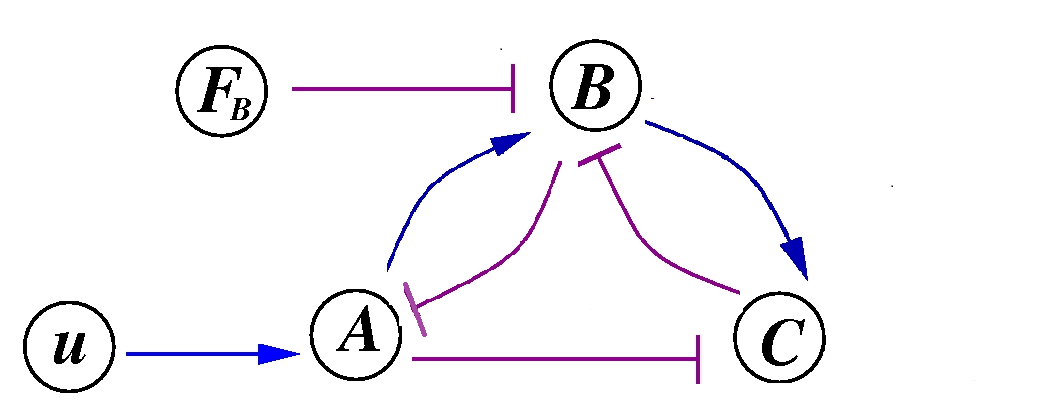}}               
  \subfloat[Ciircuit 6.]{\label{fig:f6}\includegraphics[width=0.35\textwidth]{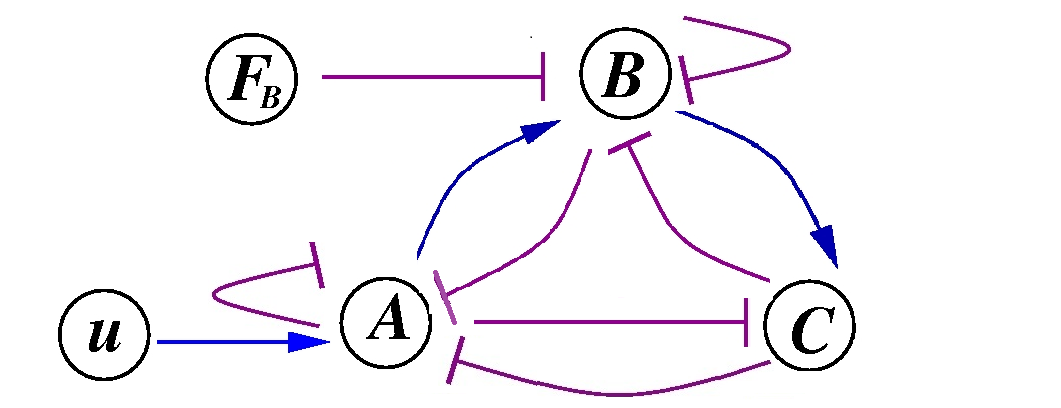}}\\
\subfloat[Circuit 7.]{\label{fig:f7}\includegraphics[width=0.35\textwidth]{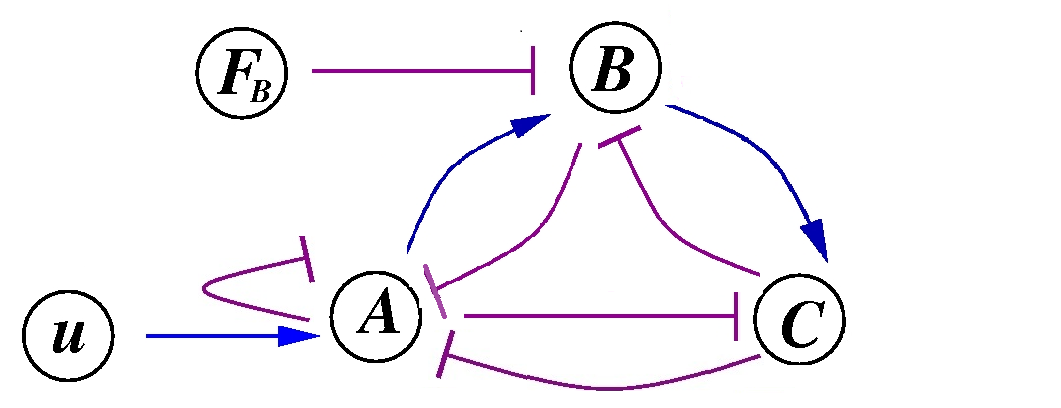}}               
  \subfloat[Circuit 8.]{\label{fig:f8}\includegraphics[width=0.35\textwidth]{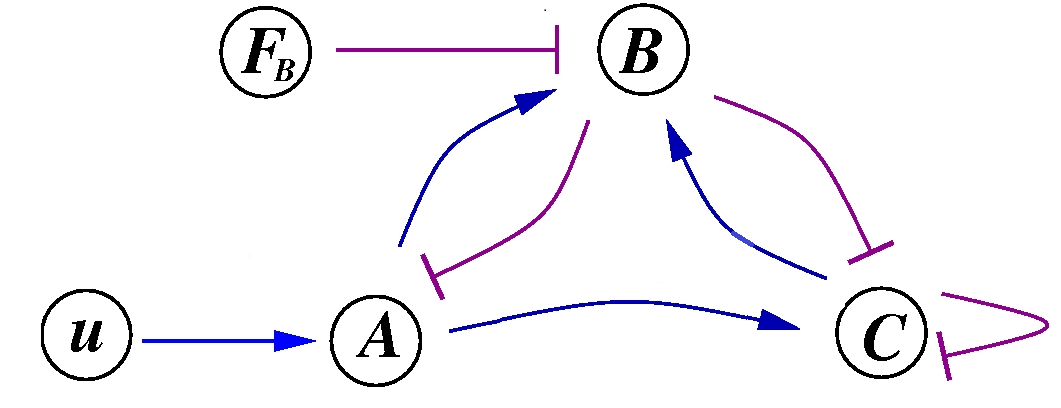}}
\subfloat[Circuit 9.]{\label{fig:f7}\includegraphics[width=0.35\textwidth]{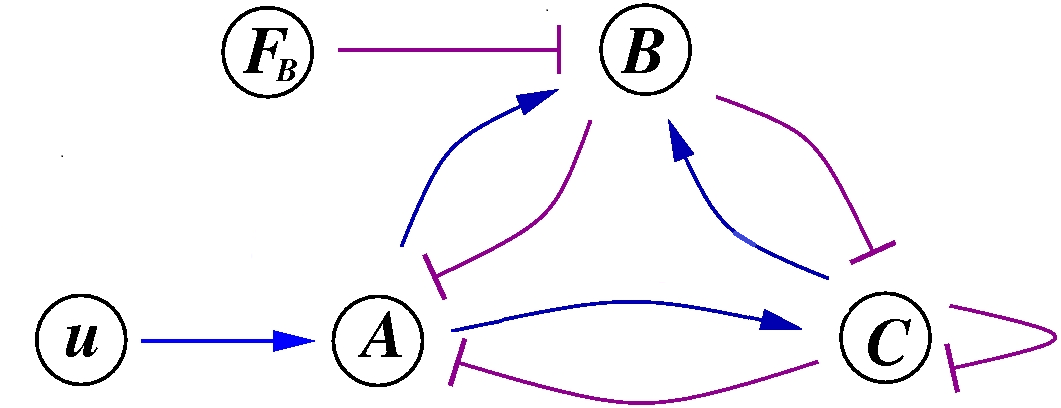}} \\              
  \subfloat[Circuit 10 ]{\label{fig:f8}\includegraphics[width=0.35\textwidth]{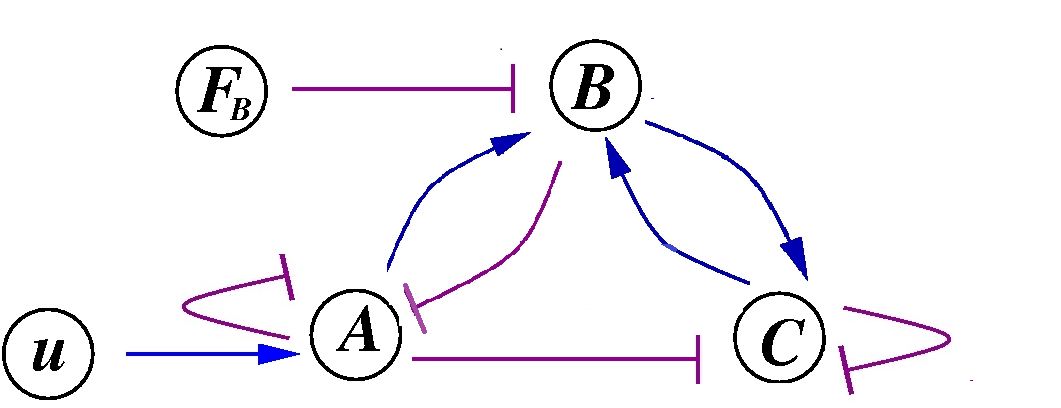}}
  \subfloat[Circuit 11. ]{\label{fig:f8}\includegraphics[width=0.35\textwidth]{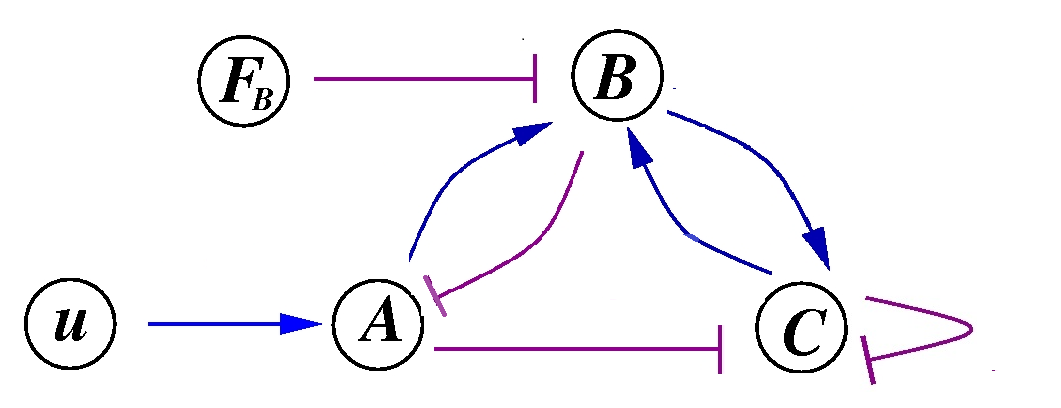}} 
    \subfloat[Circuit 12. ]{\label{fig:f8}\includegraphics[width=0.35\textwidth]{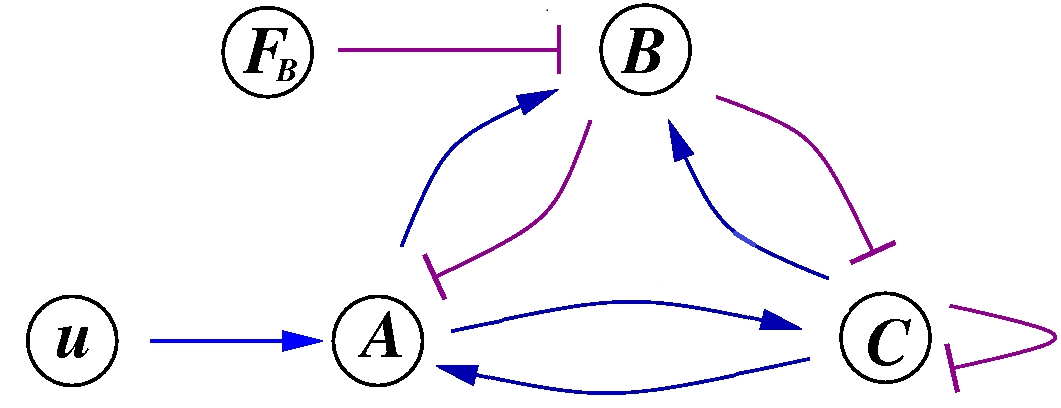}} \\
    
     \subfloat[Circuit 13.]{\label{fig:f8}\includegraphics[width=0.35\textwidth]{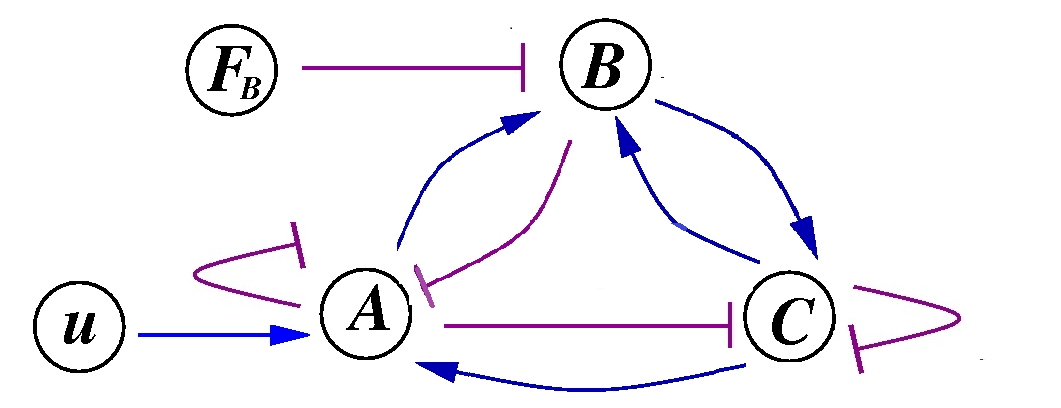}}
  \subfloat[Circuit 14.]{\label{fig:f8}\includegraphics[width=0.35\textwidth]{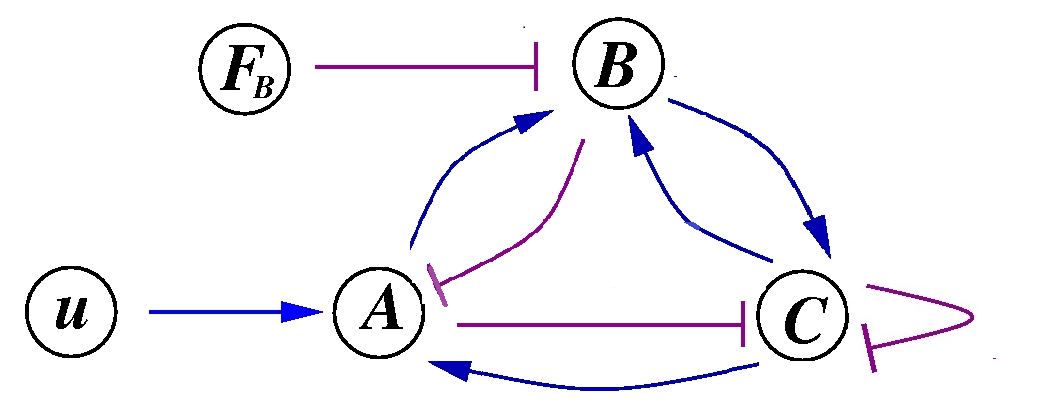}} 
    \subfloat[Circuits 15 -17 ]{\label{fig:f8}\includegraphics[width=0.35\textwidth]{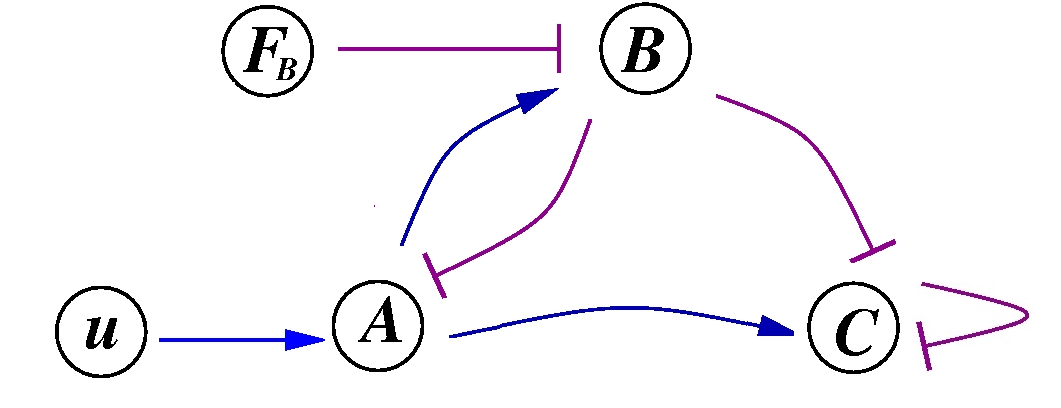}} \\ 
    
      \subfloat[Circuit 18. ]{\label{fig:f8}\includegraphics[width=0.35\textwidth]{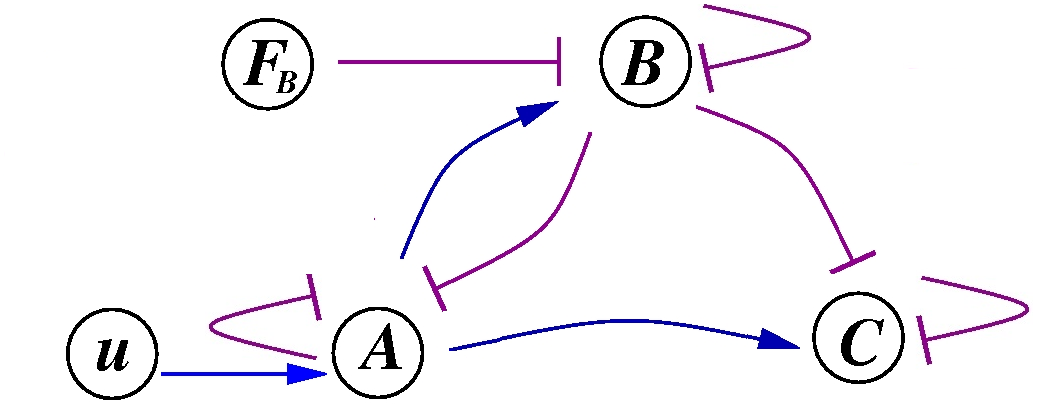}}
  \subfloat[Circuit 19. ]{\label{fig:f8}\includegraphics[width=0.35\textwidth]{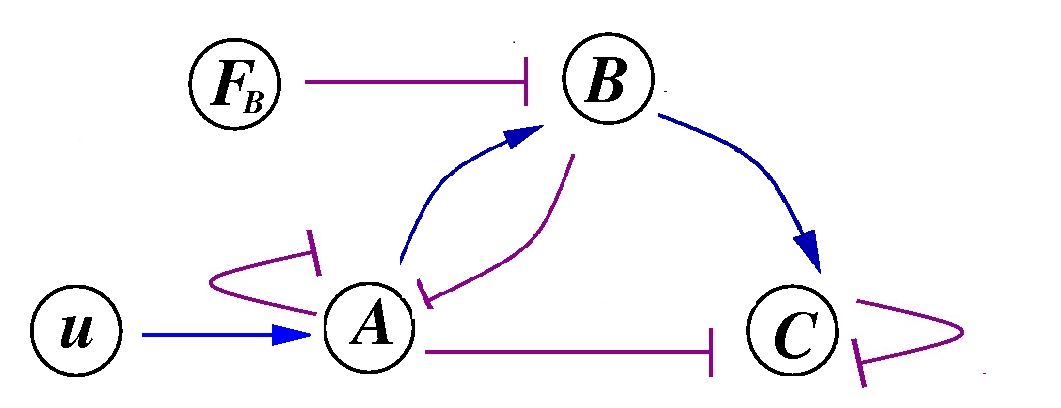}} 
    \subfloat[Circuit 20.]{\label{fig:f8}\includegraphics[width=0.35\textwidth]{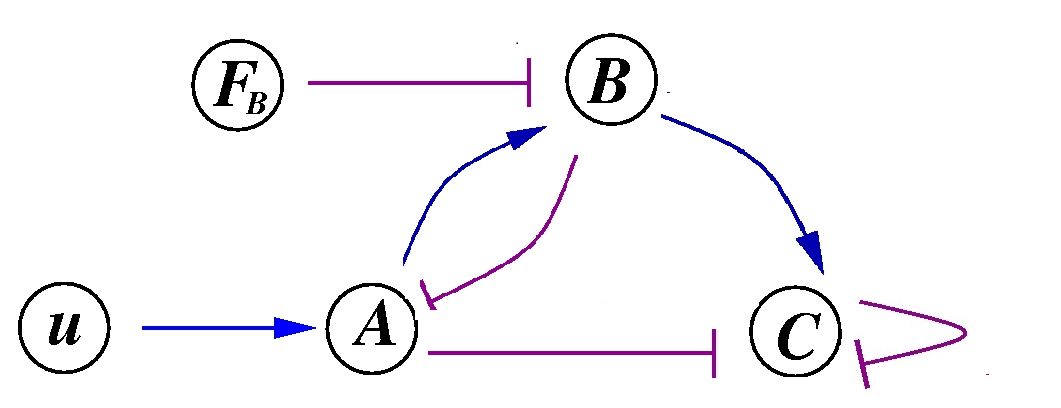}} \\

        \subfloat[Circuit 21 - 22 ]{\label{fig:f8}\includegraphics[width=0.35\textwidth]{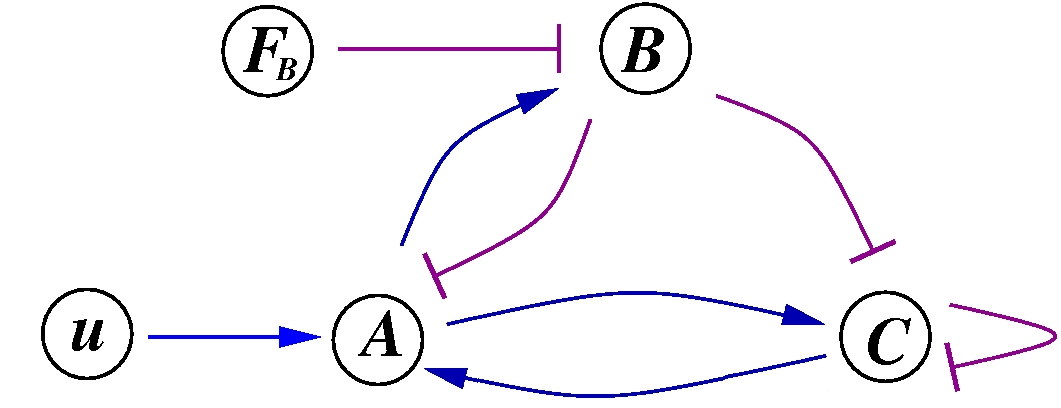}}
  \subfloat[Circuit 23. ]{\label{fig:f8}\includegraphics[width=0.35\textwidth]{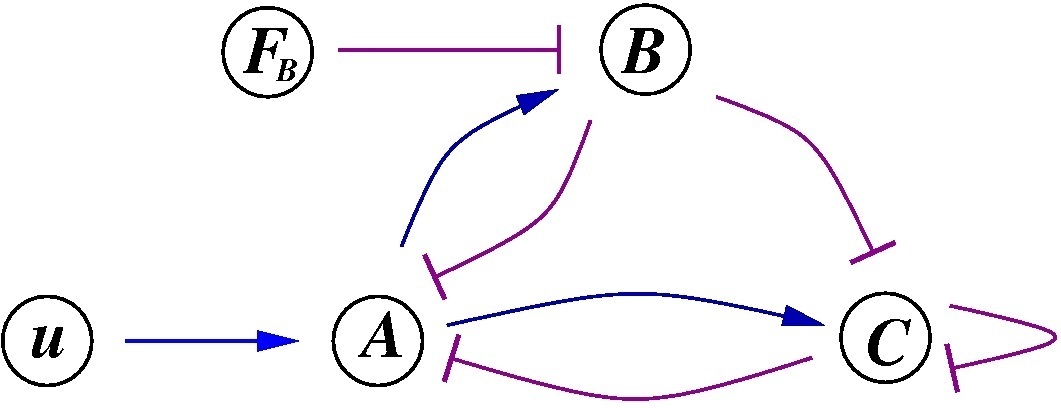}} 
    \subfloat[Circuit 24 - 25]{\label{fig:f8}\includegraphics[width=0.35\textwidth]{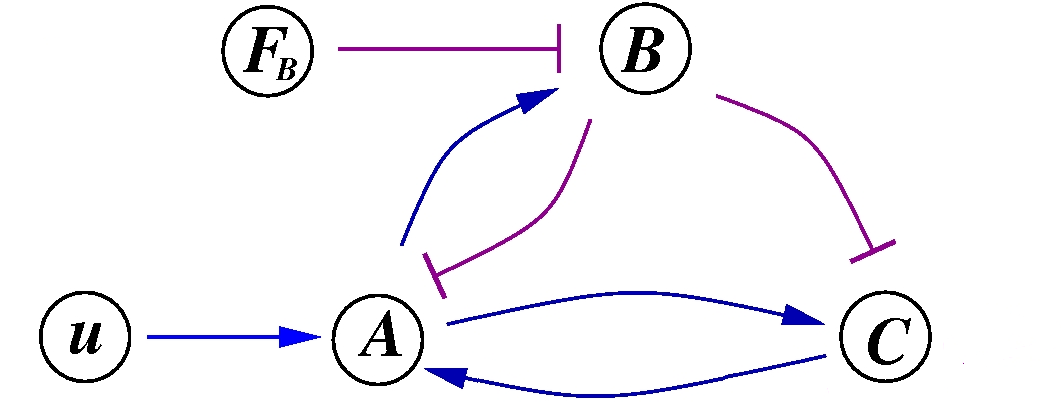}} \\
    \end{figure}

\clearpage
Circuit 1.  
\beqn
\dxA&=&k_{{\inp}A} {\inp} \frac{\txA}{\txA+K_{{\inp}A}}-k_{BA} \xB\frac{\xA}{\xA+K_{BA}}\\
\dxB&=&k_{AB}\xA\frac{\txB}{\txB+K_{AB}}-k_{F_BB} \xFB \frac{\xB}{\xB+K_{F_BB}}\\
\dxC&=&{k_{AC}}\xA\frac{\txC}{\txC+K_{AC}}- k_{BC}\xB\frac{\xC}{\xC+K_{BC}}\\
\eeqn
Parameters: $K_{AB}=0.001191;$ $k_{AB}=1.466561;$ $K_{AC}= 0.113697;$ $k_{AC}=1.211993;$
$K_{BA}= 0.001688;$ 
$k_{BA}= 44.802268;$ $K_{BC}= 0.009891;$ $k_{BC}= 7.239357;$ $K_{{\inp}A}= 0.093918;$ $k_{{\inp}A}=11.447219;$ 
$k_{AC}=1.211993;$ $K_{AC}=0.1136927;$ $K_{F_B}=9.424319;$ $k_{F_B}=22.745736$

\begin{figure}[h]
  \centering
\subfloat[Dynamics of A and B in linearized model]{\label{fig:f11}\includegraphics[width=0.55\textwidth]{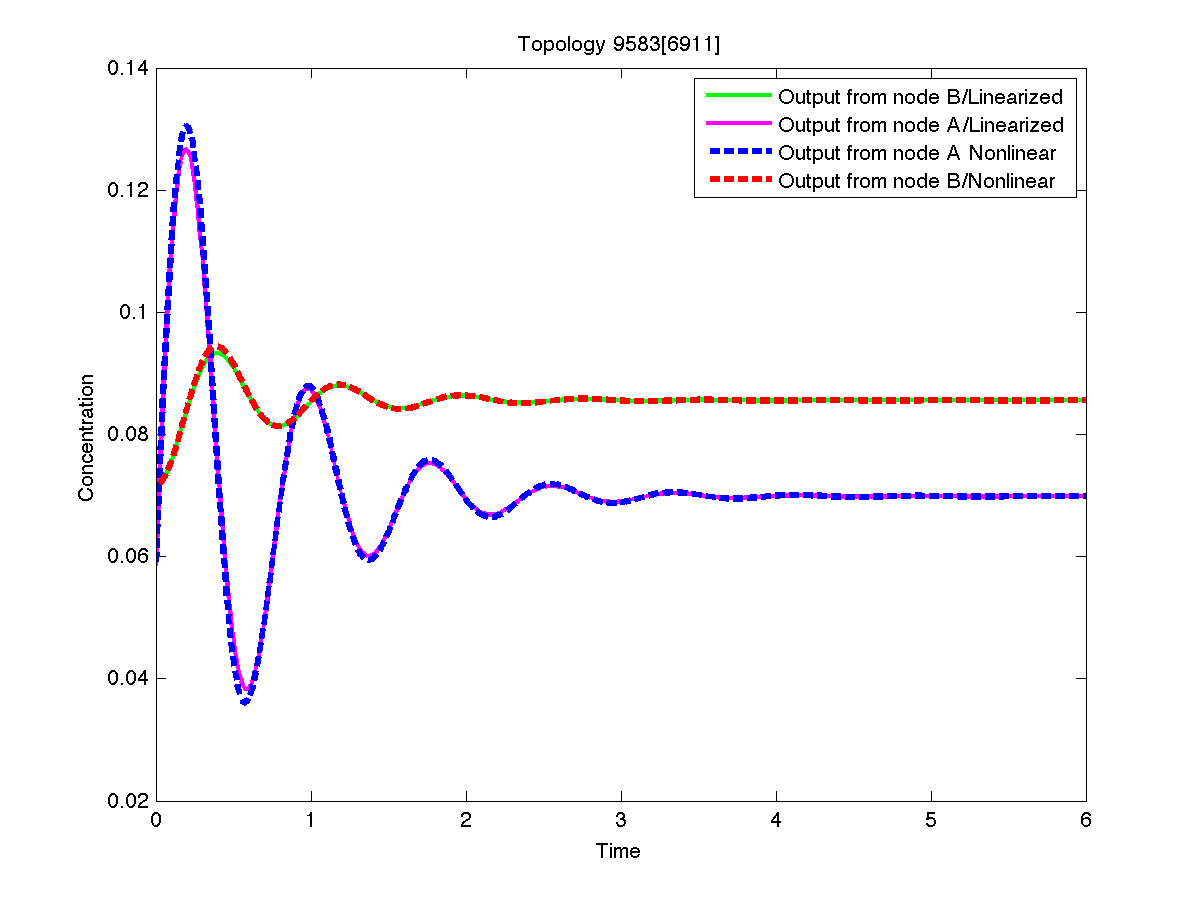}}                
  \subfloat[Ouput from C  nonlinear model]{\label{fig:f12}\includegraphics[width=0.55\textwidth]{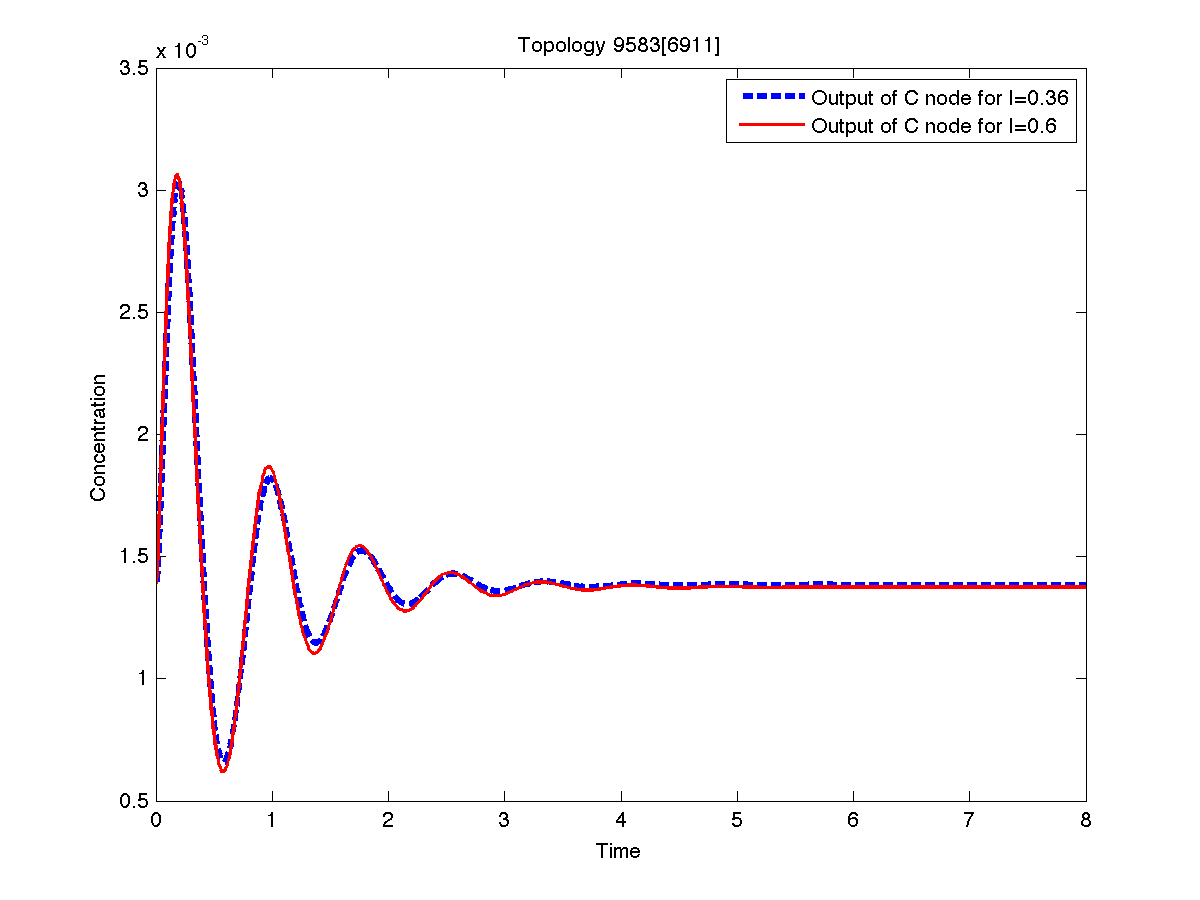}}\\
\subfloat[Quadratic approx. and output of nonlinear system]{\label{fig:f31}\includegraphics[width=0.55\textwidth]{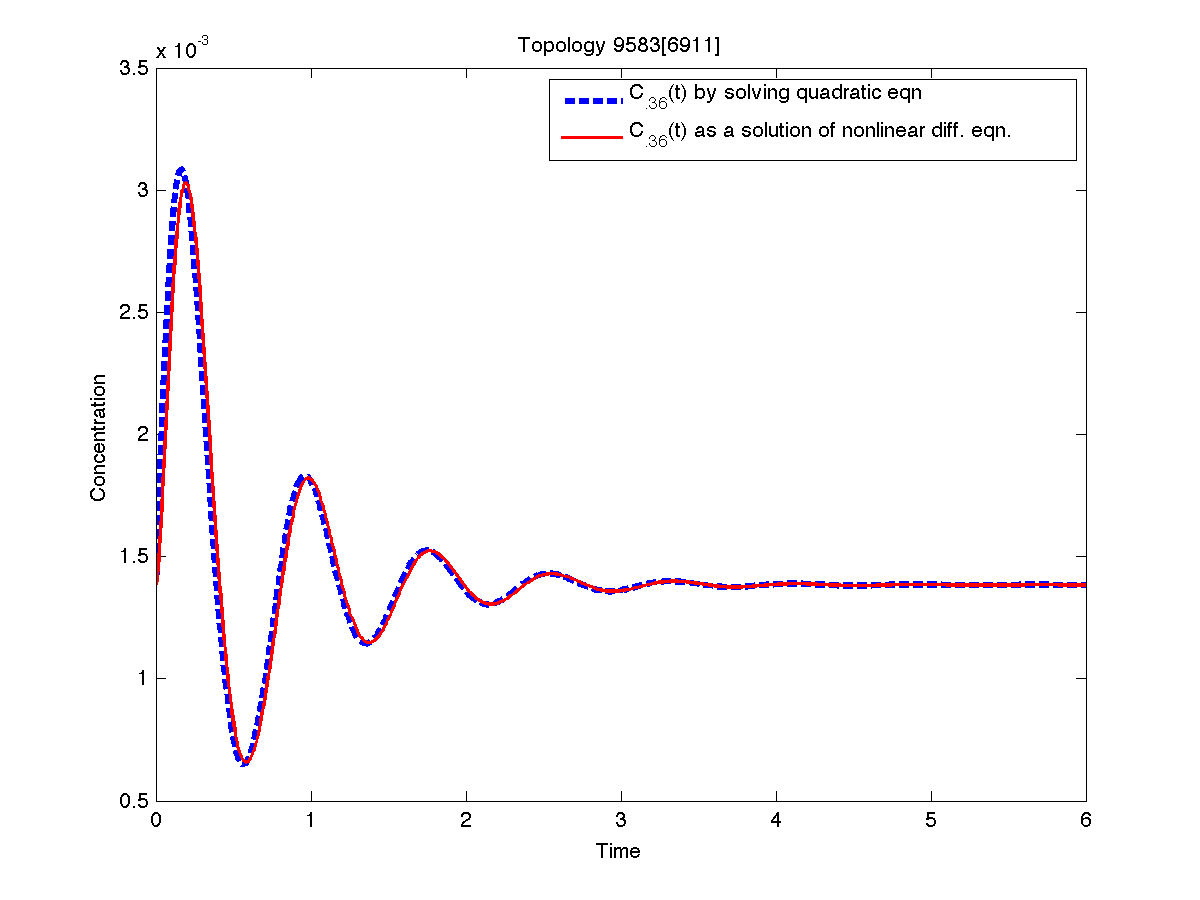}}                
  \subfloat[Quadratic approx. and output of nonlinear system]{\label{fig:f32}\includegraphics[width=0.55\textwidth]{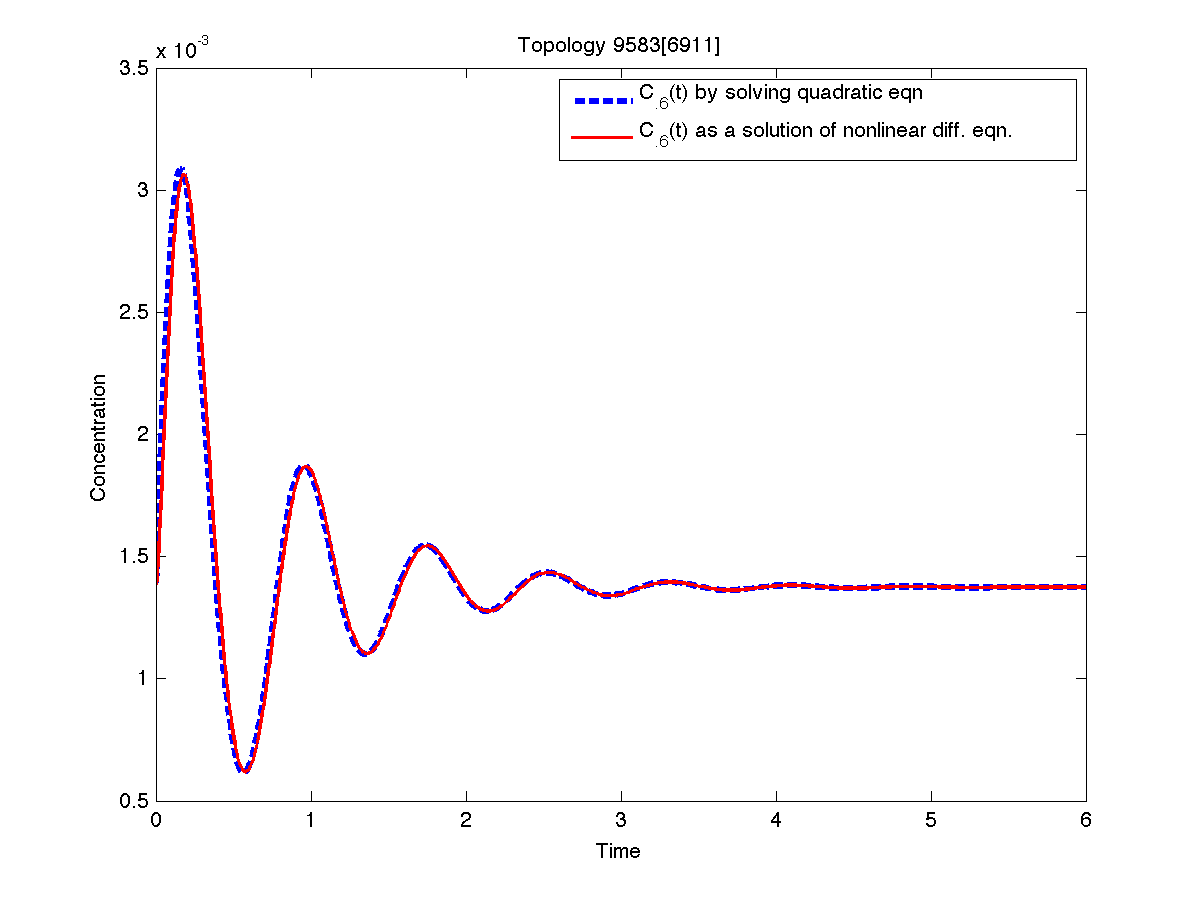}}
\end{figure}

\clearpage

Circuit 2.
\beqn
\dxA&=&k_{{\inp}A} {\inp} \frac{\txA}{\txA+K_{{\inp}A}}-k_{BA} \xB\frac{\xA}{\xA+K_{BA}}-k_{CA} \xC\frac{\xA}{\xA+K_{CA}}\\
\dxB&=&k_{AB}\xA\frac{\txB}{\txB+K_{AB}}-k_{F_BB} \xFB \frac{\xB}{\xB+K_{F_BB}}\\
\dxC&=&{k_{AC}}\xA\frac{\txC}{\txC+K_{AC}}- k_{BC}\xB\frac{\xC}{\xC+K_{BC}}\\
\eeqn
Parameters:  $K_{{\inp}A}= 0.093918;$  $k_{{\inp}A}= 11.447219;$
$K_{BA}= 0.001688;$ $k_{BA}= 44.802268;$
$K_{CA}= 90.209027;$ $k_{CA}= 96.671843;$
$K_{AB}=0.001191;$ $k_{AB}=1.466561;$
 $K_{F_B}=9.424319;$ $k_{F_B}=22.745736;$
$K_{AC}= 0.113697;$ $k_{AC}=1.211993;$
$K_{BC}=0.009891;$  $k_{BC}=7.239357$

\begin{figure}[ht]
  \centering
\subfloat[Dynamics of A and B in linearized model]{\label{fig:f41}\includegraphics[width=0.55\textwidth]{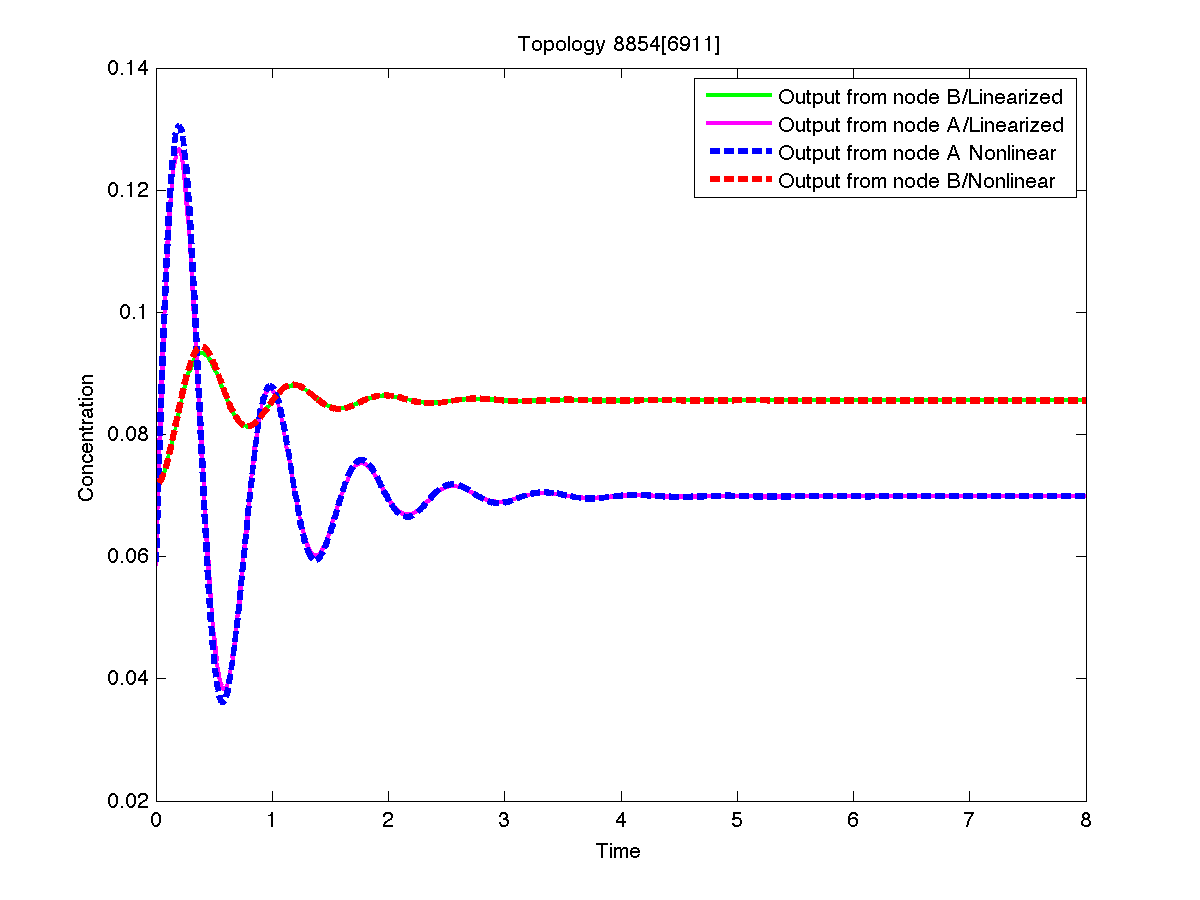}}                
  \subfloat[Ouput from C  nonlinear model]{\label{fig:f42}\includegraphics[width=0.55\textwidth]{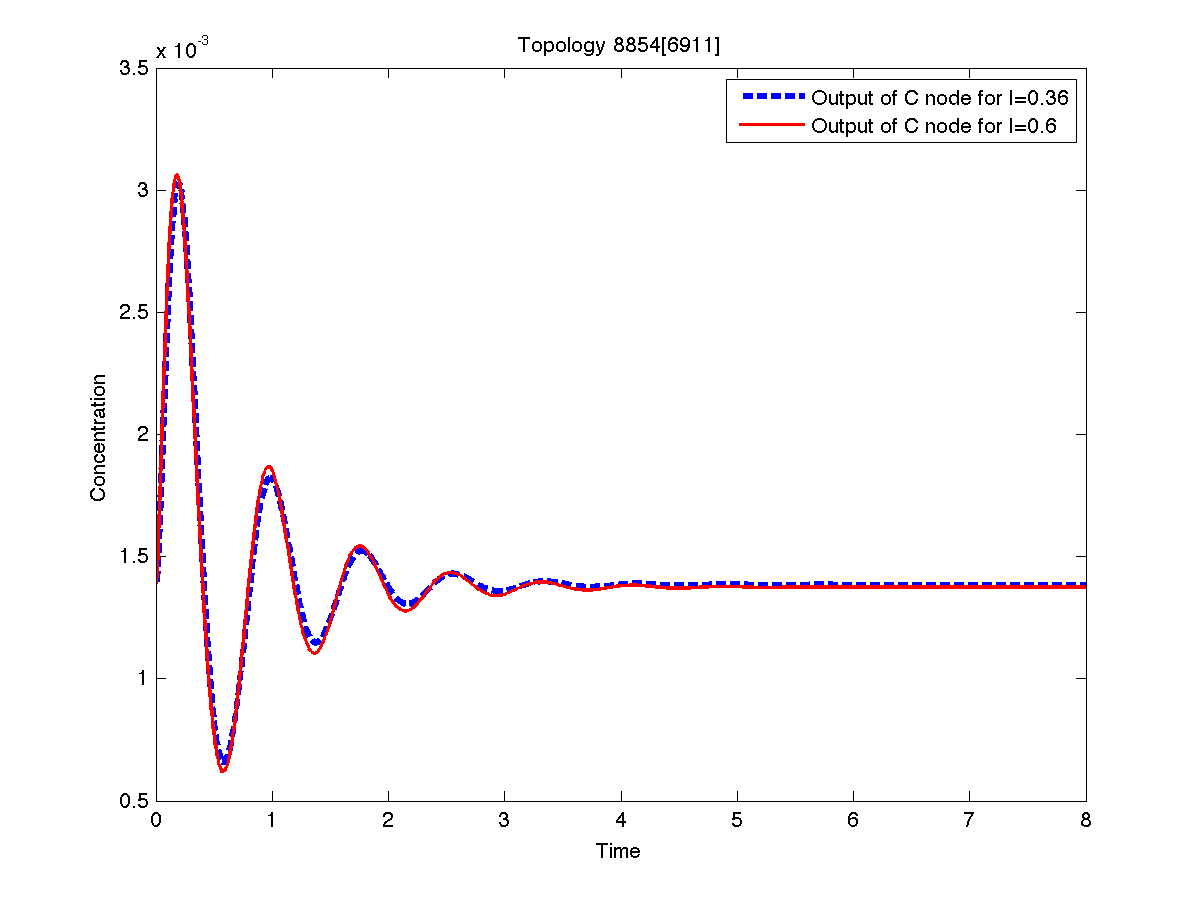}}\\
\subfloat[Quadratic approx. and output of nonlinear system]{\label{fig:f61}\includegraphics[width=0.55\textwidth]{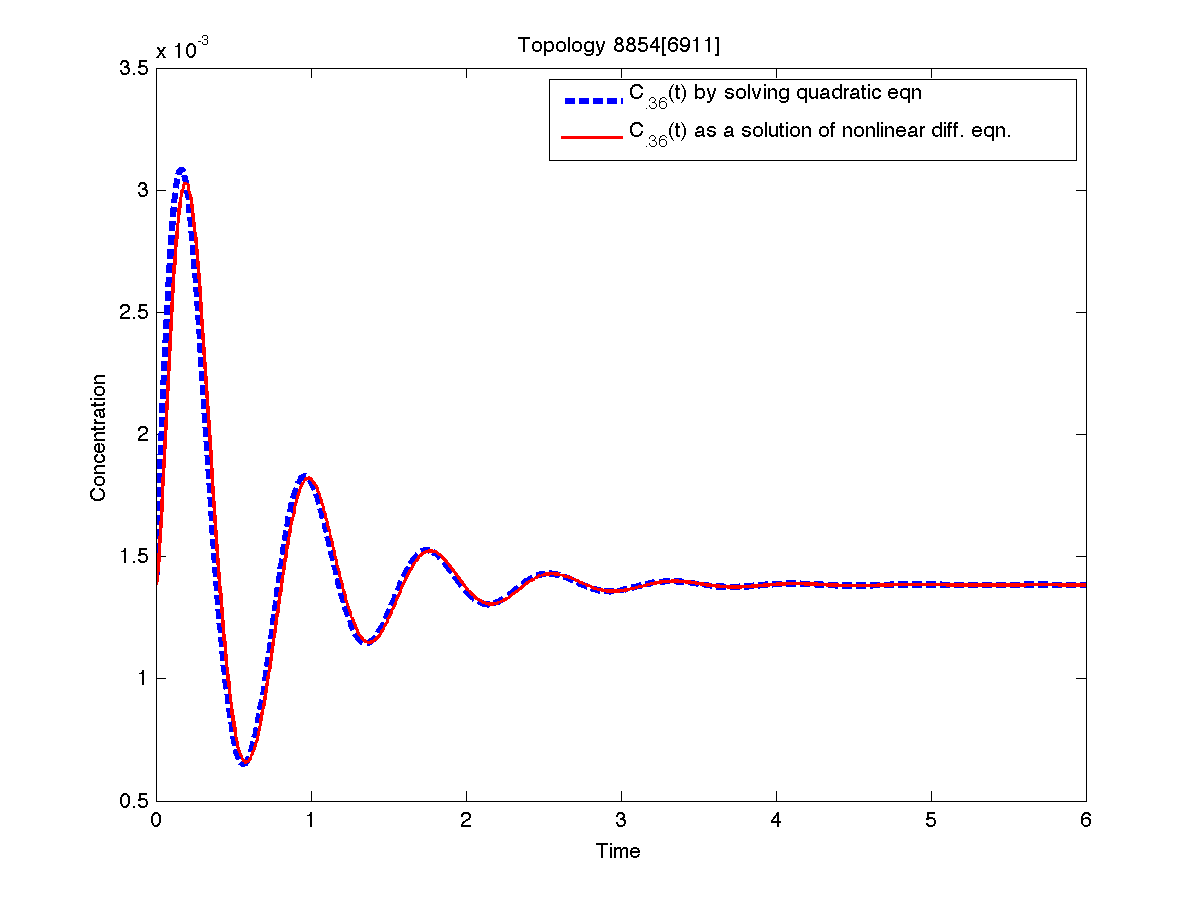}}                
  \subfloat[Quadratic approx. and output of nonlinear system]{\label{fig:f62}\includegraphics[width=0.55\textwidth]{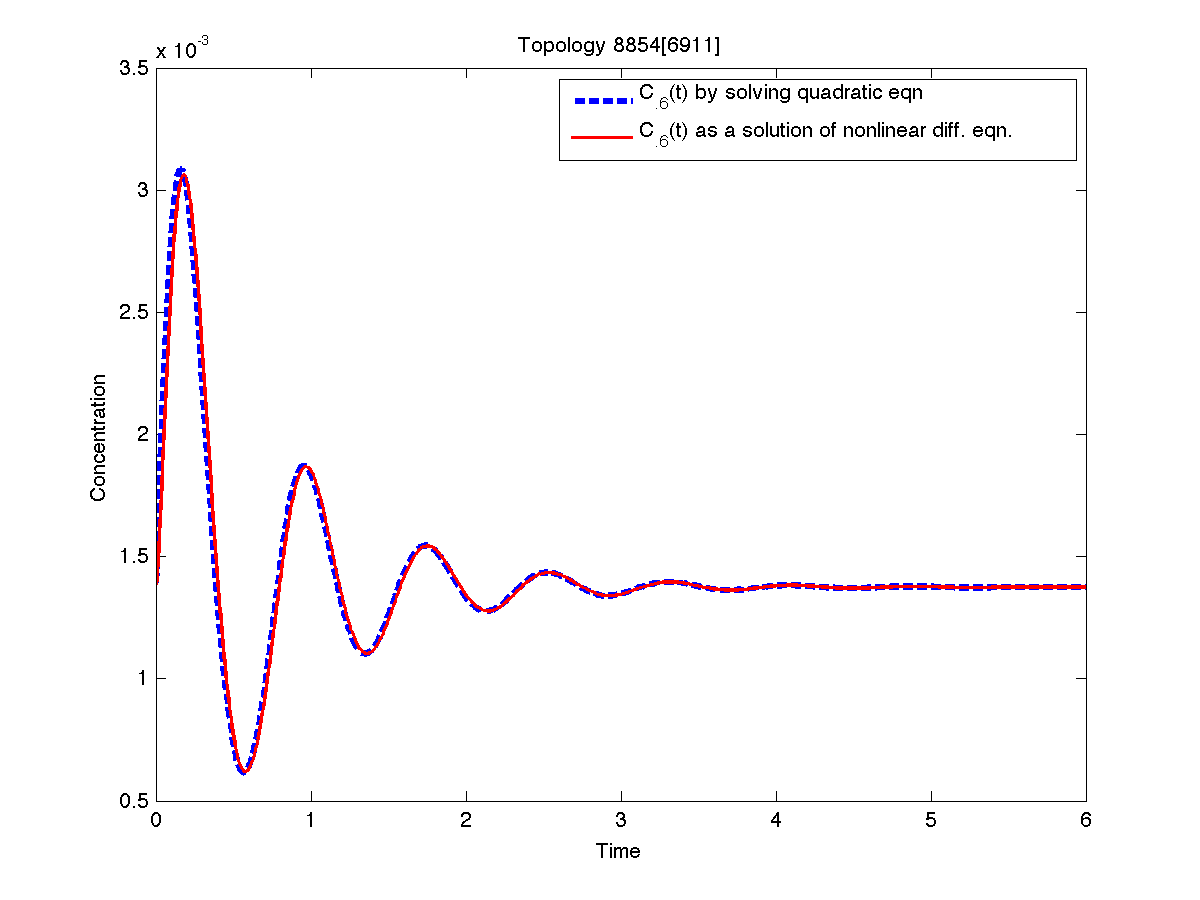}}
\end{figure}

\clearpage
Circuit 3.
\beqn
\dxA&=&k_{{\inp}A} {\inp} \frac{\txA}{\txA+K_{{\inp}A}}-k_{BA} \xB\frac{\xA}{\xA+K_{BA}}-k_{AA} \xA\frac{\xA}{\xA+K_{AA}}\\
\dxB&=&k_{AB}\xA\frac{\txB}{\txB+K_{AB}}-k_{CB}\xB \frac{\xB}{\xB+K_{CB}}-k_{BB}\xB \frac{\xB}{\xB+K_{BB}}\\
\dxC&=&{k_{BC}}\xB\frac{\txC}{\txC+K_{BC}}- k_{AC}\xA\frac{\xC}{\xC+K_{AC}}\\
\eeqn
Parameters: $K_{AA}=7.633962;$ $k_{AA}=86.238263;$
 $K_{AB}=20.265158;$ $k_{AB}=5.428752;$ $K_{AC}=0.258375; k_{AC}=62.416585;$
$K_{BA}=0.003960;$ $k_{BA}=17.705166;$
$K_{BB}=31.604578;$ $k_{BB}=3.692326;$
$K_{BC}=44.386408;$ $k_{BC}=65.027941;$
$K_{CB}=0.701052;$ $k_{CB}=26.091557;$
$K_{{\inp}A}=0.464248;$ $k_{{\inp}A}=1.882348$

\begin{figure}[ht]
  \centering
\subfloat[Dynamics of A and B in linearized model]{\label{fig:f71}\includegraphics[width=0.55\textwidth]{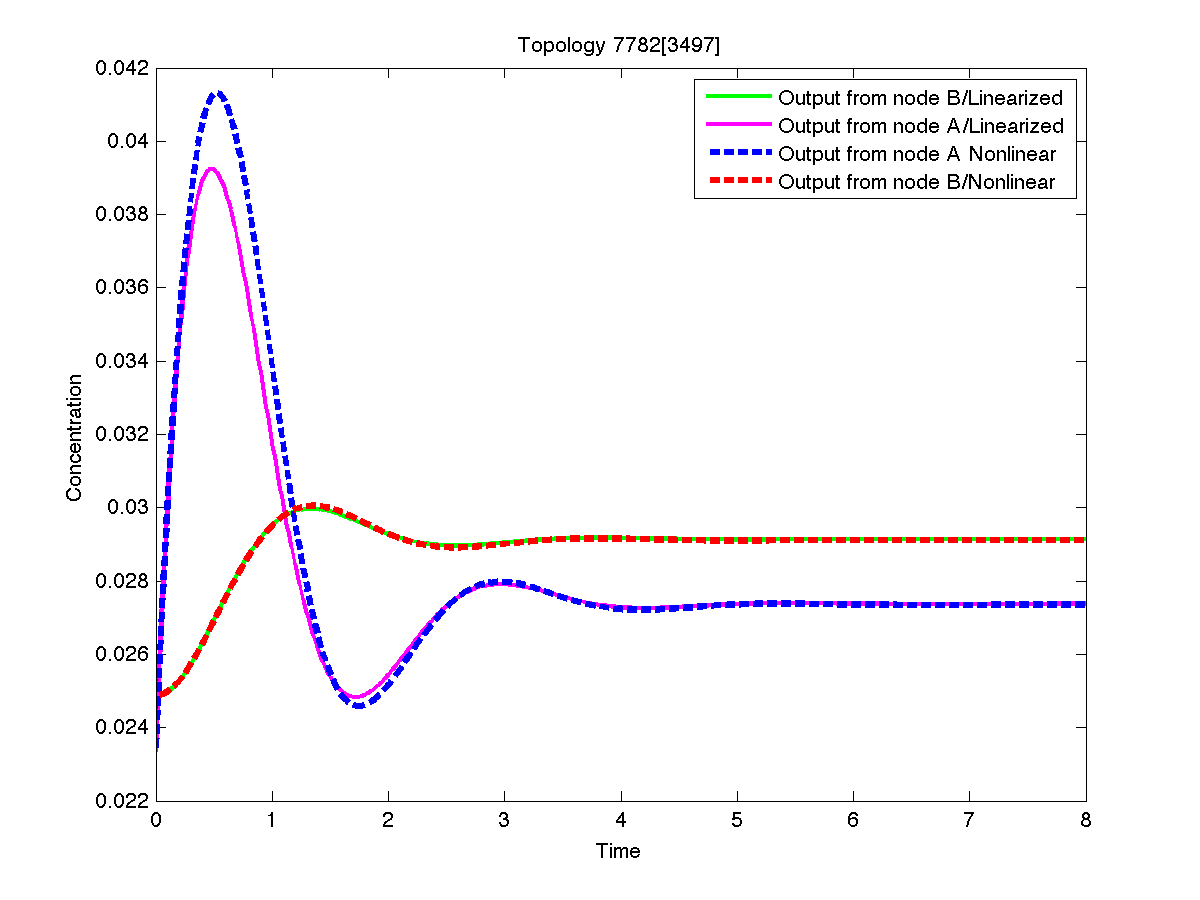}}                
  \subfloat[Ouput from C  nonlinear model]{\label{fig:f72}\includegraphics[width=0.55\textwidth]{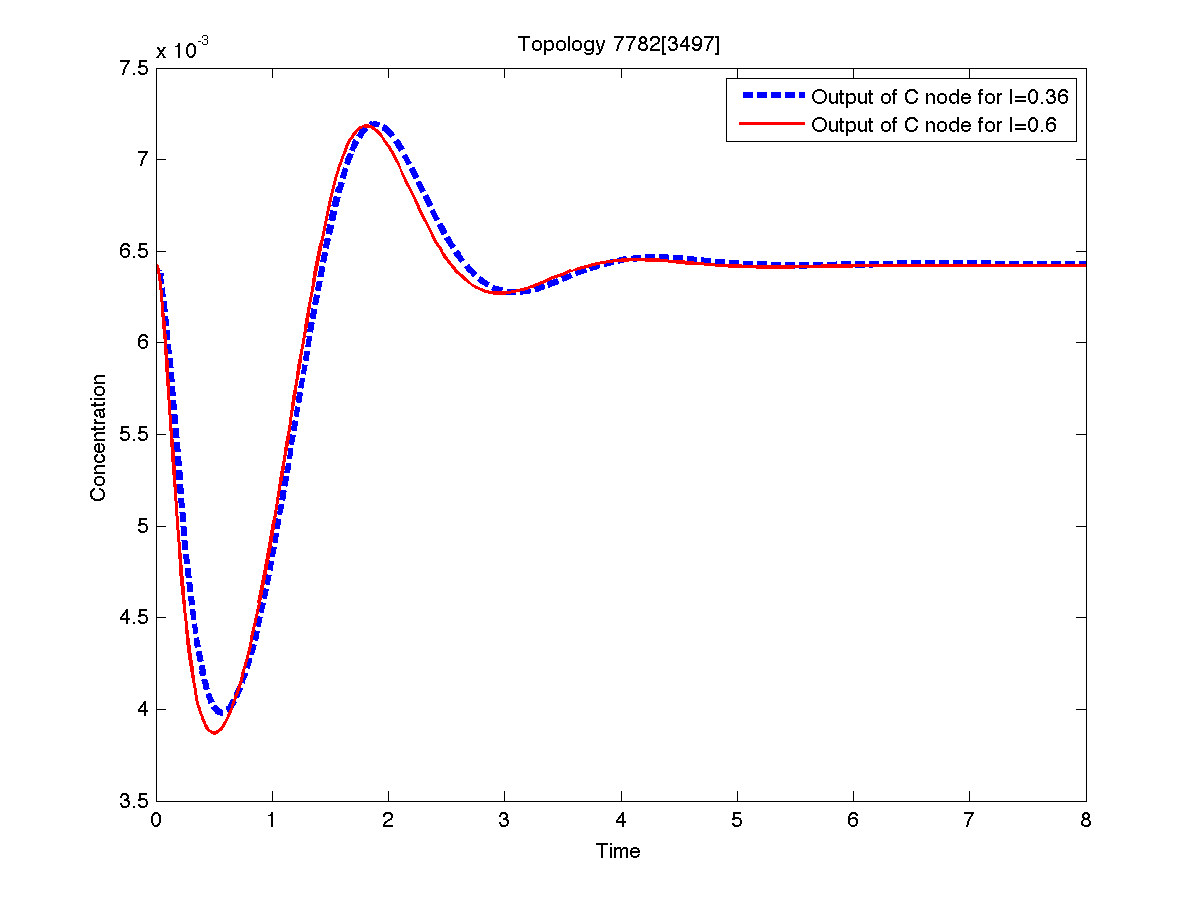}}\\
\subfloat[Quadratic approx. and output of nonlinear system]{\label{fig:f91}\includegraphics[width=0.55\textwidth]{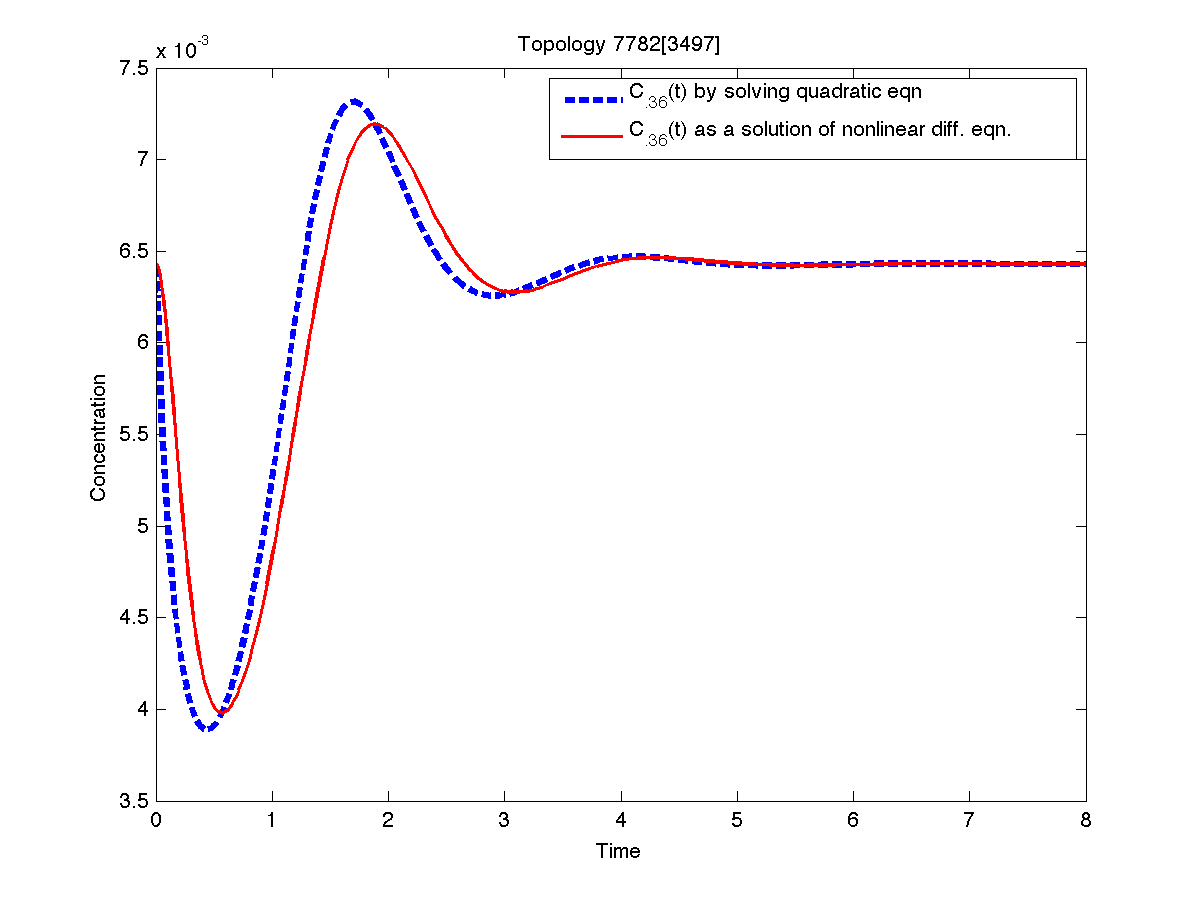}}                
  \subfloat[Quadratic approx. and output of nonlinear system]{\label{fig:f92}\includegraphics[width=0.55\textwidth]{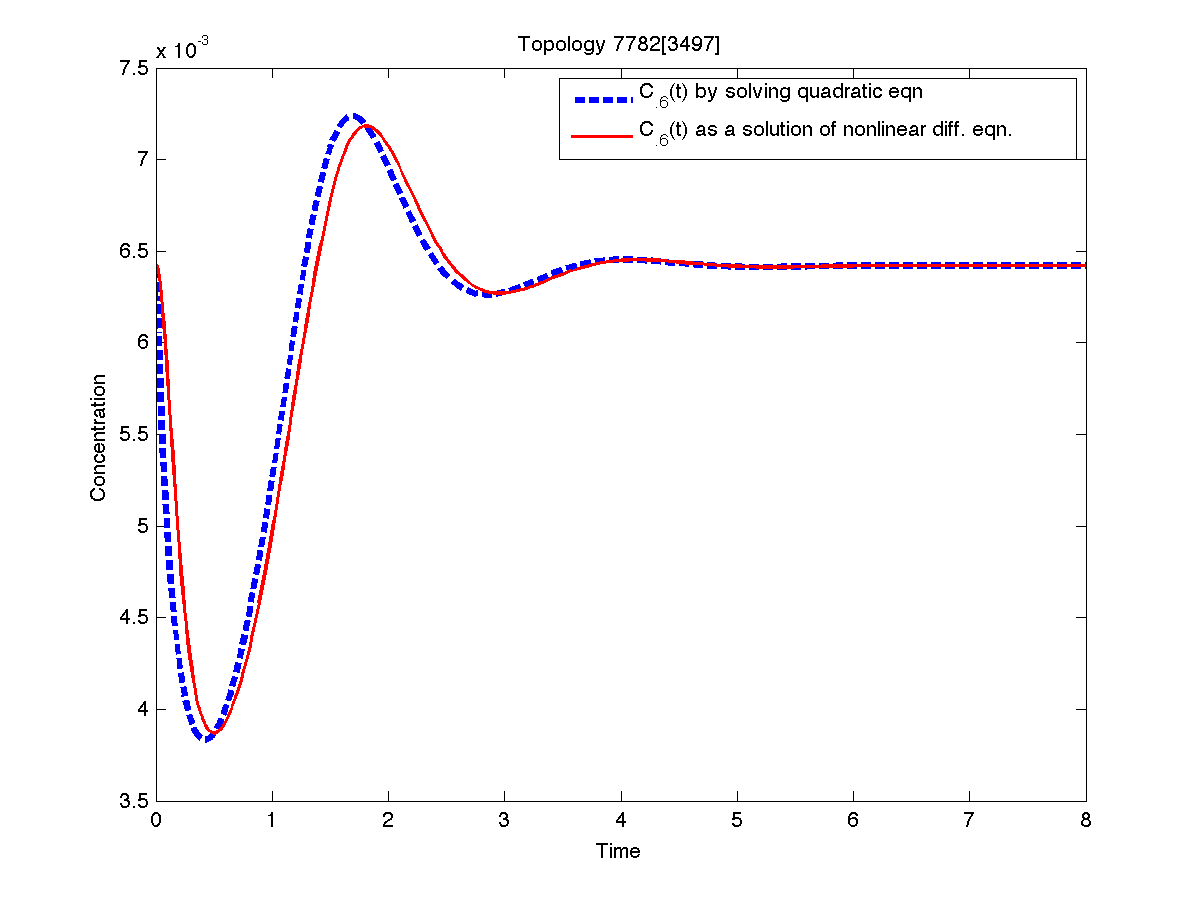}}
\end{figure}

Circuit 4.
\beqn
\dxA&=&k_{{\inp}A} {\inp} \frac{\txA}{\txA+K_{{\inp}A}}-k_{BA} \xB\frac{\xA}{\xA+K_{BA}}-k_{AA} \xA\frac{\xA}{\xA+K_{AA}}\\
\dxB&=&k_{AB}\xA\frac{\txB}{\txB+K_{AB}}-k_{CB} \xC\frac{\xB}{\xB+K_{CB}}\\
\dxC&=&{k_{BC}}\xB\frac{\txC}{\txC+K_{BC}}- k_{AC}\xA\frac{\xC}{\xC+K_{AC}}\\
\eeqn
Parameters: $K_{AA}=7.633962;$ $k_{AA}=86.238263;$
 $K_{AB}= 20.265158;$ $k_{AB}=5.428752;$
 $K_{AC}=0.258375;$ $k_{AC}=62.416585;$
 $K_{BA}=0.003960;$ $k_{BA}=17.705166;$
 $K_{BC}=44.386408;$ $k_{BC}= 65.027941;$
 $K_{CB}=0.701052;$ $k_{CB}=26.091557;$
 $K_{{\inp}A}=0.464248;$ $k_{{\inp}A}=1.882348$

\begin{figure}[ht]
  \centering
\subfloat[Dynamics of A and B in linearized model]{\label{fig:f101}\includegraphics[width=0.55\textwidth]{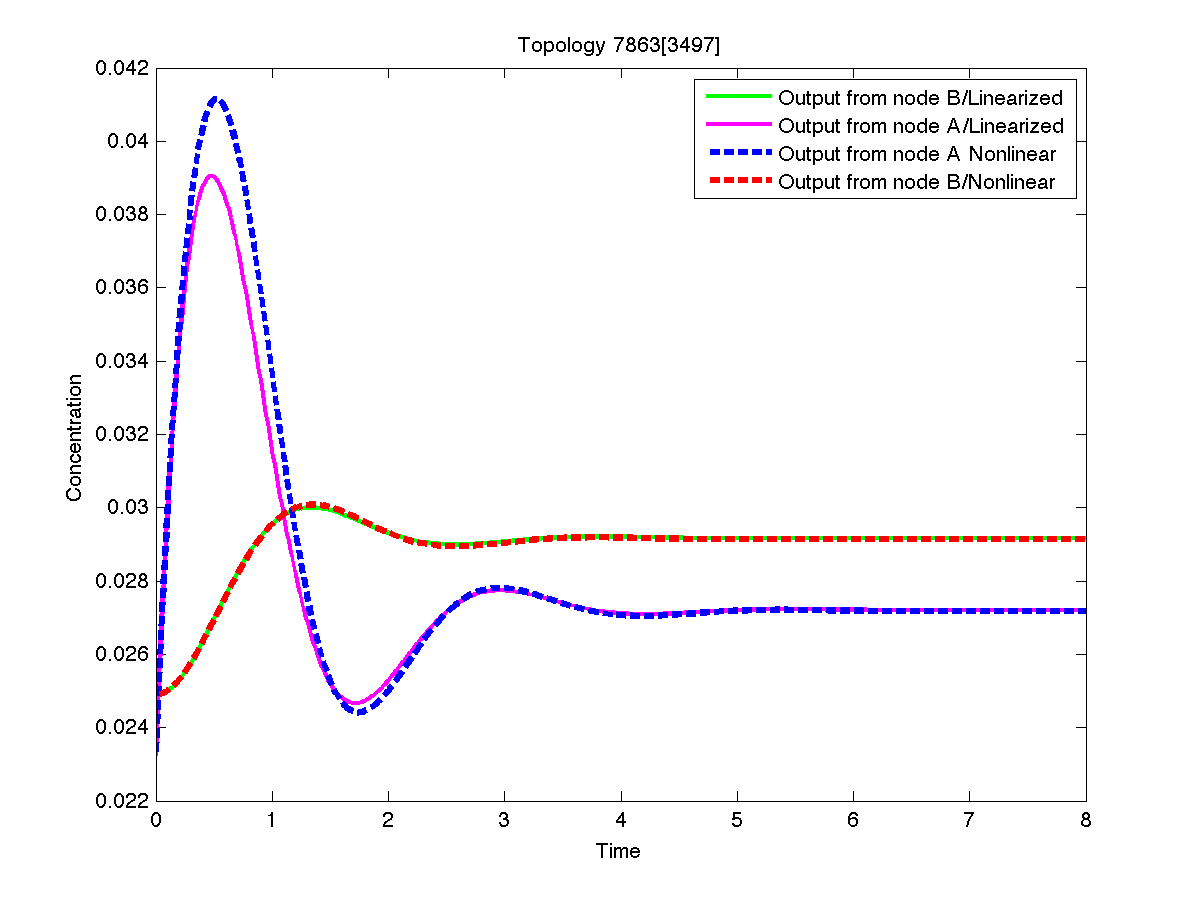}}                
  \subfloat[Ouput from C  nonlinear model]{\label{fig:f102}\includegraphics[width=0.55\textwidth]{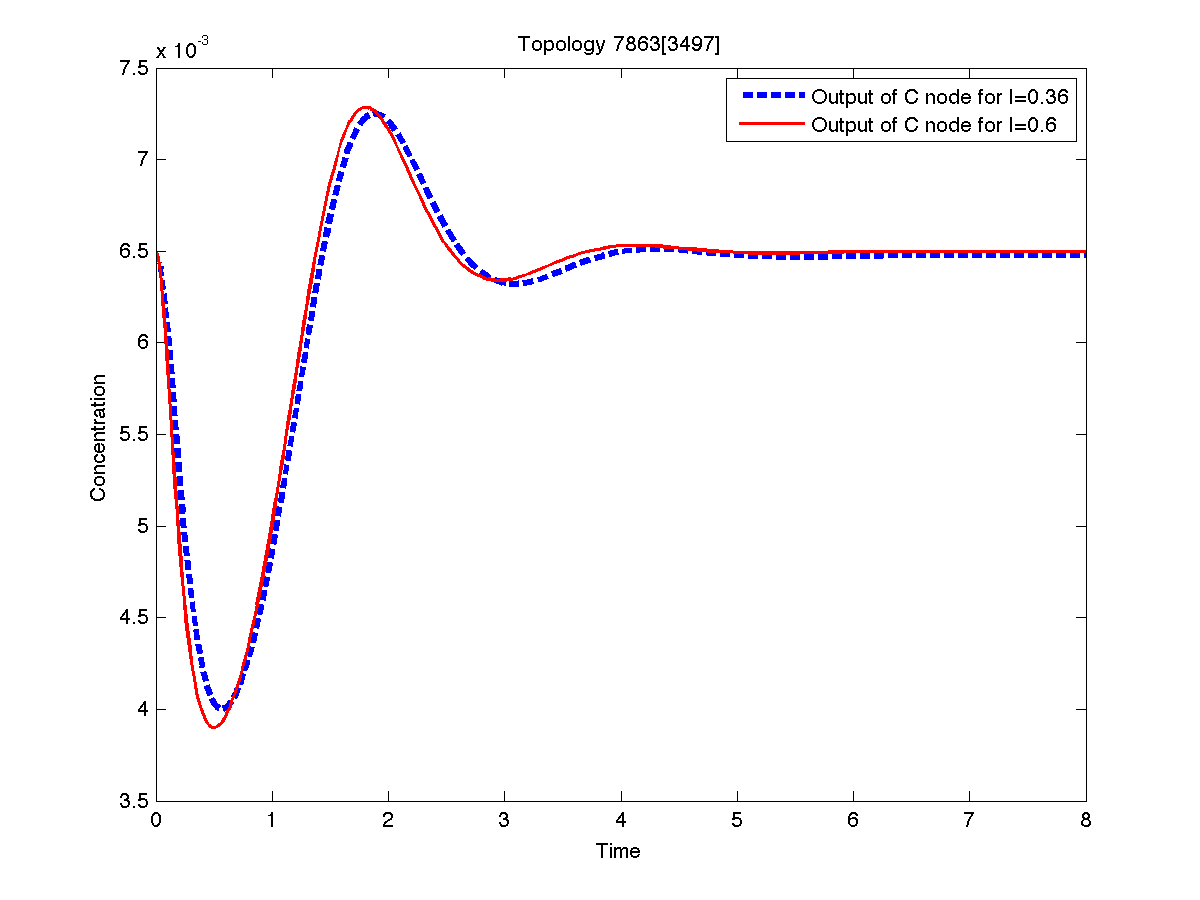}}\\
\subfloat[Quadratic approx. and output of nonlinear system]{\label{fig:f121}\includegraphics[width=0.55\textwidth]{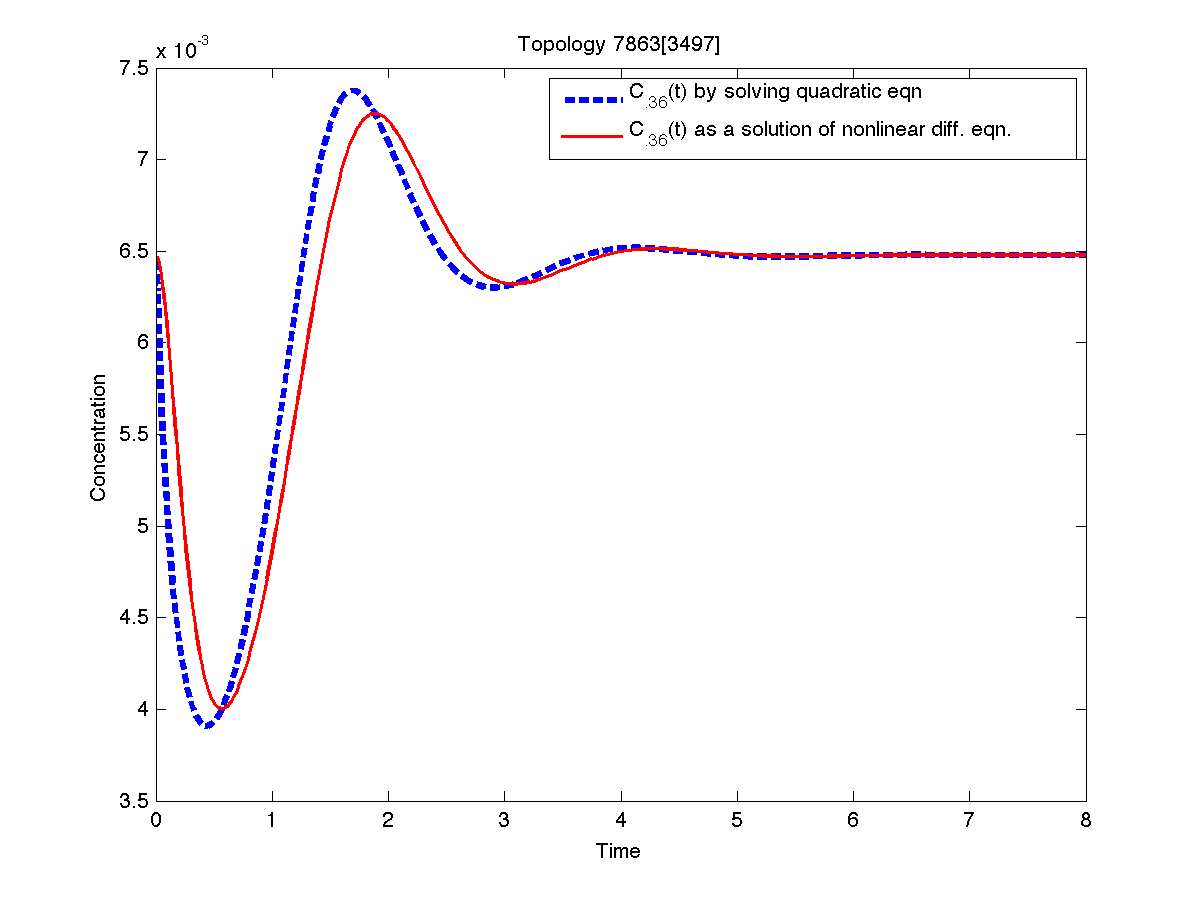}}                
  \subfloat[Quadratic approx. and output of nonlinear system]{\label{fig:f122}\includegraphics[width=0.55\textwidth]{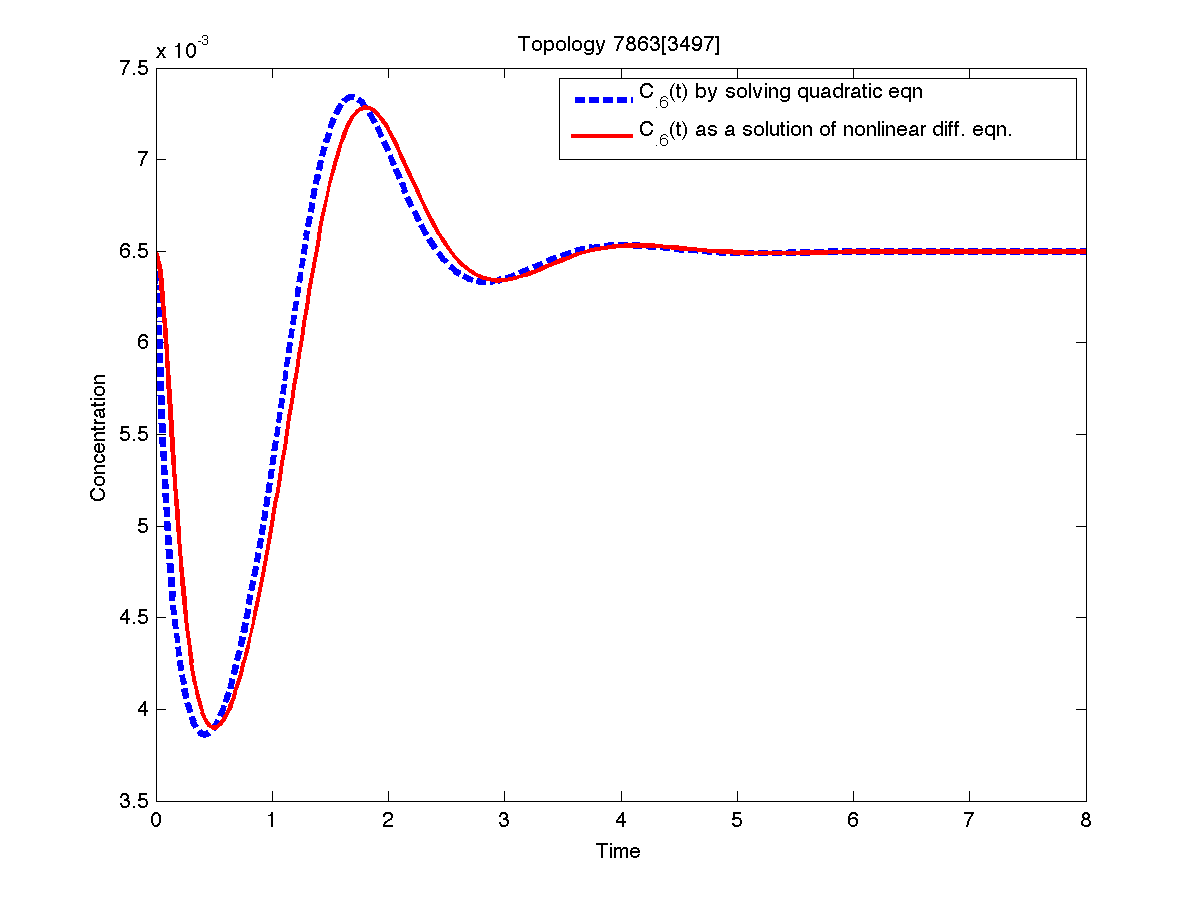}}
\end{figure}

\clearpage

Circuit 5.
\beqn
\dxA&=&k_{{\inp}A} {\inp} \frac{\txA}{\txA+K_{{\inp}A}}-k_{BA} \xB\frac{\xA}{\xA+K_{BA}}\\
\dxB&=&k_{AB}\xA\frac{\txB}{\txB+K_{AB}}-k_{CB} \xC\frac{\xB}{\xB+K_{CB}}\\
\dxC&=&{k_{BC}}\xB\frac{\txC}{\txC+K_{BC}}- k_{AC}\xA\frac{\xC}{\xC+K_{AC}}\\
\eeqn
Parameters:$K_{AB}= 63.277600;$ $k_{AB}= 6.638959;$
$K_{AC}=0.133429;$ $k_{AC}= 55.731406;$
$K_{BA}=0.011188;$ $k_{BA}= 2.749793;$
$K_{BC}= 0.013374;$ $k_{BC}= 45.175191;$
$K_{CB}= 1.457975;$ $k_{CB}= 2.114949;$
$K_{{\inp}A}= 24.589517;$ $k_{{\inp}A}= 5.346875$ 

\begin{figure}[hb]
  \centering
\subfloat[Dynamics of A and B in linearized model]{\label{fig:f131}\includegraphics[width=0.55\textwidth]{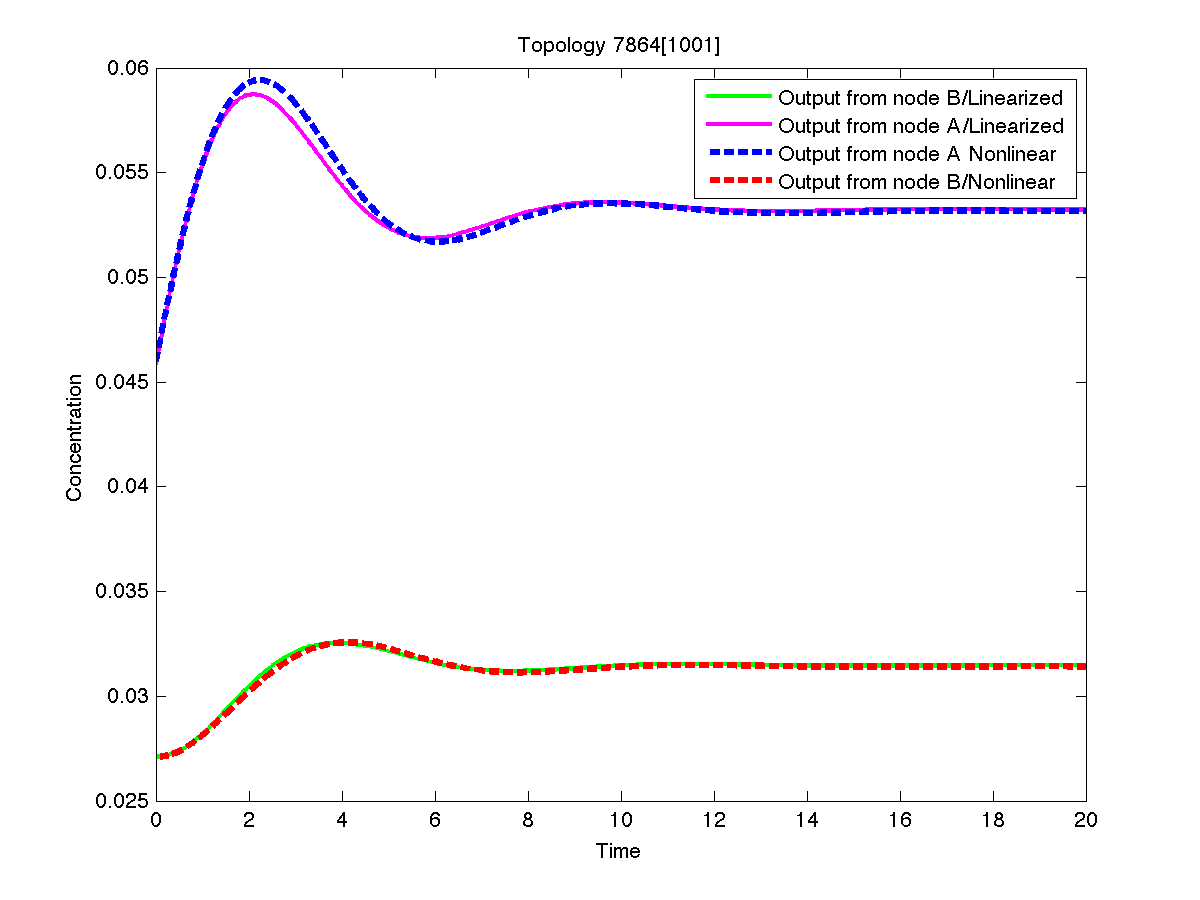}}                
  \subfloat[Ouput from C  nonlinear model]{\label{fig:f132}\includegraphics[width=0.55\textwidth]{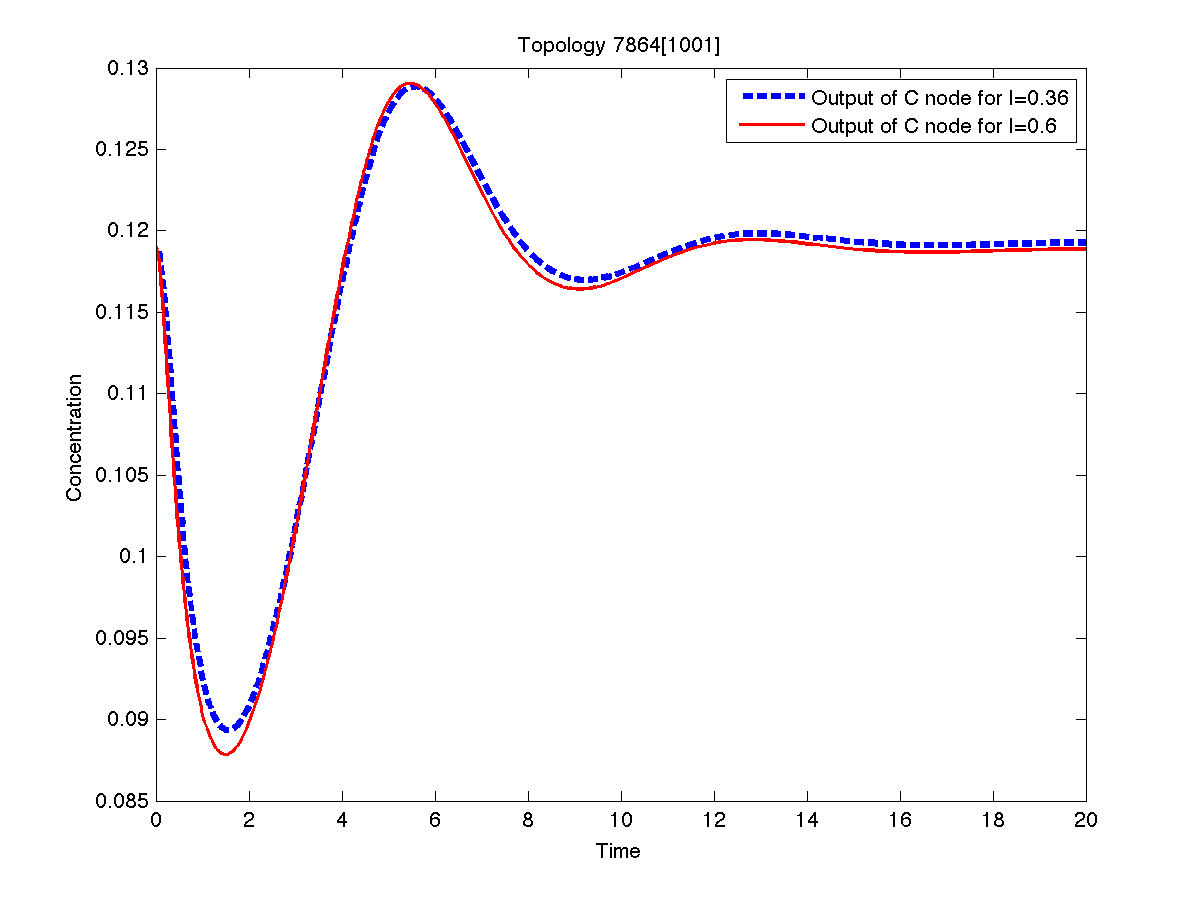}}\\
\subfloat[Quadratic approx. and output of nonlinear system]{\label{fig:f151}\includegraphics[width=0.55\textwidth]{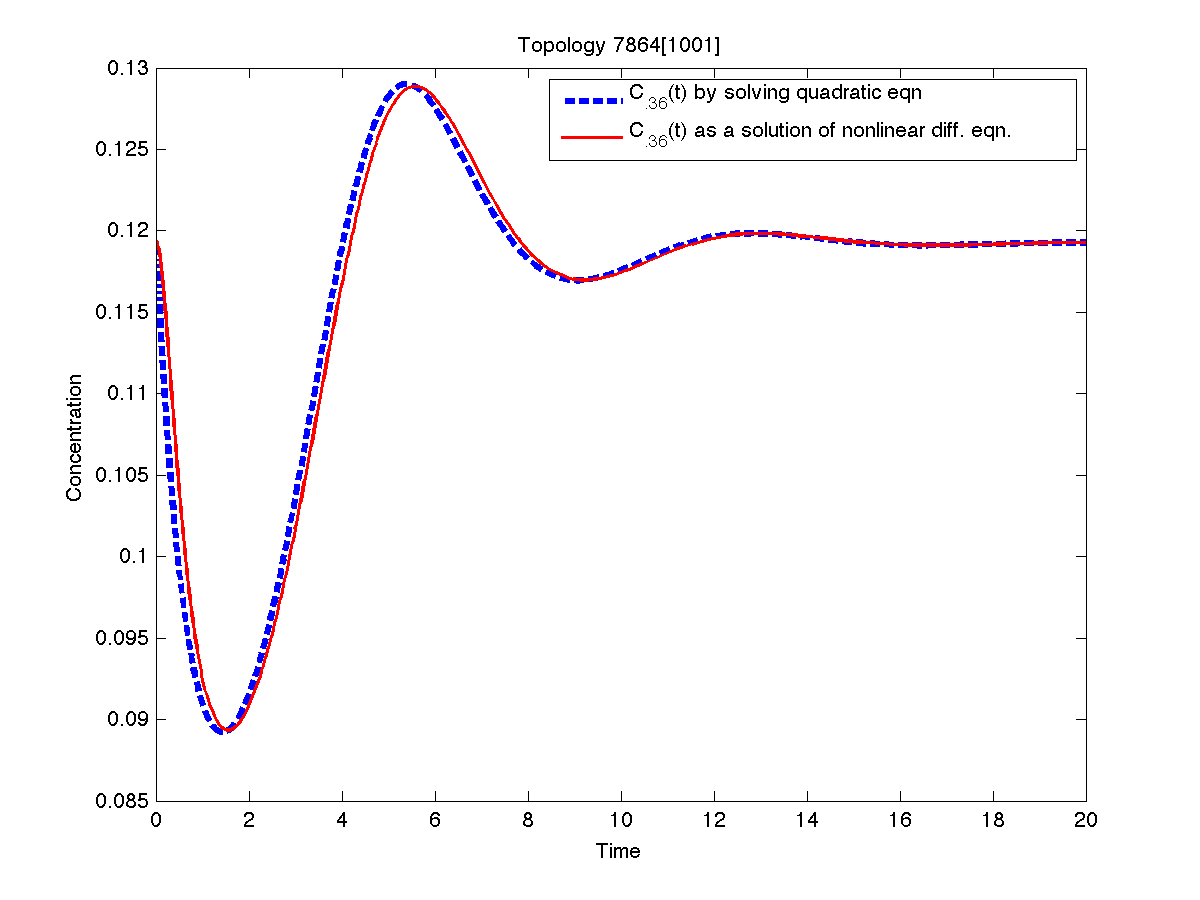}}                
  \subfloat[Quadratic approx. and output of nonlinear system]{\label{fig:f152}\includegraphics[width=0.55\textwidth]{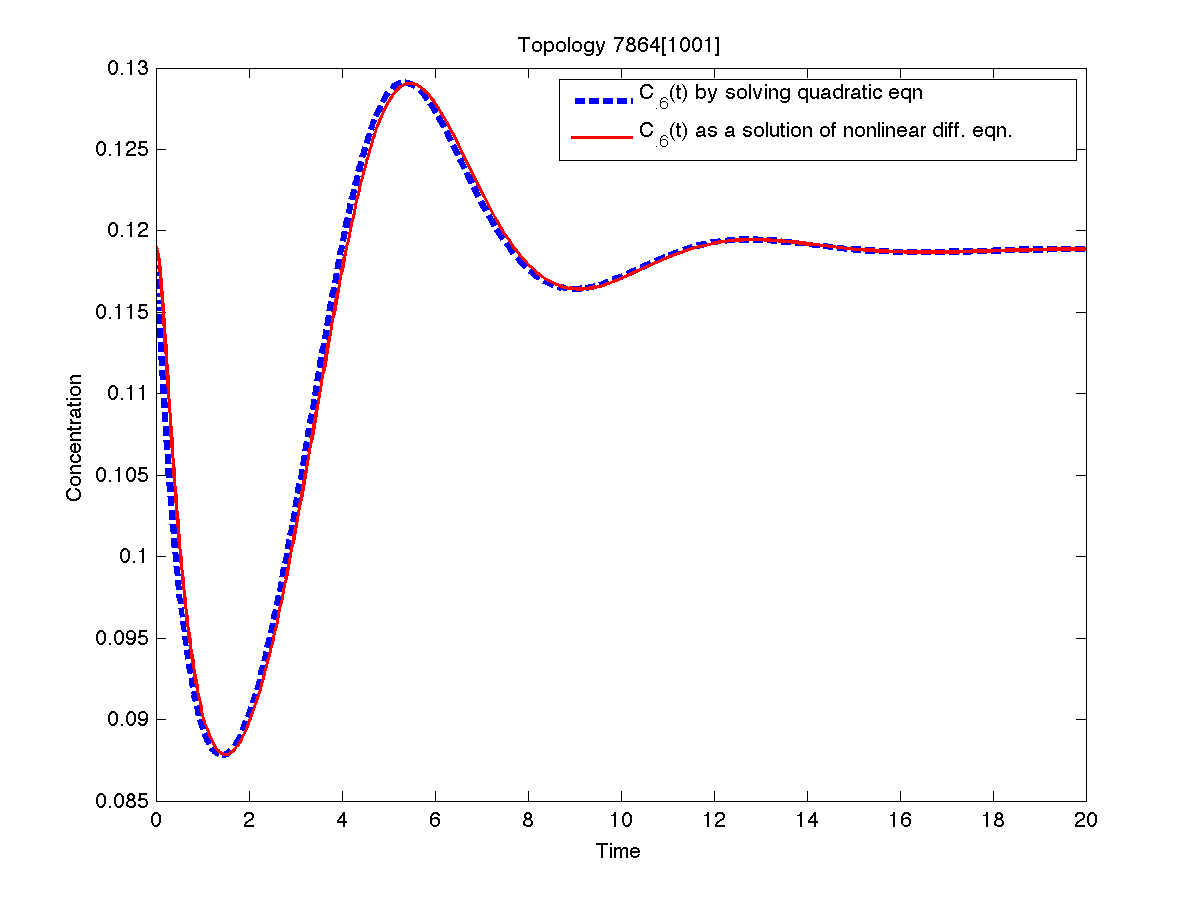}}
\end{figure}

\clearpage

Circuit 6.
\beqn
\dxA&=&k_{{\inp}A} {\inp} \frac{\txA}{\txA+K_{{\inp}A}}-k_{BA} \xB\frac{\xA}{\xA+K_{BA}}-k_{AA} \xA\frac{\xA}{\xA+K_{AA}}-k_{CA} \xC\frac{\xA}{\xA+K_{CA}}\\
\dxB&=&k_{AB}\xA\frac{\txB}{\txB+K_{AB}}-k_{CB} \xC\frac{\xB}{\xB+K_{CB}}-k_{BB} \xB\frac{\xB}{\xB+K_{BB}}\\
\dxC&=&{k_{BC}}\xB\frac{\txC}{\txC+K_{BC}}- k_{AC}\xA\frac{\xC}{\xC+K_{AC}}\\
\eeqn
Parameters:  $K_{AA}= 7.633962;$ $k_{AA}= 86.238263;$
$K_{AB}= 20.265158;$ $k_{AB}= 5.428752;$
 $K_{AC}= 0.258375;$ $k_{AC}= 62.416585;$
 $K_{BA}= 0.003960;$ $k_{BA}= 17.705166;$
 $K_{BB}= 31.604578;$ $k_{BB}= 3.692326;$
$K_{BC}= 44.386408;$ $k_{BC}= 65.027941;$
$K_{CA}= 26.714681;$ $k_{CA}= 2.806080;$
 $K_{CB}=0.701052;$ $k_{CB}= 26.091557;$
 $K_{{\inp}A}= 0.464248;$ $k_{{\inp}A}= 1.882348$

\begin{figure}[ht]
  \centering
\subfloat[Dynamics of A and B in linearized model]{\label{fig:f161}\includegraphics[width=0.55\textwidth]{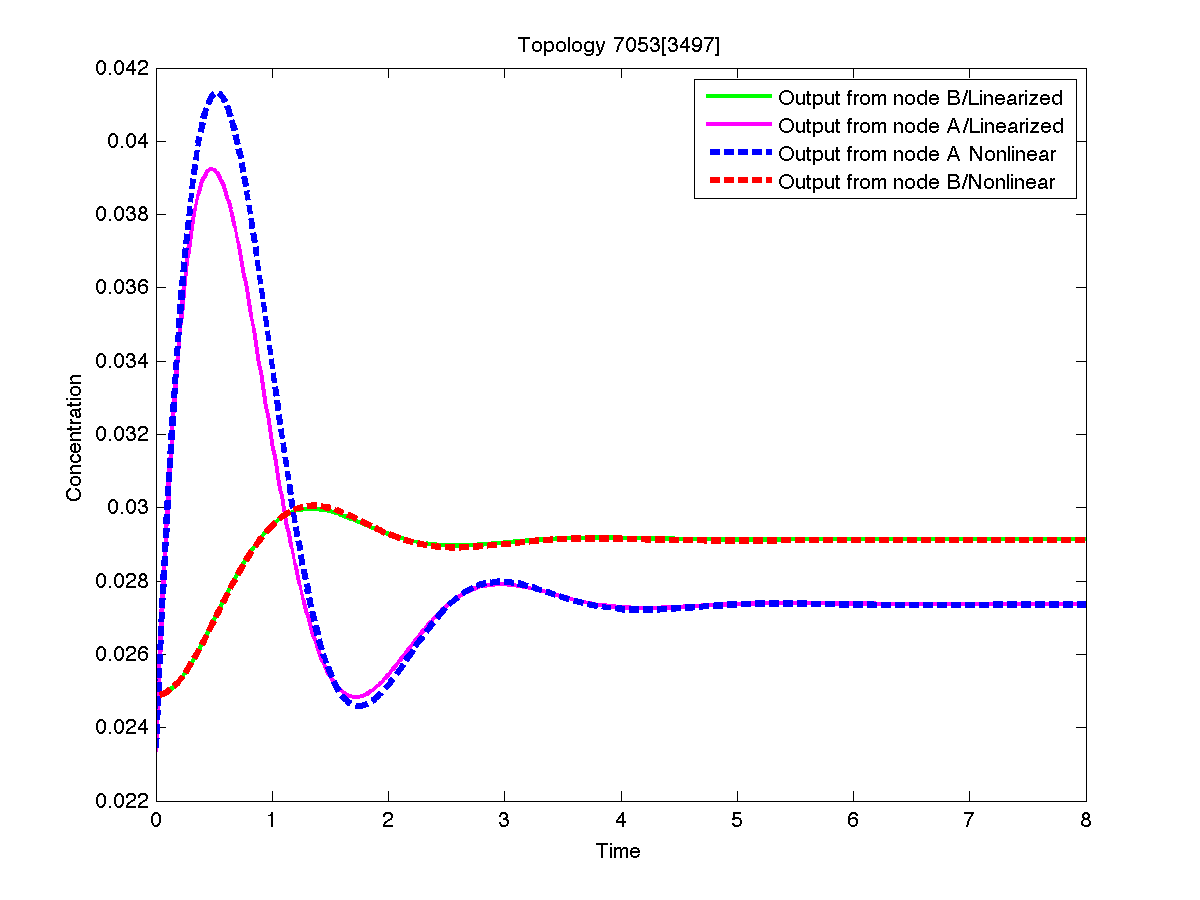}}                
  \subfloat[Ouput from C  nonlinear model]{\label{fig:f162}\includegraphics[width=0.55\textwidth]{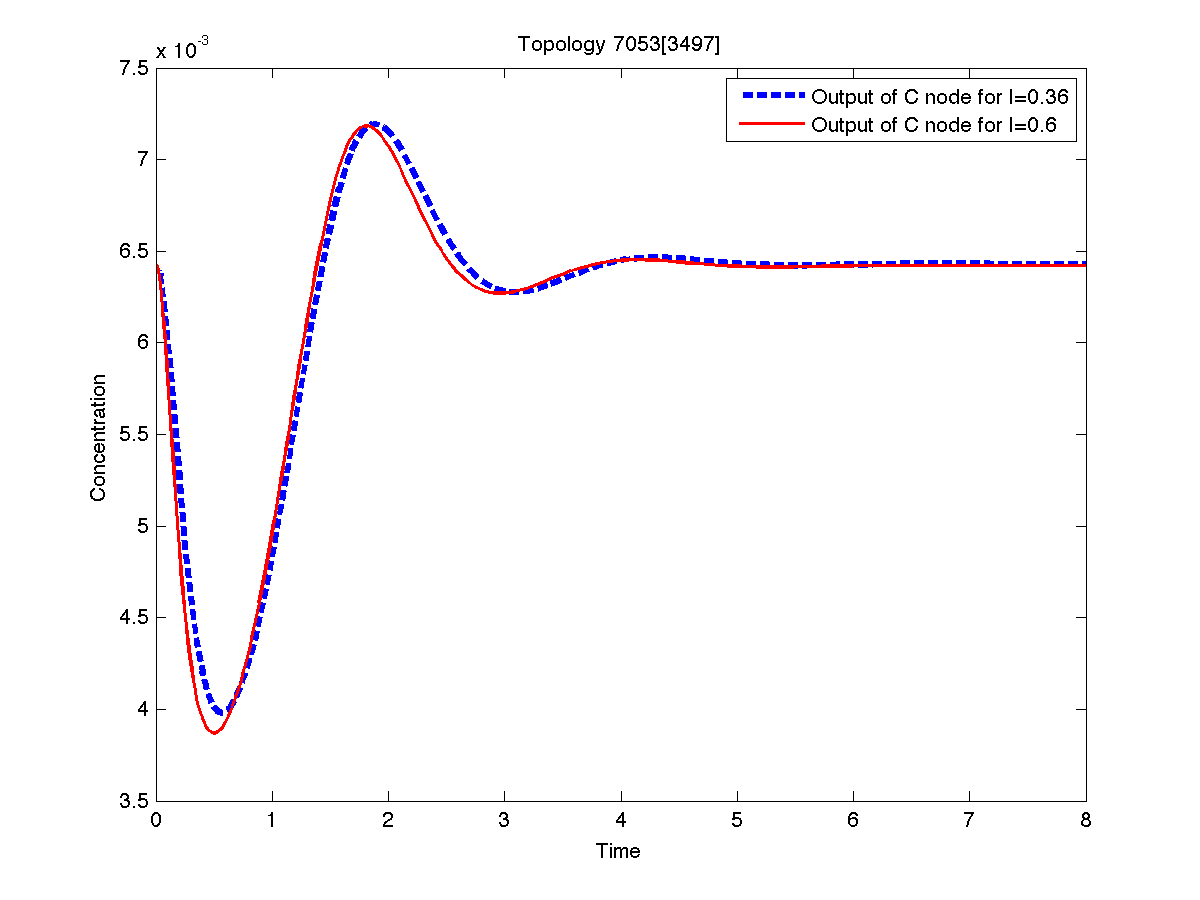}}\\
\subfloat[Quadratic approx. and output of nonlinear system]{\label{fig:f181}\includegraphics[width=0.55\textwidth]{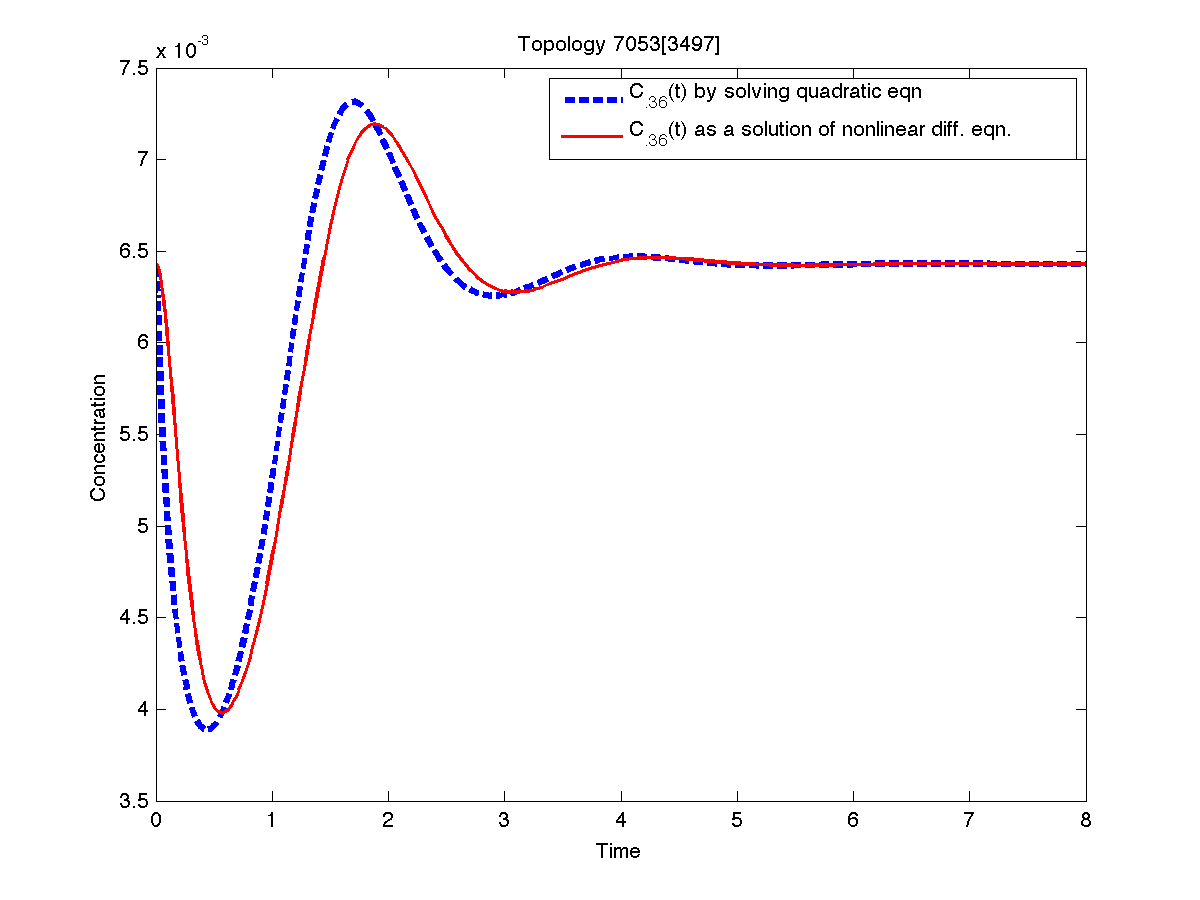}}                
  \subfloat[Quadratic approx. and output of nonlinear system]{\label{fig:f182}\includegraphics[width=0.55\textwidth]{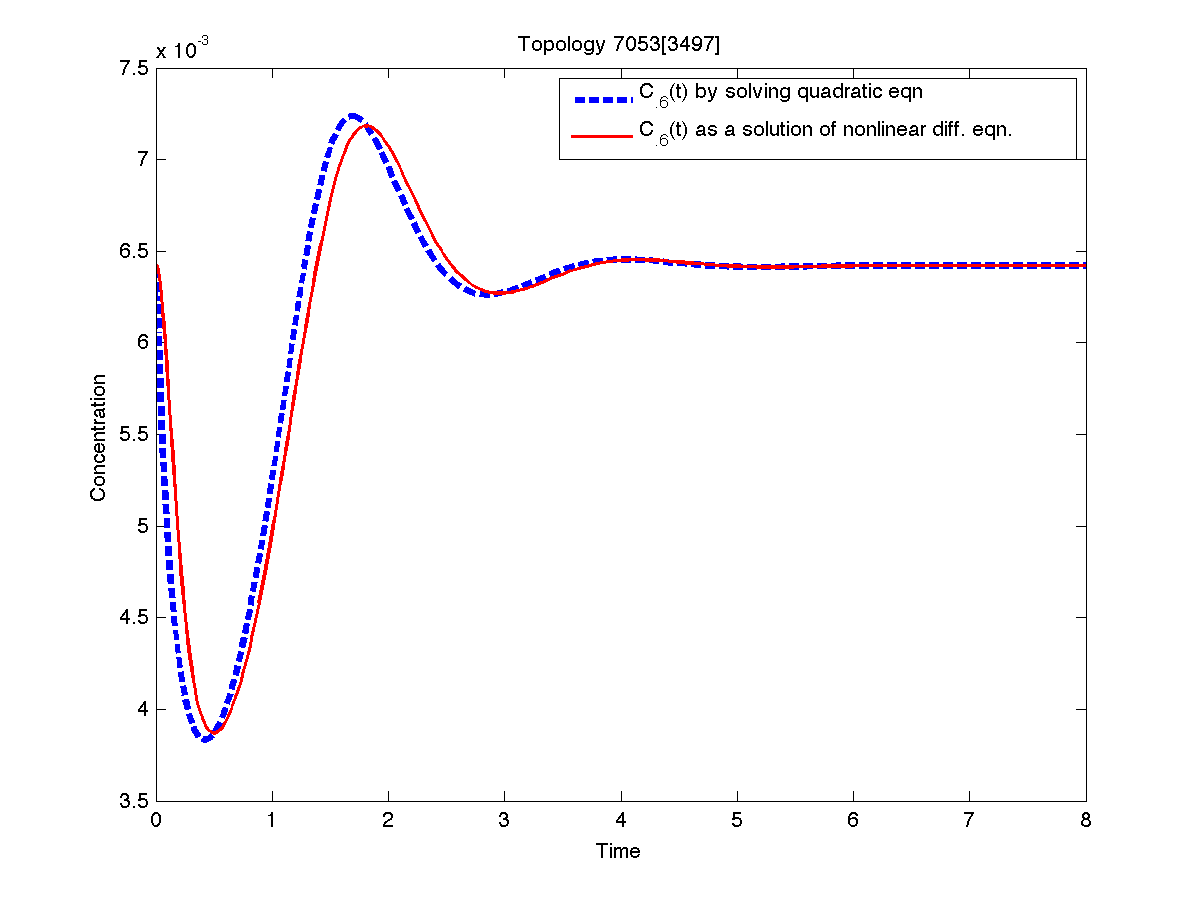}}
\end{figure}

\clearpage

Circuit 7.
\beqn
\dxA&=&k_{{\inp}A} {\inp} \frac{\txA}{\txA+K_{{\inp}A}}-k_{BA} \xB\frac{\xA}{\xA+K_{BA}}-k_{AA} \xA\frac{\xA}{\xA+K_{AA}}-k_{CA} \xC\frac{\xA}{\xA+K_{CA}}\\
\dxB&=&k_{AB}\xA\frac{\txB}{\txB+K_{AB}}-k_{CB} \xC \frac{\xB}{\xB+K_{CB}}\\
\dxC&=&{k_{BC}}\xB\frac{\txC}{\txC+K_{BC}}- k_{AC}\xA\frac{\xC}{\xC+K_{AC}}\\
\eeqn
Parameters: \  $K_{AA}= 7.633962;$ $k_{AA}= 86.238263;$
$K_{AB}= 20.265158;$ $k_{AB}= 5.428752;$
 $K_{AC}= 0.258375;$ $k_{AC}= 62.416585;$
 $K_{BA}= 0.003960;$ $k_{BA}= 17.705166;$
$K_{BC}= 44.386408;$ $k_{BC}= 65.027941;$
$K_{CA}= 26.714681;$ $k_{CA}= 2.806080;$
 $K_{CB}=0.701052;$ $k_{CB}= 26.091557;$
 $K_{{\inp}A}= 0.464248;$ $k_{{\inp}A}= 1.882348$

\begin{figure}[hb]
  \centering
\subfloat[Dynamics of A and B in linearized model]{\label{fig:f191}\includegraphics[width=0.55\textwidth]{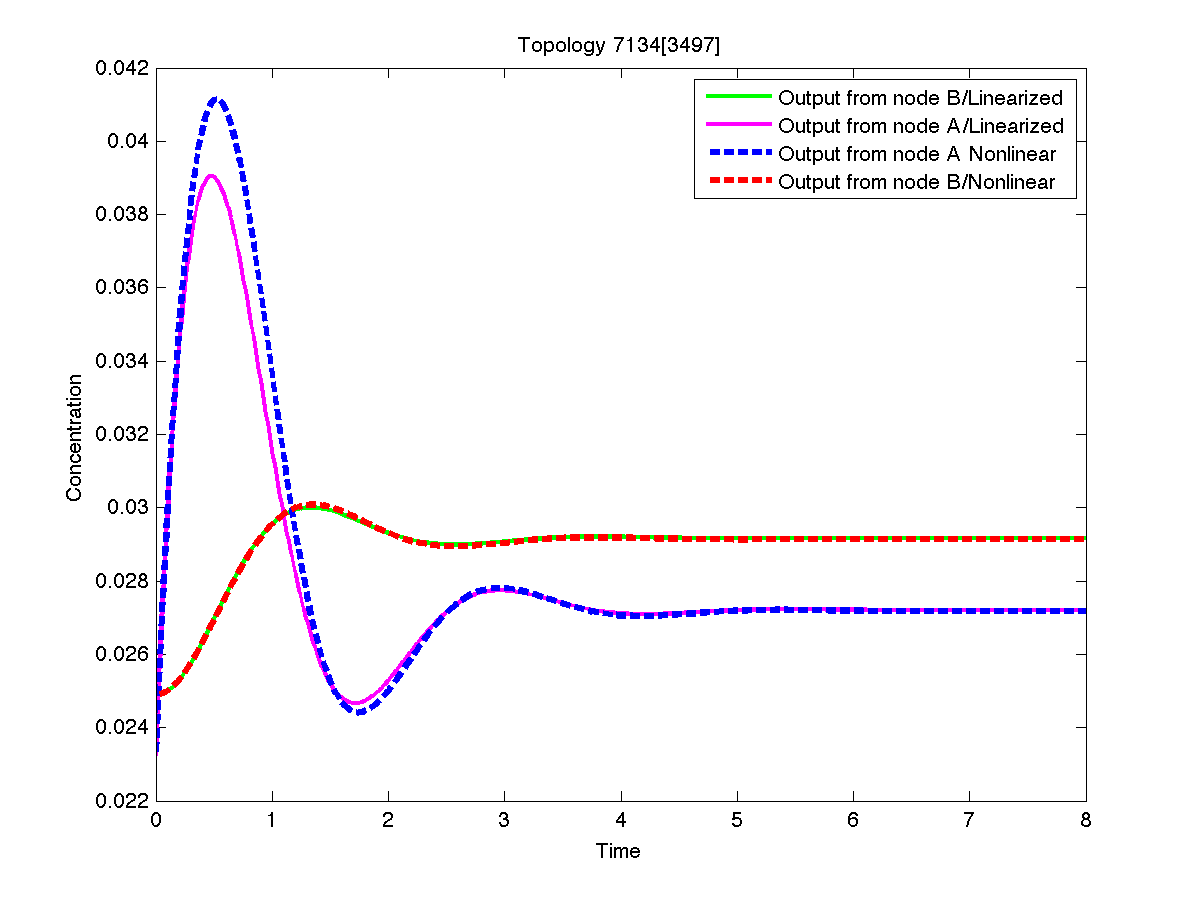}}                
  \subfloat[Ouput from C  nonlinear model]{\label{fig:f192}\includegraphics[width=0.55\textwidth]{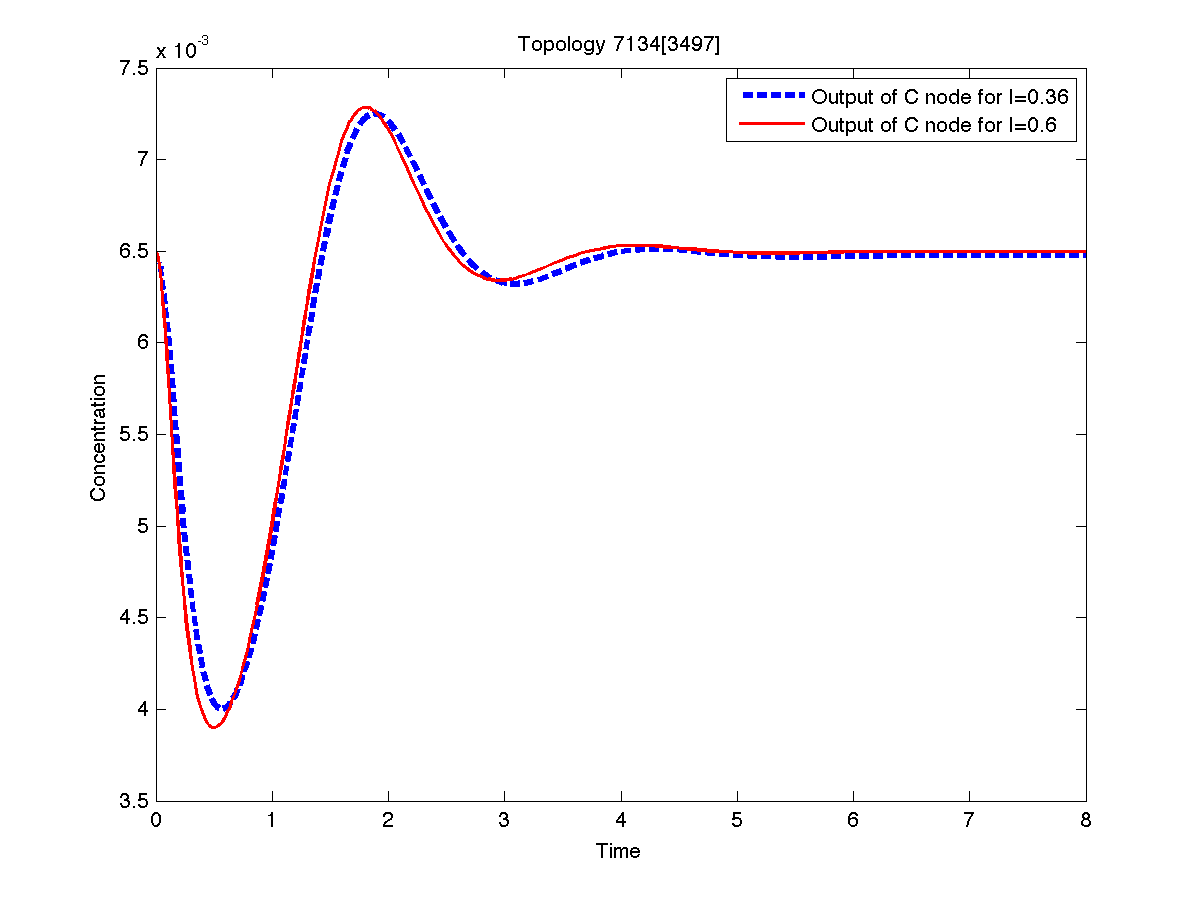}}\\
\subfloat[Quadratic approx. and output of nonlinear system]{\label{fig:211}\includegraphics[width=0.55\textwidth]{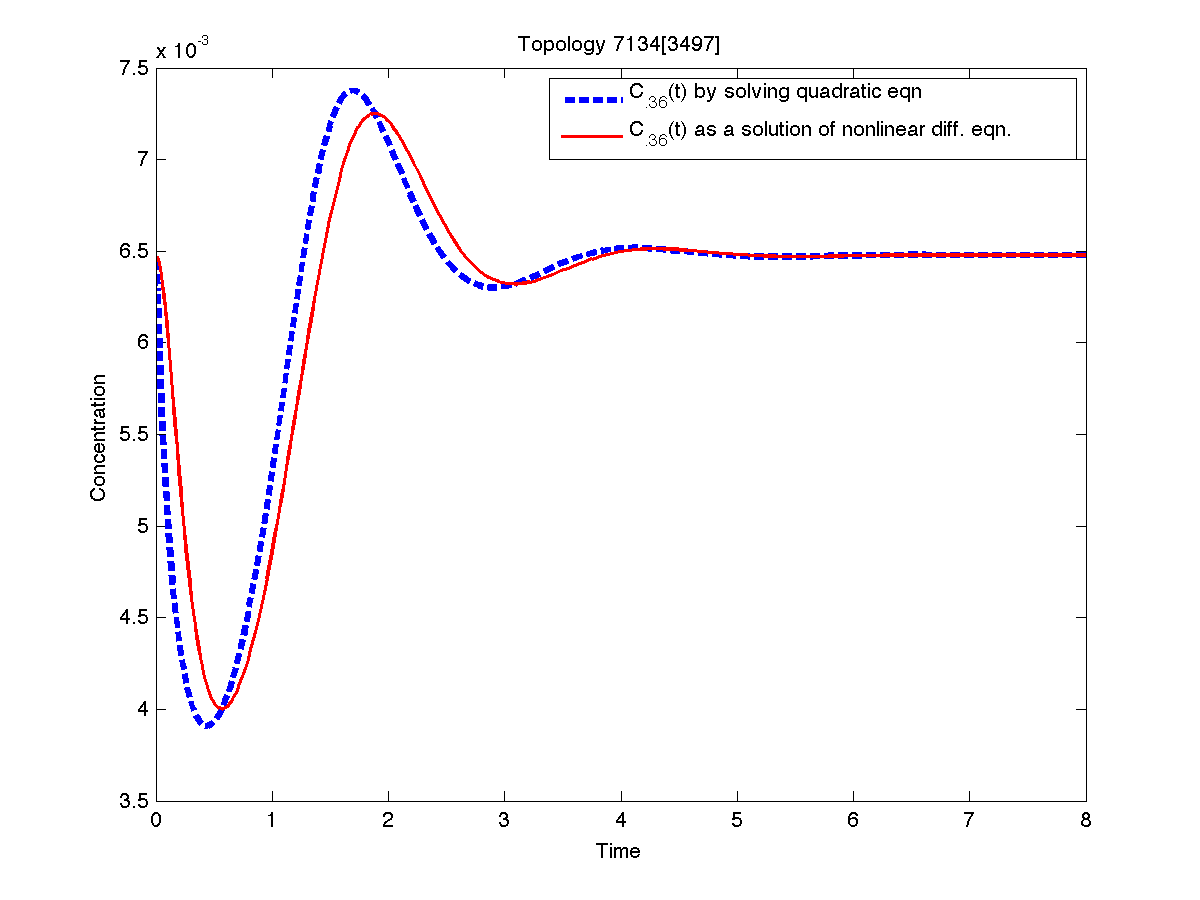}}                
  \subfloat[Quadratic approx. and output of nonlinear system]{\label{fig:f212}\includegraphics[width=0.55\textwidth]{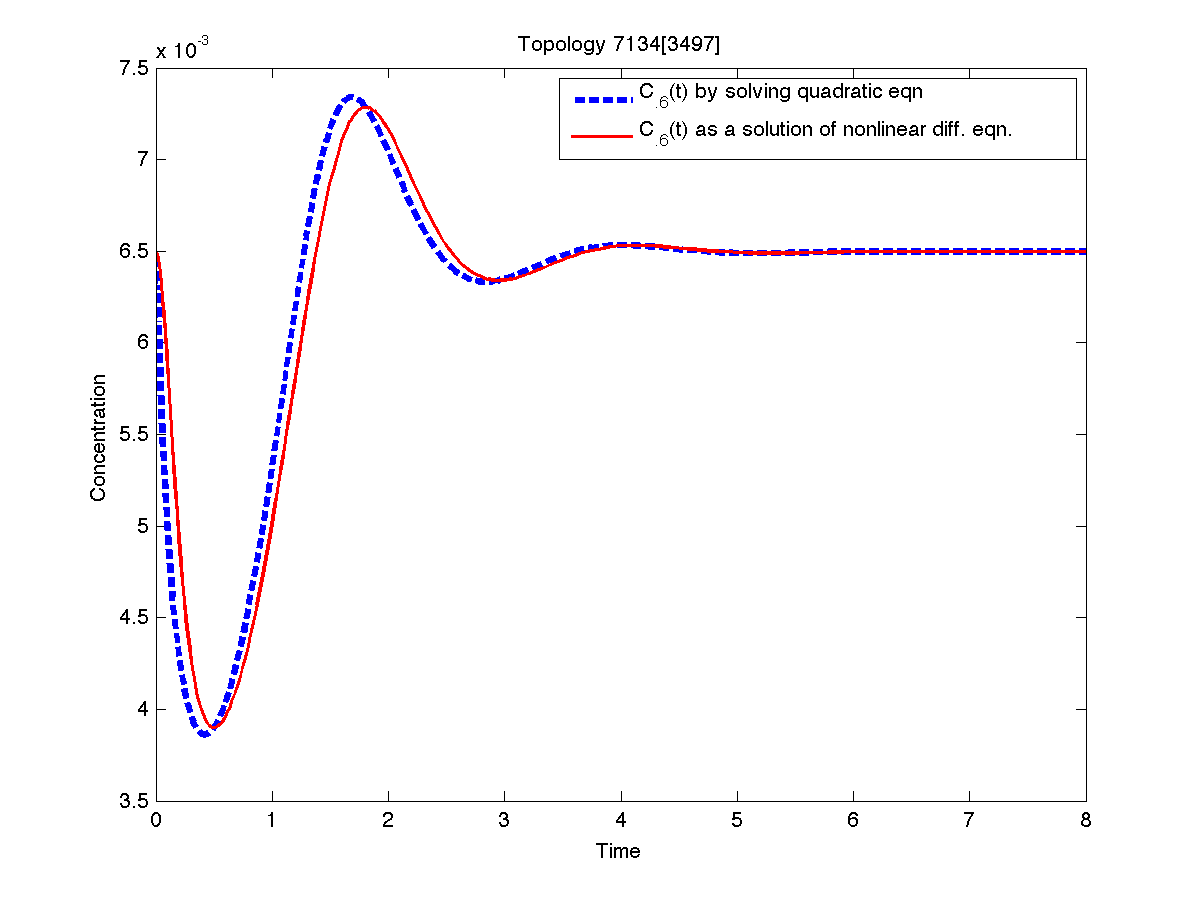}}
\end{figure}

Circuit 8.
\beqn
\dxA&=&k_{{\inp}A} {\inp} \frac{\txA}{\txA+K_{{\inp}A}}-k_{BA} \xB\frac{\xA}{\xA+K_{BA}}\\
\dxB&=&k_{AB}\xA\frac{\txB}{\txB+K_{AB}}-k_{F_BB} \xFB \frac{\xB}{\xB+K_{F_BB}}+k_{CB} \xC \frac{\txB}{\txB+K_{CB}}\\
\dxC&=&{k_{AC}}\xA\frac{\txC}{\txC+K_{AC}}- k_{BC}\xB\frac{\xC}{\xC+K_{BC}}-k_{CC}\xC\frac{\xC}{\xC+K_{CC}}\\
\eeqn
Parameters: \  $K_{{\inp}A}= 0.093918;$ $k_{{\inp}A}= 11.447219;$ 
$K_{BA}= 0.001688;$ $k_{BA}= 44.802268;$
$K_{AB}=0.001191;$ $k_{AB}=1.466561;$
 $K_{F_B}=9.424319;$ $k_{F_B}=22.745736;$
$K_{AC}= 0.113697;$ $k_{AC}=1.211993;$
$K_{BC}=0.009891;$  $k_{BC}=7.239357;$
$K_{CB}=30.602013;$ $k_{CB}= 3.811536;$
$K_{CC}=0.189125;$ $k_{CC}= 17.910182$

\begin{figure}[ht]
  \centering
\subfloat[Dynamics of A and B in linearized model]{\label{fig:f221}\includegraphics[width=0.55\textwidth]{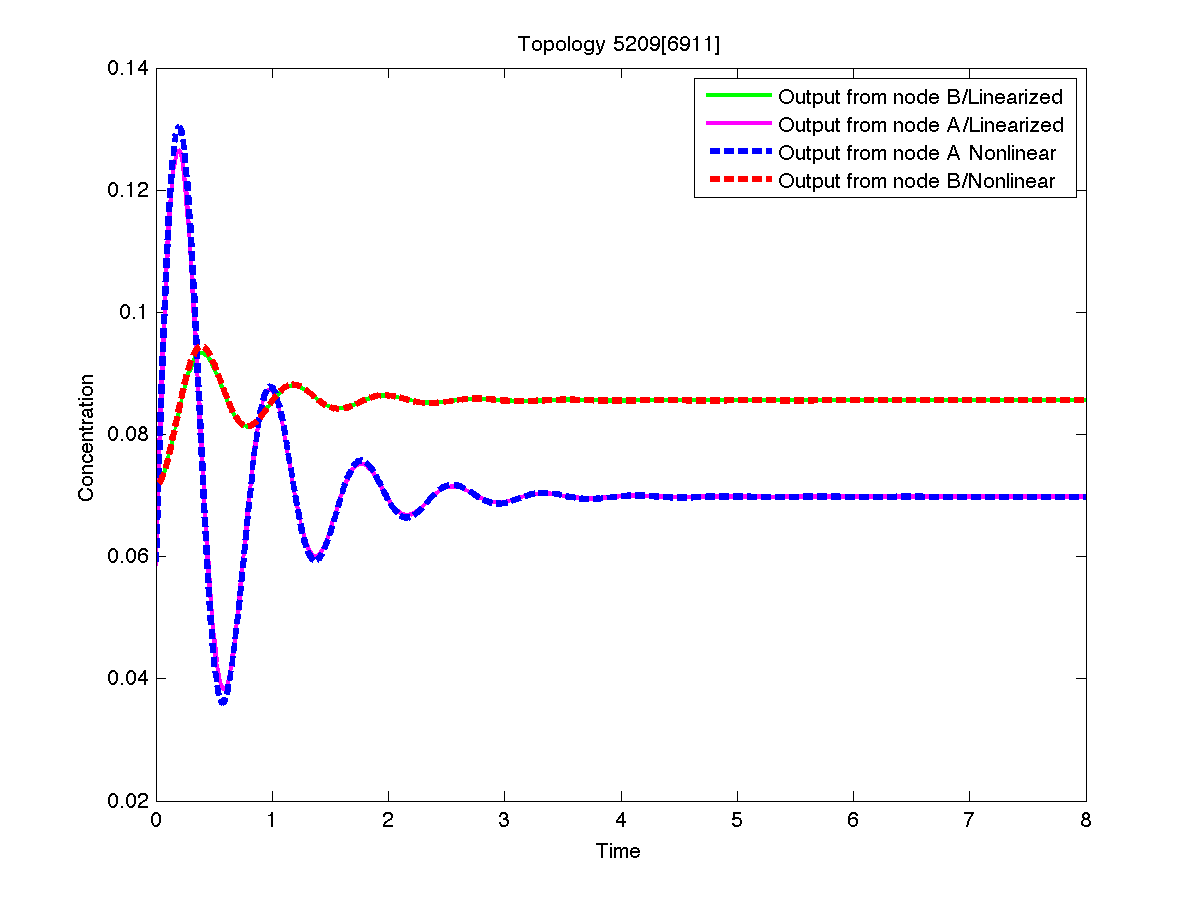}}                
  \subfloat[Ouput from C  nonlinear model]{\label{fig:f222}\includegraphics[width=0.55\textwidth]{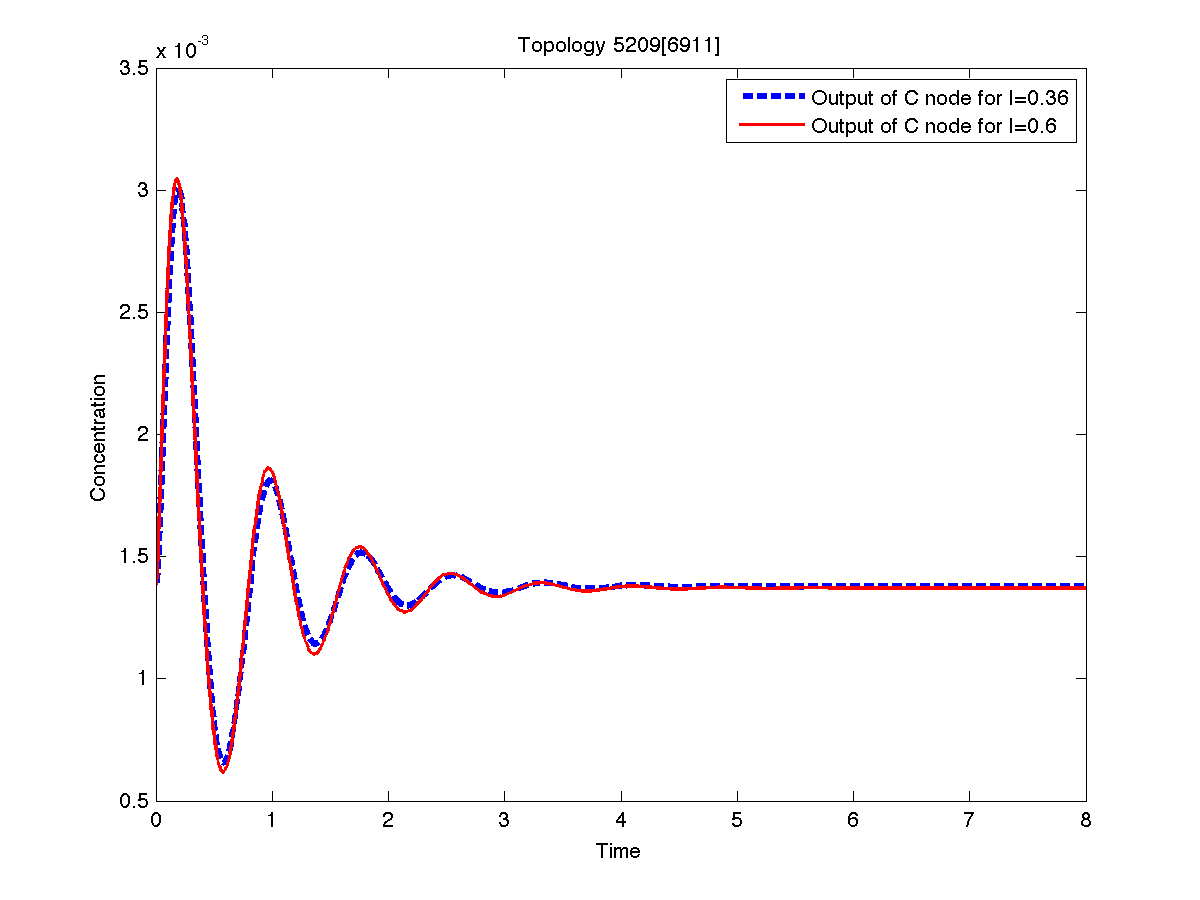}}\\
\subfloat[Quadratic approx. and output of nonlinear system]{\label{fig:f241}\includegraphics[width=0.55\textwidth]{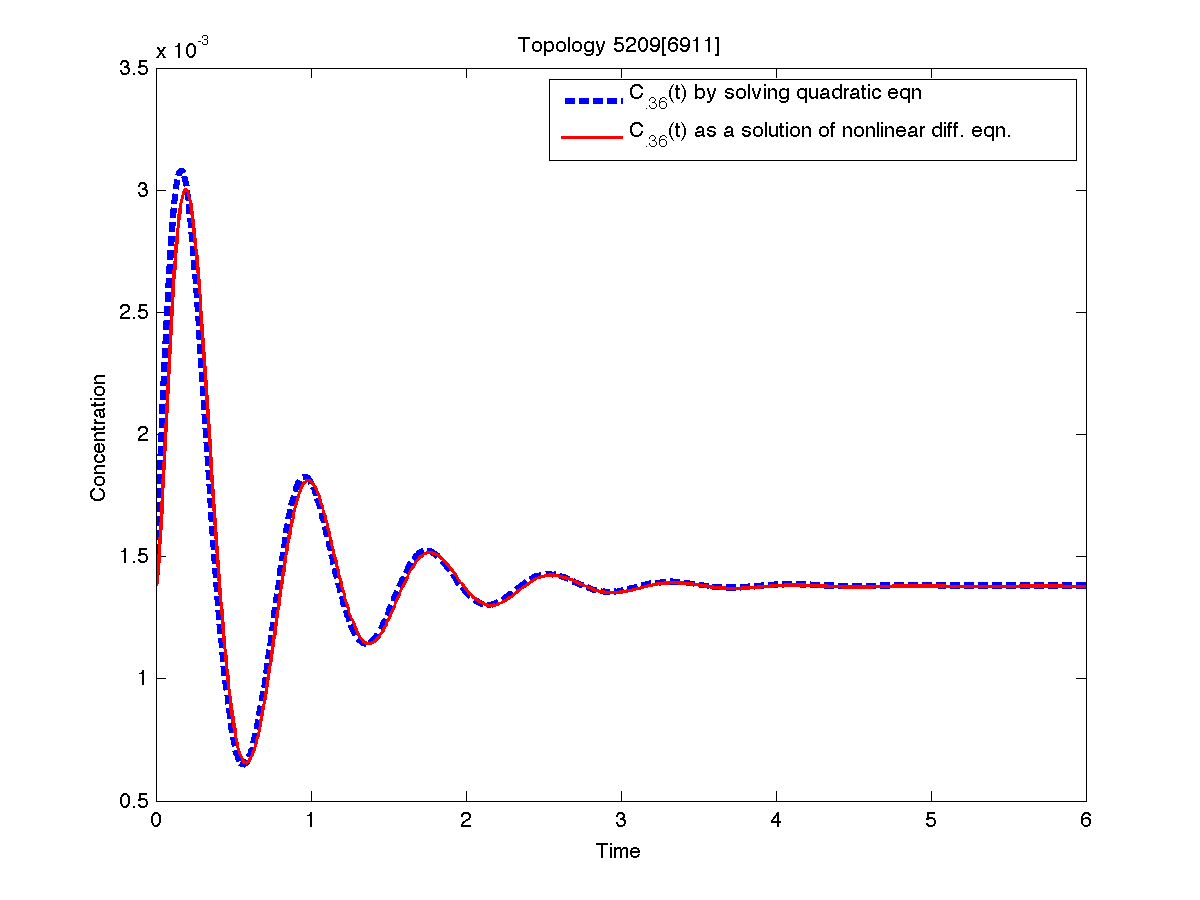}}                
  \subfloat[Quadratic approx. and output of nonlinear system]{\label{fig:f242}\includegraphics[width=0.55\textwidth]{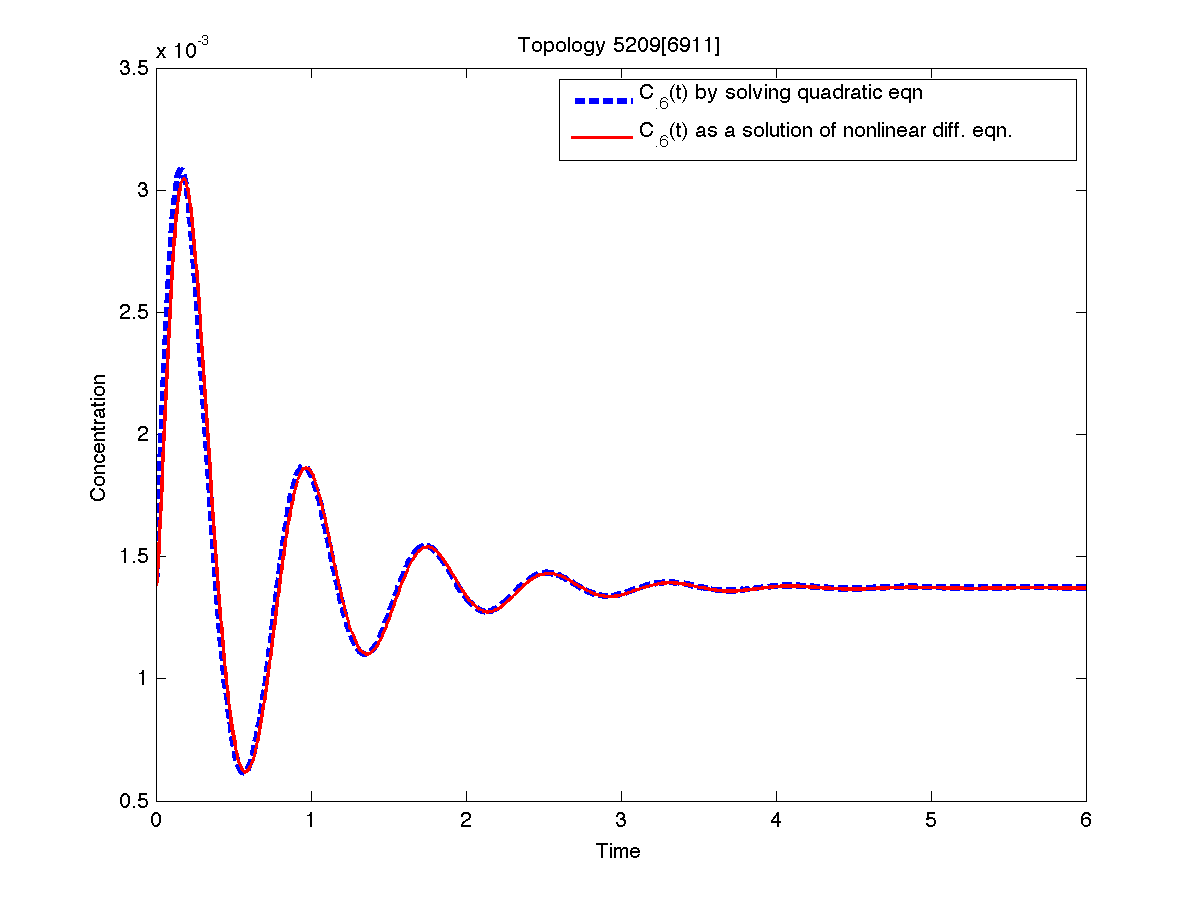}}
\end{figure}

\clearpage

Circuit 9.
\beqn
\dxA&=&k_{{\inp}A} {\inp} \frac{\txA}{\txA+K_{{\inp}A}}-k_{BA} \xB\frac{\xA}{\xA+K_{BA}}-k_{CA} \xC\frac{\xA}{\xA+K_{CA}}\\
\dxB&=&k_{AB}\xA\frac{\txB}{\txB+K_{AB}}+k_{CB} \xC \frac{\txB}{\txB+K_{CB}}-k_{F_BB} \xFB \frac{\xB}{\xB+K_{F_BB}}\\
\dxC&=&{k_{AC}}\xA\frac{\txC}{\txC+K_{AC}}- k_{BC}\xB\frac{\xC}{\xC+K_{BC}}-k_{CC}\xC\frac{\xC}{\xC+K_{CC}}\\
\eeqn
Parameters: \  $K_{{\inp}A}= 0.093918;$ $k_{{\inp}A}= 11.447219;$ 
$K_{BA}= 0.001688;$ $k_{BA}= 44.802268;$
$K_{CA}= 90.209027;$ $k_{CA}= 96.671843;$
$K_{AB}=0.001191;$ $k_{AB}=1.466561;$
$ K_{F_B}=9.424319;$ $k_{F_B}=22.745736;$
$K_{Ac}= 0.113697;$ $k_{AC}=1.211993;$
$K_{BC}=0.009891;$ $k_{BC}=7.239357;$
$K_{CB}=30.602013;$ $k_{CB}= 3.811536;$
$K_{CC}=0.189125;$ $k_{CC}= 17.910182$

\begin{figure}[hb]
  \centering
\subfloat[Dynamics of A and B in linearized model]{\label{fig:f251}\includegraphics[width=0.55\textwidth]{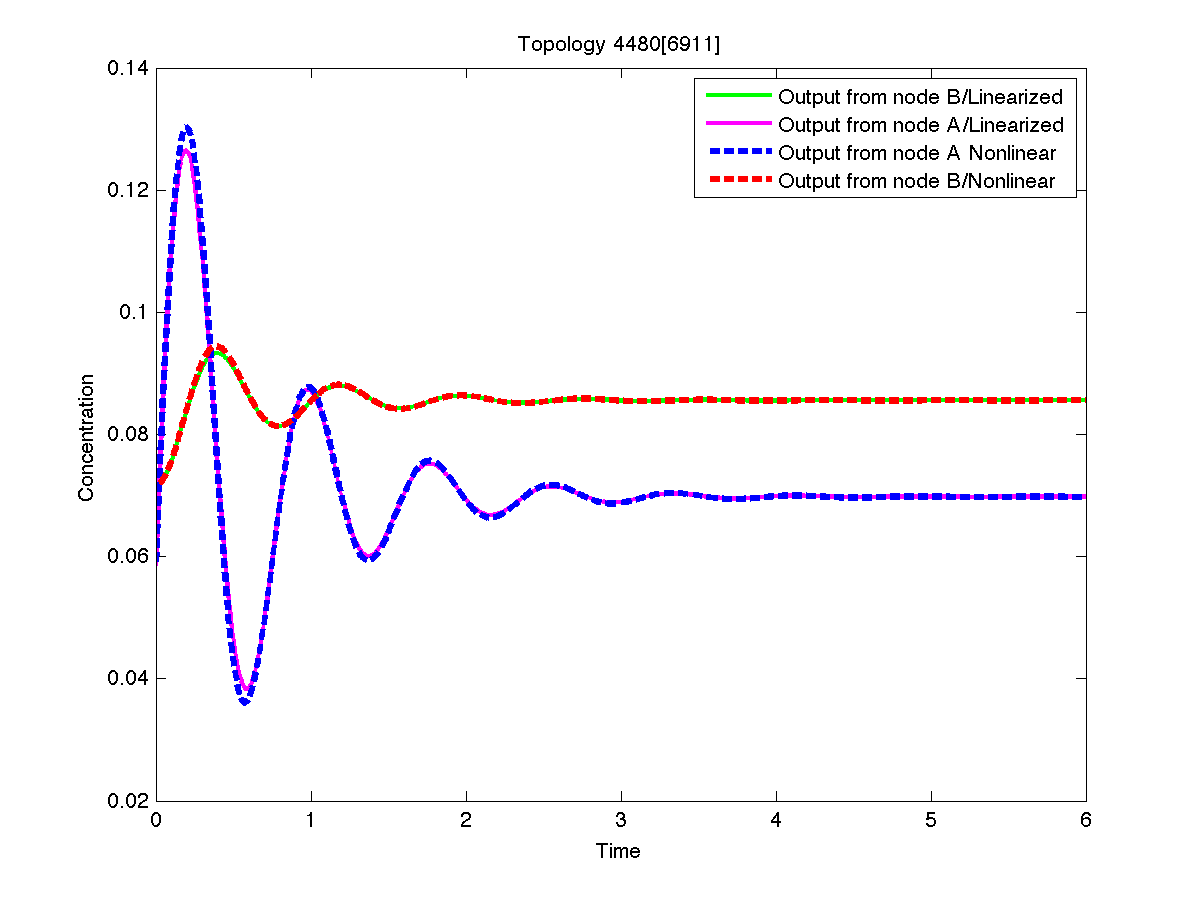}}                
  \subfloat[Ouput from C  nonlinear model]{\label{fig:f252}\includegraphics[width=0.55\textwidth]{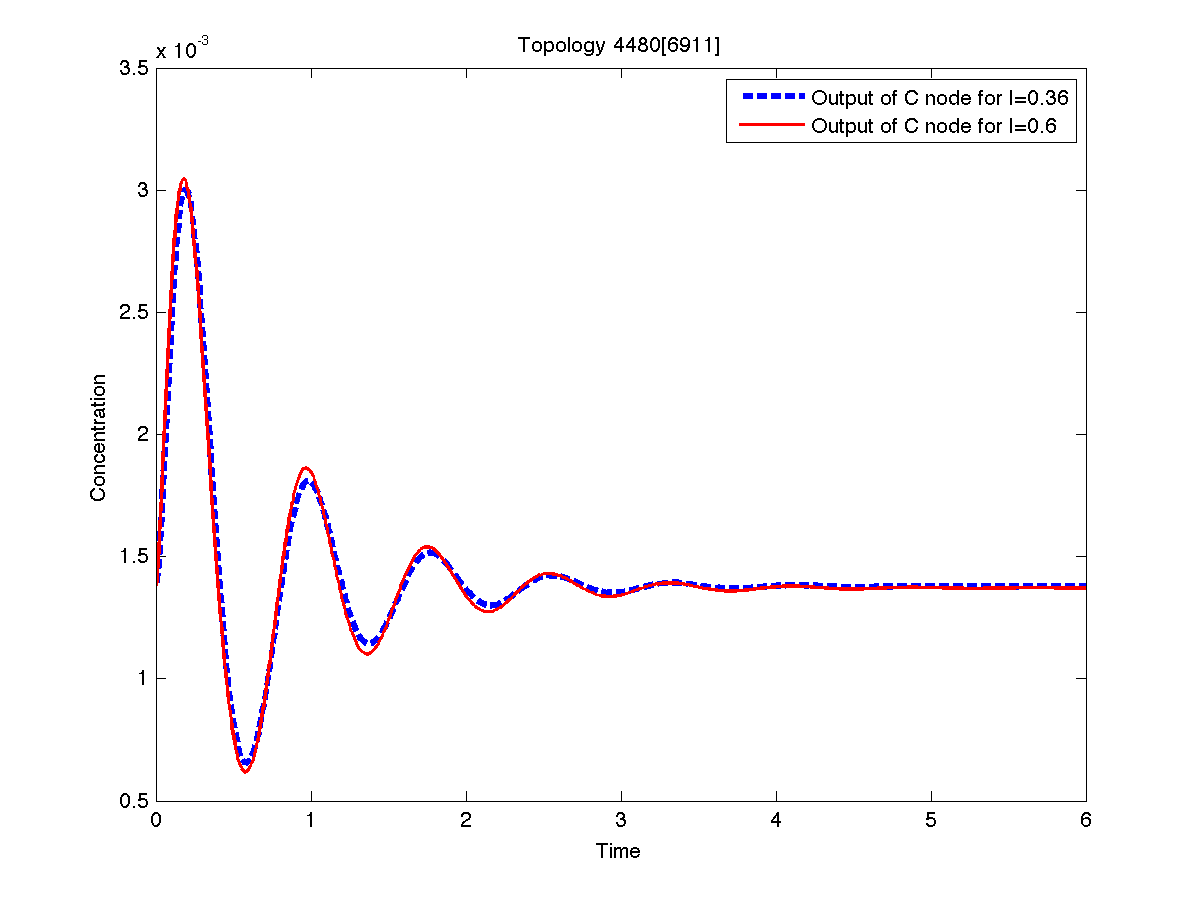}}\\
\subfloat[Quadratic approx. and output of nonlinear system]{\label{fig:f271}\includegraphics[width=0.55\textwidth]{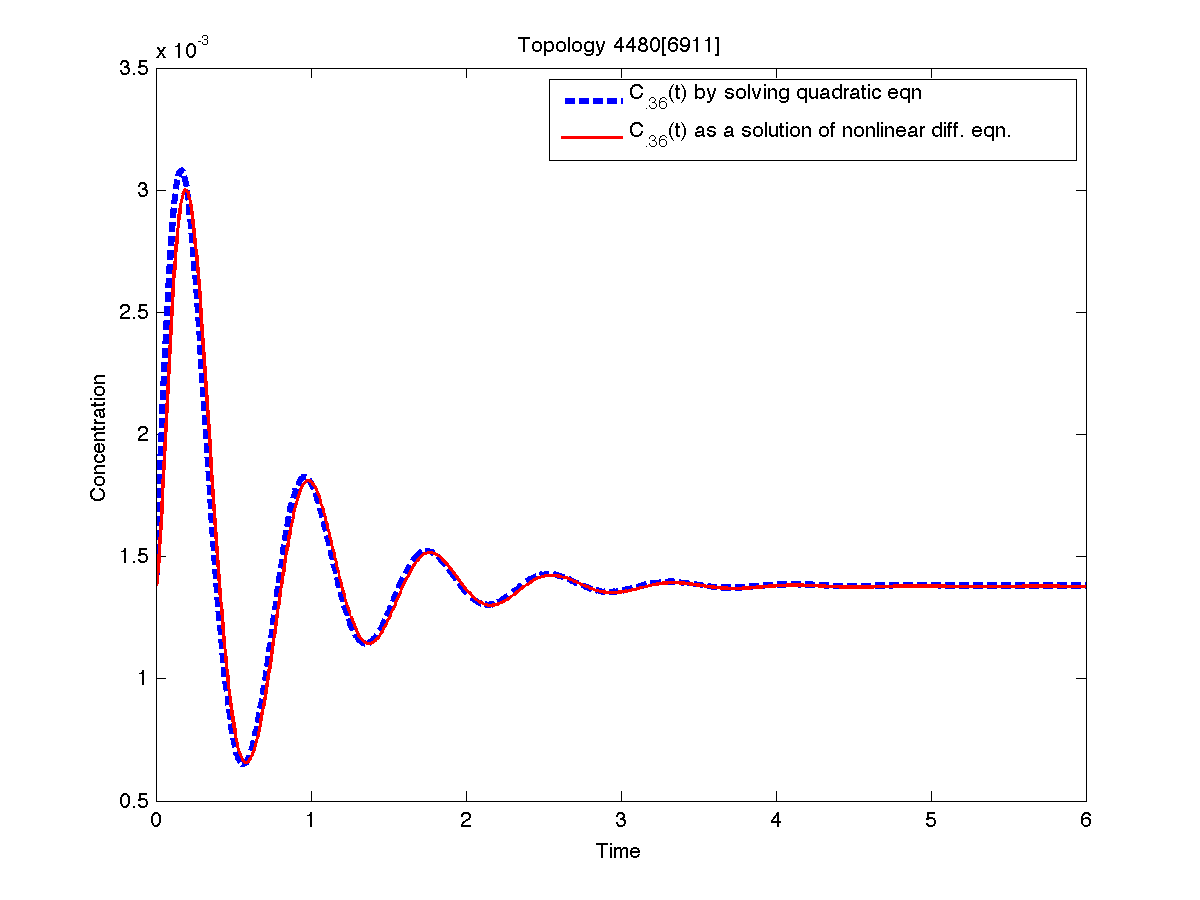}}                
  \subfloat[Quadratic approx. and output of nonlinear system]{\label{fig:f272}\includegraphics[width=0.55\textwidth]{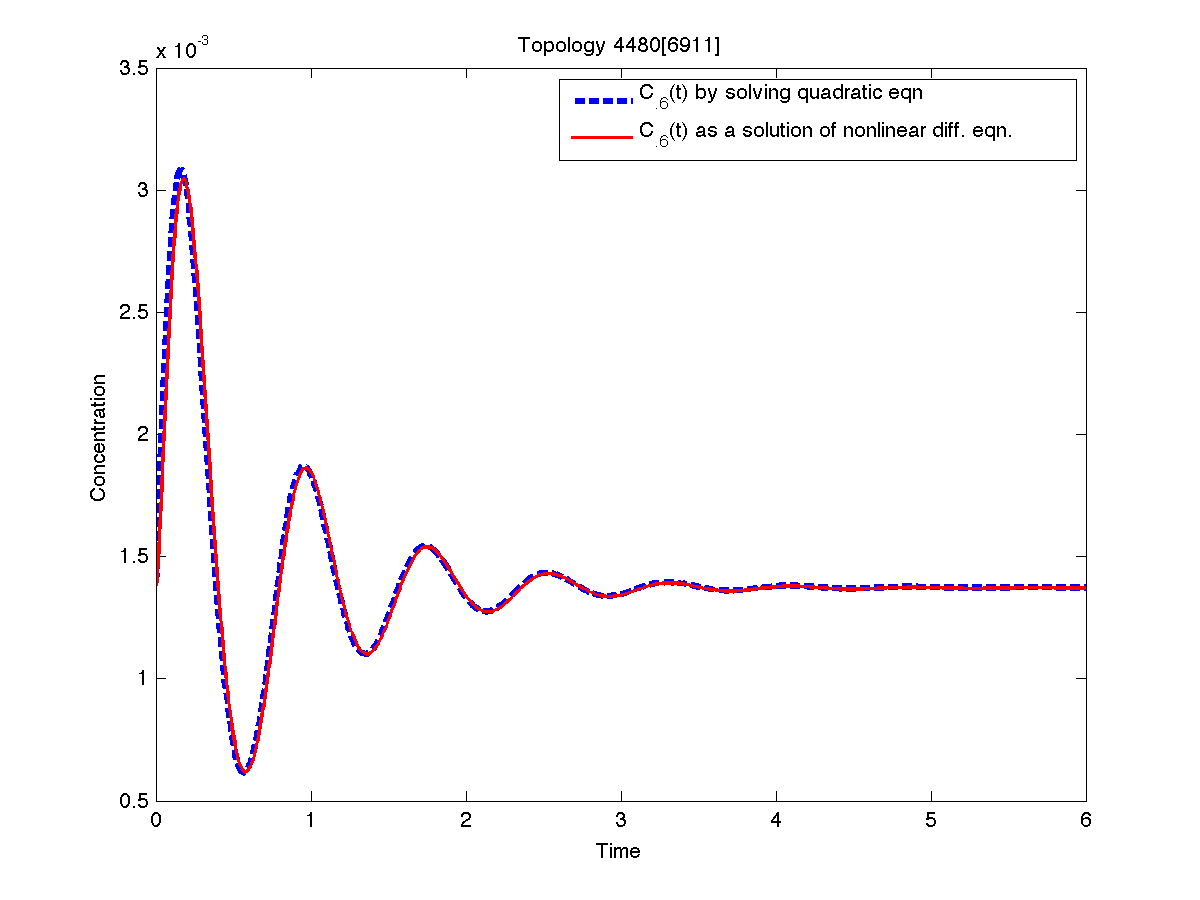}}
\end{figure}

Circuit 10.
\beqn
\dxA&=&k_{{\inp}A} {\inp} \frac{\txA}{\txA+K_{{\inp}A}}-k_{BA} \xB\frac{\xA}{\xA+K_{BA}}-k_{AA} \xA\frac{\xA}{\xA+K_{AA}}\\
\dxB&=&k_{AB}\xA\frac{\txB}{\txB+K_{AB}}+k_{CB} \xC \frac{\txB}{\txB+K_{CB}}-k_{F_BB} \xFB \frac{\xB}{\xB+K_{F_BB}}\\
\dxC&=&{k_{BC}}\xB\frac{\txC}{\txC+K_{BC}}- k_{AC}\xA\frac{\xC}{\xC+K_{AC}}-k_{CC}\xC\frac{\xC}{\xC+K_{CC}}\\
\eeqn
Parameters: \  $K_{AA}= 24.989065;$ $k_{AA}= 53.174082;$
 $K_{AB}= 0.444375;$ $k_{AB}= 12.053134;$
 $K_{F_B}= 1.716920;$ $k_{F_B}= 11.601122;$
 $K_{AC}= 0.013988;$ $k_{AC}= 8.521185;$
 $K_{BA}= 0.005461;$ $k_{BA}= 7.103952;$
 $K_{BC}=51.850148;$ $k_{BC}= 80.408137;$
 $K_{CB}=5.392001;$ $k_{CB}= 3.086740;$
$K_{CC}= 1.962230;$ $k_{CC}= 17.382010;$
$K_{{\inp}A}= 4.387832;$ $k_{{\inp}A}= 19.638124$ 

\begin{figure}[ht]
  \centering
\subfloat[Dynamics of A and B in linearized model]{\label{fig:f281}\includegraphics[width=0.55\textwidth]{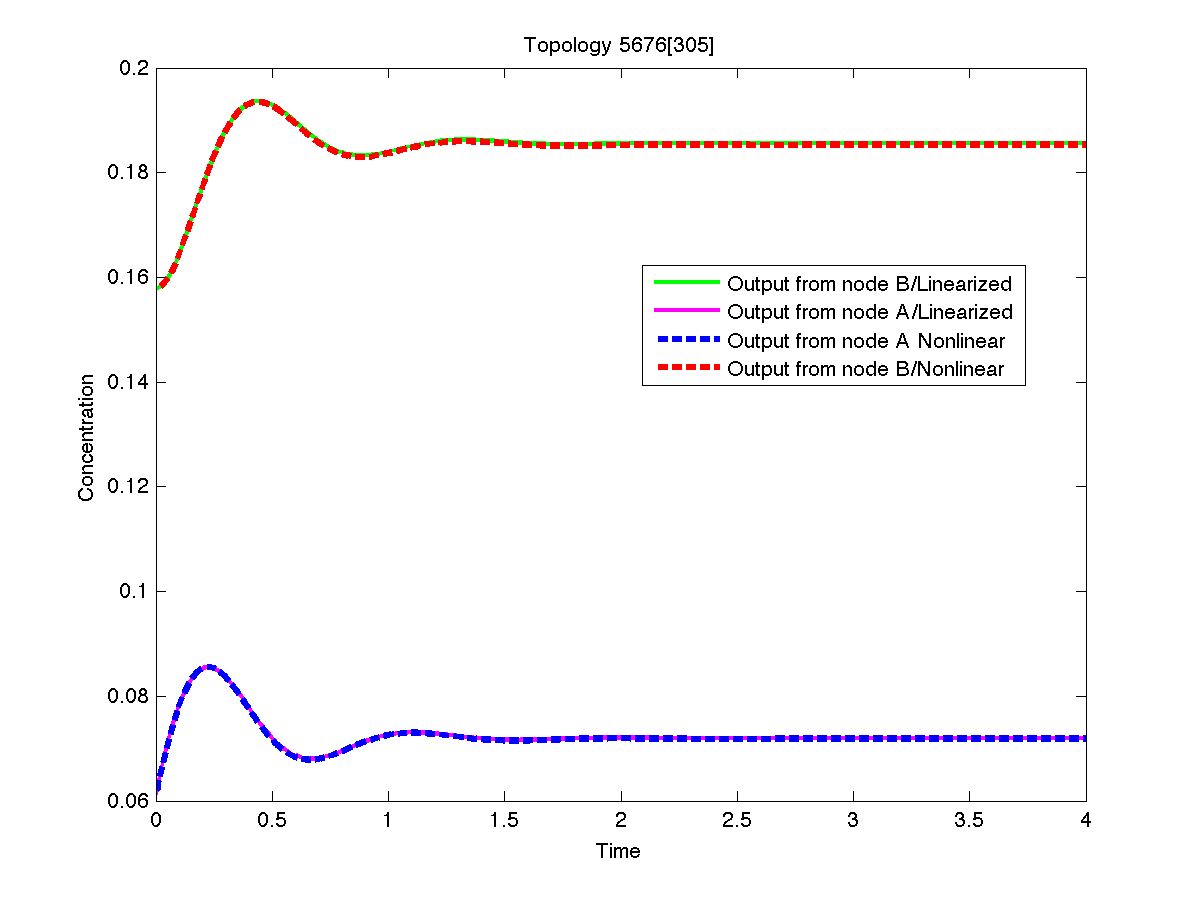}}                
  \subfloat[Ouput from C  nonlinear model]{\label{fig:f282}\includegraphics[width=0.55\textwidth]{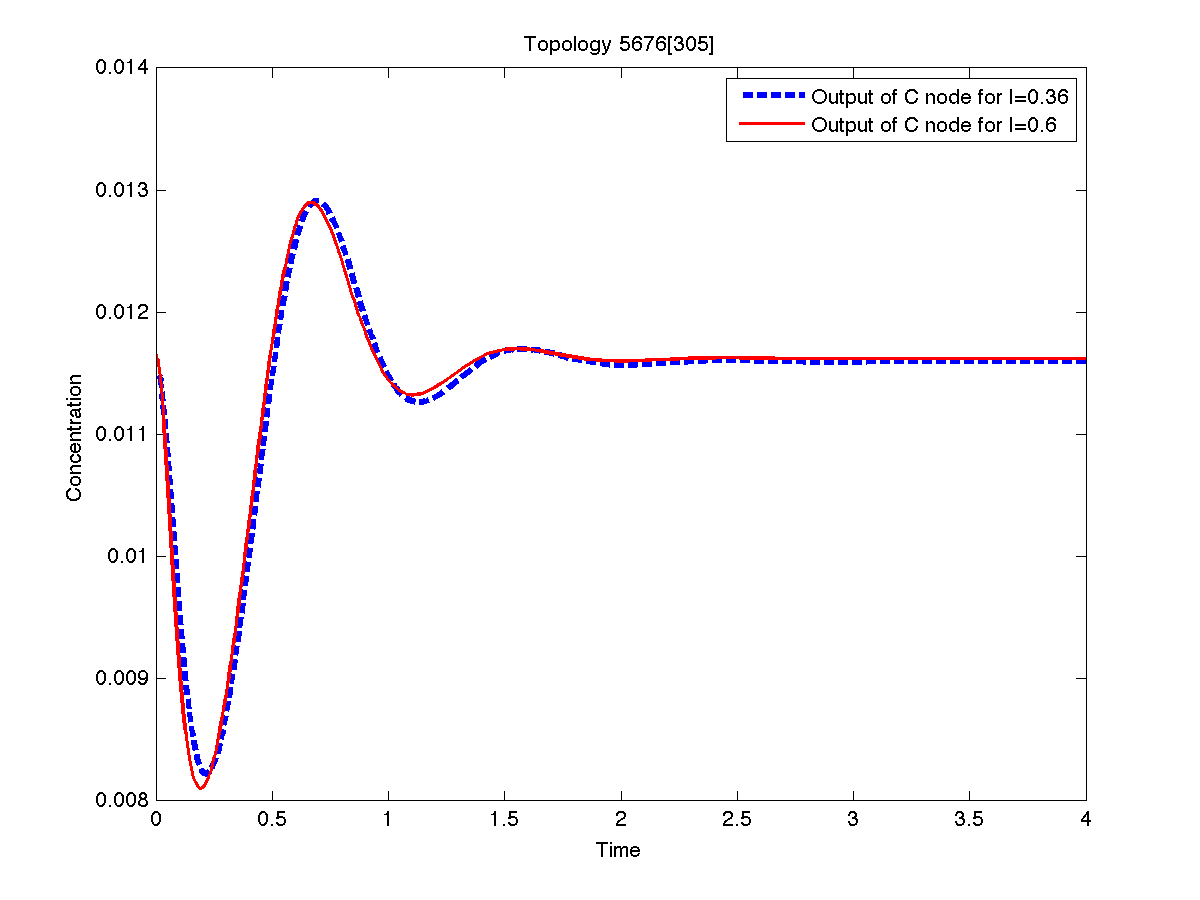}}\\
\subfloat[Quadratic approx. and output of nonlinear system]{\label{fig:f301}\includegraphics[width=0.55\textwidth]{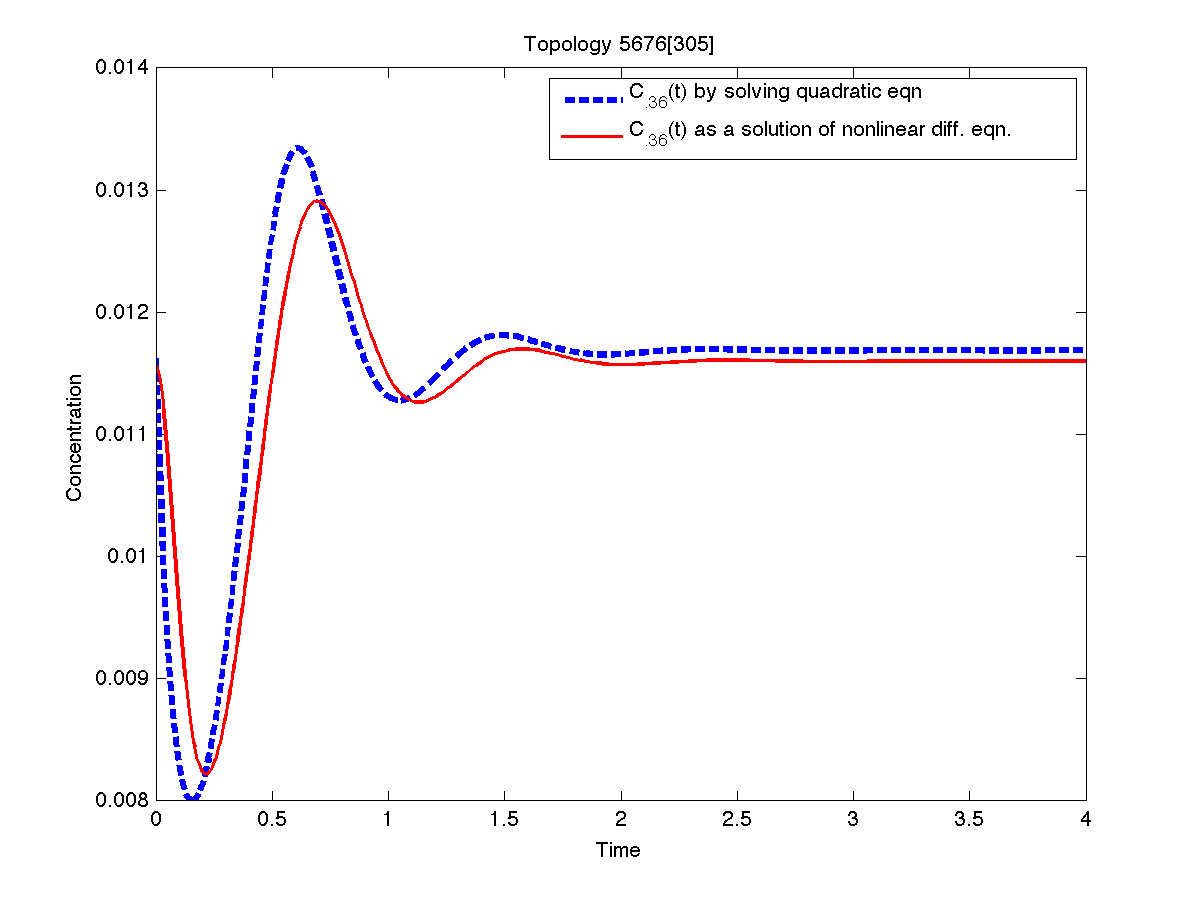}}                
  \subfloat[Quadratic approx. and output of nonlinear system]{\label{fig:f302}\includegraphics[width=0.55\textwidth]{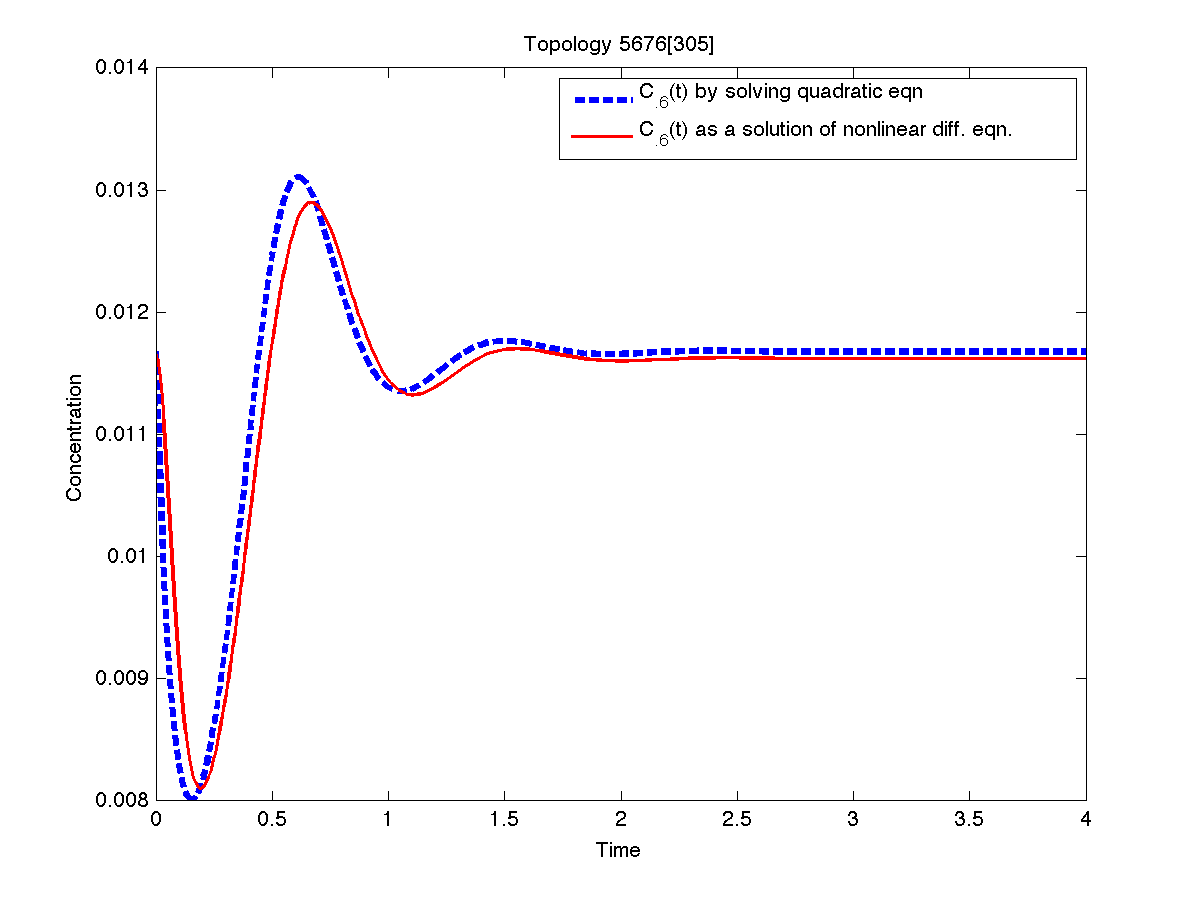}}
\end{figure}

\clearpage

Circuit 11.
\beqn
\dxA&=&k_{{\inp}A} {\inp} \frac{\txA}{\txA+K_{{\inp}A}}-k_{BA} \xB\frac{\xA}{\xA+K_{BA}}\\
\dxB&=&k_{AB}\xA\frac{\txB}{\txB+K_{AB}}+k_{CB} \xC \frac{\txB}{\txB+K_{CB}}-k_{F_BB} \xFB \frac{\xB}{\xB+K_{F_BB}}\\
\dxC&=&{k_{BC}}\xB\frac{\txC}{\txC+K_{BC}}- k_{AC}\xA\frac{\xC}{\xC+K_{AC}}-k_{CC}\xC\frac{\xC}{\xC+K_{CC}}\\
\eeqn
Parameters: \   $K_{AB}= 0.444375;$ $k_{AB}= 12.053134;$
 $K_{F_B}= 1.716920;$ $k_{F_B}= 11.601122;$
 $K_{AC}= 0.013988;$ $k_{AC}= 8.521185;$
 $K_{BA}= 0.005461;$ $k_{BA}= 7.103952;$
 $K_{BC}=51.850148;$ $k_{BC}= 80.408137;$
 $K_{CB}=5.392001;$ $k_{CB}= 3.086740;$
$K_{CC}= 1.962230;$ $k_{CC}= 17.382010;$
$K_{{\inp}A}= 4.387832;$ $k_{{\inp}A}= 19.638124$

\begin{figure}[hb]
  \centering
\subfloat[Dynamics of A and B in linearized model]{\label{fig:f311}\includegraphics[width=0.55\textwidth]{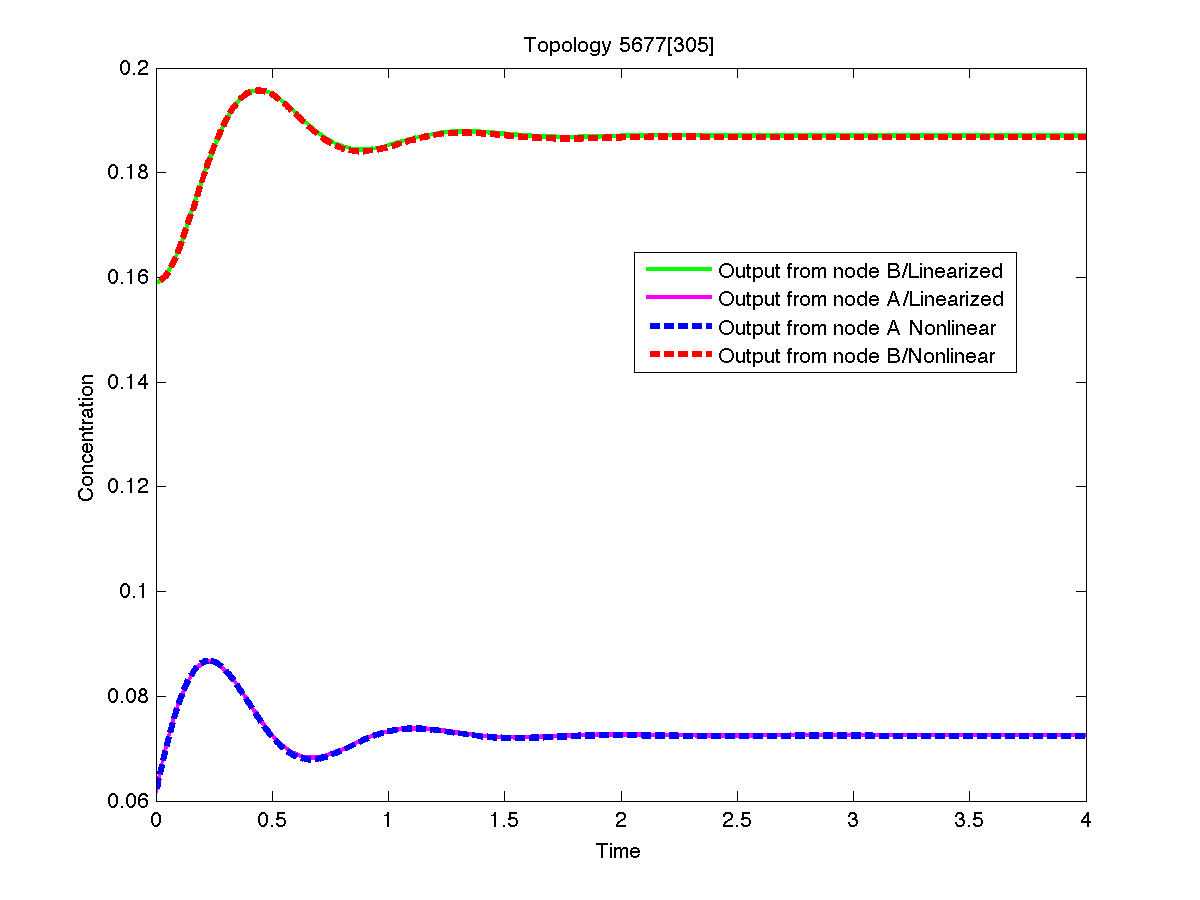}}                
  \subfloat[Ouput from C  nonlinear model]{\label{fig:f312}\includegraphics[width=0.55\textwidth]{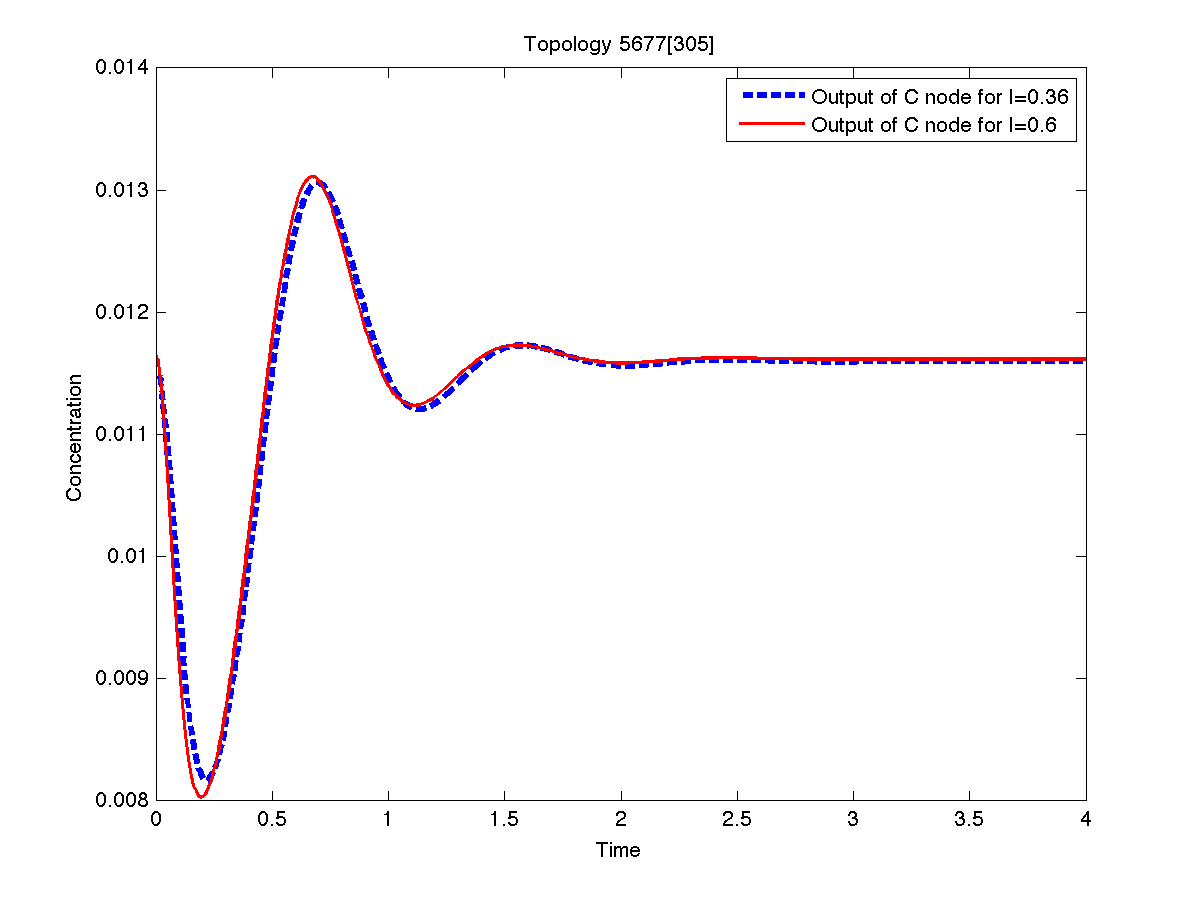}}\\
\subfloat[Quadratic approx. and output of nonlinear system]{\label{fig:331}\includegraphics[width=0.55\textwidth]{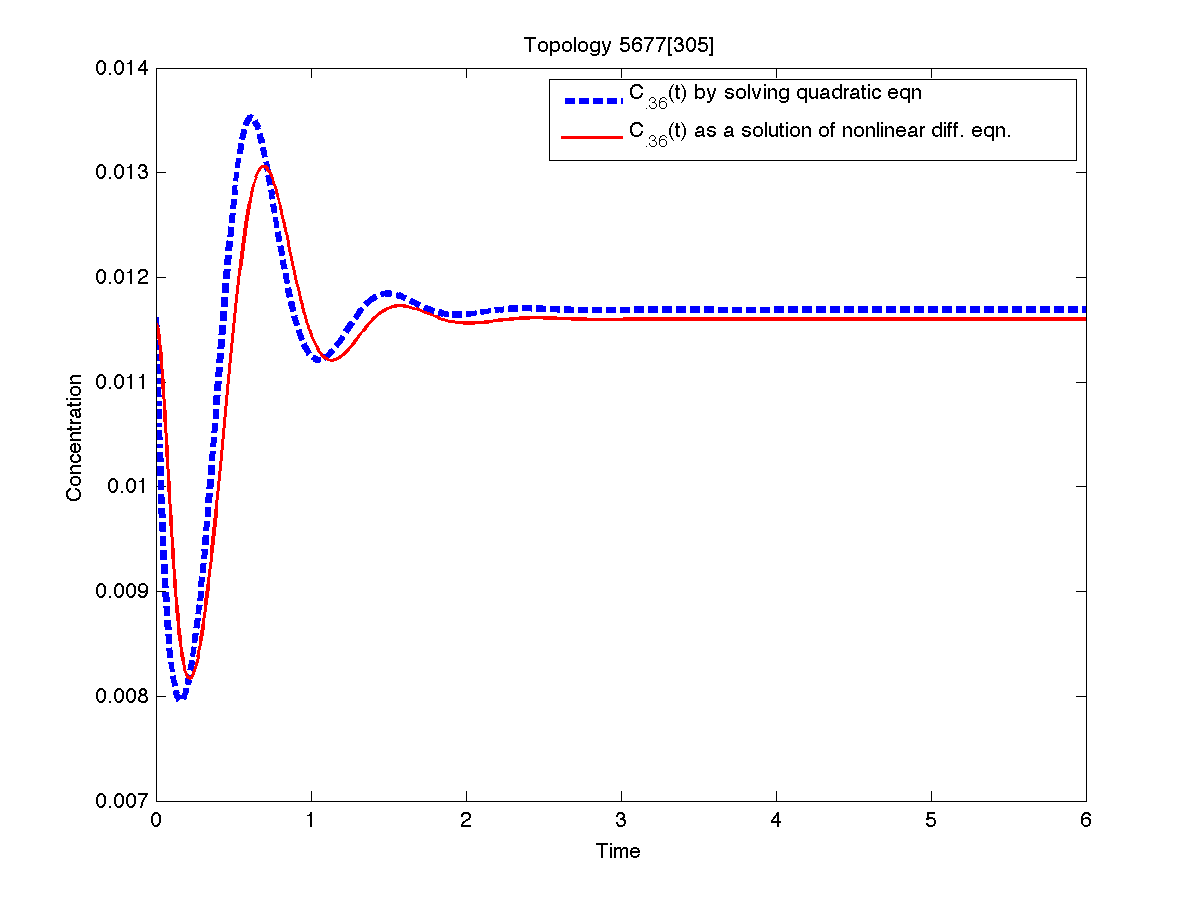}}                
  \subfloat[Quadratic approx. and output of nonlinear system]{\label{fig:f332}\includegraphics[width=0.55\textwidth]{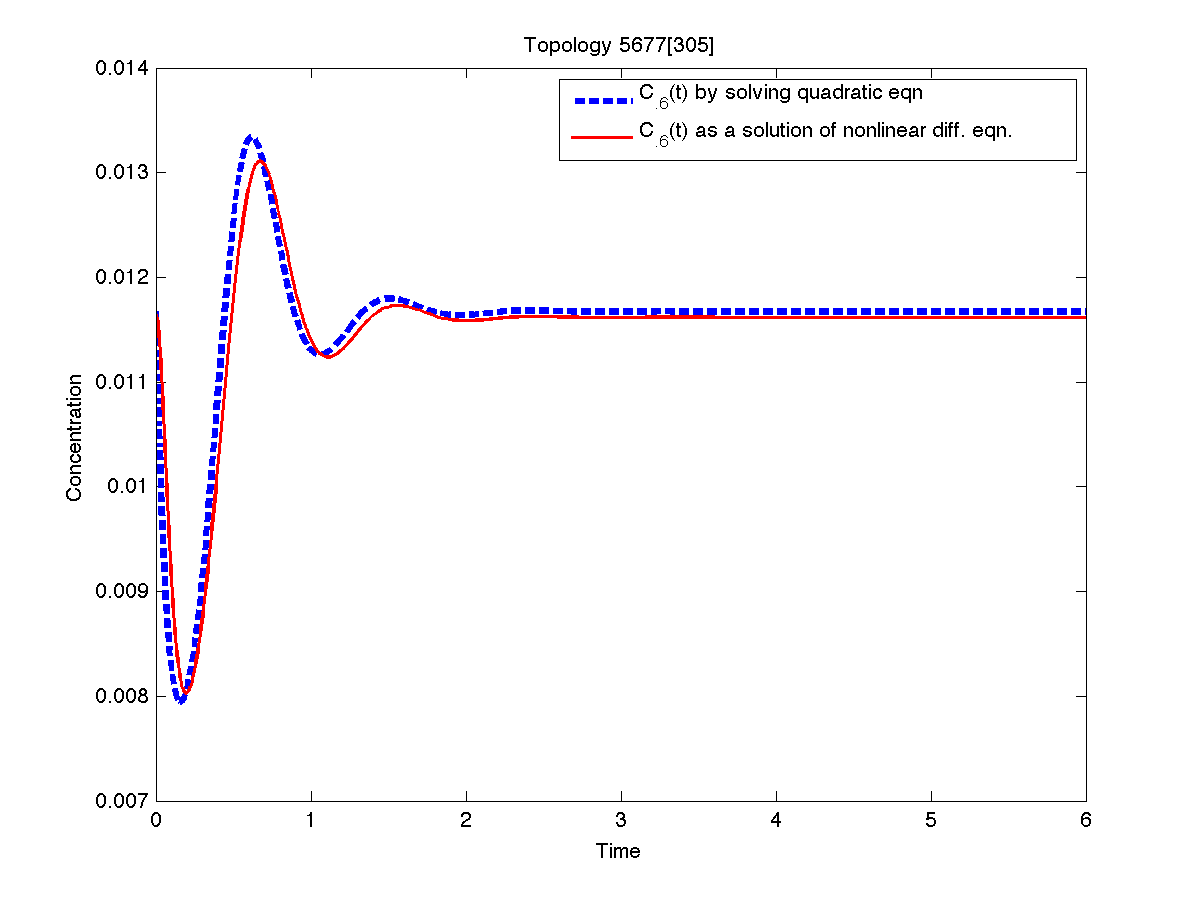}}
\end{figure}

Circuit 12.
\beqn
\dxA&=&k_{{\inp}A} {\inp} \frac{\txA}{\txA+K_{{\inp}A}}-k_{BA} \xB\frac{\xA}{\xA+K_{BA}}+k_{CA} \xC\frac{\txA}{\txA+K_{CA}}\\
\dxB&=&k_{AB}\xA\frac{\txB}{\txB+K_{AB}}+k_{CB} C \frac{\txB}{\txB+K_{CB}}-k_{F_BB} \xFB \frac{\xB}{\xB+K_{F_BB}}\\
\dxC&=&{k_{AC}}\xA\frac{\txC}{\txC+K_{AC}}- k_{BC}\xB\frac{\xC}{\xC+K_{BC}}-k_{CC}\xC\frac{\xC}{\xC+K_{CC}}\\
\eeqn
Parameters: \ $K_{{\inp}A}= 0.093918;$ $k_{{\inp}A}= 11.447219;$ 
$K_{BA}= 0.001688;$ $k_{BA}= 44.802268;$
$K_{CA}= 5.026318;$ $k_{CA}= 45.803641;$
$K_{AB}=0.001191;$ $k_{AB}=1.466561;$
 $K_{F_B}=9.424319;$ $k_{F_B}=22.745736;$
$K_{AC}= 0.113697;$ $k_{AC}=1.211993;$
$K_{BC}=0.009891;$ $k_{BC}=7.239357;$
$K_{CB}=30.602013;$ $k_{CB}= 3.811536;$
$K_{CC}=0.189125;$ $k_{CC}= 17.910182$

\begin{figure}[ht]
  \centering
\subfloat[Dynamics of A and B in linearized model]{\label{fig:f341}\includegraphics[width=0.55\textwidth]{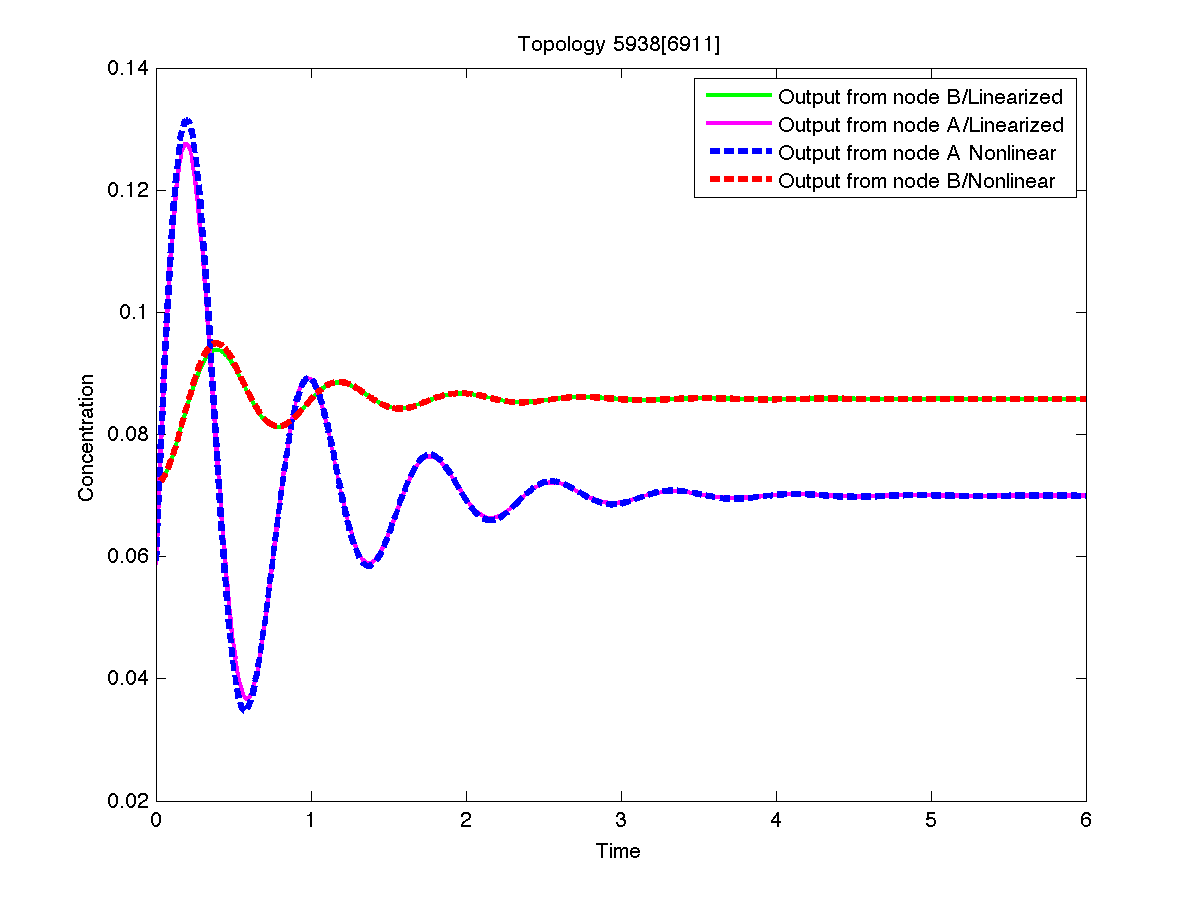}}                
  \subfloat[Ouput from C  nonlinear model]{\label{fig:f342}\includegraphics[width=0.55\textwidth]{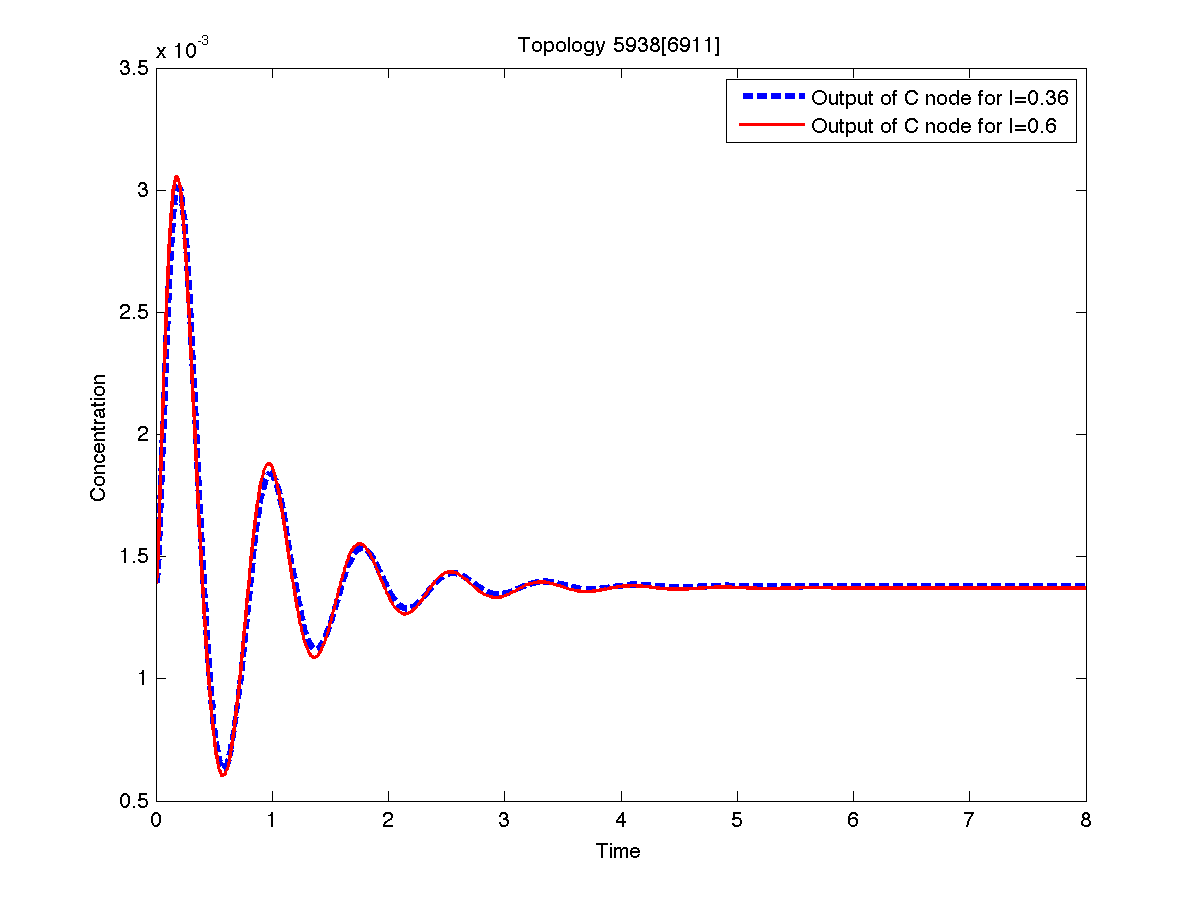}}\\
\subfloat[Quadratic approx. and output of nonlinear system]{\label{fig:f361}\includegraphics[width=0.55\textwidth]{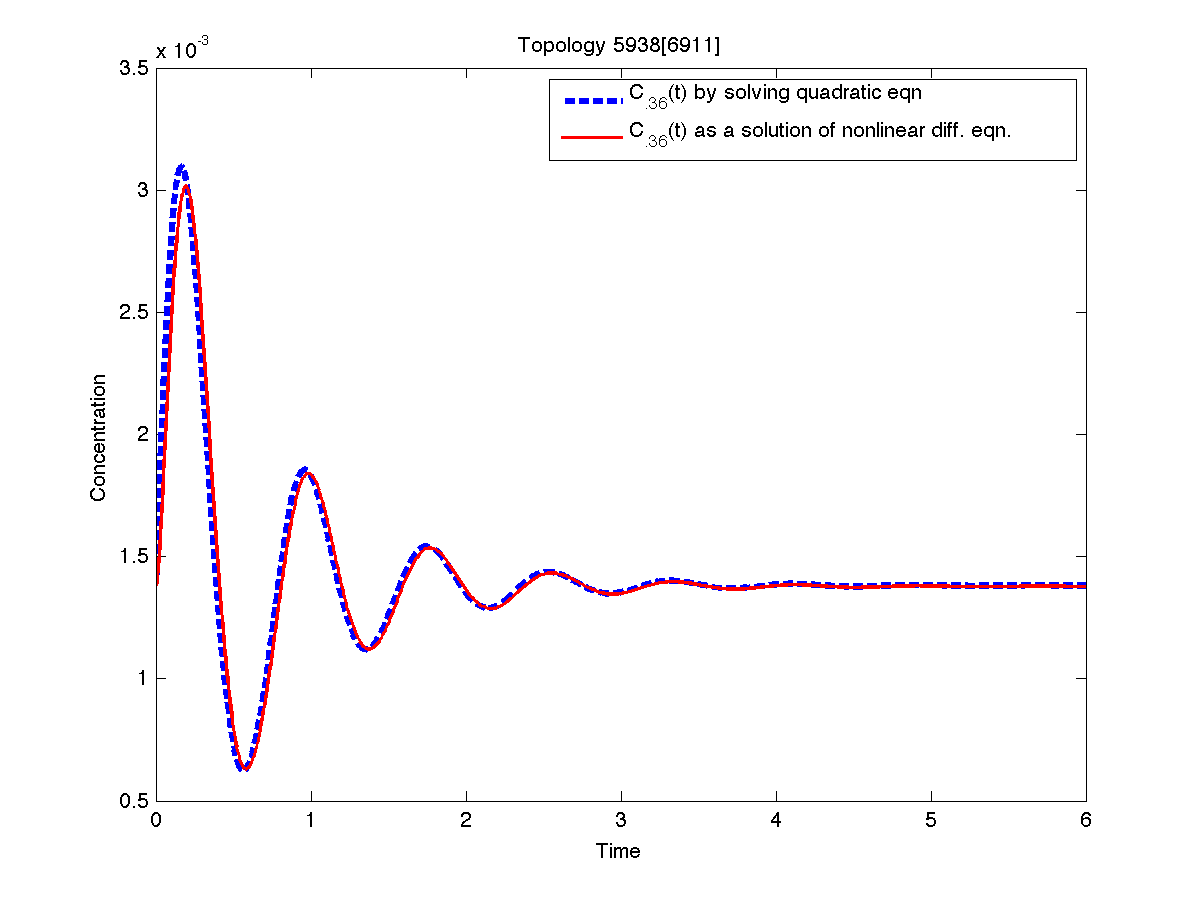}}                
  \subfloat[Quadratic approx. and output of nonlinear system]{\label{fig:f362}\includegraphics[width=0.55\textwidth]{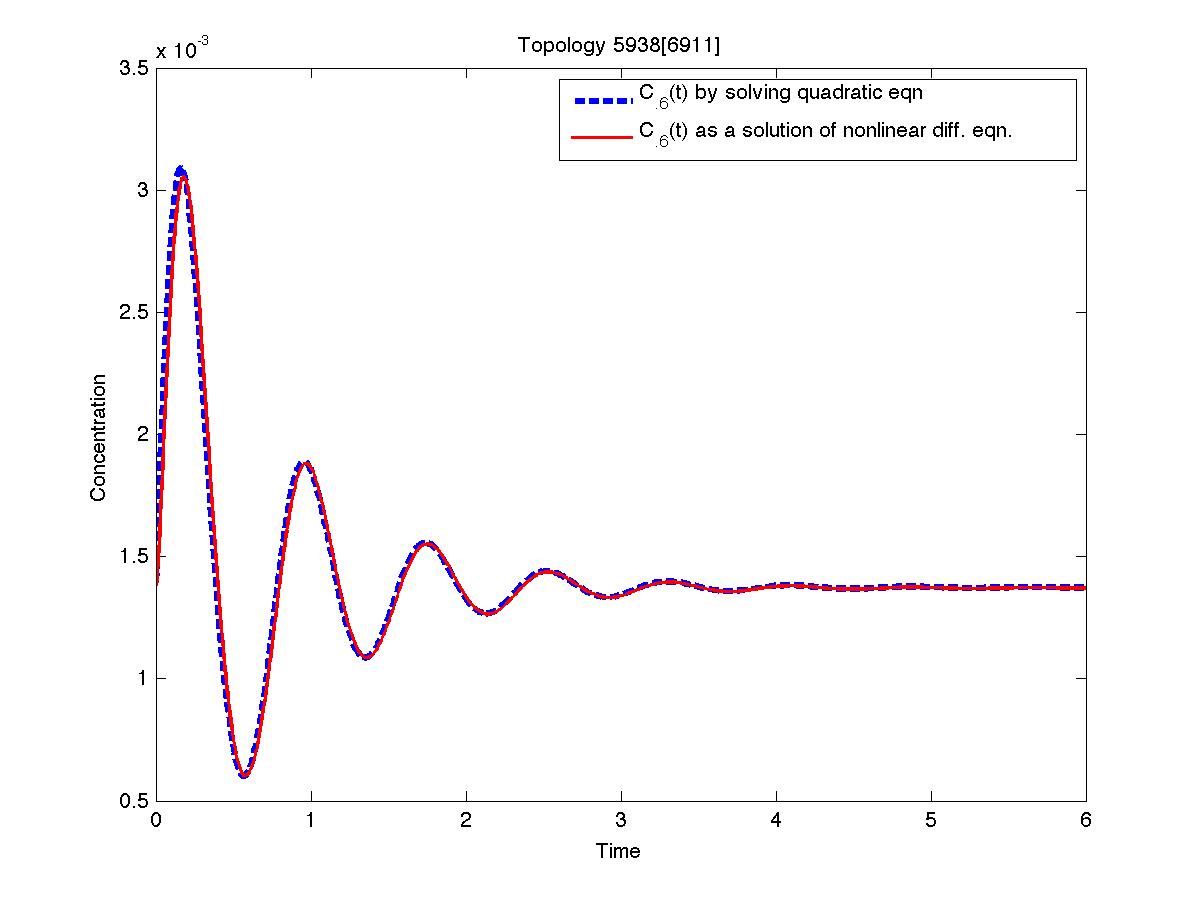}}
\end{figure}

\clearpage
Circuit 13. 
\beqn
\dxA&=&k_{{\inp}A} {\inp} \frac{\txA}{\txA+K_{{\inp}A}}-k_{BA} \xB\frac{\xA}{\xA+K_{BA}}-k_{AA} \xA\frac{\xA}{\xA+K_{AA}}+k_{CA} \xC\frac{\txA}{\txA+K_{CA}}\\
\dxB&=&k_{AB}\xA\frac{\txB}{\txB+K_{AB}}+k_{CB} \xC \frac{\txB}{\txB+K_{CB}}-k_{BB} \xB\frac{\xB}{\xB+K_{BB}}\\
\dxC&=&{k_{BC}}\xB\frac{\txC}{\txC+K_{BC}}- k_{AC}\xA\frac{\xC}{\xC+K_{AC}}-k_{CC}\xA\frac{\xC}{\xC+K_{CC}}\\
\eeqn
Parameters: \ $K_{AA}= 24.989065;$ $k_{AA}= 53.174082;$
 $K_{AB}= 0.444375;$ $k_{AB}= 12.053134;$
 $K_{F_B}= 1.716920;$ $k_{F_B}= 11.601122;$
 $K_{AC}= 0.013988;$ $k_{AC}= 8.521185;$
 $K_{BA}= 0.005461;$ $k_{BA}= 7.103952;$
 $K_{BC}=51.850148;$ $k_{BC}= 80.408137;$
 $K_{CB}=5.392001;$ $k_{CB}= 3.086740;$
$K_{CC}= 1.962230;$ $k_{CC}= 17.382010;$
$K_{{\inp}A}= 4.387832;$ $k_{{\inp}A}= 19.638124;$
$K_{CA}= 15.479253;$ $k_{CA}= 4.903430$

\begin{figure}[hb]
  \centering
\subfloat[Dynamics of A and B in linearized model]{\label{fig:f371}\includegraphics[width=0.55\textwidth]{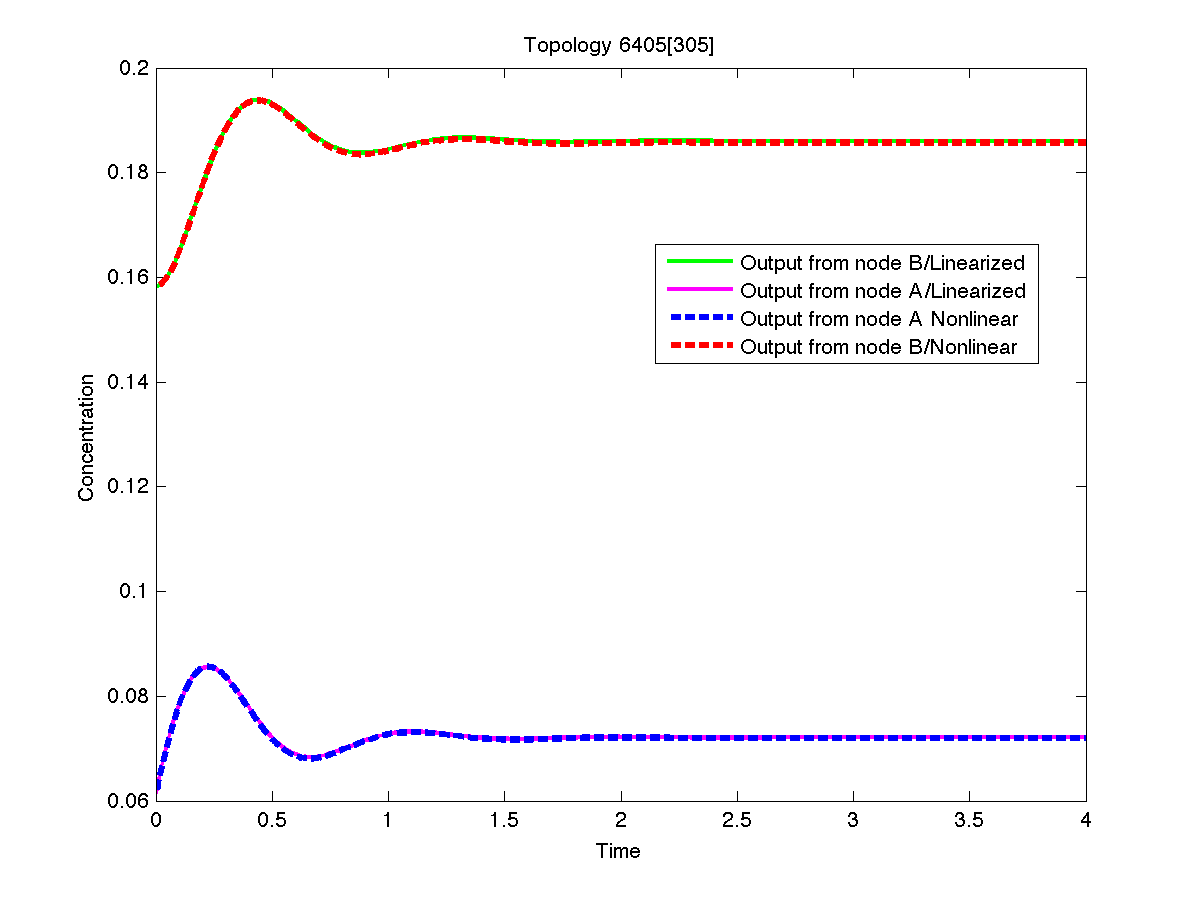}}                
  \subfloat[Ouput from C  nonlinear model]{\label{fig:f372}\includegraphics[width=0.55\textwidth]{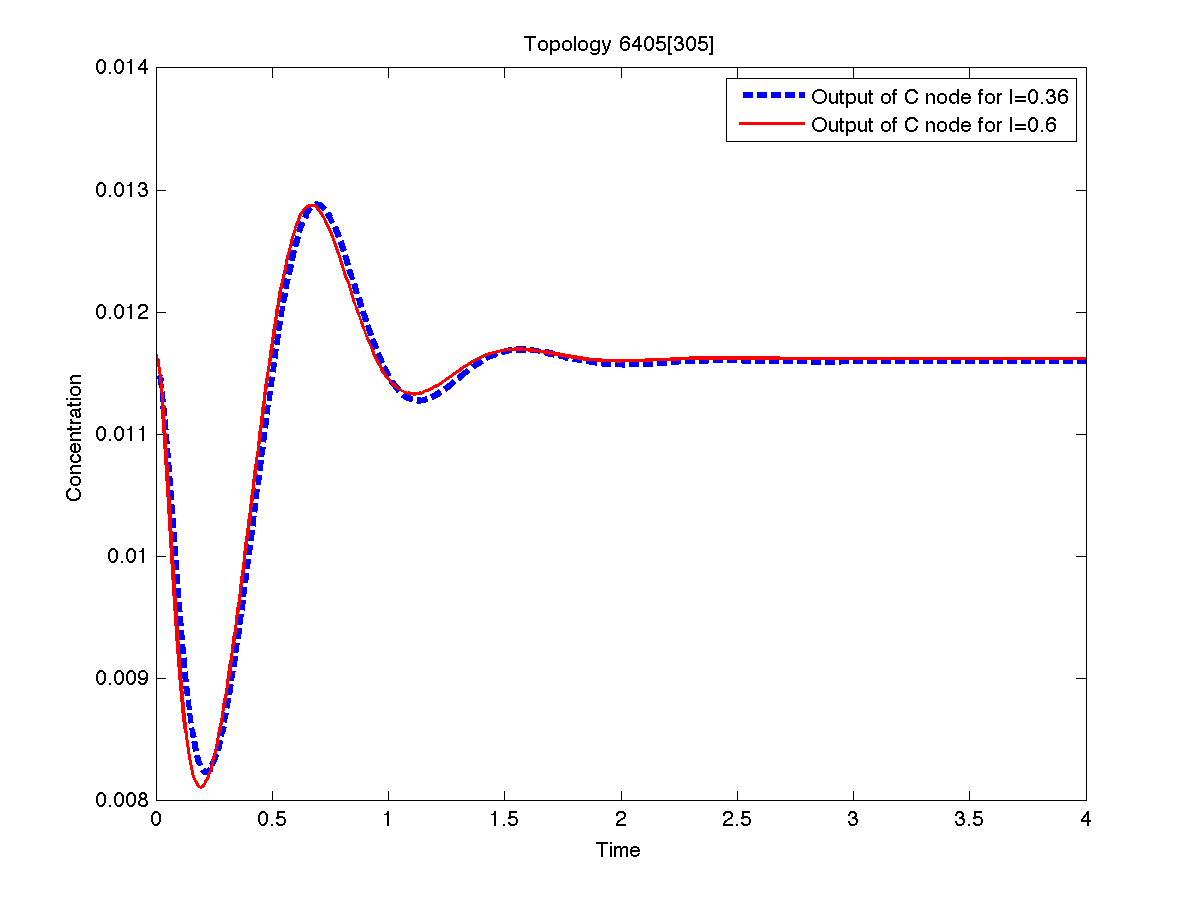}}\\
\subfloat[Quadratic approx. and output of nonlinear system]{\label{fig:f391}\includegraphics[width=0.55\textwidth]{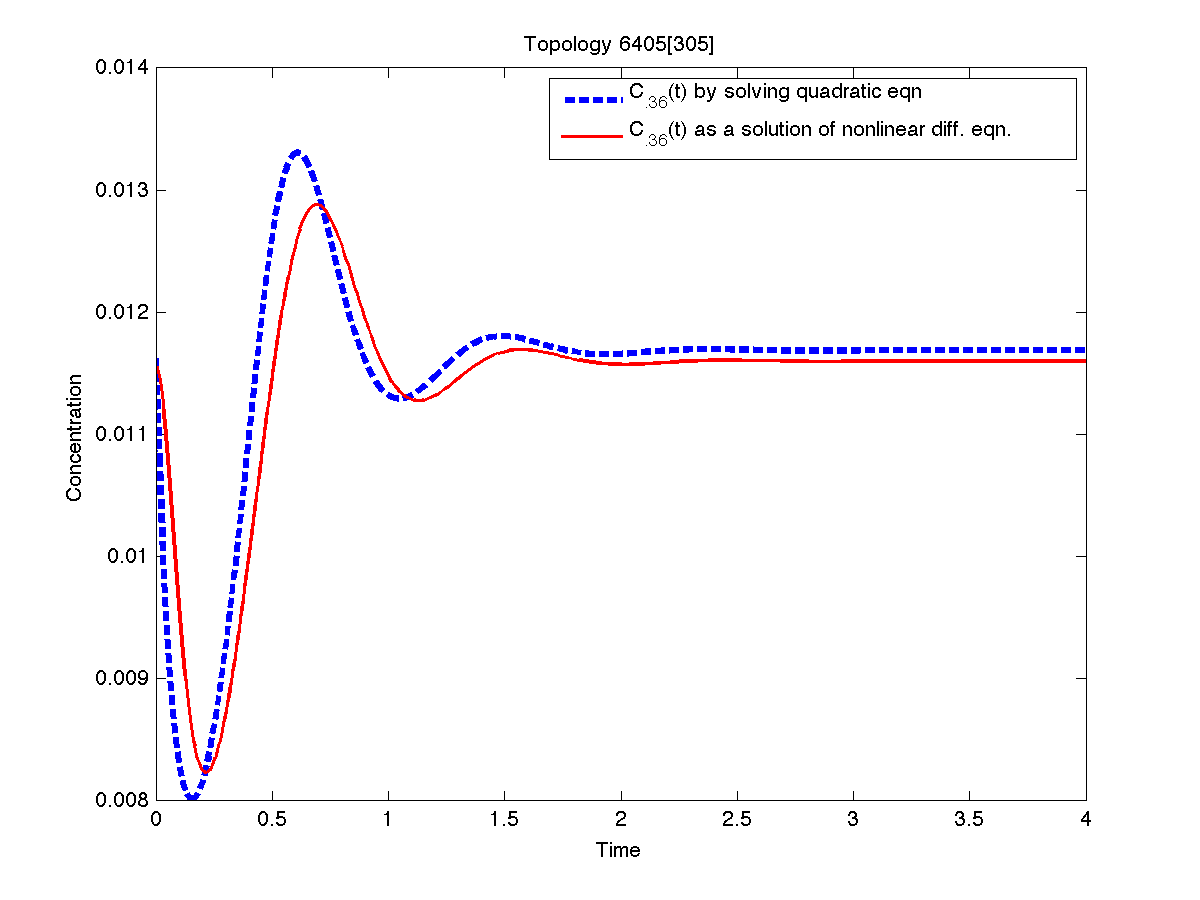}}                
  \subfloat[Quadratic approx. and output of nonlinear system]{\label{fig:f392}\includegraphics[width=0.55\textwidth]{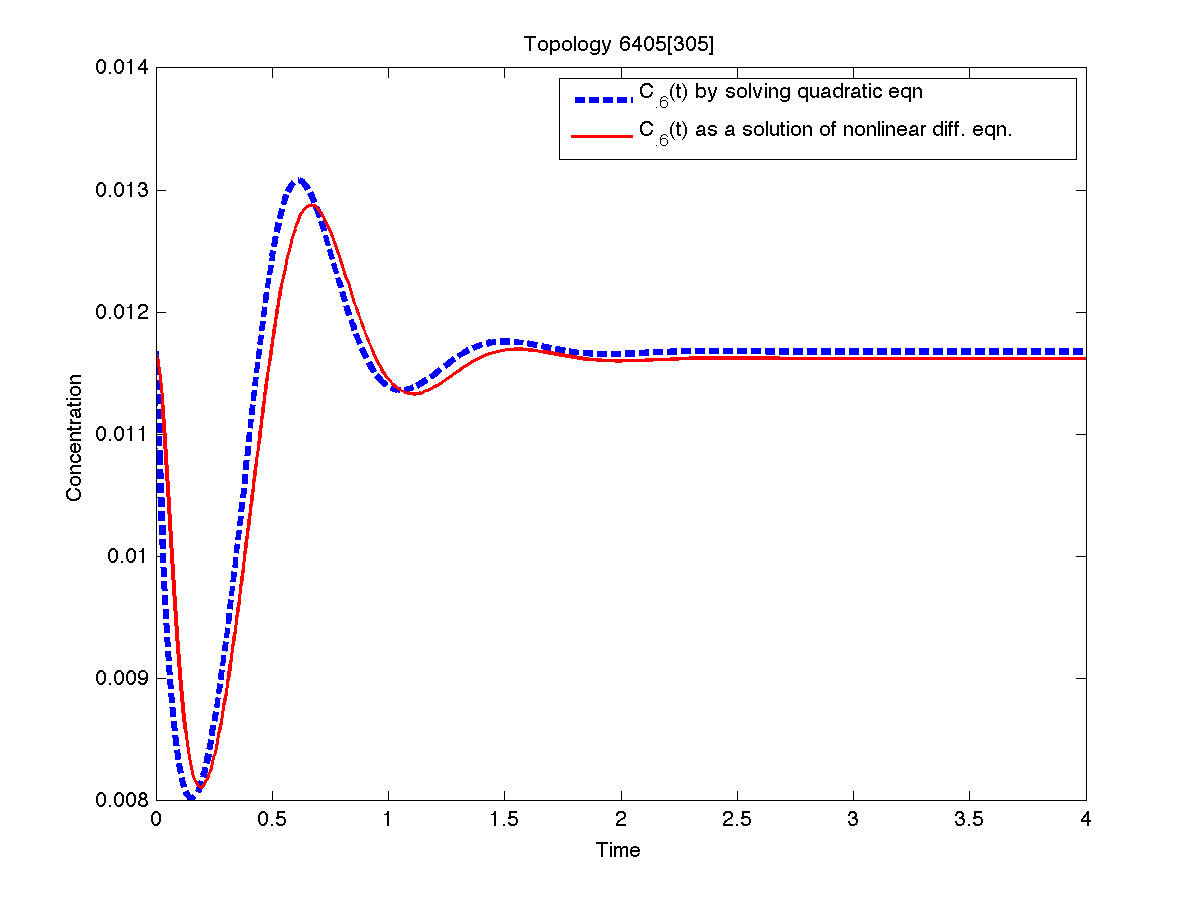}}
\end{figure}

Circuit 14.
\beqn
\dxA&=&k_{{\inp}A} {\inp} \frac{\txA}{\txA+K_{{\inp}A}}-k_{BA} \xB\frac{\xA}{\xA+K_{BA}}+k_{CA} \xC\frac{\txA}{\txA+K_{CA}}\\
\dxB&=&k_{AB}\xA\frac{\txB}{\txB+K_{AB}}+k_{CB} \xC \frac{\txB}{\txB+K_{CB}}-k_{BB} \xB\frac{\xB}{\xB+K_{BB}}\\
\dxC&=&{k_{BC}}\xB\frac{\txC}{\txC+K_{BC}}- k_{AC}\xA\frac{\xC}{\xC+K_{AC}}-k_{CC}\xA\frac{\xC}{\xC+K_{CC}}\\
\eeqn
Parameters: \  $K_{AB}= 0.444375;$ $k_{AB}= 12.053134;$
 $K_{F_B} 1.716920;$ $k_{F_B}= 11.601122;$
 $K_{AC}= 0.013988;$ $k_{AC}= 8.521185;$
 $K_{BA}= 0.005461;$ $k_{BA}= 7.103952;$
 $K_{BC}=51.850148;$ $k_{BC}= 80.408137;$
 $K_{CB}=5.392001;$ $k_{CB}= 3.086740;$
$K_{CC}= 1.962230;$ $k_{CC}= 17.382010;$
$K_{{\inp}A}= 4.387832;$ $k_{{\inp}A}= 19.638124;$
$K_{CA}= 15.479253;$ $k_{CA}= 4.903430$

\begin{figure}[ht]
  \centering
\subfloat[Dynamics of A and B in linearized model]{\label{fig:f401}\includegraphics[width=0.55\textwidth]{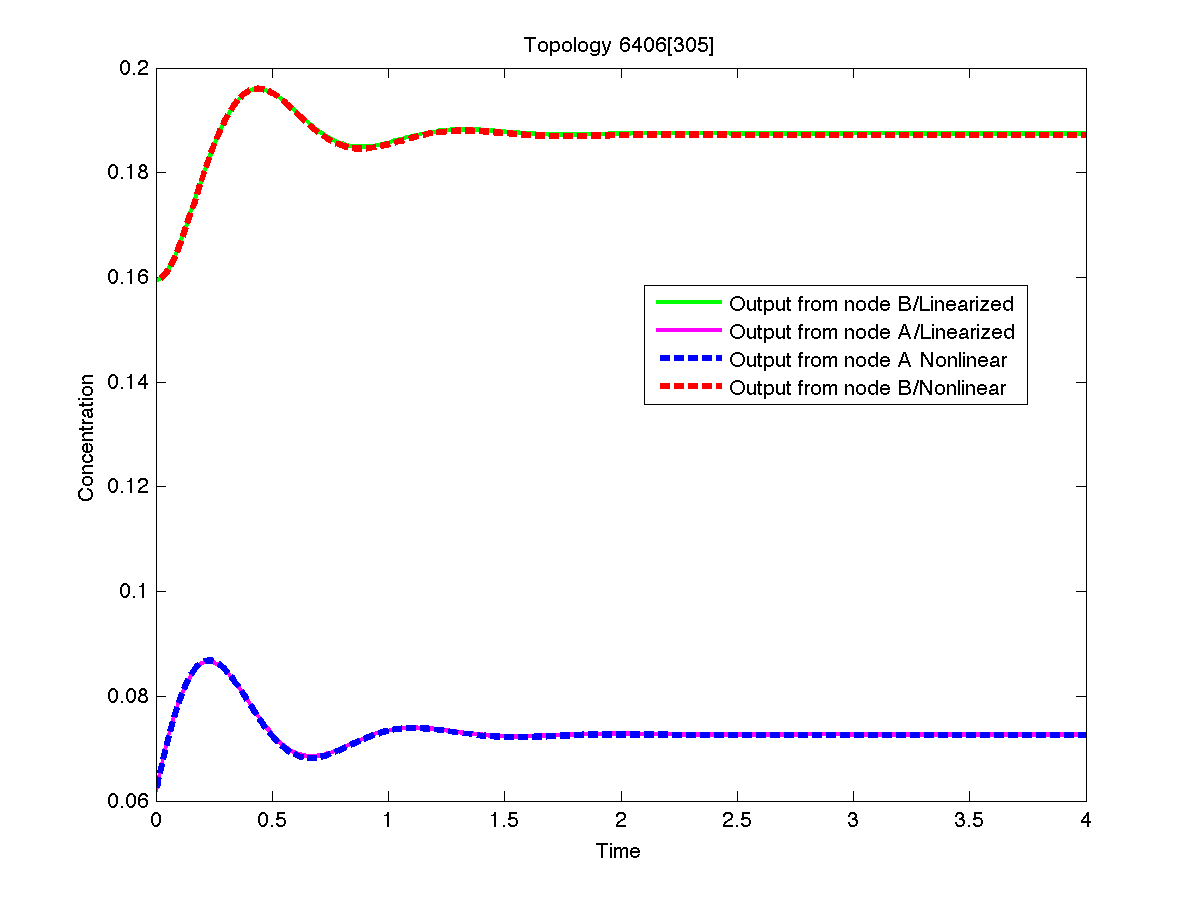}}                
  \subfloat[Ouput from C  nonlinear model]{\label{fig:f402}\includegraphics[width=0.55\textwidth]{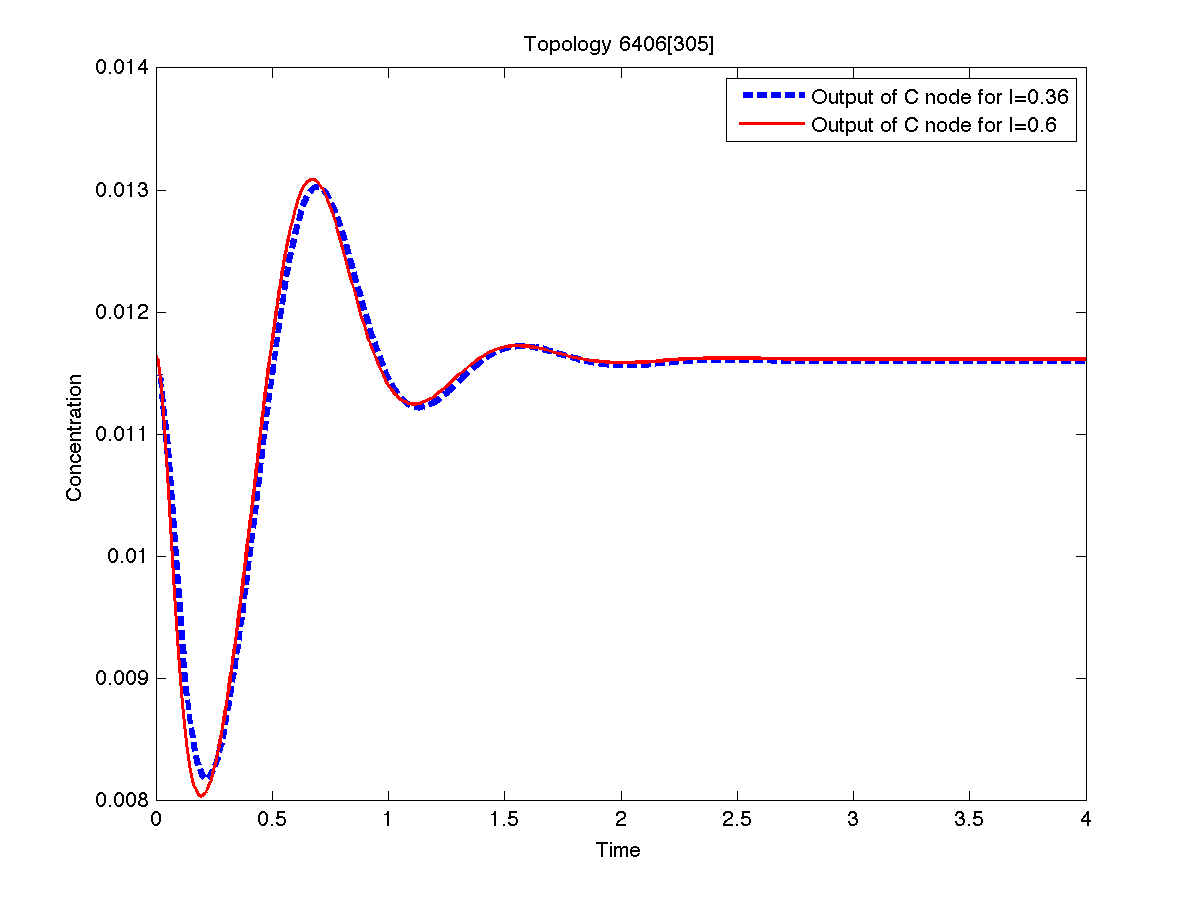}}\\
\subfloat[Quadratic approx. and output of nonlinear system]{\label{fig:f421}\includegraphics[width=0.55\textwidth]{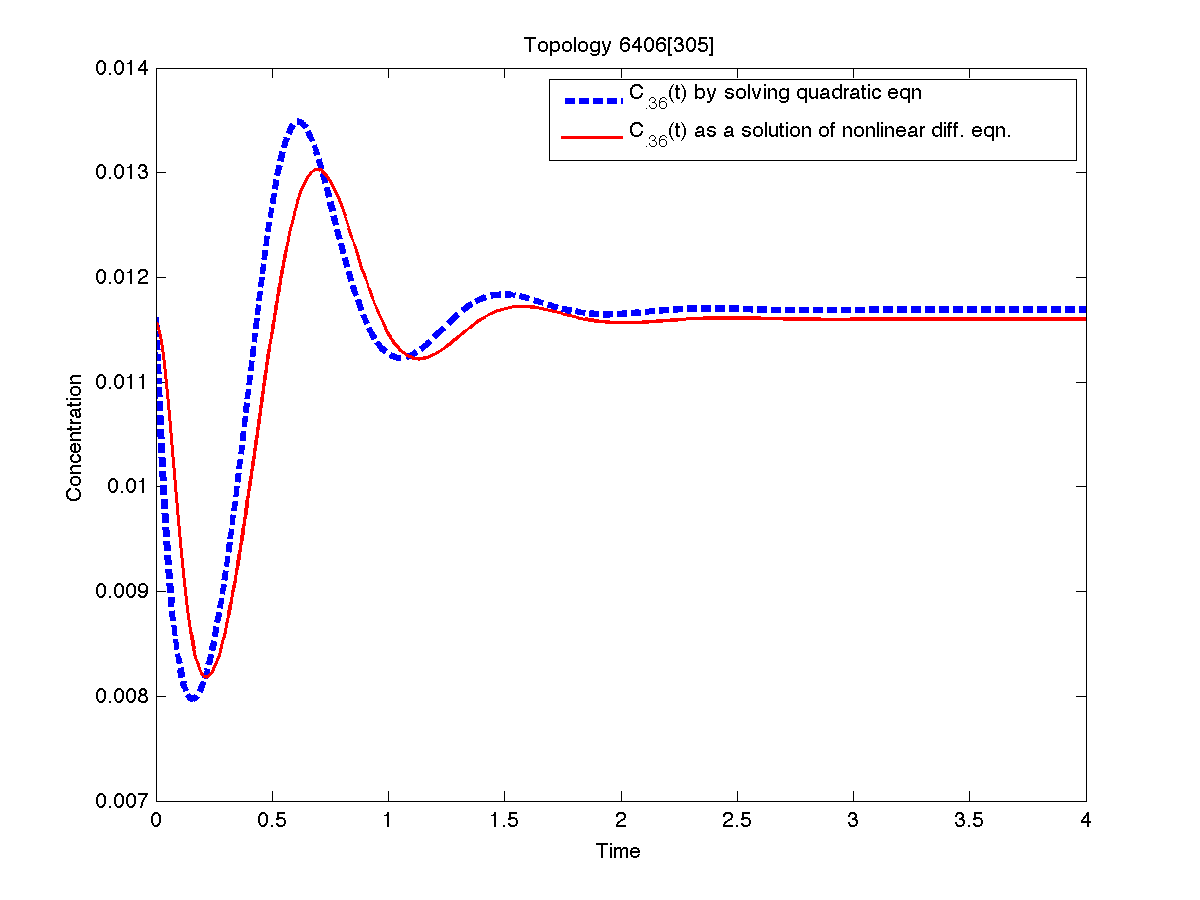}}                
  \subfloat[Quadratic approx. and output of nonlinear system]{\label{fig:f422}\includegraphics[width=0.55\textwidth]{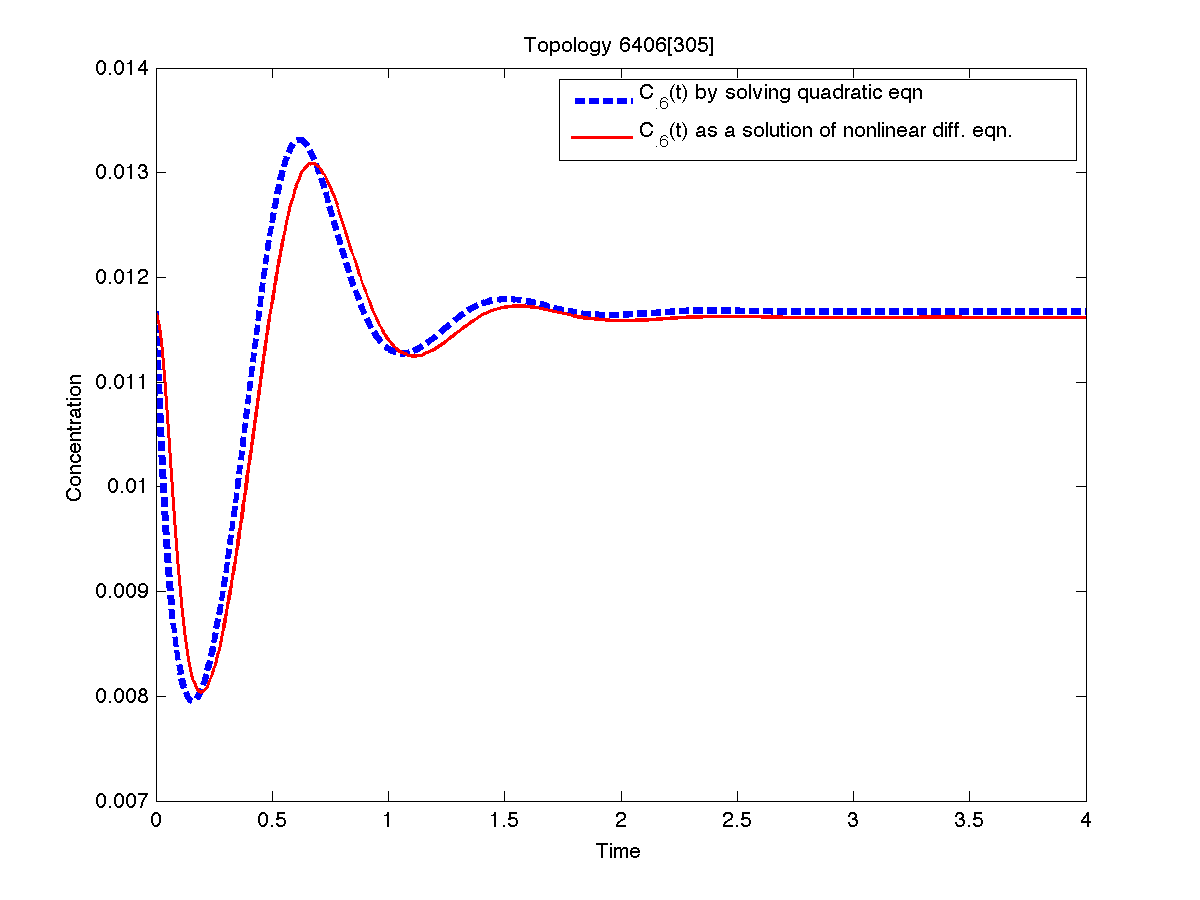}}
\end{figure}

\clearpage
Circuit 15.
\beqn
\dxA&=&k_{{\inp}A} {\inp} \frac{\txA}{\txA+K_{{\inp}A}}-k_{BA} \xB\frac{\xA}{\xA+K_{BA}}\\
\dxB&=&k_{AB}\xA\frac{\txB}{\txB+K_{AB}}-k_{F_BB} \xFB \frac{\xB}{\xB+K_{F_BB}}\\
\dxC&=&{k_{AC}}\xA\frac{\txC}{\txC+K_{AC}}- k_{BC}\xB\frac{\xC}{\xC+K_{BC}}-k_{CC}\xA\frac{\xC}{\xC+K_{CC}}\\
\eeqn
Parameters: \ $K_{AB}= 0.709169;$ $k_{AB}= 7.445605;$
 $K_{F_B}= 1.495375;$ $k_{F_B}= 7.282827;$
 $K_{AC}= 0.002566;$ $k_{AC}= 1.115065;$
 $K_{BA}= 0.002522;$ $k_{BA}= 5.753075;$
 $K_{BC}= 0.017051;$ $k_{BC}= 2.777794;$
 $K_{CC} =0.195997;$ $k_{CC}= 1.480130;$
$K_{{\inp}A}= 0.225814;$ $k_{{\inp}A}= 2.492872$

\begin{figure}[hb]
  \centering
\subfloat[Dynamics of A and B in linearized model]{\label{fig:f431}\includegraphics[width=0.55\textwidth]{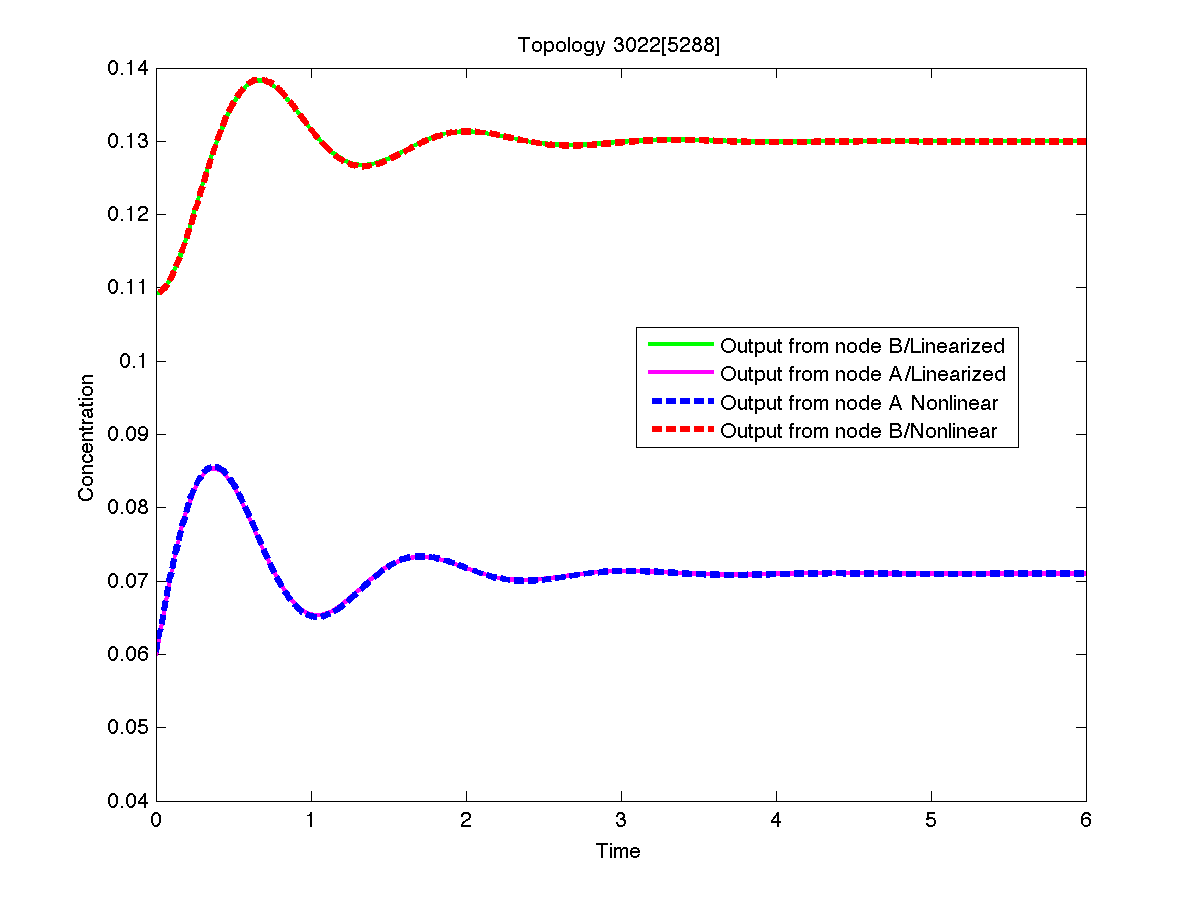}}                
  \subfloat[Ouput from C  nonlinear model]{\label{fig:f432}\includegraphics[width=0.55\textwidth]{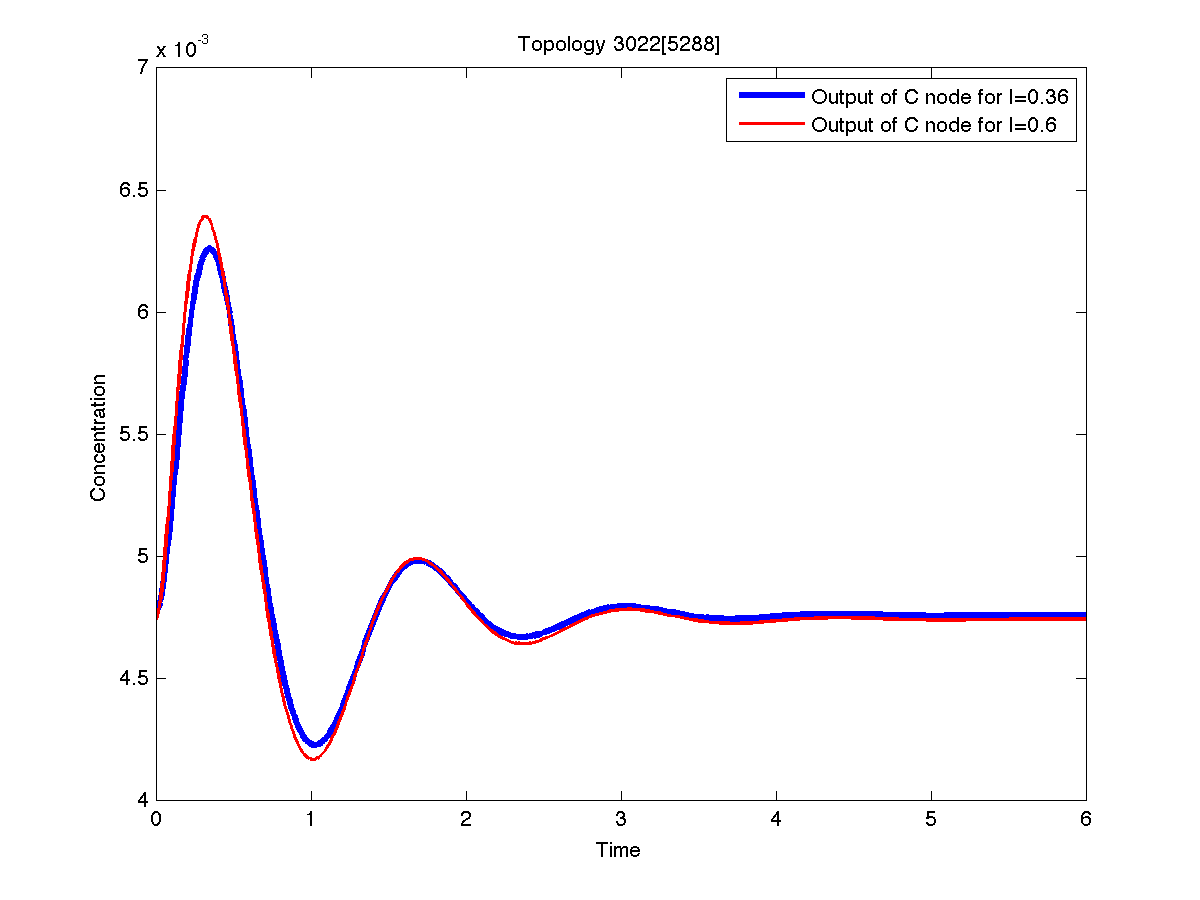}}\\
\subfloat[Quadratic approx. and output of nonlinear system]{\label{fig:f451}\includegraphics[width=0.55\textwidth]{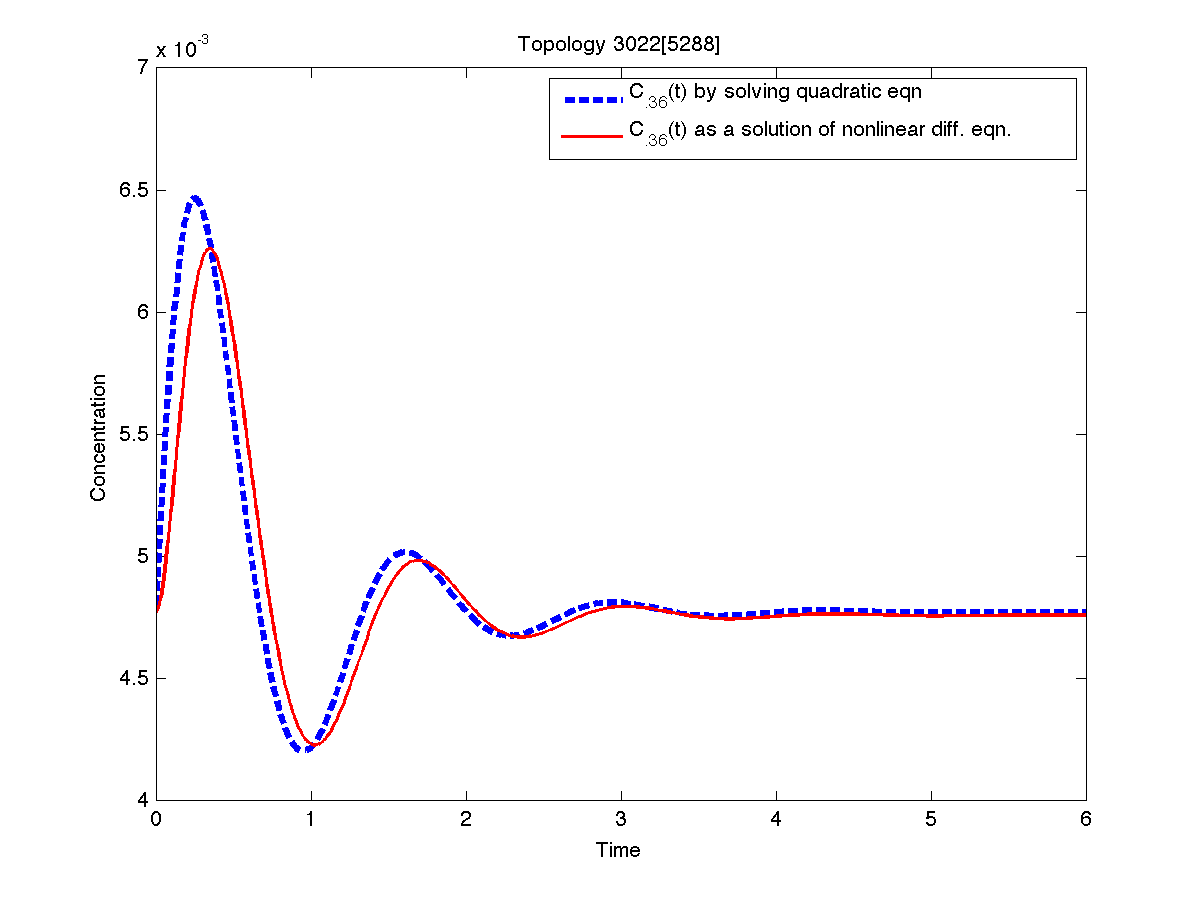}}                
  \subfloat[Quadratic approx. and output of nonlinear system]{\label{fig:f452}\includegraphics[width=0.55\textwidth]{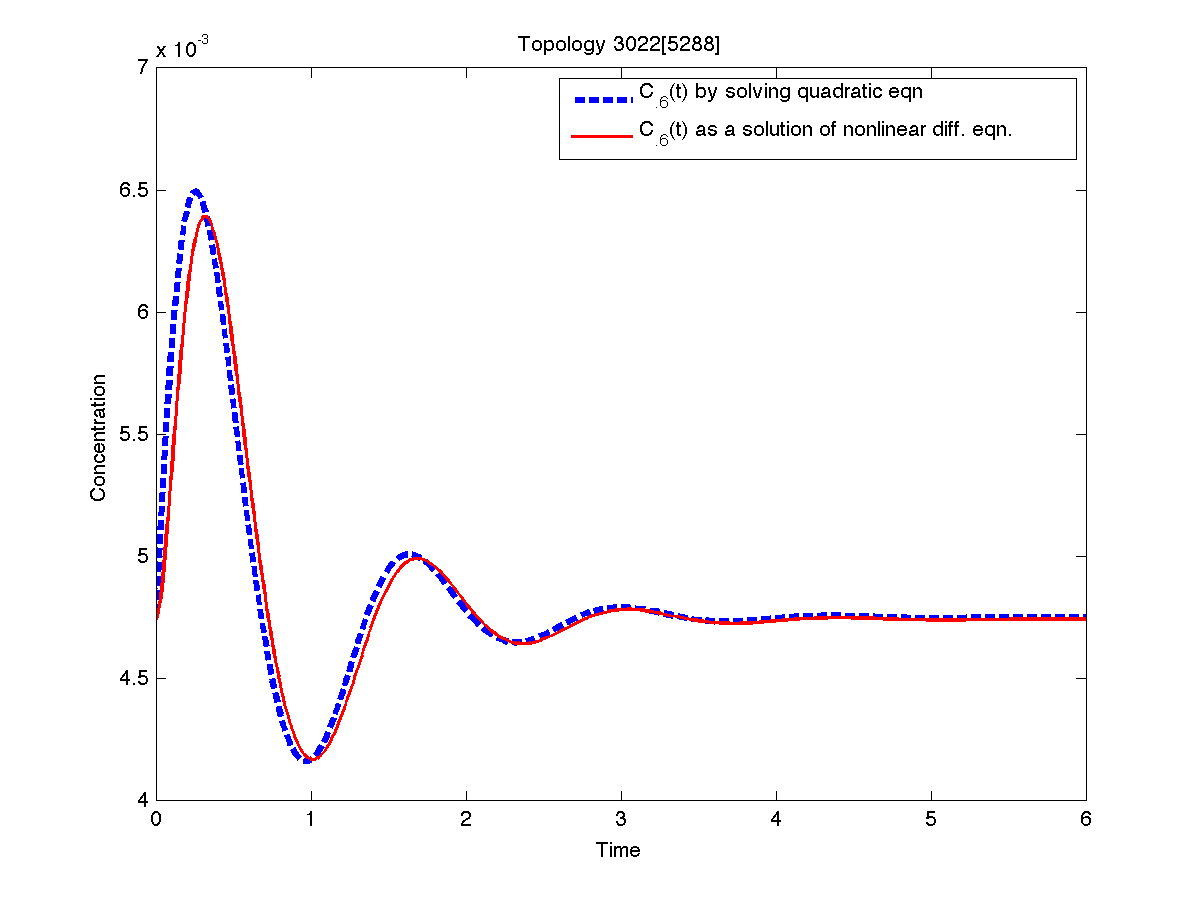}}
\end{figure}
\
\
\
\clearpage
Circuit 16. 

This is the same topology as in the previous case, only a different parameter set was used:

Parameters: \ $K_{AB}= 0.001191;$ $k_{AB}= 1.466561;$
$K_{F_B}= 9.424319;$ $k_{F_B}= 22.745736;$
 $K_{AC}= 0.113697;$ $k_{AC}= 1.211993;$
$K_{BA}= 0.001688;$ $k_{BA}= 44.802268;$
 $K_{BC}= 0.009891;$ $k_{BC}= 7.239357;$
 $K_{CC}= 0.189125;$ $k_{CC}= 17.910182;$
$K_{{\inp}A}= 0.093918;$ $k_{{\inp}A}= 11.447219$

\begin{figure}[hb]
  \centering
\subfloat[Dynamics of A and B in linearized model]{\label{fig:f731}\includegraphics[width=0.55\textwidth]{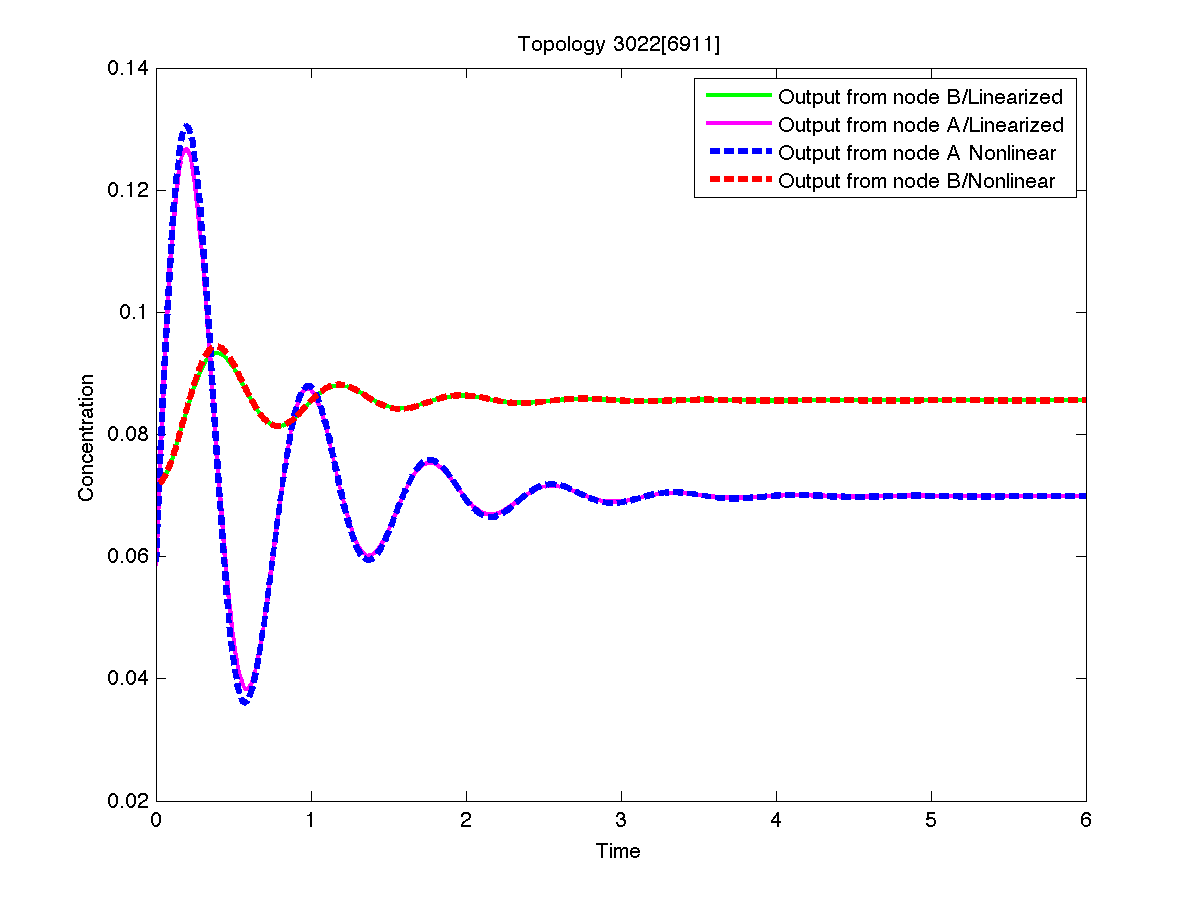}}
  \subfloat[Ouput from C  nonlinear model]{\label{fig:f732}\includegraphics[width=0.55\textwidth]{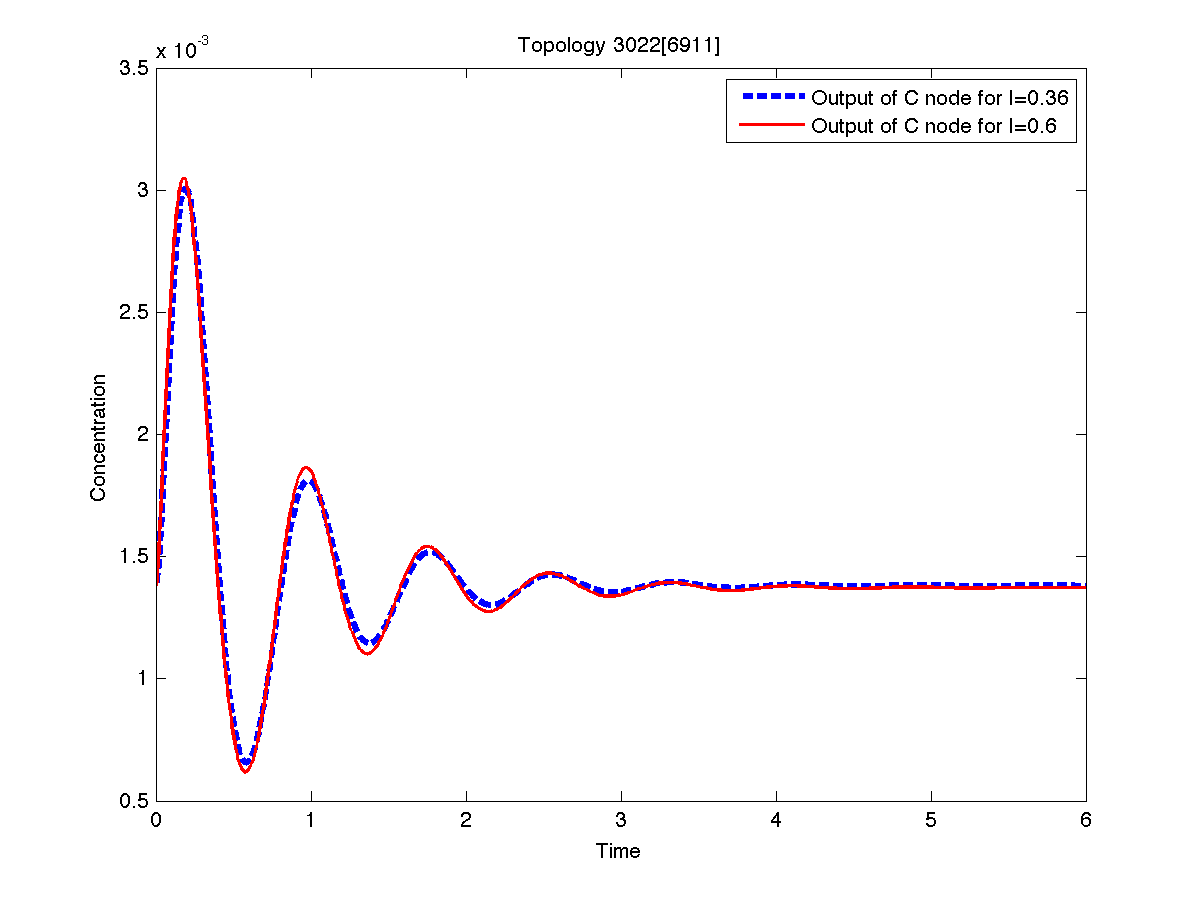}}\\
\subfloat[Quadratic approx. and output of nonlinear system]{\label{fig:f751}\includegraphics[width=0.55\textwidth]{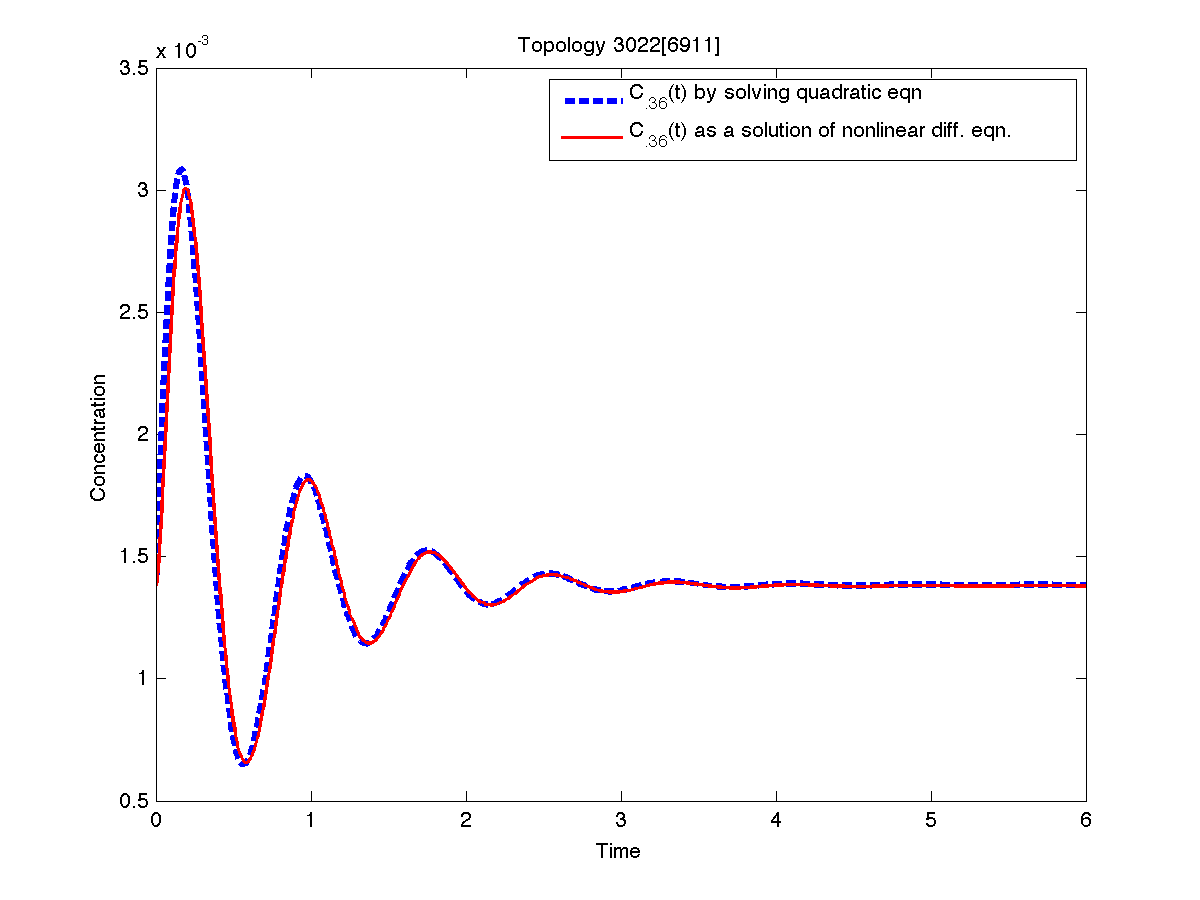}}  
  \subfloat[Quadratic approx. and output of nonlinear system]{\label{fig:f752}\includegraphics[width=0.55\textwidth]{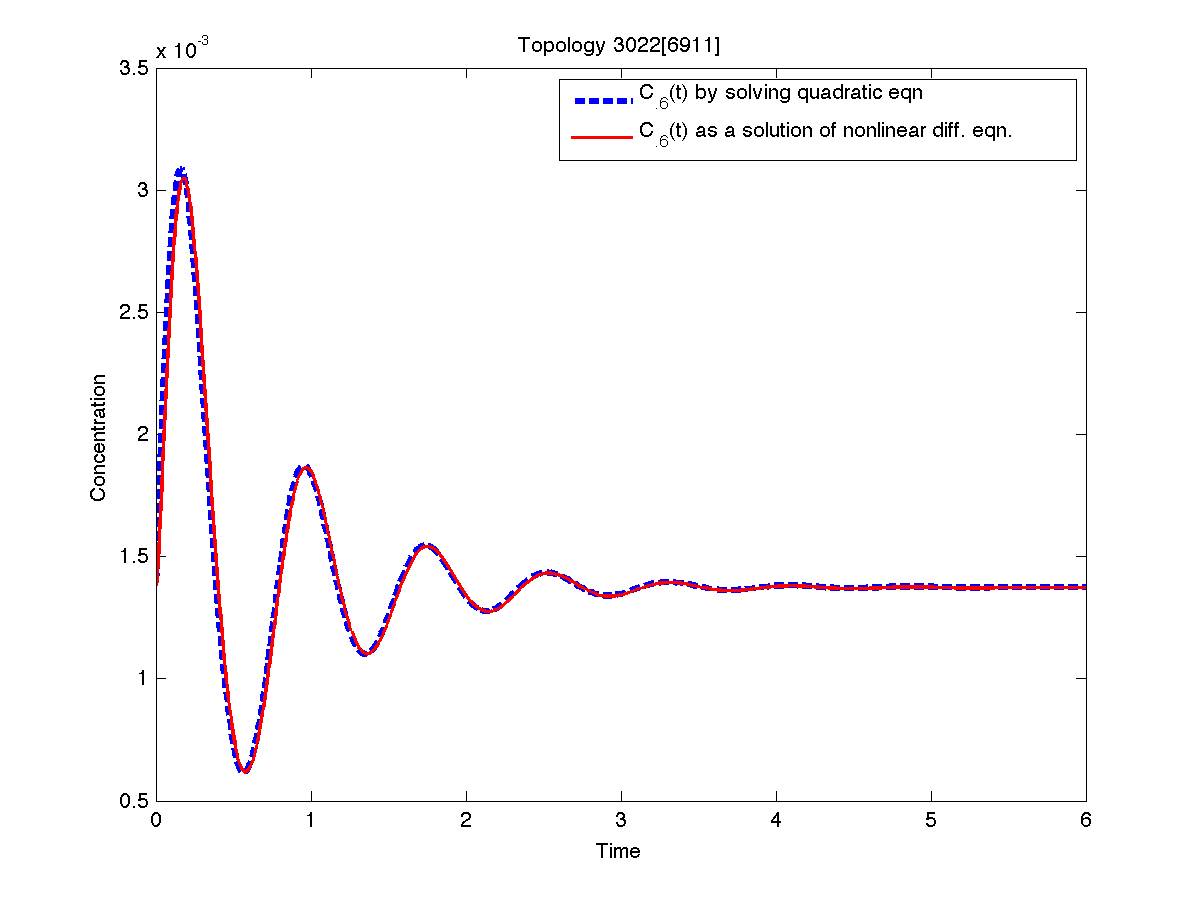}}
\end{figure}

\clearpage
Circuit 17.

This is the same topology as in the previous case, only a different parameter set was used:

Parameters: \  $K_{AB}= 1.620877;$ $k_{AB}= 2.306216;$
 $K_{F_B}= 2.012565;$ $k_{F_B}= 2.700847;$
 $K_{AC}= 0.010933;$ $k_{AC}= 8.968091;$
 $K_{BA}= 0.001812;$ $k_{BA}= 10.039221;$
 $K_{BC}= 0.014199;$ $k_{BC}= 17.762333;$
 $K_{CC}=2.686891;$ $k_{CC}= 4.139044;$
$K_{{\inp}A}= 0.161715;$ $k_{{\inp}A}= 1.933303$
\
\

\begin{figure}[hb]
  \centering
\subfloat[Dynamics of A and B in linearized model]{\label{fig:f731}\includegraphics[width=0.55\textwidth]{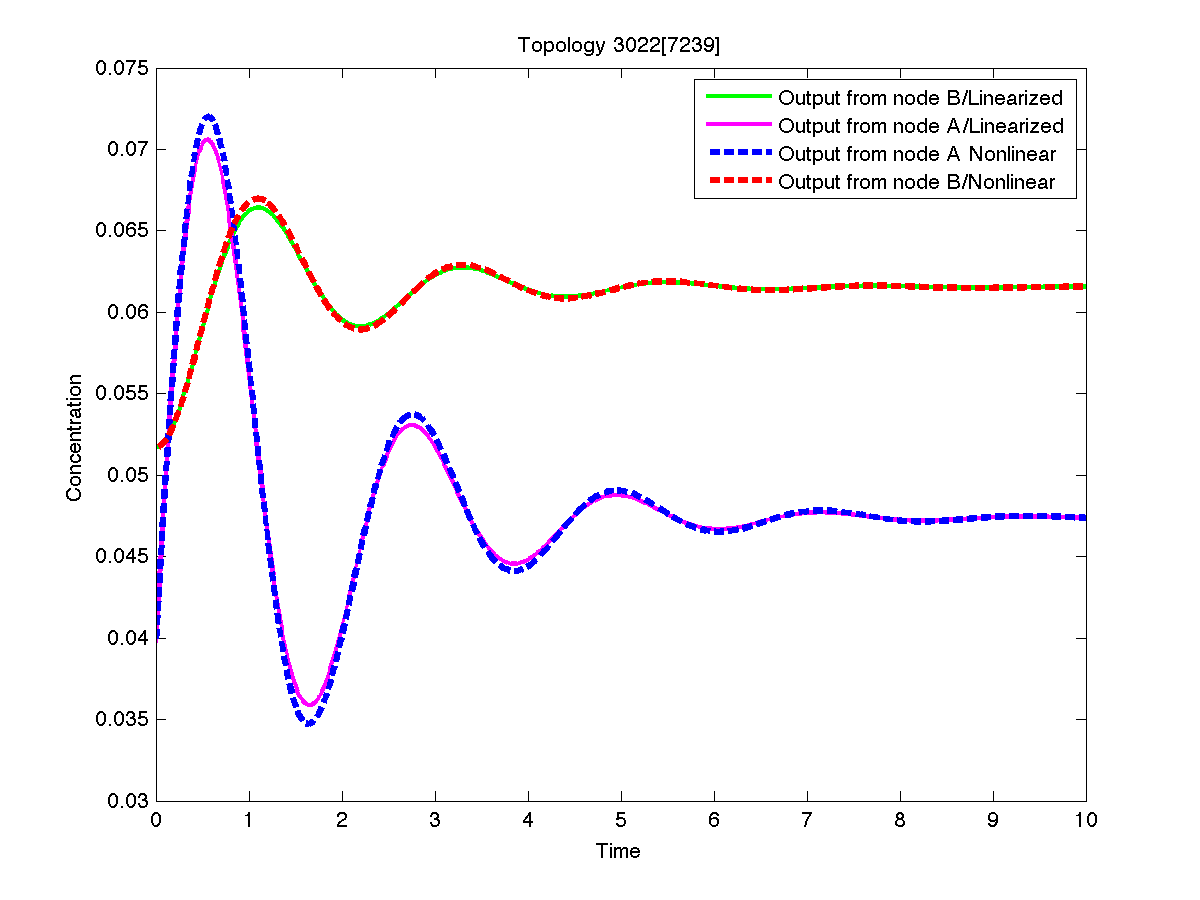}}
  \subfloat[Ouput from C  nonlinear model]{\label{fig:f732}\includegraphics[width=0.55\textwidth]{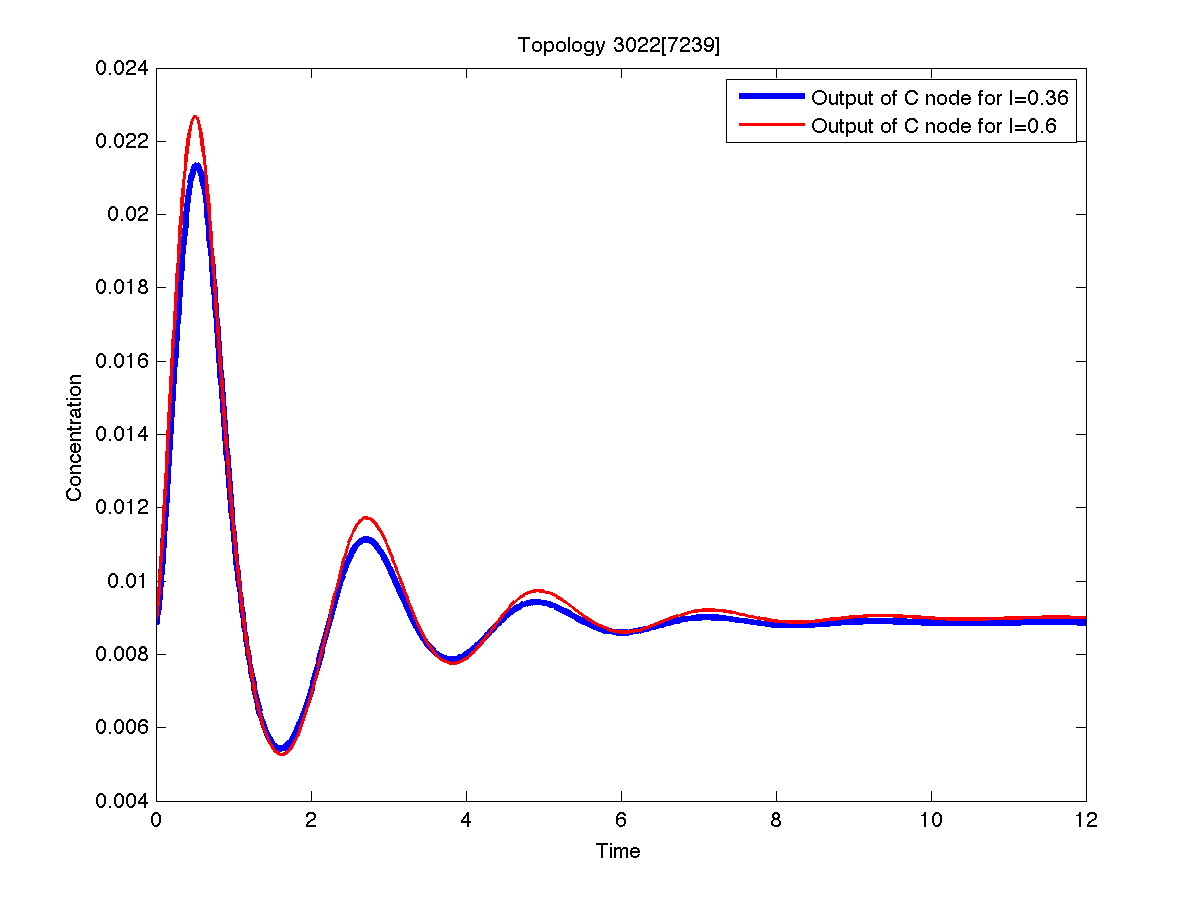}}\\
\subfloat[Quadratic approx. and output of nonlinear system]{\label{fig:f751}\includegraphics[width=0.55\textwidth]{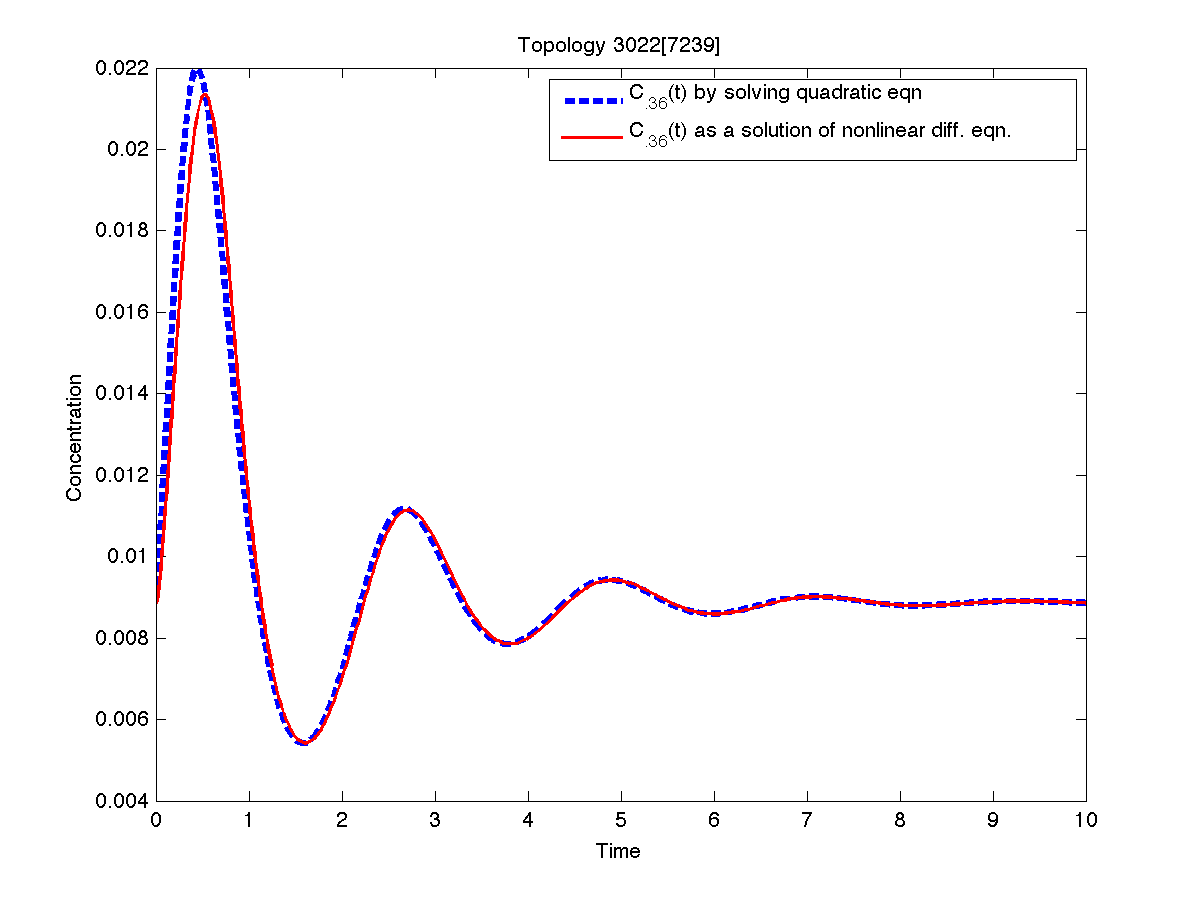}}  
  \subfloat[Quadratic approx. and output of nonlinear system]{\label{fig:f752}\includegraphics[width=0.55\textwidth]{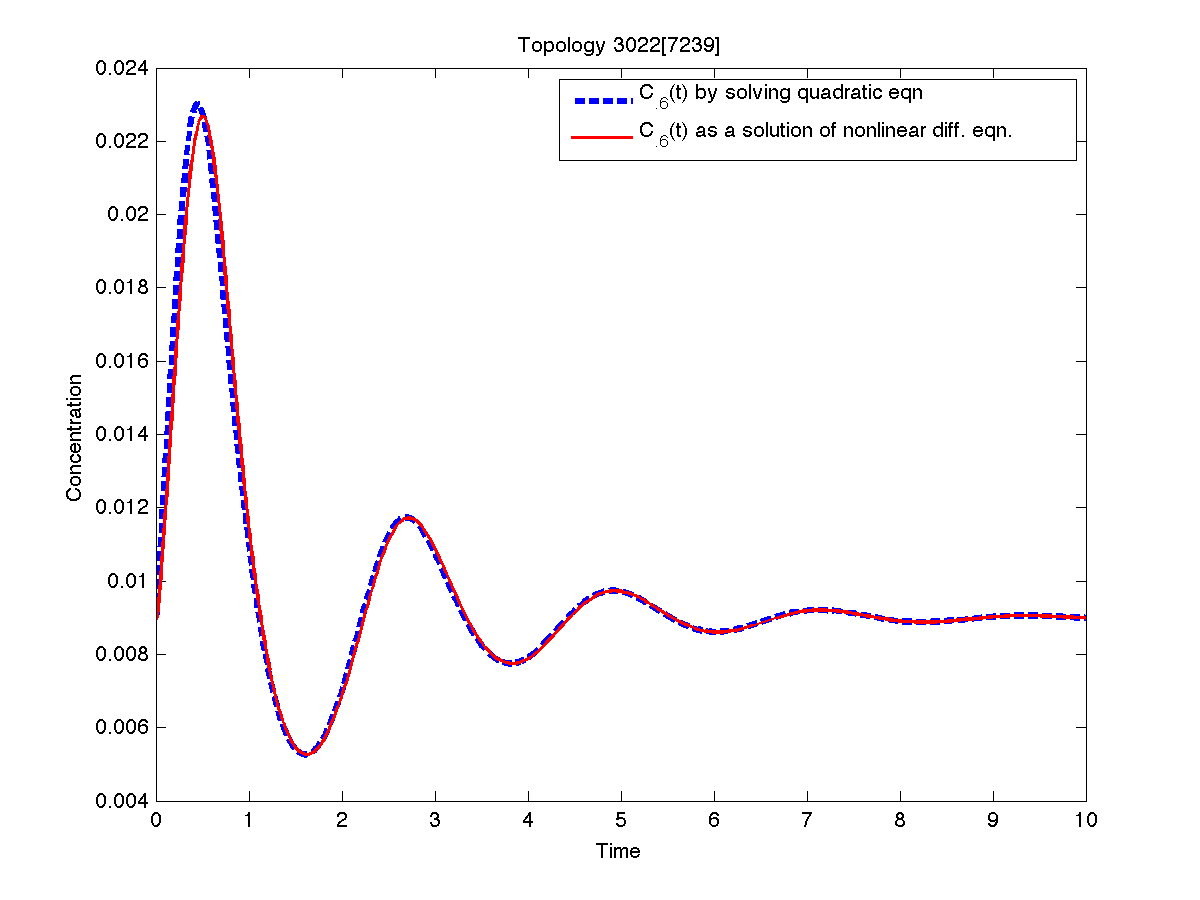}}
\end{figure}

Circuit 18. 
\beqn
\dxA&=&k_{{\inp}A} {\inp} \frac{\txA}{\txA+K_{{\inp}A}}-k_{BA} \xB\frac{\xA}{\xA+K_{BA}}-k_{AA} \xA\frac{\xA}{\xA+K_{AA}}\\
\dxB&=&k_{AB}\xA\frac{\txB}{\txB+K_{AB}}+k_{BB} \xB\frac{\txB}{\txB+K_{BB}}-k_{F_BB} \xFB \frac{\xB}{\xB+K_{F_BB}}\\
\dxC&=&{k_{AC}}\xA\frac{\txC}{\txC+K_{AC}}- k_{BC}\xB\frac{\xC}{\xC+K_{BC}}-k_{CC}\xC\frac{\xC}{\xC+K_{CC}}\\
\eeqn
Parameters: \   $K_{AA}= 17.569120;$ $k_{AA}= 2.198366;$
 $K_{AB}= 9.435176;$ $k_{AB}= 3.134007;$
 $K_{F_B}= 0.469083;$ $k_{F_B}= 1.934194;$
$K_{AC}= 0.062914;$ $k_{AC}= 2.742206;$
$K_{BA}= 0.003245;$ $k_{BA}= 75.352905;$
 $K_{BB}= 27.463128;$ $k_{BB}= 10.551155;$
$K_{BC}= 0.041615;$ $k_{BC}= 61.333818;$
 $K_{CC}= 0.039332;$ $k_{CC}= 4.756637;$
$K_{{\inp}A}= 0.005167;$ $k_{{\inp}A}= 8.186533$

\begin{figure}[ht]
  \centering
\subfloat[Dynamics of A and B in linearized model]{\label{fig:f521}\includegraphics[width=0.55\textwidth]{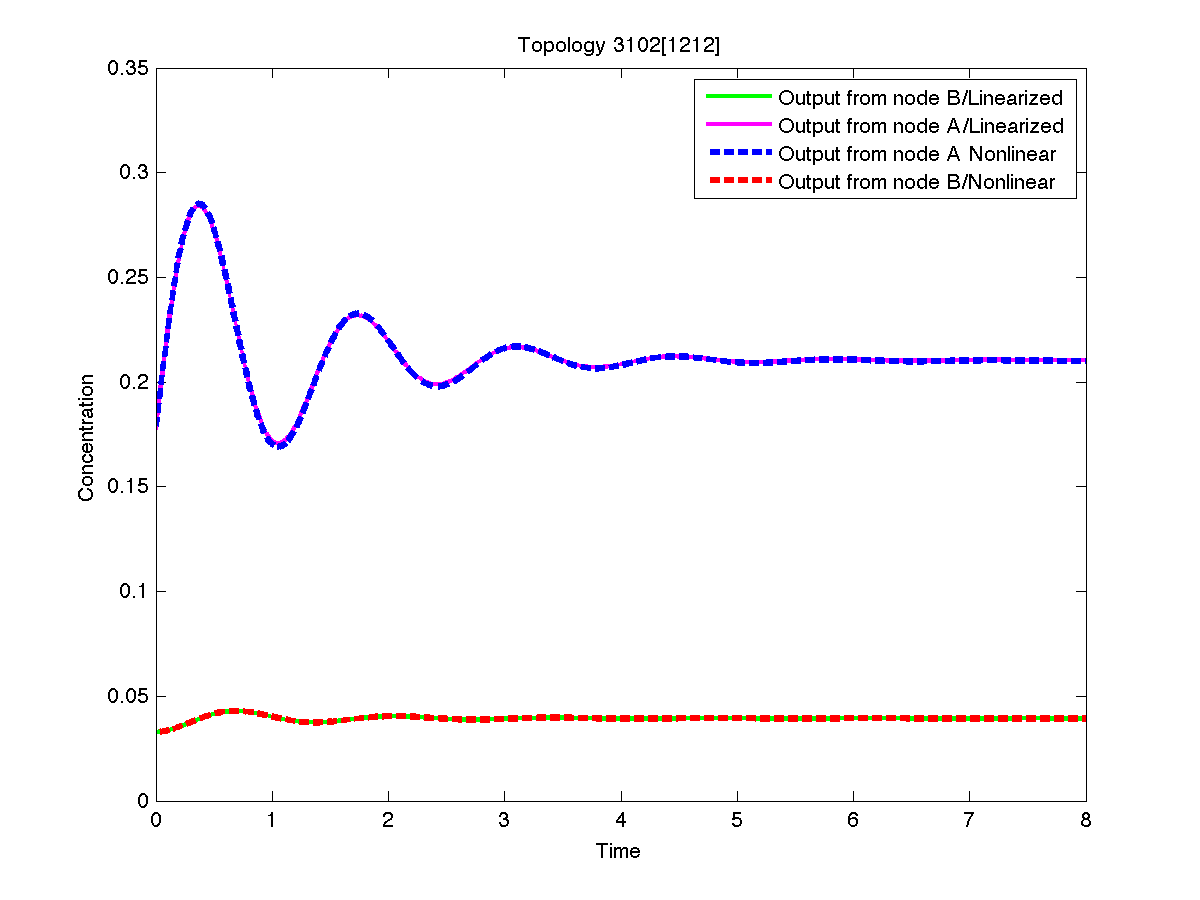}}                
  \subfloat[Ouput from C  nonlinear model]{\label{fig:f522}\includegraphics[width=0.55\textwidth]{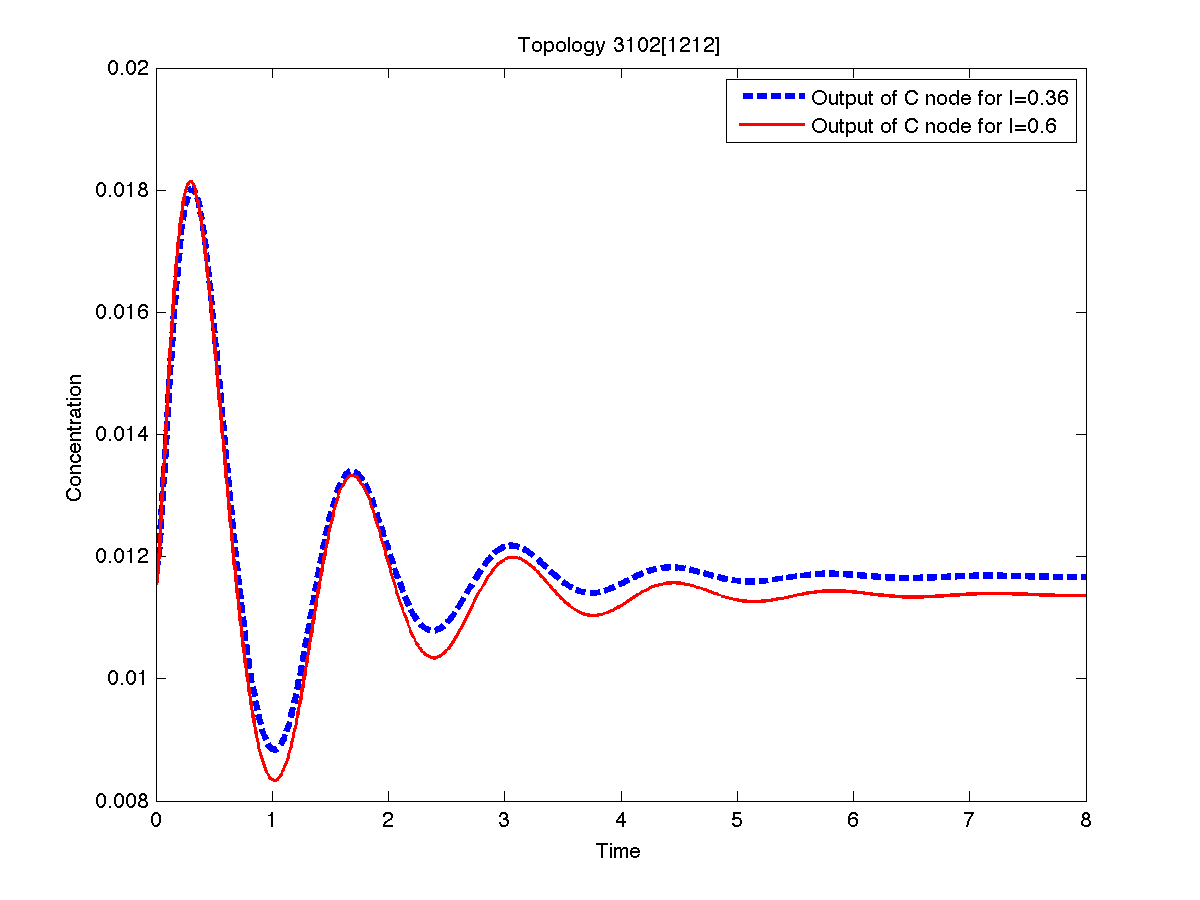}}\\
\subfloat[Quadratic approx. and output of nonlinear system]{\label{fig:f541}\includegraphics[width=0.55\textwidth]{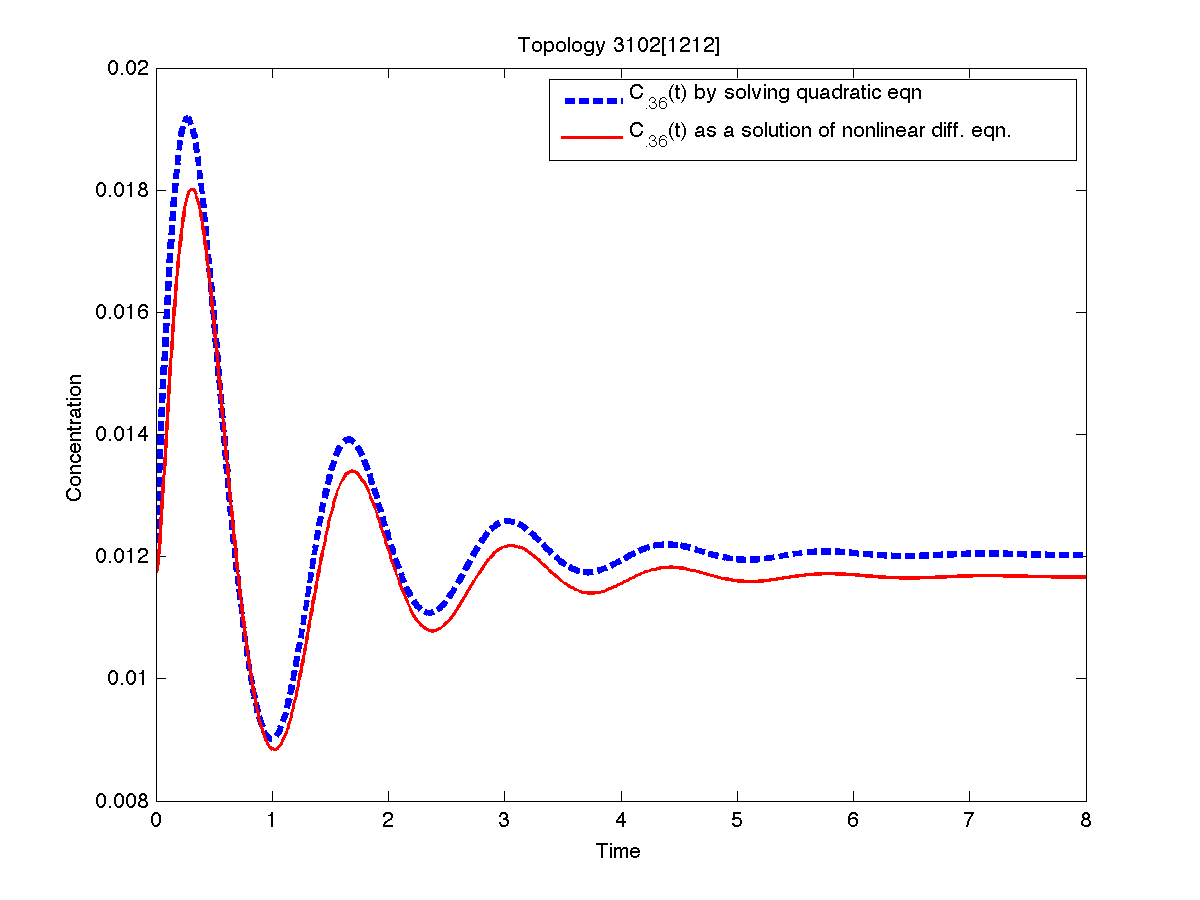}}                
  \subfloat[Quadratic approx. and output of nonlinear system]{\label{fig:f542}\includegraphics[width=0.55\textwidth]{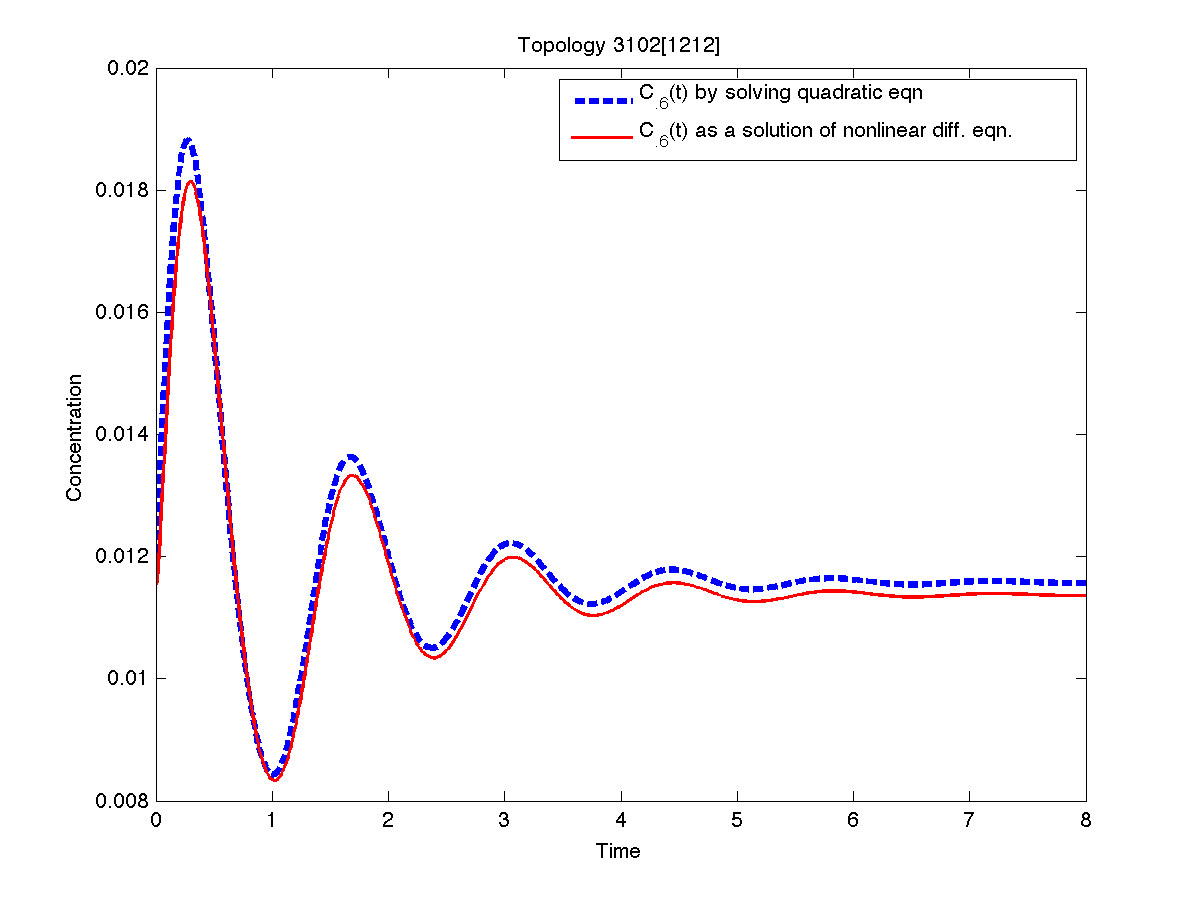}}
\end{figure}

\clearpage
Circuit 19.
\beqn
\dxA&=&k_{{\inp}A} {\inp} \frac{\txA}{\txA+K_{{\inp}A}}-k_{BA} \xB\frac{\xA}{\xA+K_{BA}}-k_{AA} \xA\frac{\xA}{\xA+K_{AA}}\\
\dxB&=&k_{AB}\xA\frac{\txB}{\txB+K_{AB}}-k_{F_BB} \xFB \frac{\xB}{\xB+K_{F_BB}}\\
\dxC&=&{k_{BC}}\xB\frac{\txC}{\txC+K_{BC}}- k_{AC}\xA\frac{\xC}{\xC+K_{AC}}-k_{CC}\xC\frac{\xC}{\xC+K_{CC}}\\
\eeqn
Parameters: \  $K_{{\inp}A}= 4.387832;$ $k_{{\inp}A}= 19.638124;$
 $K_{BA}= 0.005461;$ $k_{BA}= 7.103952;$
$K_{AA}= 24.989065;$ $k_{AA}= 53.174082;$
 $K_{AB}= 0.444375;$ $k_{AB}= 12.053134;$
 $K_{F_B}= 1.716920;$ $k_{F_B}= 11.601122;$
 $K_{BC}=51.850148;$ $k_{BC}= 80.408137;$
 $K_{AC}= 0.013988;$ $k_{AC}= 8.521185;$
$K_{CC}= 1.962230;$ $k_{CC}= 17.382010$

\begin{figure}[hb]
  \centering
\subfloat[Dynamics of A and B in linearized model]{\label{fig:f551}\includegraphics[width=0.55\textwidth]{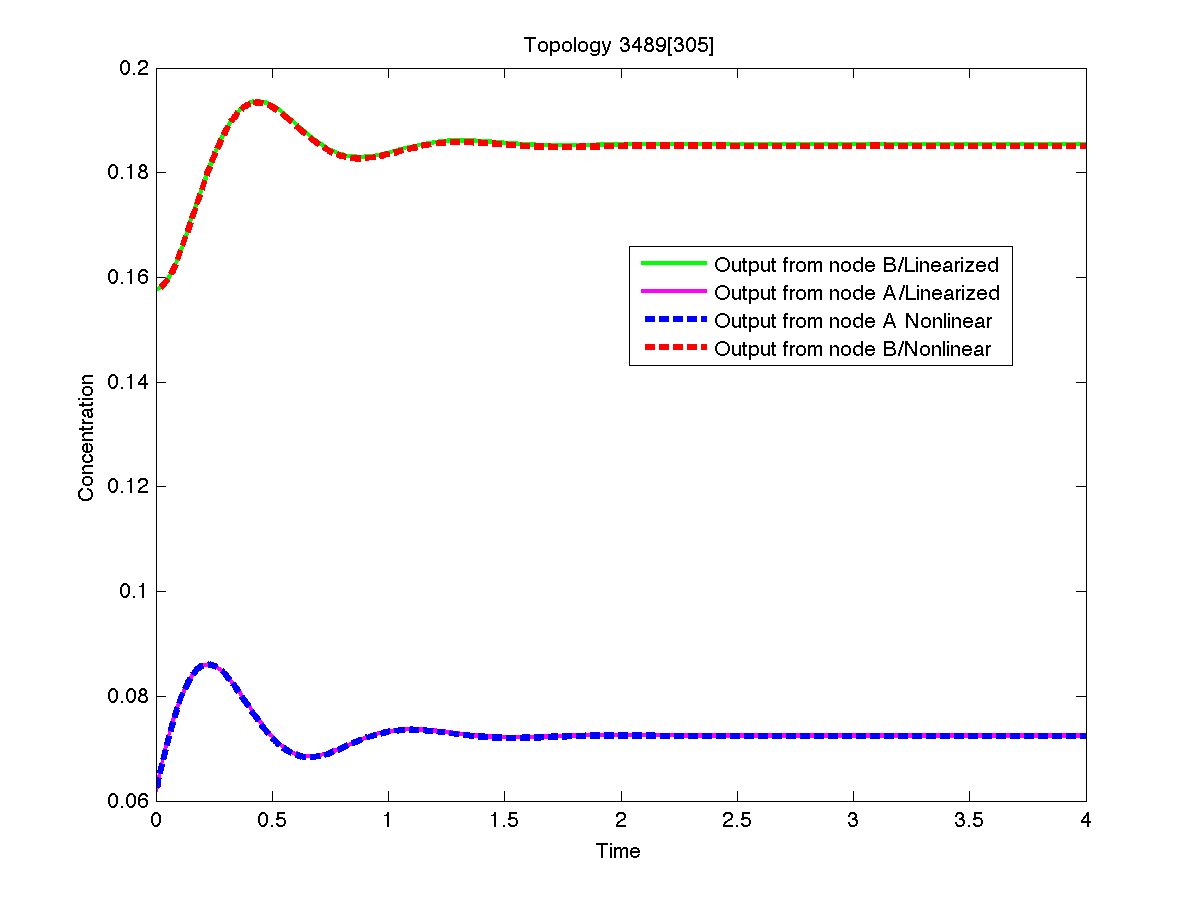}}                
  \subfloat[Ouput from C  nonlinear model]{\label{fig:f552}\includegraphics[width=0.55\textwidth]{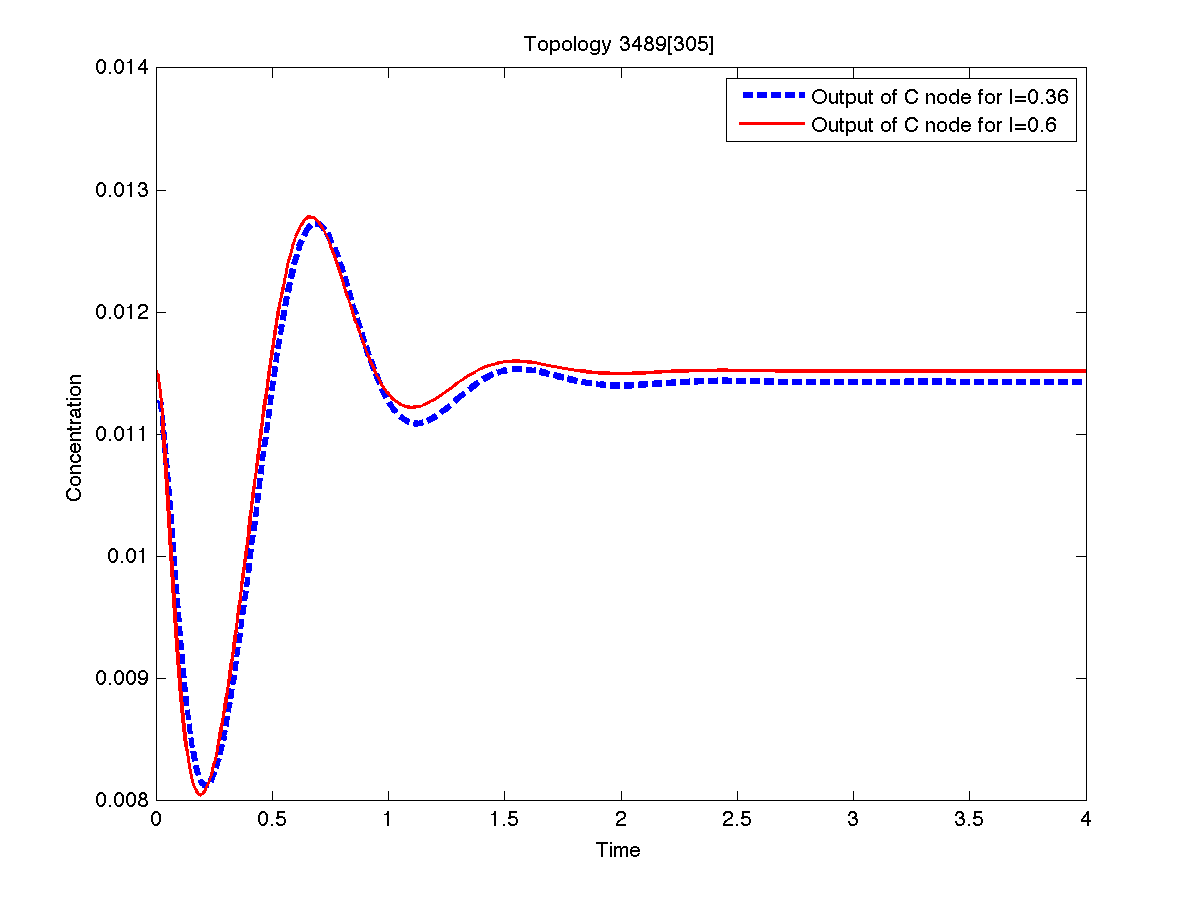}}\\
\subfloat[Quadratic approx. and output of nonlinear system]{\label{fig:f571}\includegraphics[width=0.55\textwidth]{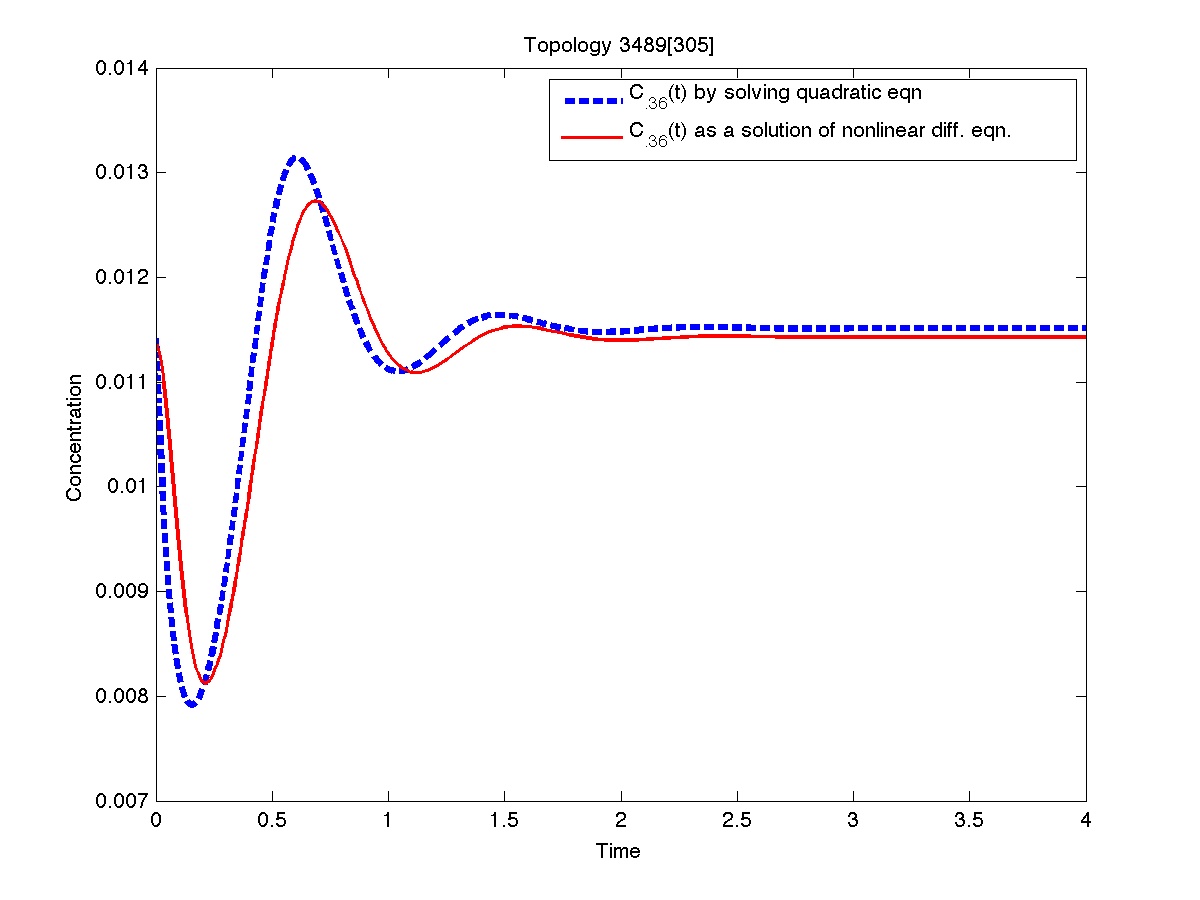}}                
  \subfloat[Quadratic approx. and output of nonlinear system]{\label{fig:f572}\includegraphics[width=0.55\textwidth]{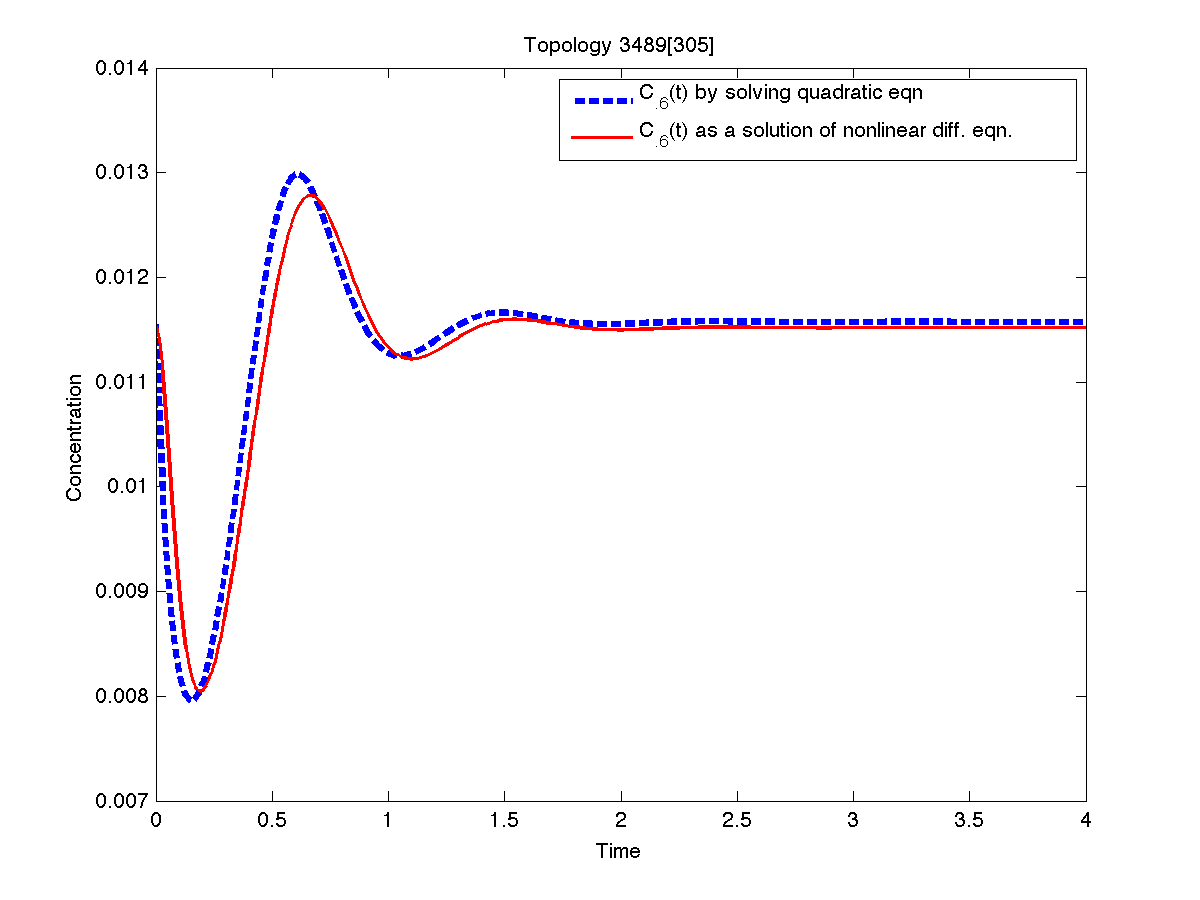}}
\end{figure}

Circuit 20. 
\beqn
\dxA&=&k_{{\inp}A} {\inp} \frac{\txA}{\txA+K_{{\inp}A}}-k_{BA} \xB\frac{\xA}{\xA+K_{BA}}\\
\dxB&=&k_{AB}\xA\frac{\txB}{\txB+K_{AB}}-k_{F_BB} \xFB \frac{\xB}{\xB+K_{F_BB}}\\
\dxC&=&{k_{BC}}\xB\frac{\txC}{\txC+K_{BC}}- k_{AC}\xA\frac{\xC}{\xC+K_{AC}}-k_{CC}\xC\frac{\xC}{\xC+K_{CC}}\\
\eeqn
Parameters: \  $K_{{\inp}A}= 4.387832;$ $k_{{\inp}A}= 19.638124;$
 $K_{BA}= 0.005461;$ $k_{BA}= 7.103952;$
 $K_{AB}= 0.444375;$ $k_{AB}= 12.053134;$
 $K_{F_B}= 1.716920;$ $k_{F_B}= 11.601122;$
 $K_{BC}=51.850148;$ $k_{BC}= 80.408137;$
 $K_{AC}= 0.013988;$ $k_{AC}= 8.521185;$
$K_{CC}= 1.962230;$ $k_{CC}= 17.382010$

\begin{figure}[ht]
  \centering
\subfloat[Dynamics of A and B in linearized model]{\label{fig:f581}\includegraphics[width=0.55\textwidth]{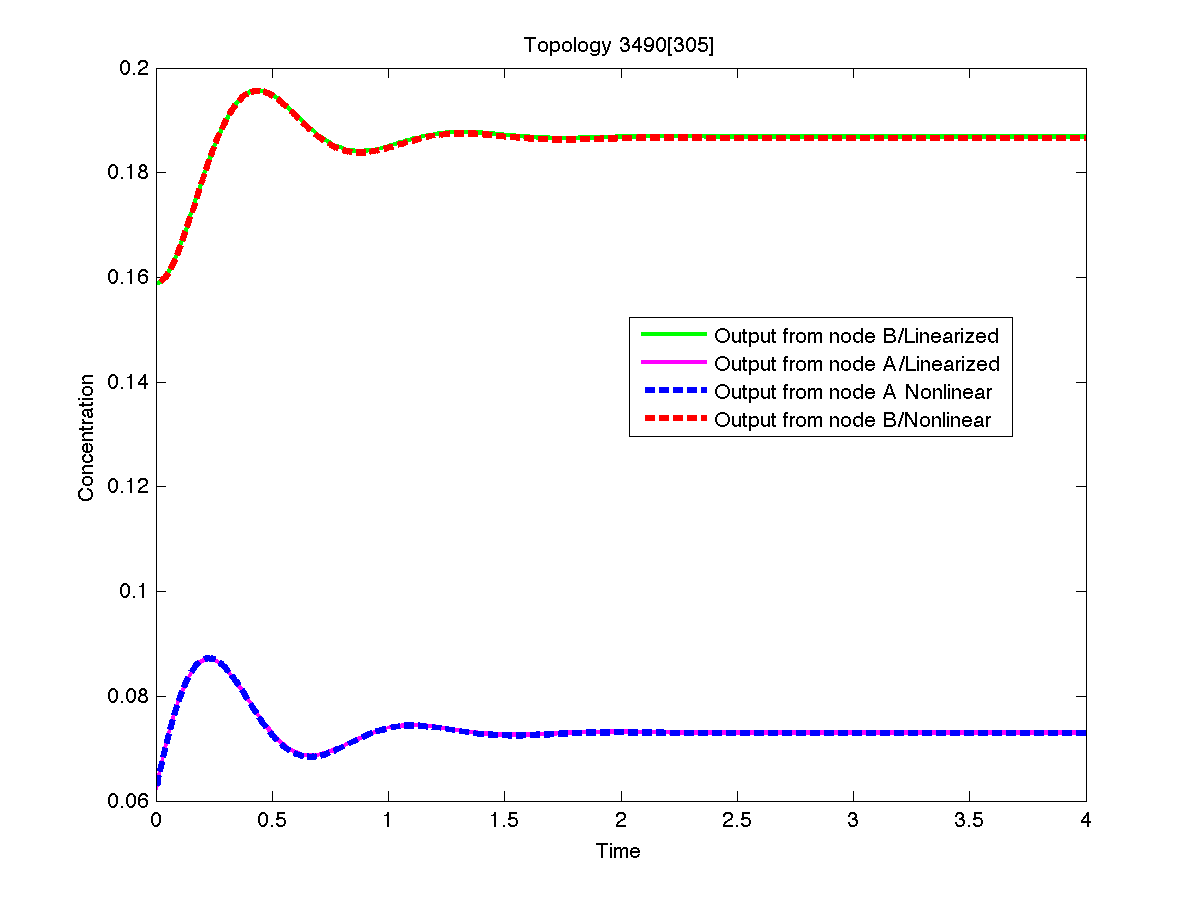}}                
  \subfloat[Ouput from C  nonlinear model]{\label{fig:f582}\includegraphics[width=0.55\textwidth]{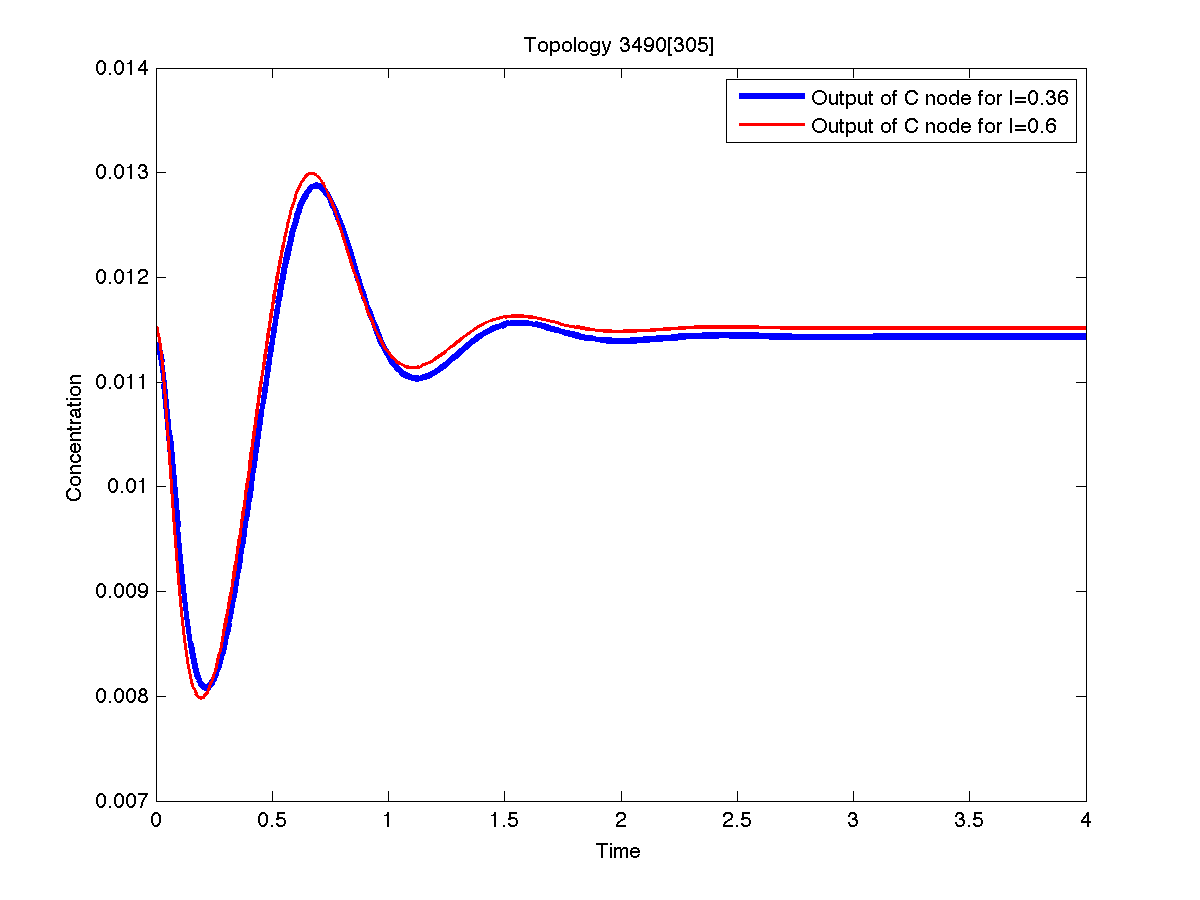}}\\
\subfloat[Quadratic approx. and output of nonlinear system]{\label{fig:f601}\includegraphics[width=0.55\textwidth]{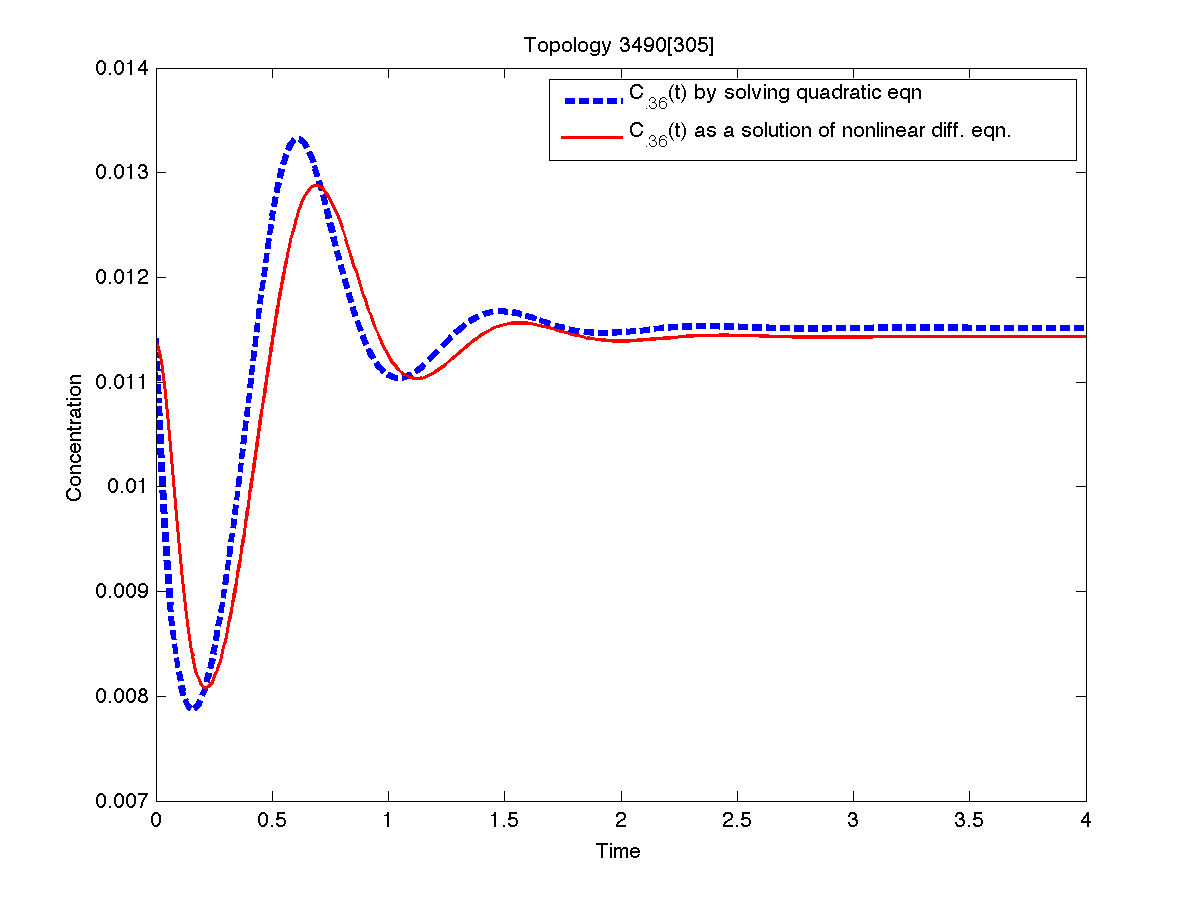}}                
  \subfloat[Quadratic approx. and output of nonlinear system]{\label{fig:f602}\includegraphics[width=0.55\textwidth]{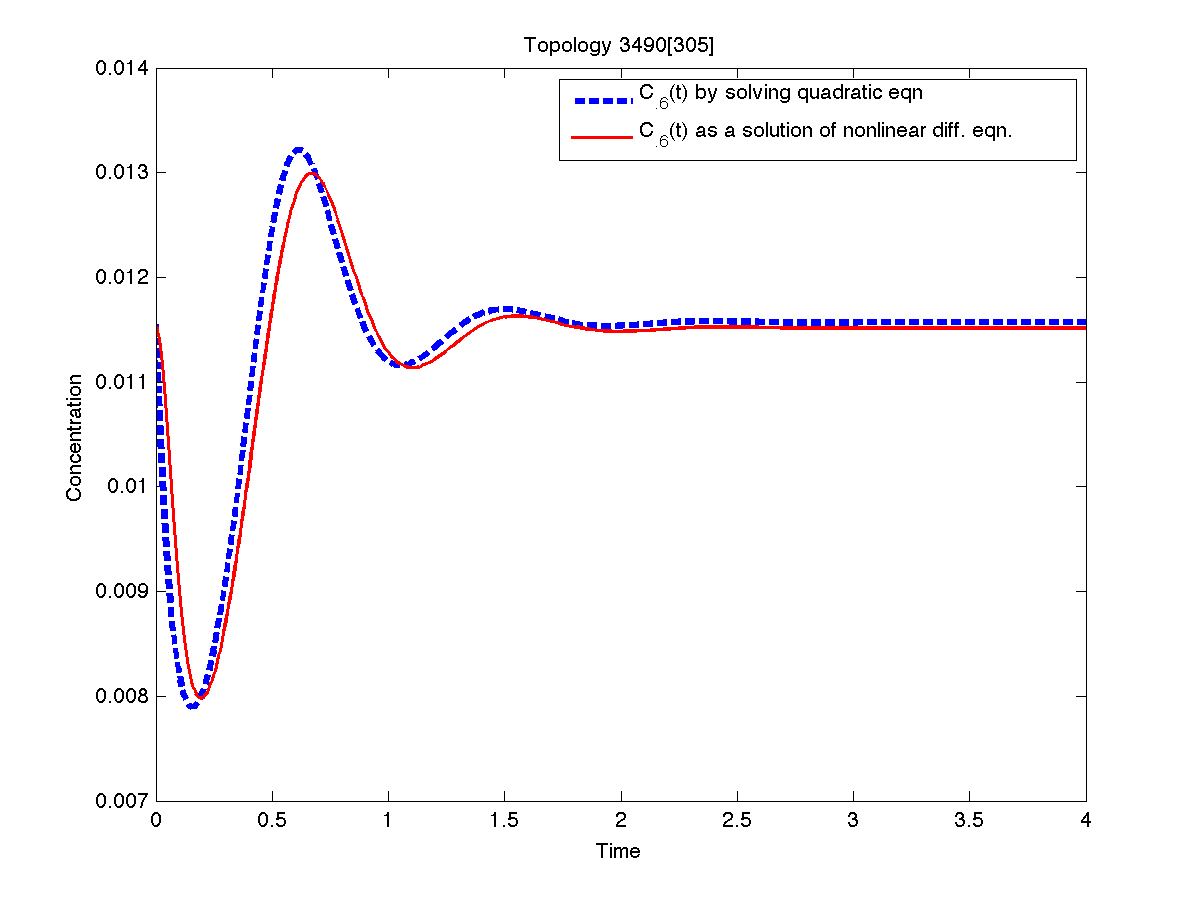}}
\end{figure}

\clearpage
Circuit 21.
\beqn
\dxA&=&k_{{\inp}A} {\inp} \frac{\txA}{\txA+K_{{\inp}A}}-k_{BA} \xB\frac{\xA}{\xA+K_{BA}}+k_{CA} \xC\frac{\txA}{\txA+K_{CA}}\\
\dxB&=&k_{AB}\xA\frac{\txB}{\txB+K_{AB}}-k_{F_BB} \xFB \frac{\xB}{\xB+K_{F_BB}}\\
\dxC&=&{k_{AC}}\xA\frac{\txC}{\txC+K_{AC}}- k_{BC}\xB\frac{\xC}{\xC+K_{BC}}-k_{CC}\xC\frac{\xC}{\xC+K_{CC}}\\
\eeqn
Parameters: \  $K_{{\inp}A}= 0.093918;$ $k_{{\inp}A}= 11.447219;$ 
$K_{BA}= 0.001688;$ $k_{BA}= 44.802268;$
$K_{CA}= 5.026318;$ $k_{CA}= 45.803641;$
$K_{AB}=0.001191;$ $k_{AB}=1.466561;$
 $K_{F_B}=9.424319;$ $k_{F_B}=22.745736;$
$K_{AC}= 0.113697;$ $k_{AC}=1.211993;$
$K_{BC}=0.009891;$  $k_{BC}=7.239357;$
$K_{CC}=0.189125;$ $k_{CC}= 17.910182$

\begin{figure}[hb]
  \centering
\subfloat[Dynamics of A and B in linearized model]{\label{fig:f611}\includegraphics[width=0.55\textwidth]{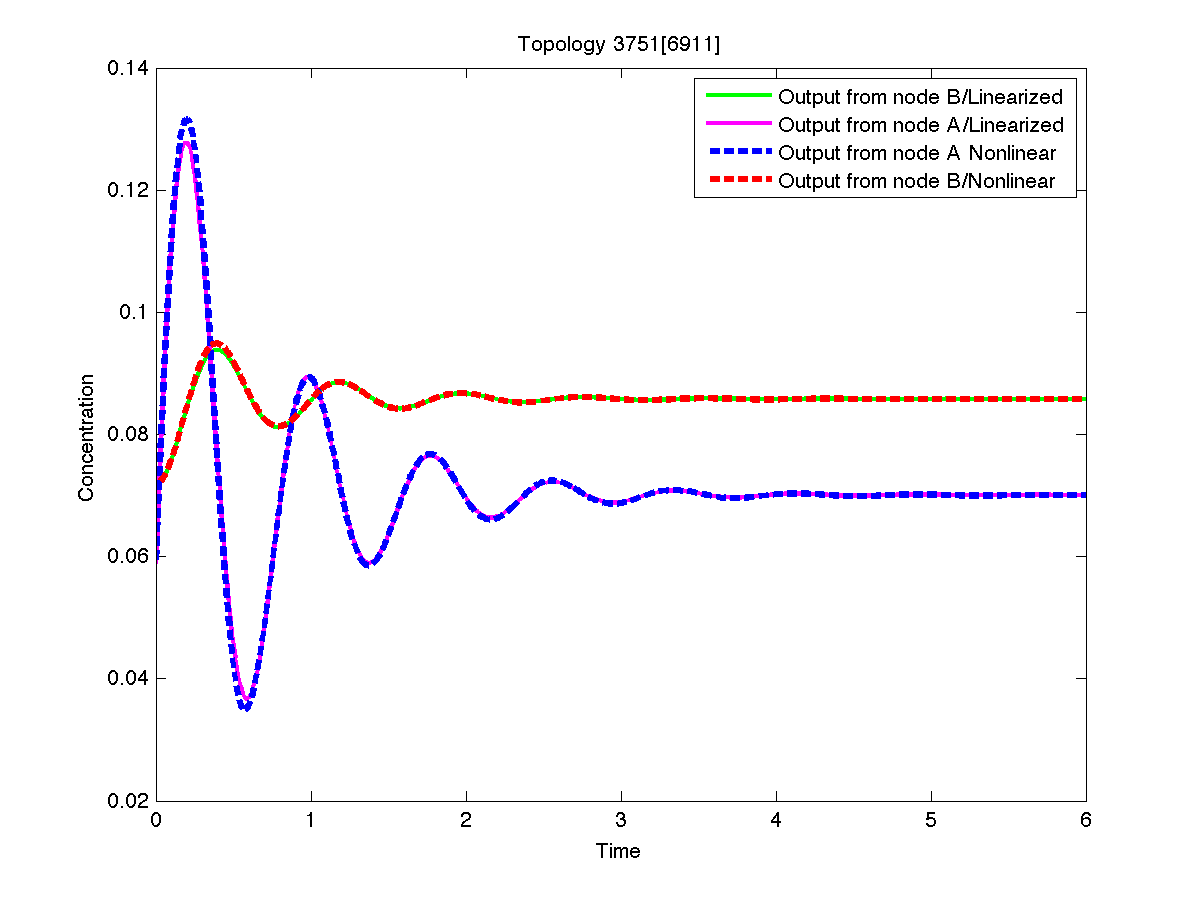}}                
  \subfloat[Ouput from C  nonlinear model]{\label{fig:f612}\includegraphics[width=0.55\textwidth]{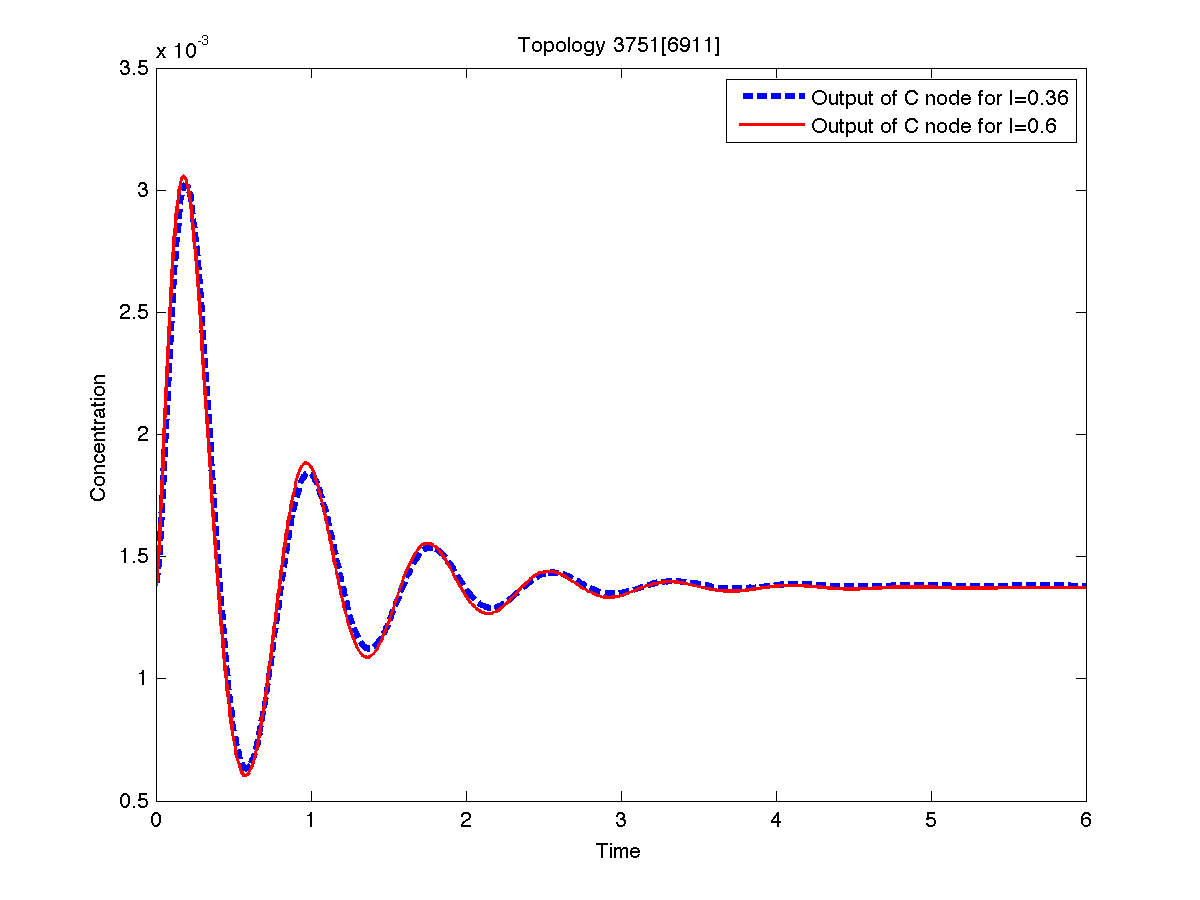}}\\
\subfloat[Quadratic approx. and output of nonlinear system]{\label{fig:f631}\includegraphics[width=0.55\textwidth]{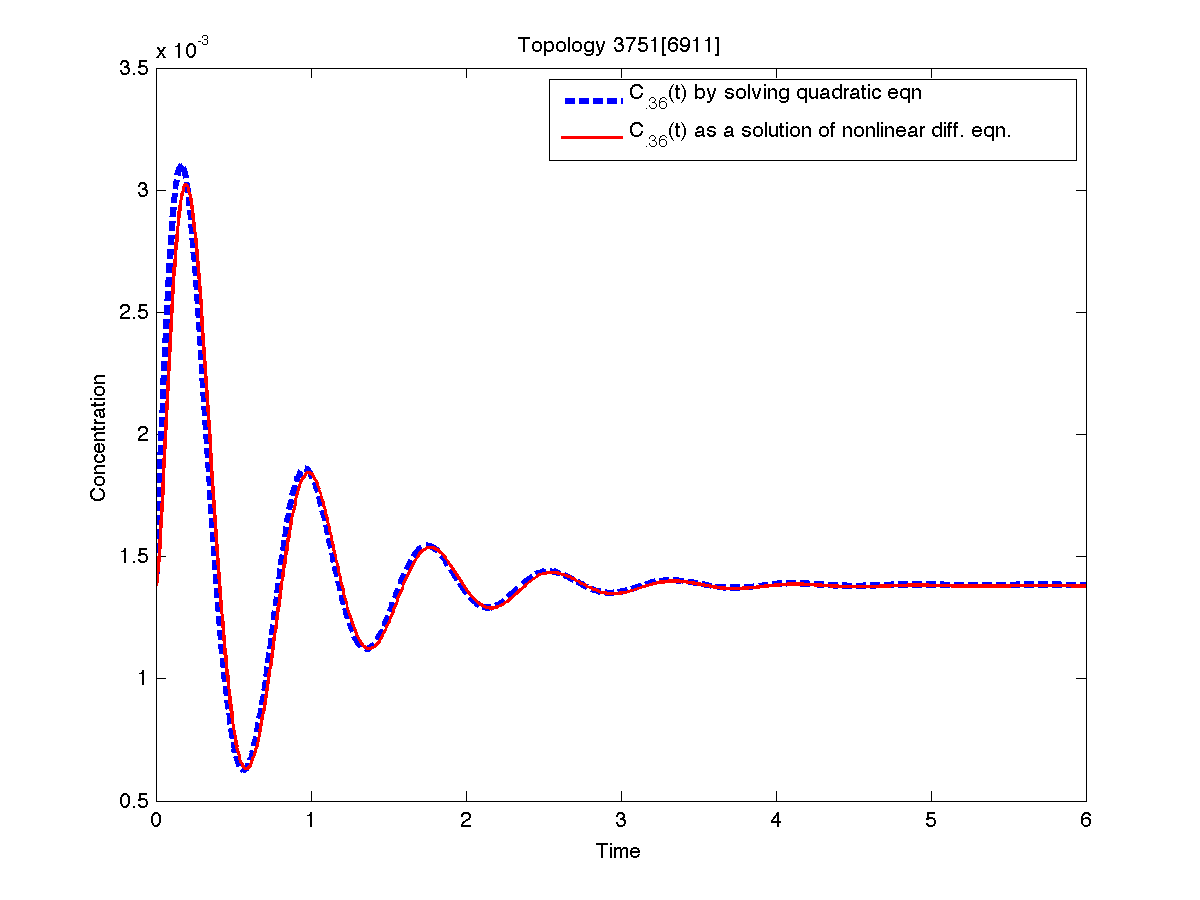}}                
  \subfloat[Quadratic approx. and output of nonlinear system]{\label{fig:f632}\includegraphics[width=0.55\textwidth]{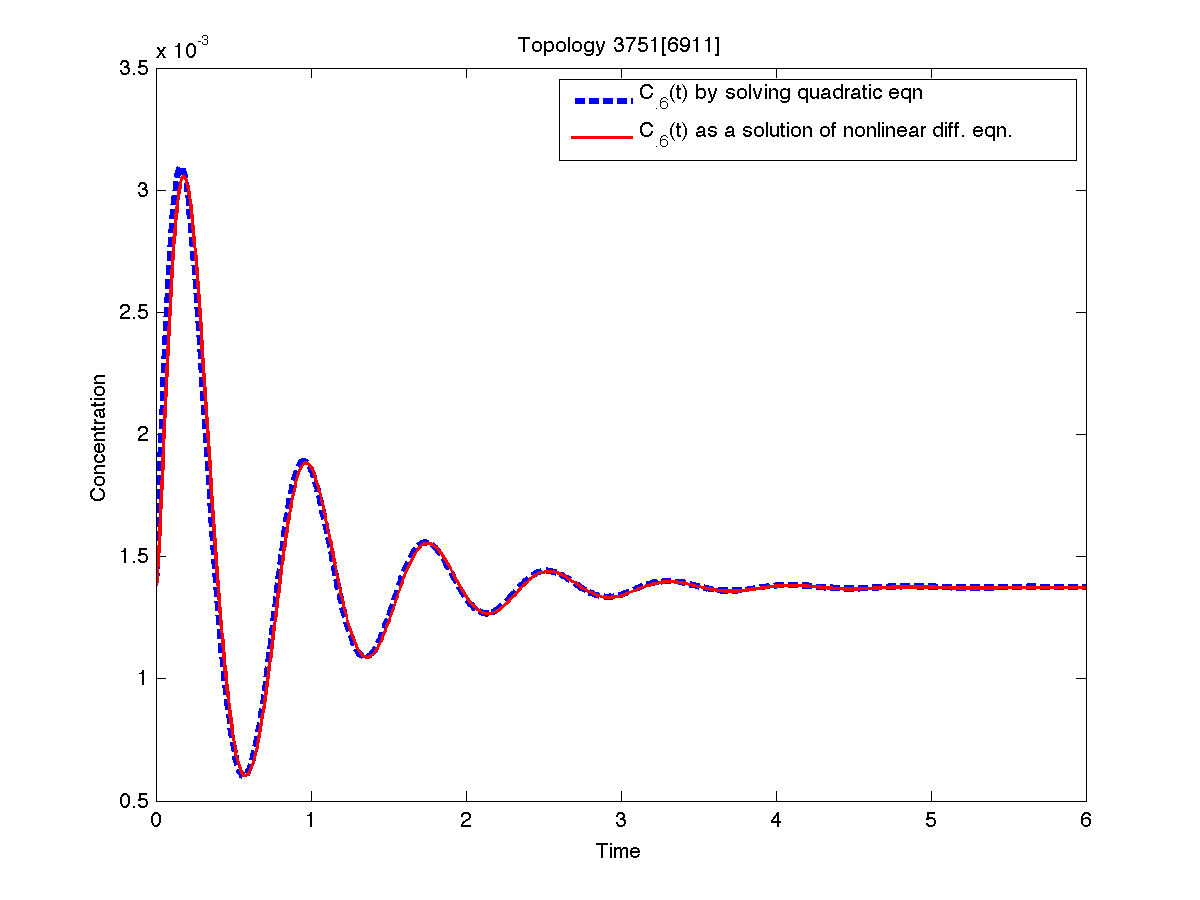}}
\end{figure}

\clearpage
Circuit 22.

This is the same topology as in the previous case, only a different parameter set was used:

Parameters: \  $K_{AB}= 1.620877;$ $k_{AB}= 2.306216;$
 $K_{F_B}= 2.012565;$ $k_{F_B}= 2.700847;$
 $K_{AC}= 0.010933;$ $k_{AC}= 8.968091;$
 $K_{BA}= 0.001812;$ $k_{BA}= 10.039221;$
 $K_{BC}= 0.014199;$ $k_{BC}= 17.762333;$
 $K_{CA}= 0.002690;$ $k_{CA}= 1.506954;$
 $K_{CC}= 2.686891;$ $k_{CC}= 4.139044;$
$K_{{\inp}A}= 0.161715;$ $ k_{{\inp}A}= 1.933303$
\

\begin{figure}[hb]
  \centering
\subfloat[Dynamics of A and B in linearized model]{\label{fig:f611}\includegraphics[width=0.55\textwidth]{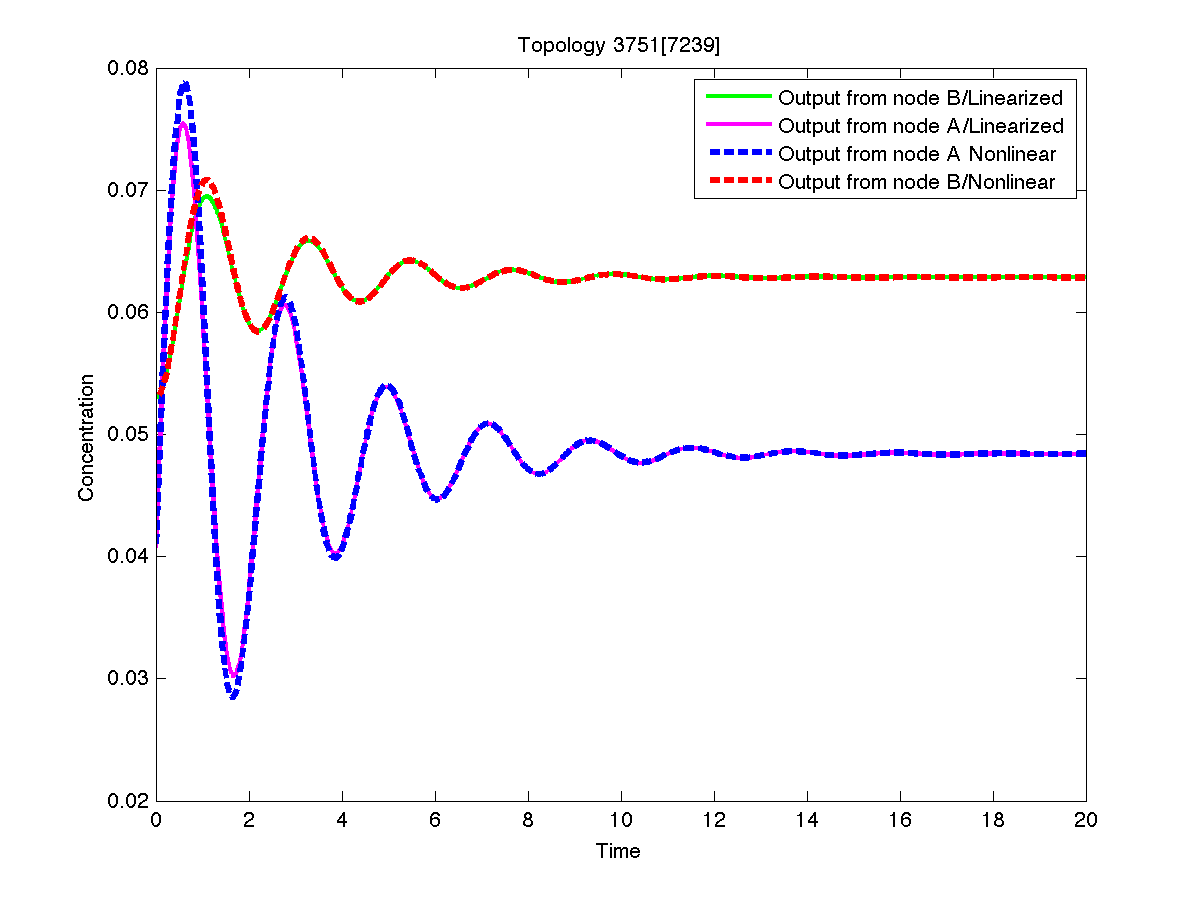}}                
  \subfloat[Ouput from C  nonlinear model]{\label{fig:f612}\includegraphics[width=0.55\textwidth]{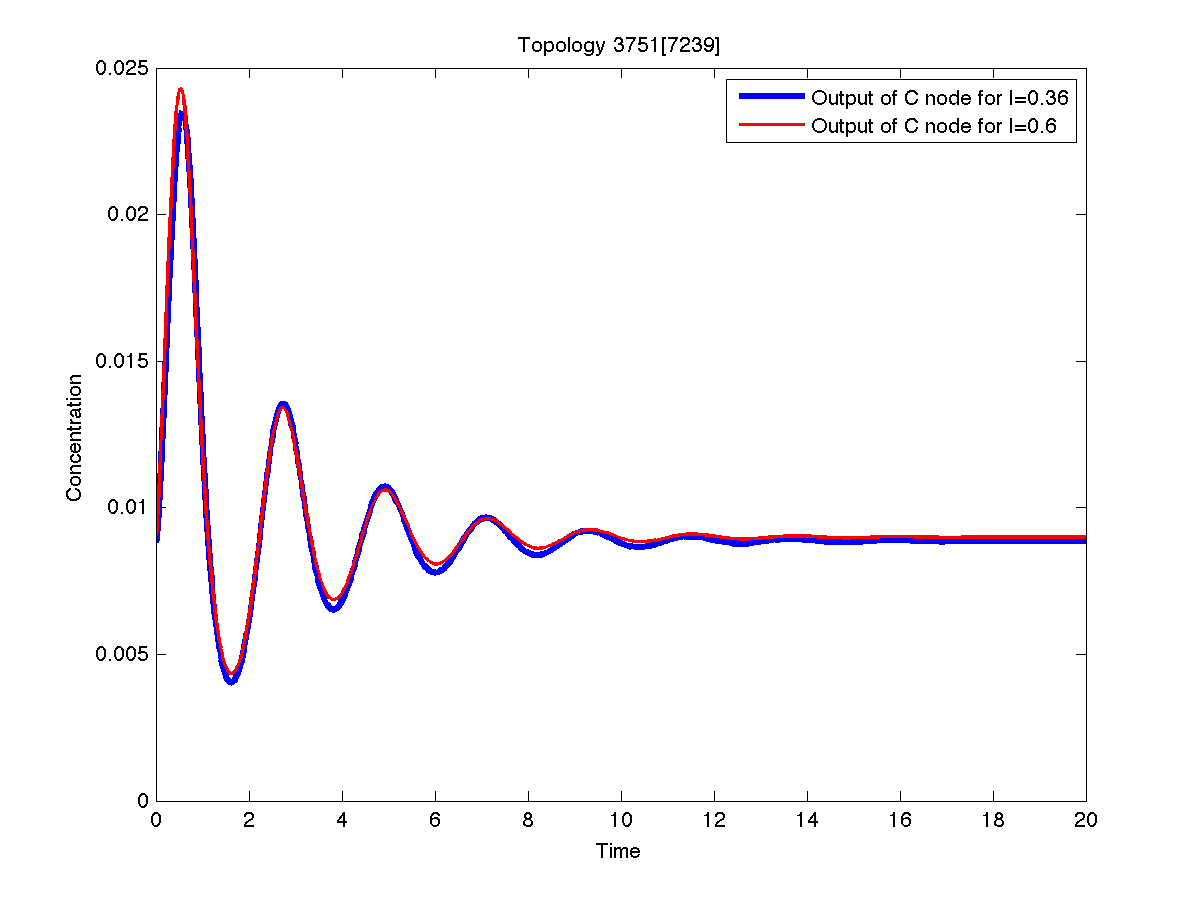}}\\
\subfloat[Quadratic approx. and output of nonlinear system]{\label{fig:f631}\includegraphics[width=0.55\textwidth]{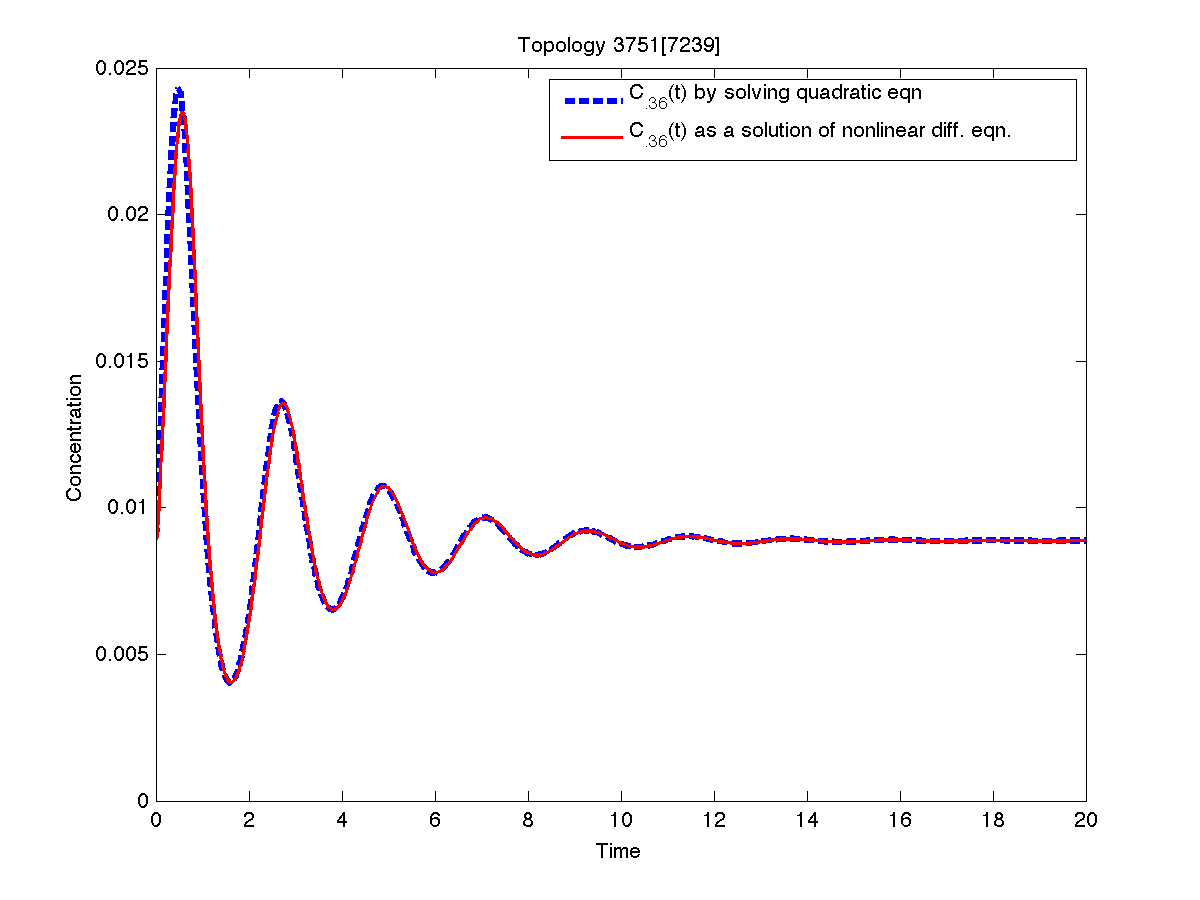}}                
  \subfloat[Quadratic approx. and output of nonlinear system]{\label{fig:f632}\includegraphics[width=0.55\textwidth]{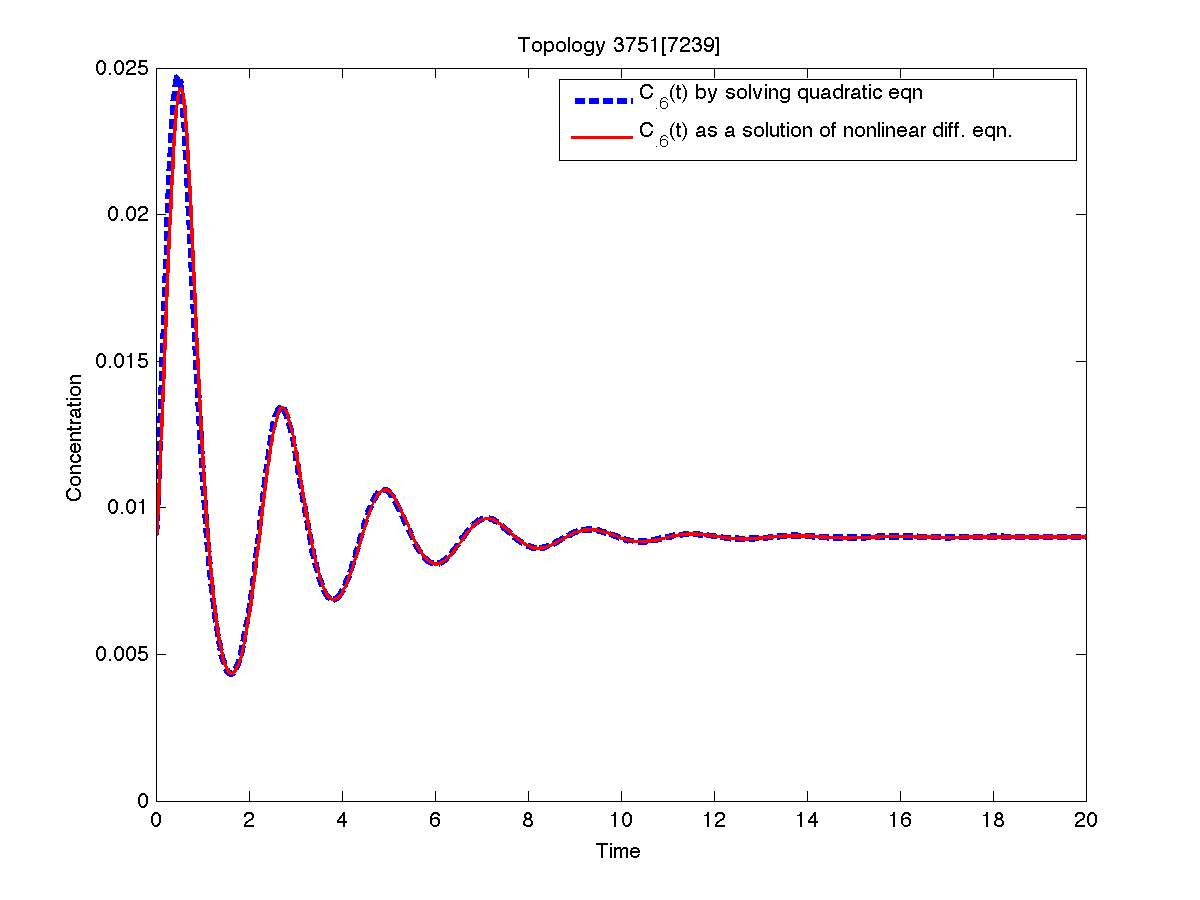}}
\end{figure}

Circuit 23.
\beqn
\dxA&=&k_{{\inp}A} {\inp} \frac{\txA}{\txA+K_{{\inp}A}}-k_{BA} \xB\frac{\xA}{\xA+K_{BA}}-k_{CA} \xC\frac{\xA}{\xA+K_{CA}}\\
\dxB&=&k_{AB}\xA\frac{\txB}{\txB+K_{AB}}-k_{F_BB} \xFB \frac{\xB}{\xB+K_{F_BB}}\\
\dxC&=&{k_{AC}}\xA\frac{\txC}{\txC+K_{AC}}-k_{BC}\xB\frac{\xC}{\xC+K_{BC}}-k_{CC}\xC\frac{\xC}{\xC+K_{CC}}\\
\eeqn
Parameters: \  $K_{{\inp}A}= 0.093918;$ $k_{{\inp}A}= 11.447219;$ 
$K_{BA}= 0.001688;$ $k_{BA}= 44.802268;$
 $K_{CA}= 90.209027;$ $k_{CA}= 96.671843;$
$K_{AB}=0.001191;$ $k_{AB}=1.466561;$
 $K_{F_B}=9.424319;$ $k_{F_B}=22.745736;$
 $K_{AC}= 0.113697;$ $k_{AC}=1.211993;$
$K_{BC}=0.009891;$ $k_{BC}=7.239357;$
$K_{CC}=0.189125;$ $k_{CC}= 17.910182$

\begin{figure}[ht]
  \centering
\subfloat[Dynamics of A and B in linearized model]{\label{fig:f671}\includegraphics[width=0.55\textwidth]{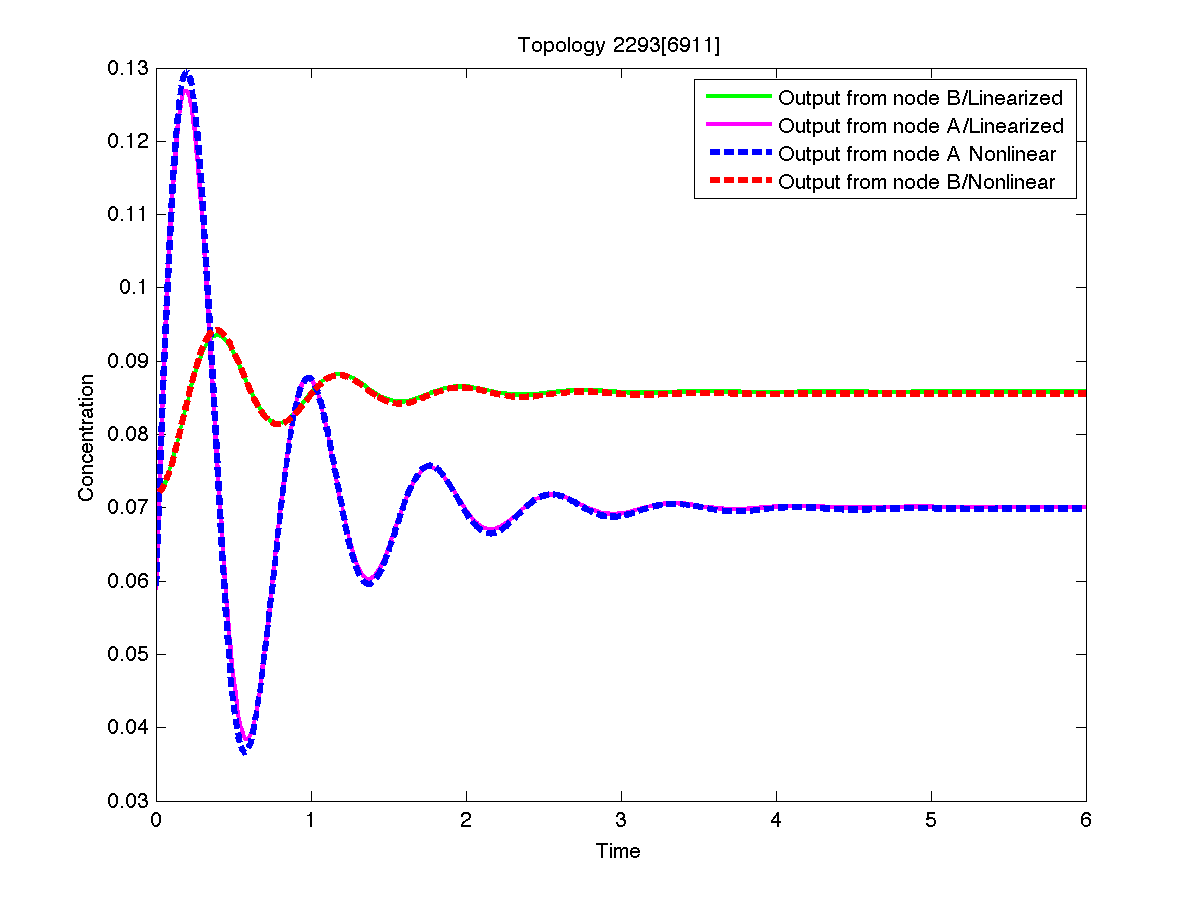}}                
  \subfloat[Ouput from C  nonlinear model]{\label{fig:f672}\includegraphics[width=0.55\textwidth]{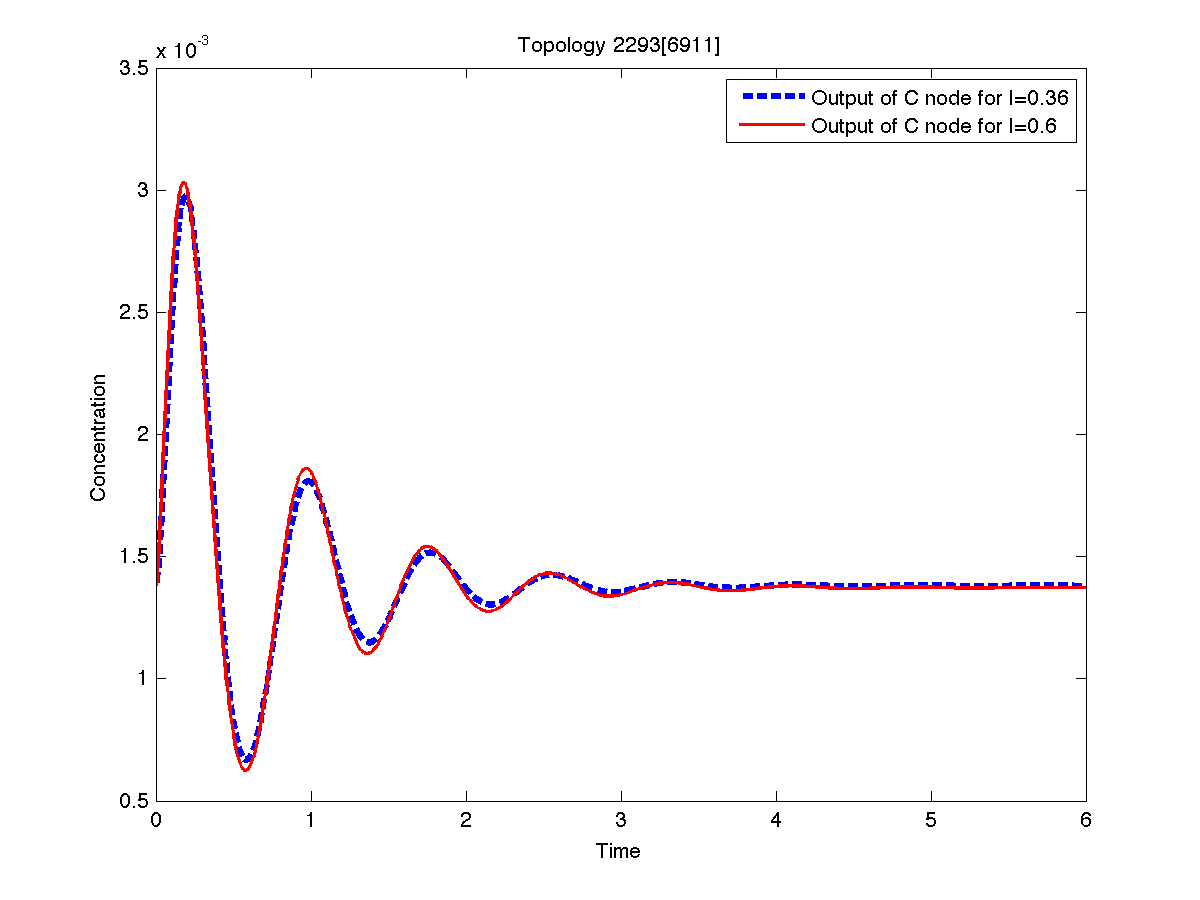}}\\
\subfloat[Quadratic approx. and output of nonlinear system]{\label{fig:f691}\includegraphics[width=0.55\textwidth]{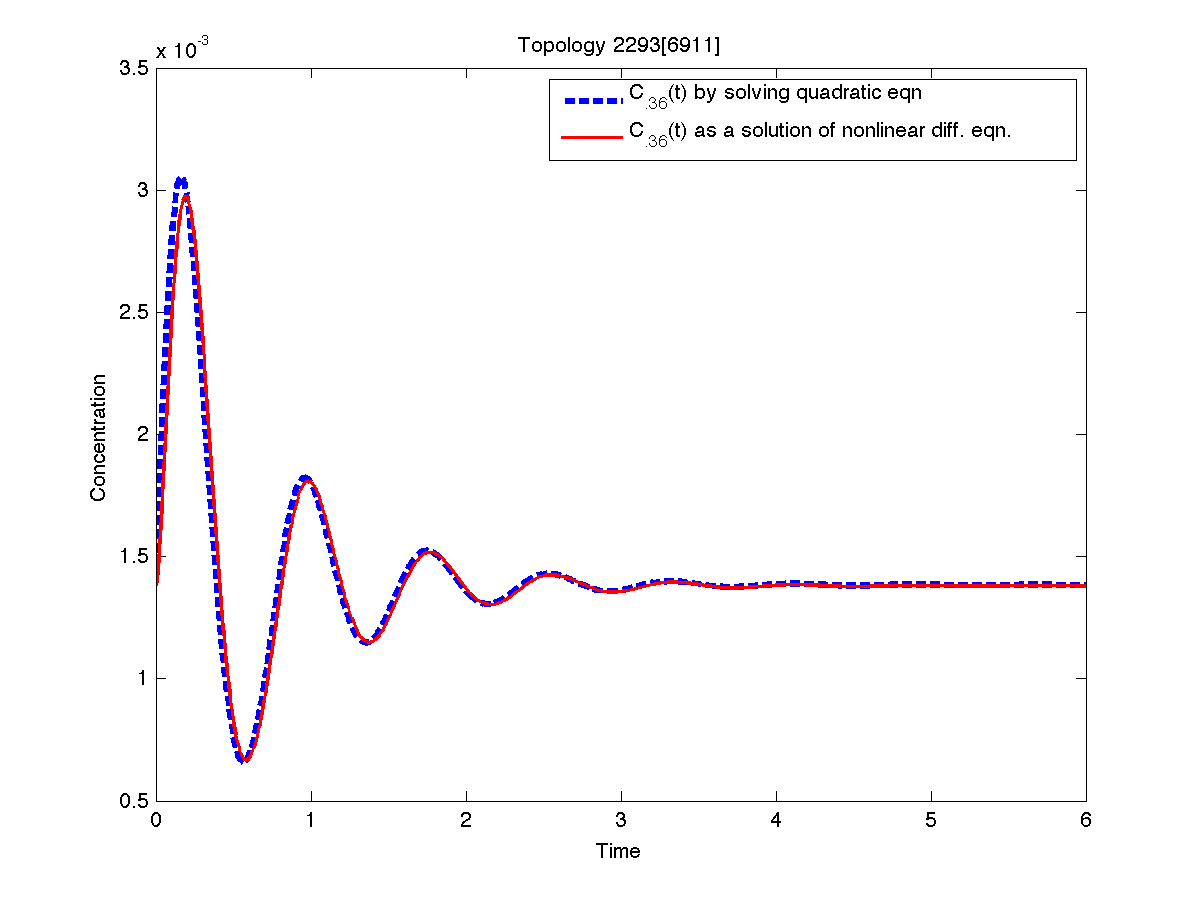}}                
  \subfloat[Quadratic approx. and output of nonlinear system]{\label{fig:f692}\includegraphics[width=0.55\textwidth]{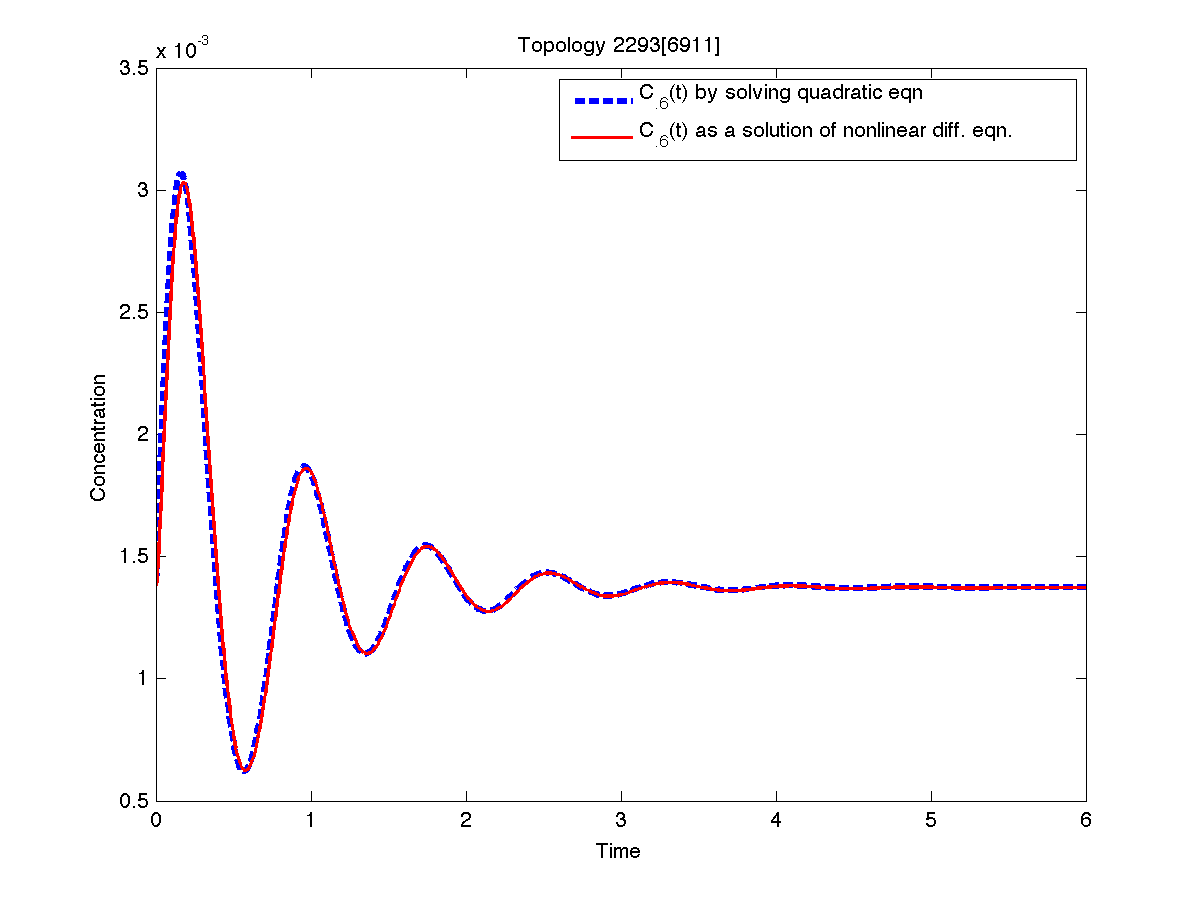}}
\end{figure}

\clearpage

Circuit 24.
\beqn
\dxA&=&k_{{\inp}A} {\inp} \frac{\txA}{\txA+K_{{\inp}A}}-k_{BA}
\xB\frac{\xA}{\xA+K_{BA}}+k_{CA} \xC\frac{\txA}{\txA+K_{CA}}\\
\dxB&=&k_{AB}\xA\frac{\txB}{\txB+K_{AB}}-k_{F_BB} \xFB \frac{\xB}{\xB+K_{F_BB}}\\
\dxC&=&{k_{AC}}\xA\frac{\txC}{\txC+K_{AC}}- k_{BC}\xB\frac{\xC}{\xC+K_{BC}}\\
\eeqn
Parameters: \   $K_{{\inp}A}= 0.093918;$ $k_{{\inp}A}= 11.447219;$ 
$K_{BA}= 0.001688;$ $k_{BA}= 44.802268;$
$K_{CA}= 5.026318;$ $k_{CA}= 45.803641;$
$K_{AB}=0.001191;$ $k_{AB}=1.466561;$
 $K_{F_B}=9.424319;$ $k_{F_B}=22.745736;$
 $K_{AC}= 0.113697;$ $k_{AC}=1.211993;$
$K_{BC}=0.009891;$  $k_{BC}=7.239357$

\begin{figure}[hb]
  \centering
\subfloat[Dynamics of A and B in linearized model]{\label{fig:f701}\includegraphics[width=0.55\textwidth]{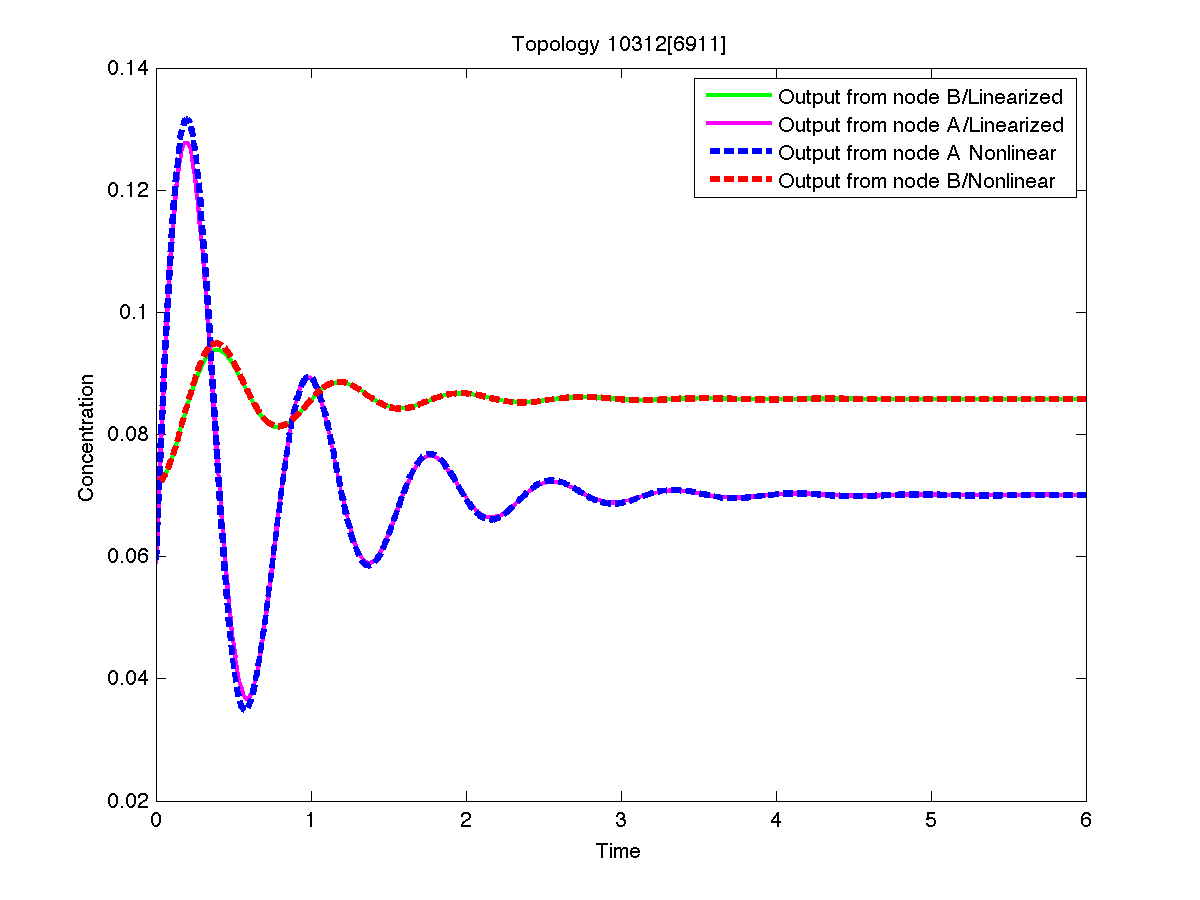}}                
  \subfloat[Ouput from C  nonlinear model]{\label{fig:f702}\includegraphics[width=0.55\textwidth]{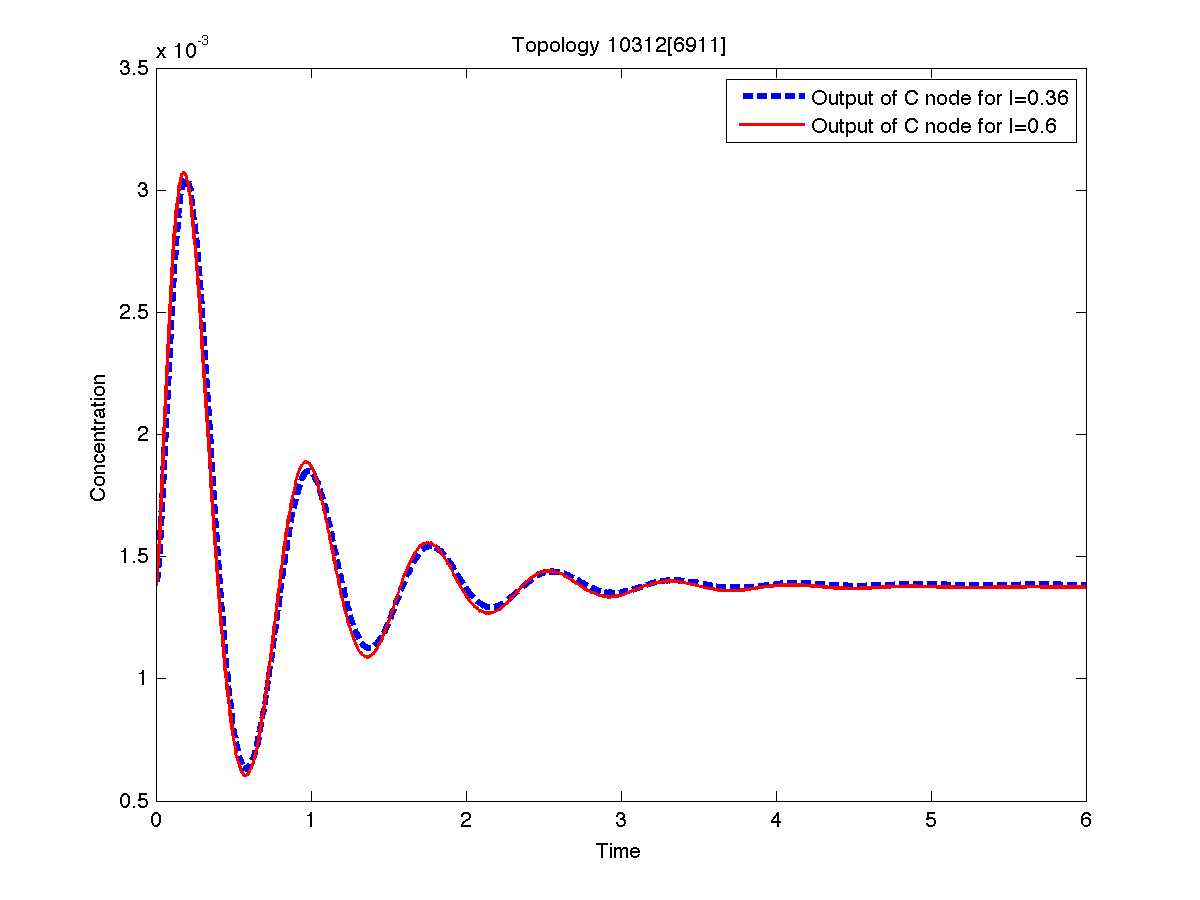}}\\
\subfloat[Quadratic approx. and output of nonlinear system]{\label{fig:f721}\includegraphics[width=0.55\textwidth]{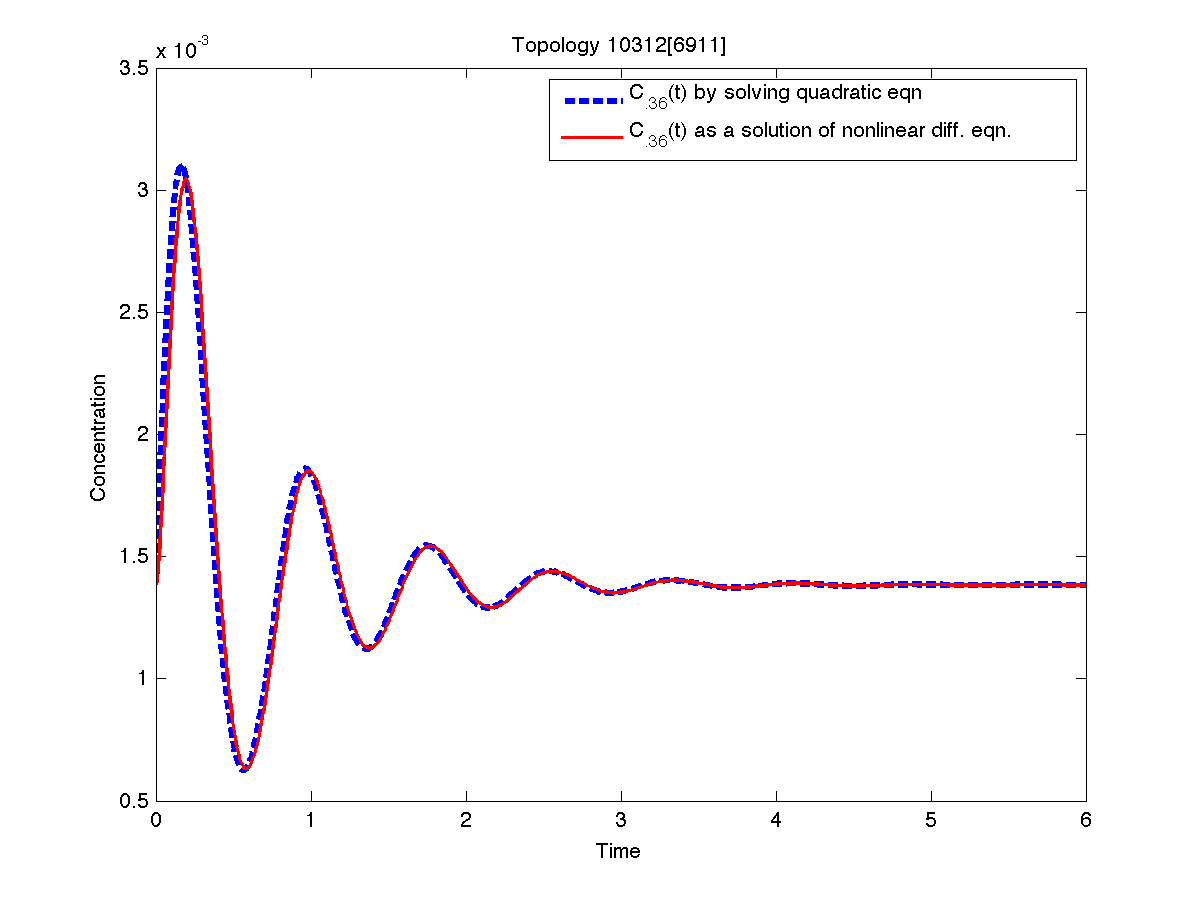}}                
  \subfloat[Quadratic approx. and output of nonlinear system]{\label{fig:f722}\includegraphics[width=0.55\textwidth]{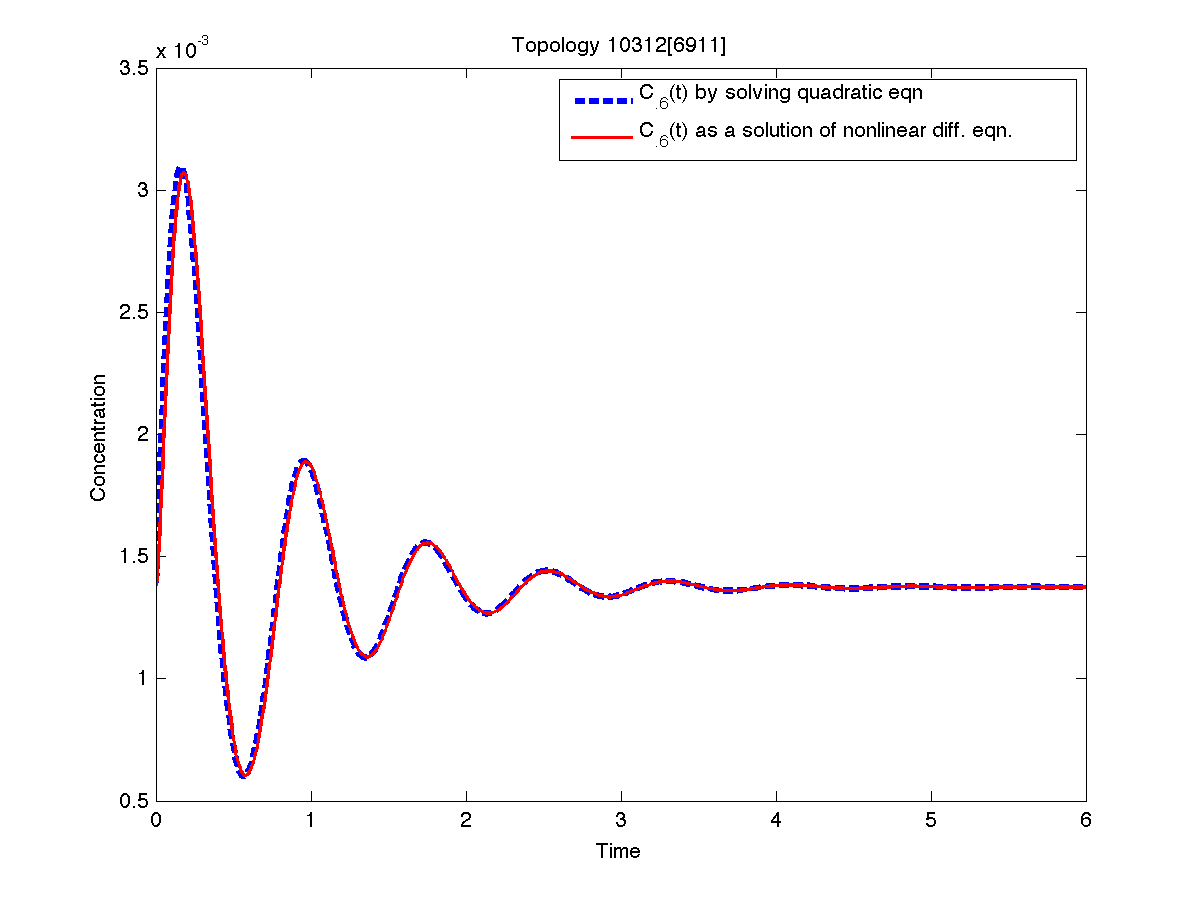}}
\end{figure}

\clearpage

Circuit 25. 

This is the same topology as in the previous case, only a different parameter set was used:

 $K_{AB}= 1.620877;$ $k_{AB}= 2.306216;$
 $K_{F_B}= 2.012565;$ $k_{F_B}= 2.700847;$
 $K_{AC}= 0.010933;$ $k_{AC}= 8.968091;$
 $K_{BA}= 0.001812;$ $k_{BA}= 10.039221;$
 $K_{BC}= 0.014199;$ $k_{BC}= 17.762333;$
 $K_{CA}= 0.002690;$ $k_{CA}= 1.506954;$
 $K_{{\inp}A}= 0.161715;$ $k_{{\inp}A}= 1.93330$

\begin{figure}[hb]
  \centering
\subfloat[Dynamics of A and B in linearized model]{\label{fig:f731}\includegraphics[width=0.55\textwidth]{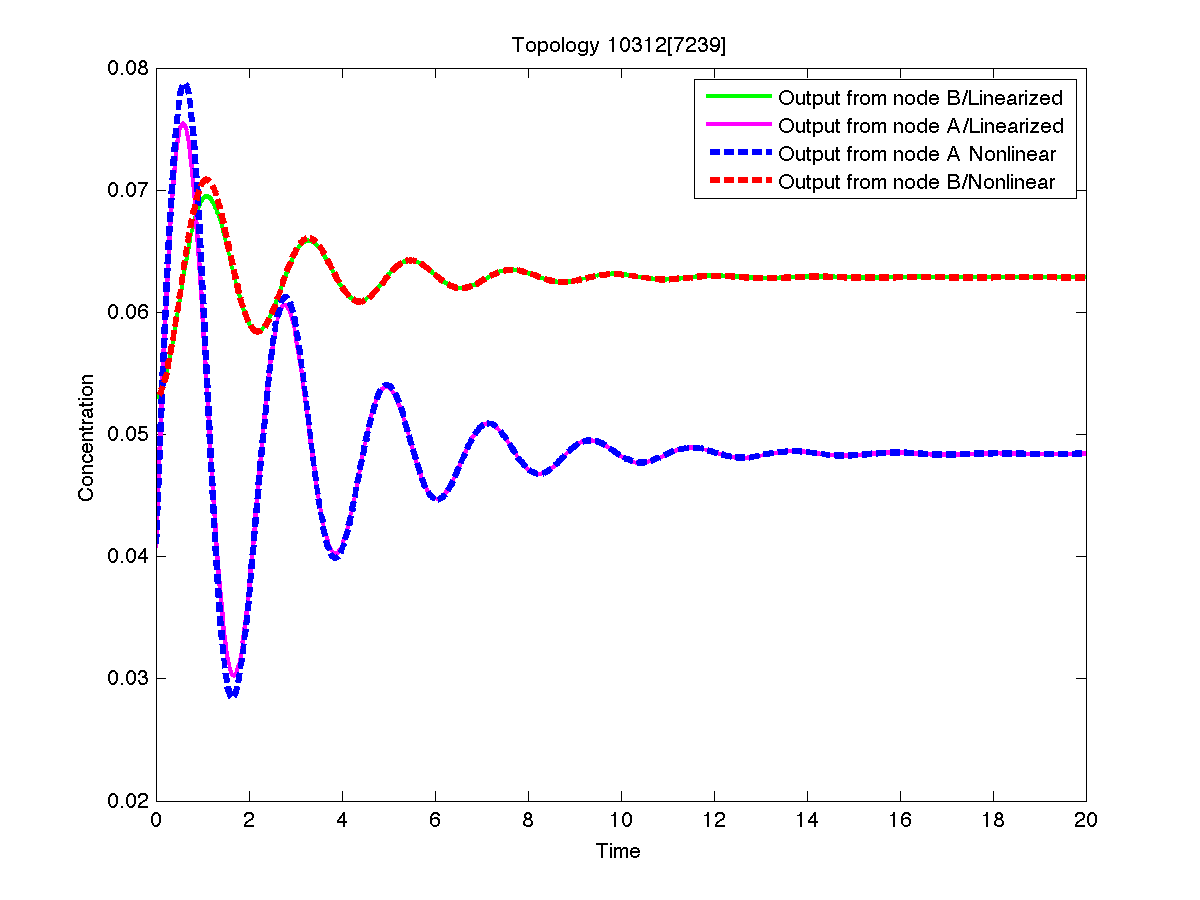}}
  \subfloat[Ouput from C  nonlinear model]{\label{fig:f732}\includegraphics[width=0.55\textwidth]{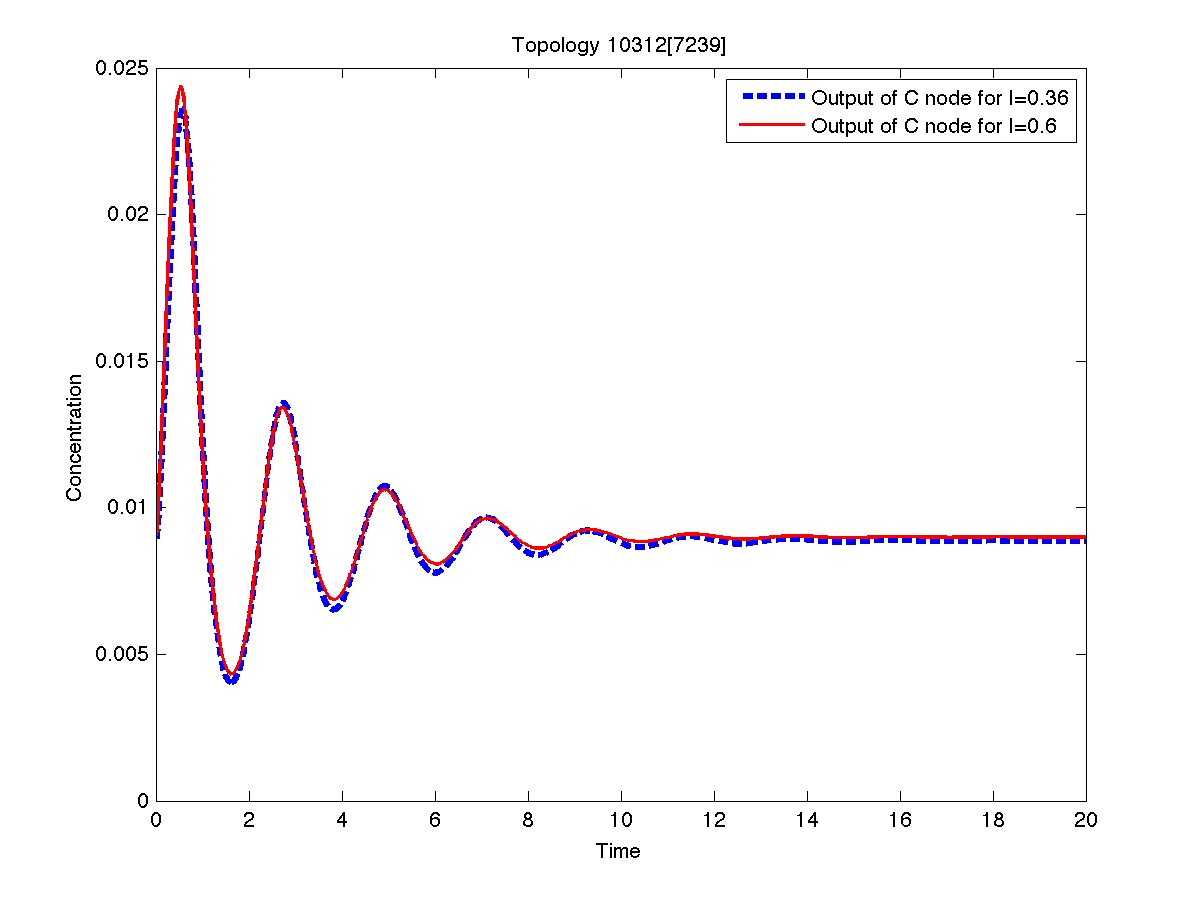}}\\
\subfloat[Quadratic approx. and output of nonlinear system]{\label{fig:f751}\includegraphics[width=0.55\textwidth]{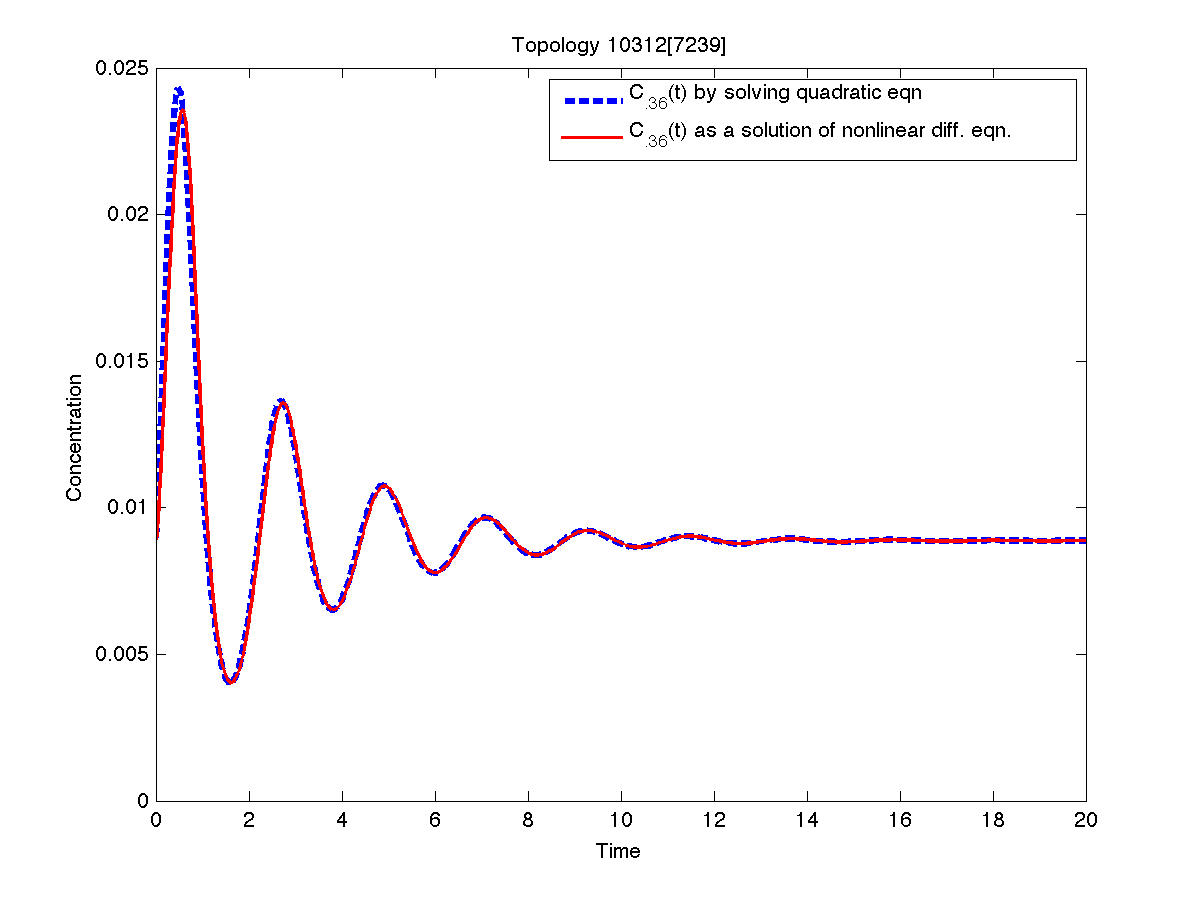}}  
  \subfloat[Quadratic approx. and output of nonlinear system]{\label{fig:f752}\includegraphics[width=0.55\textwidth]{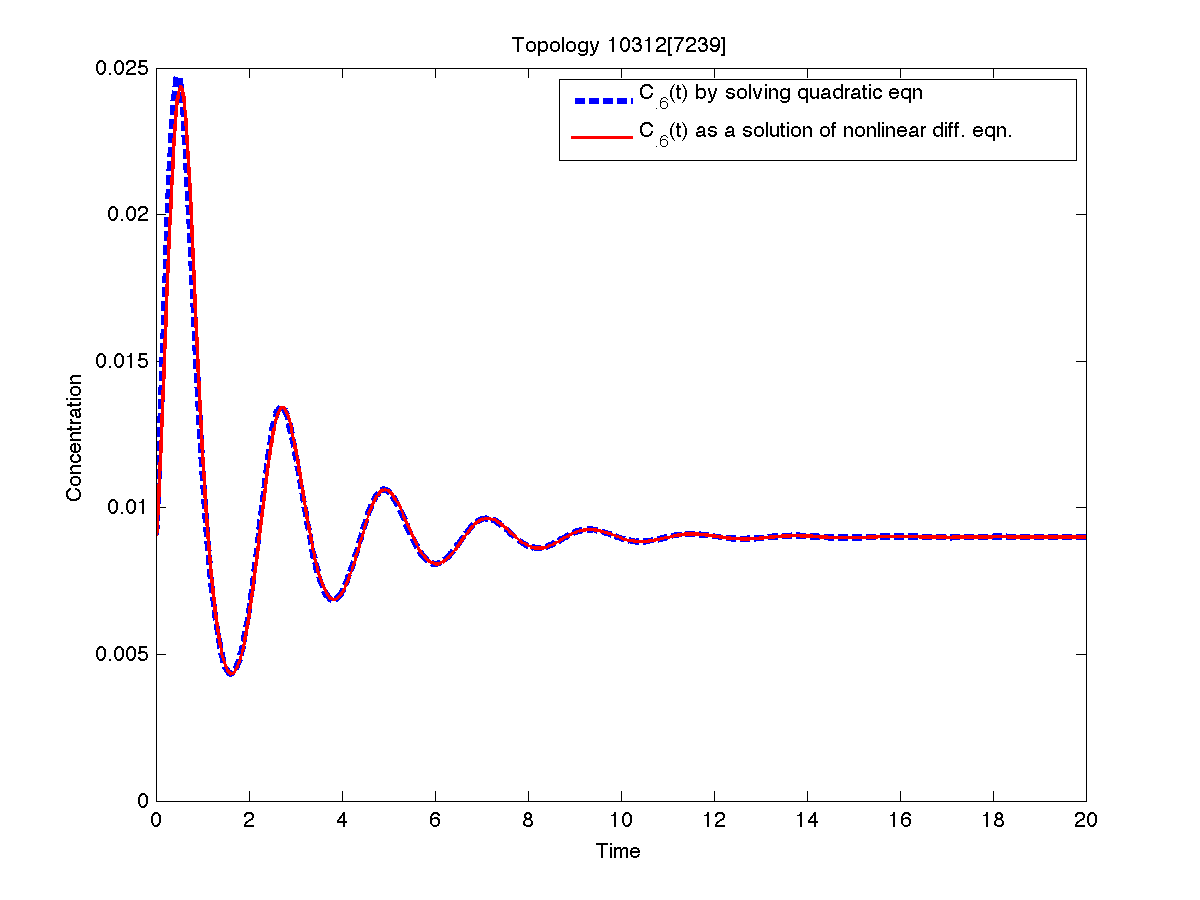}}
\end{figure}

\clearpage

\section{Ratios $\xA(t)/ \xB(t)$}

In this section, for each ASI circuit, we show that the ratio $\xA(t)/ \xB(t)$
is  approximately invariant when inputs are scaled, as discussed in the Main Text.

\clearpage

\begin{figure}[ht]
  \centering
\subfloat{\label{fig:f21}\includegraphics[width=0.55\textwidth]{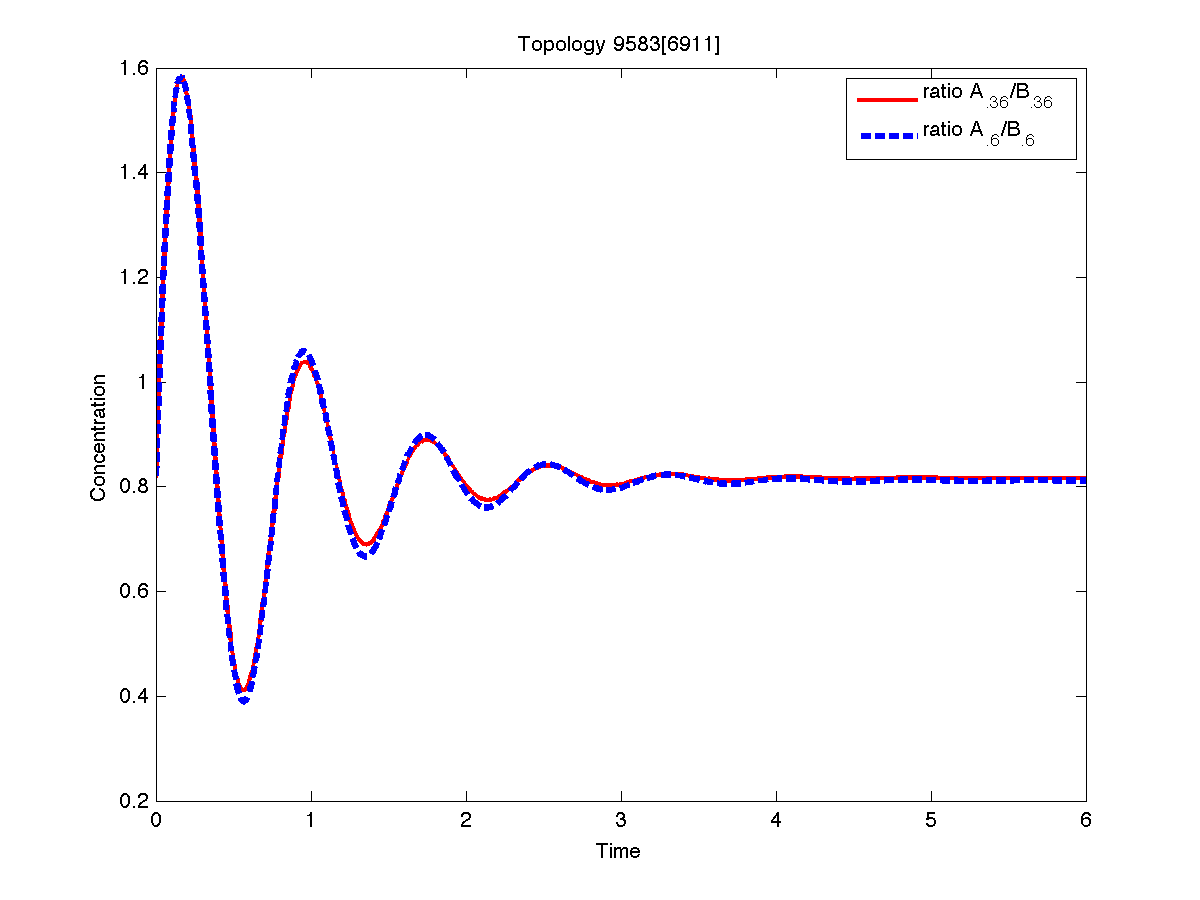}}        
\subfloat{\label{fig:f22}\includegraphics[width=0.55\textwidth]{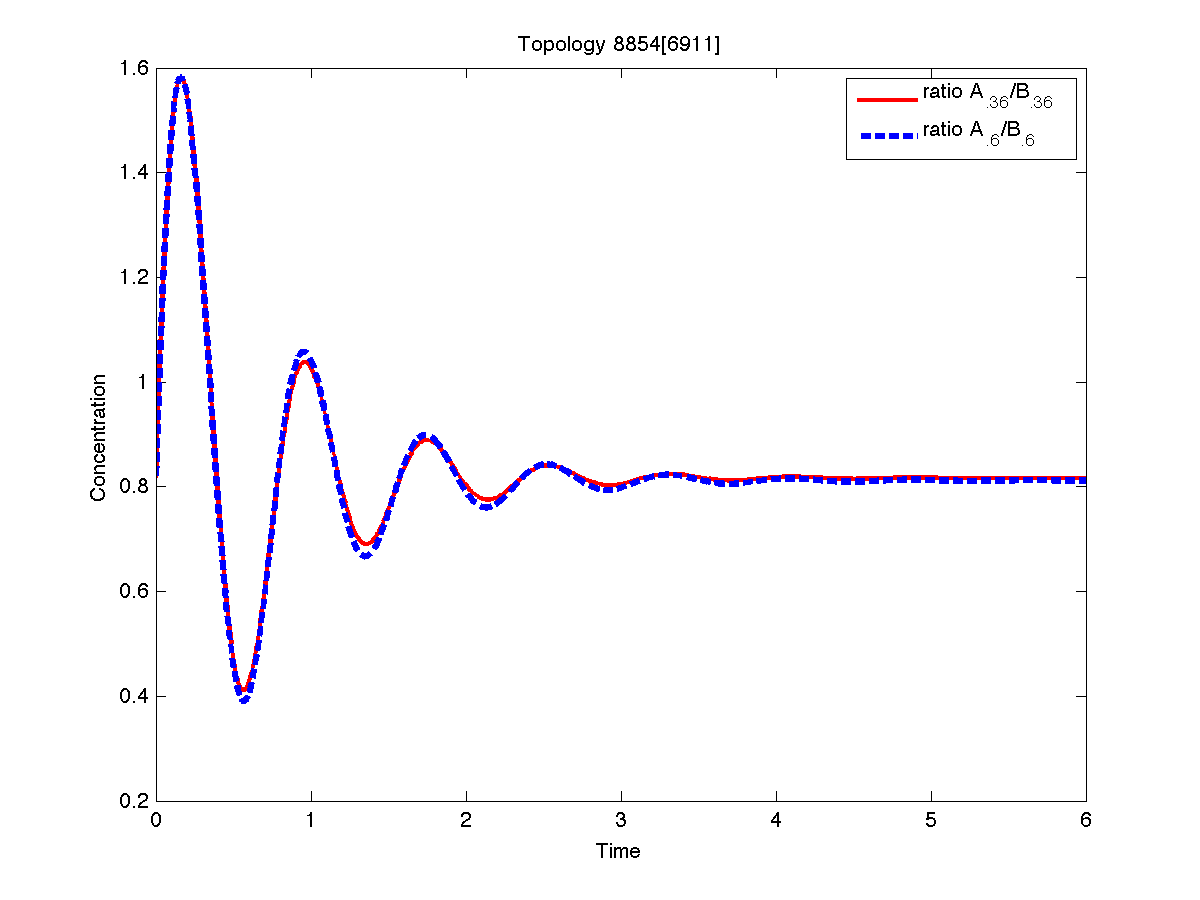}}\\
\subfloat{\label{fig:f31}\includegraphics[width=0.55\textwidth]{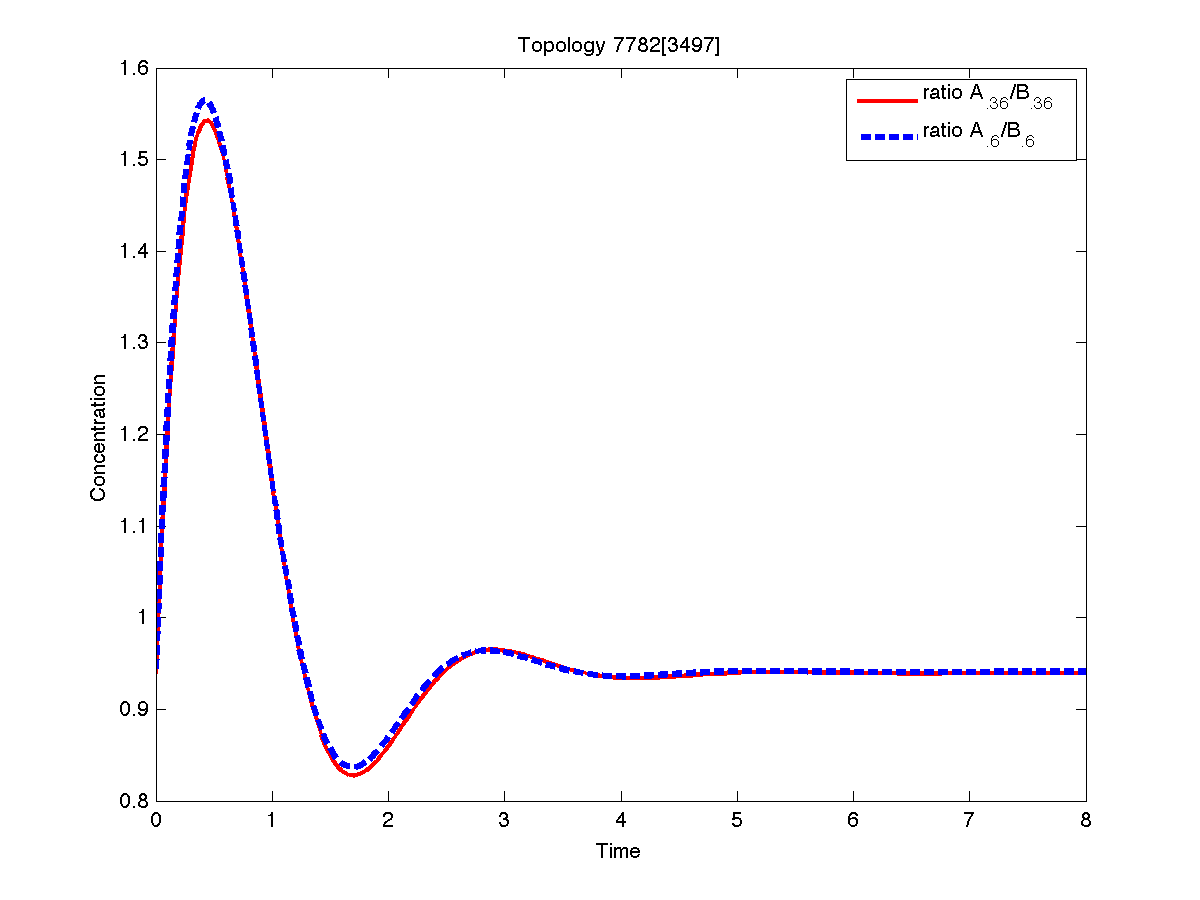}}         
\subfloat{\label{fig:f32}\includegraphics[width=0.55\textwidth]{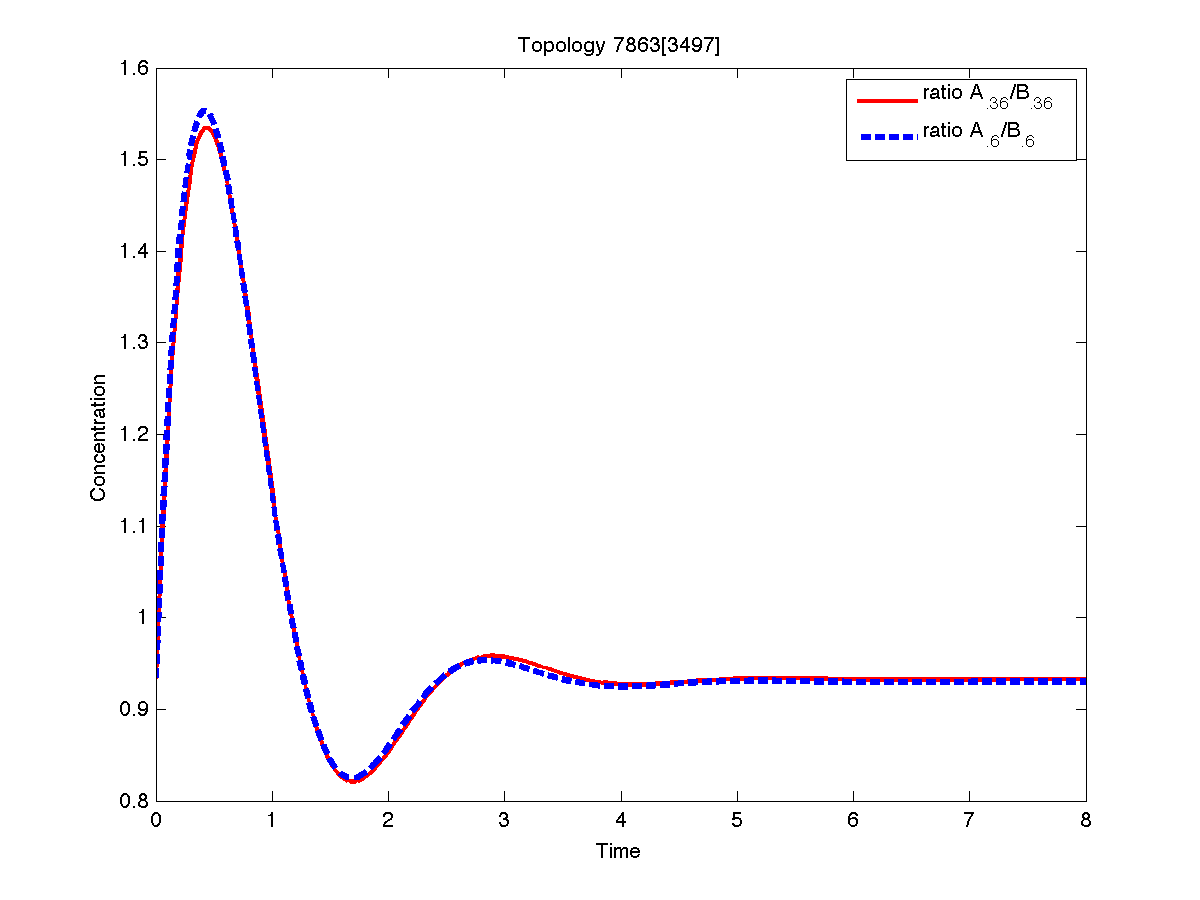}}\\
\subfloat{\label{fig:f141}\includegraphics[width=0.55\textwidth]{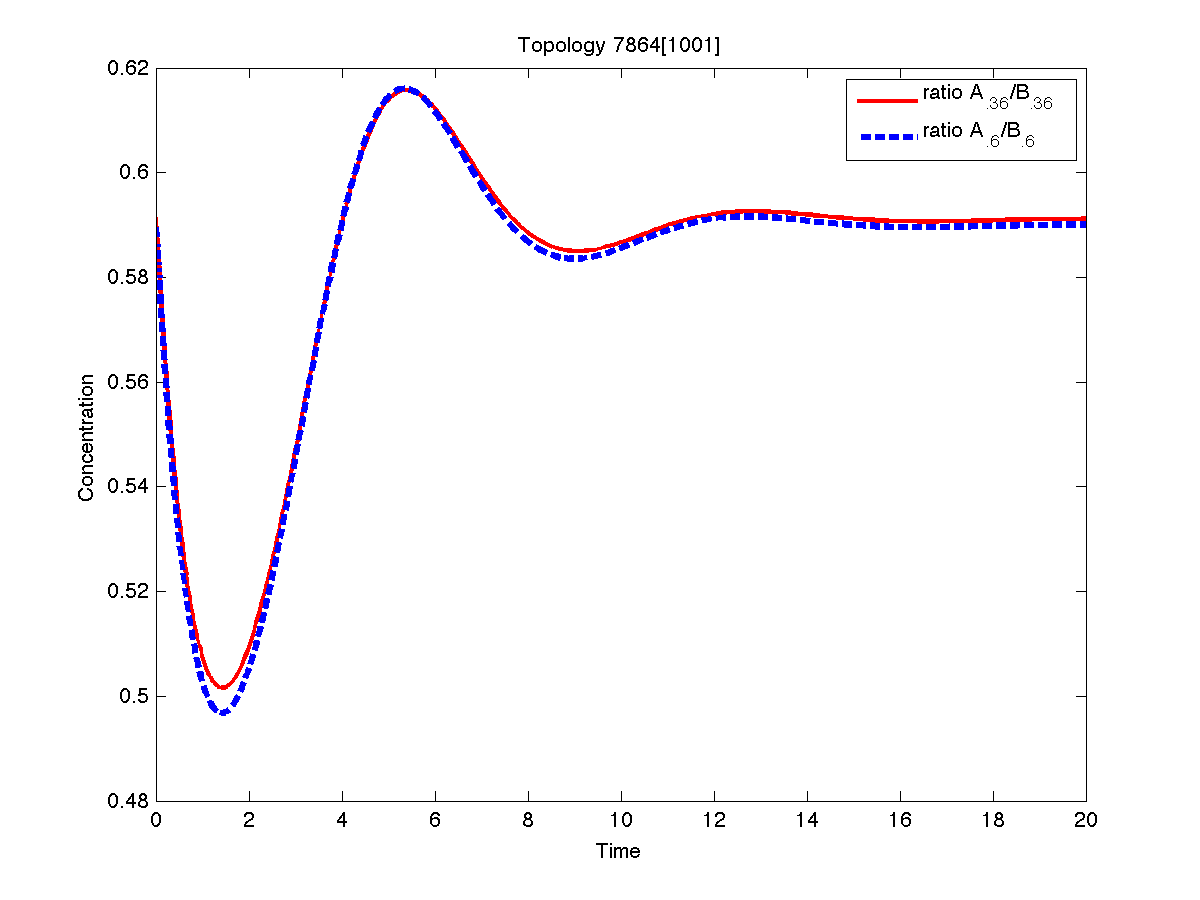}}                
 \subfloat{\label{fig:f142}\includegraphics[width=0.55\textwidth]{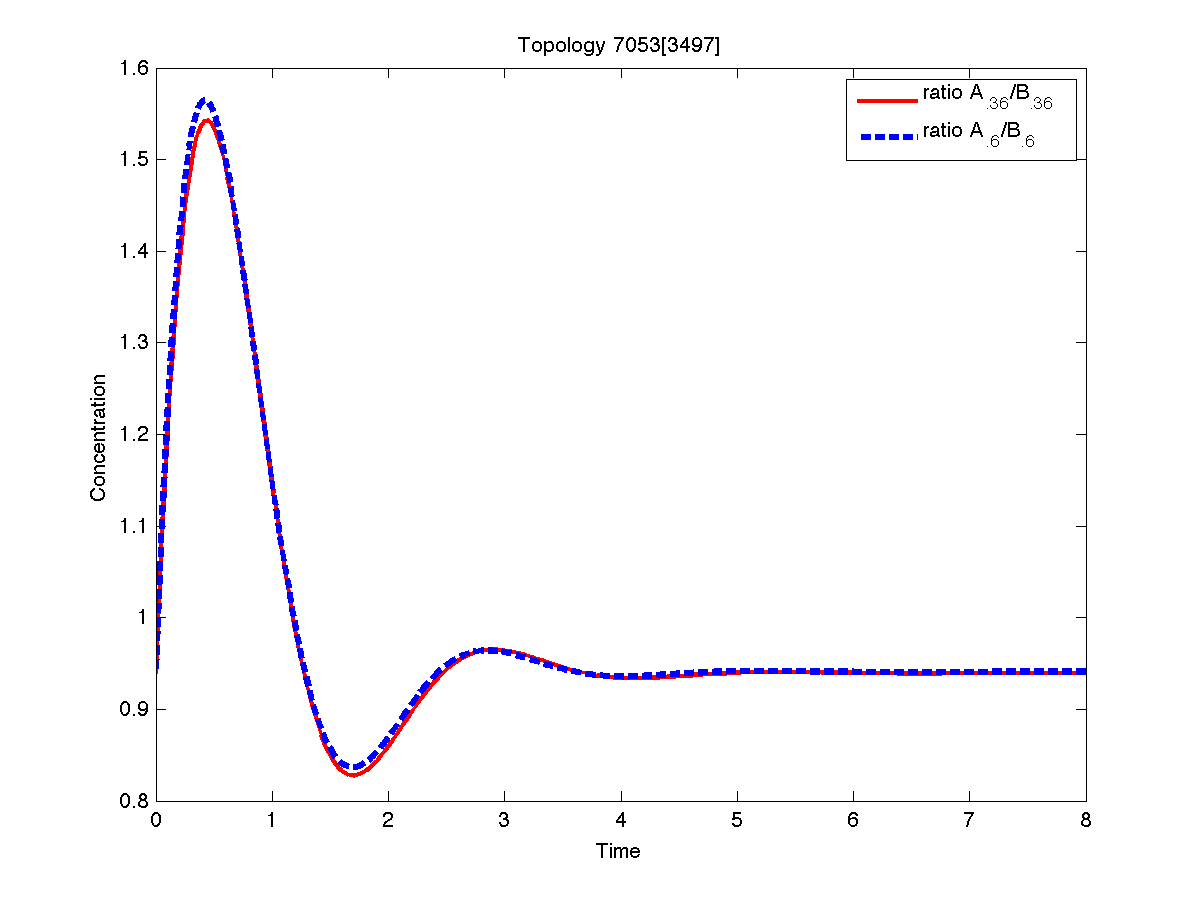}}\\
\caption{$\xA(t)/ \xB(t)$ for circuits 1-6}
\end{figure}

\clearpage

\begin{figure}[ht]
 \centering
\subfloat{\label{fig:f51}\includegraphics[width=0.55\textwidth]{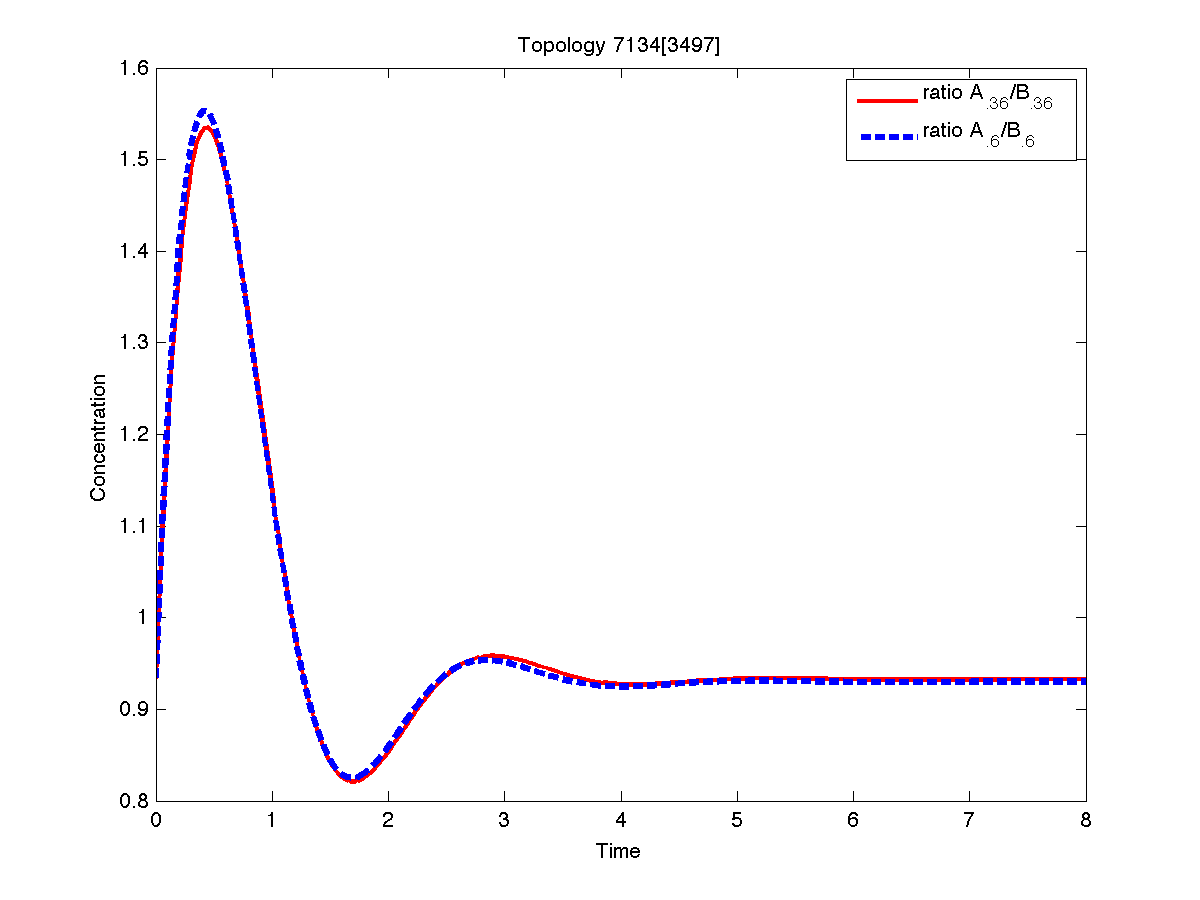}}                
\subfloat{\label{fig:f52}\includegraphics[width=0.55\textwidth]{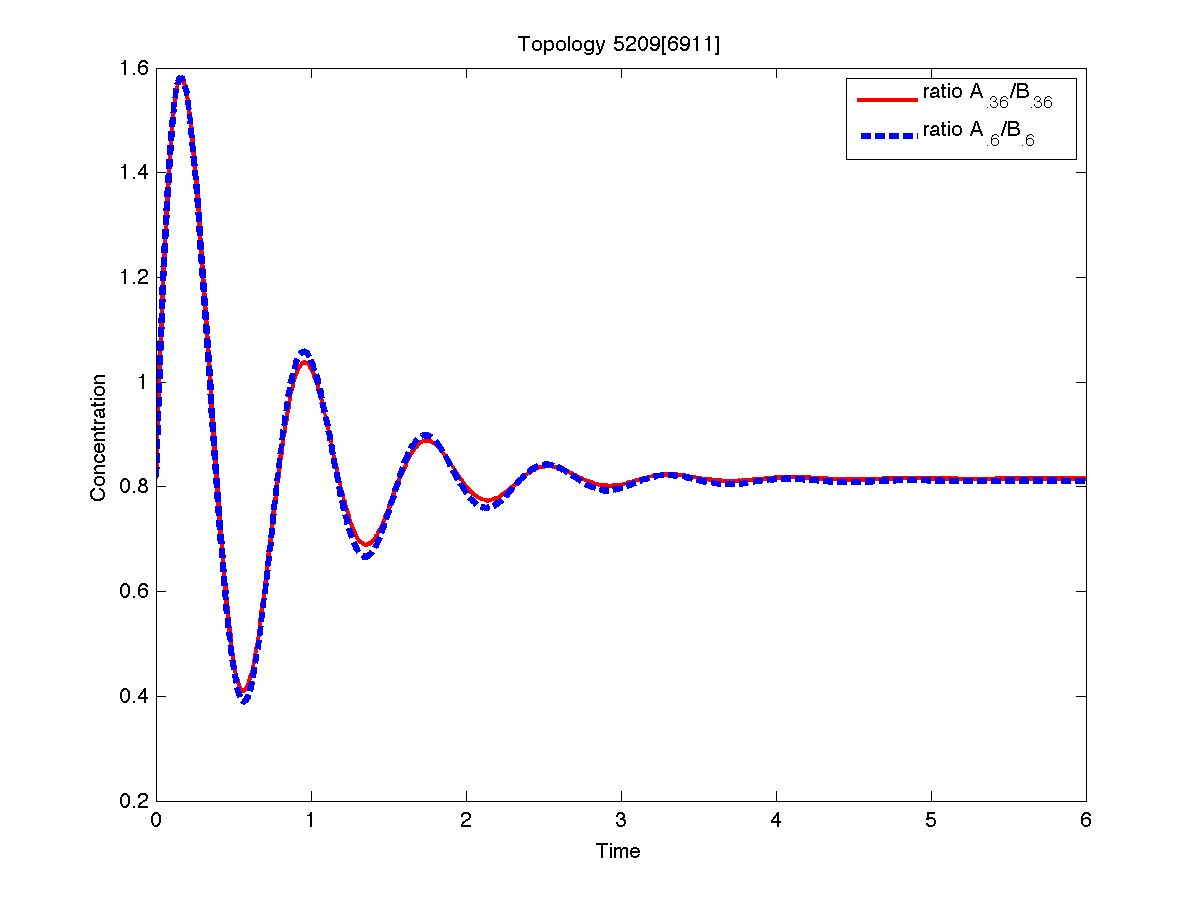}}\\   
\subfloat{\label{fig:f61}\includegraphics[width=0.55\textwidth]{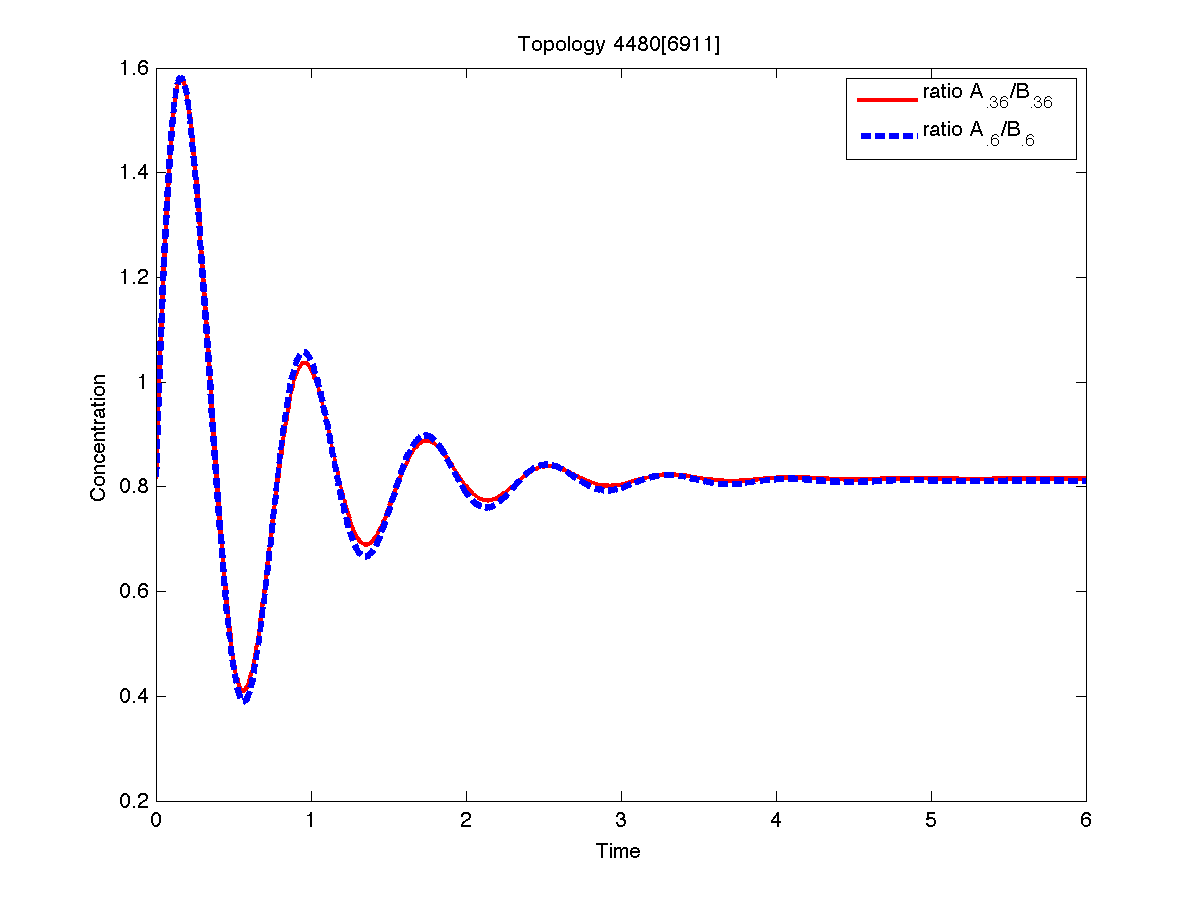}}                
 \subfloat{\label{fig:f62}\includegraphics[width=0.55\textwidth]{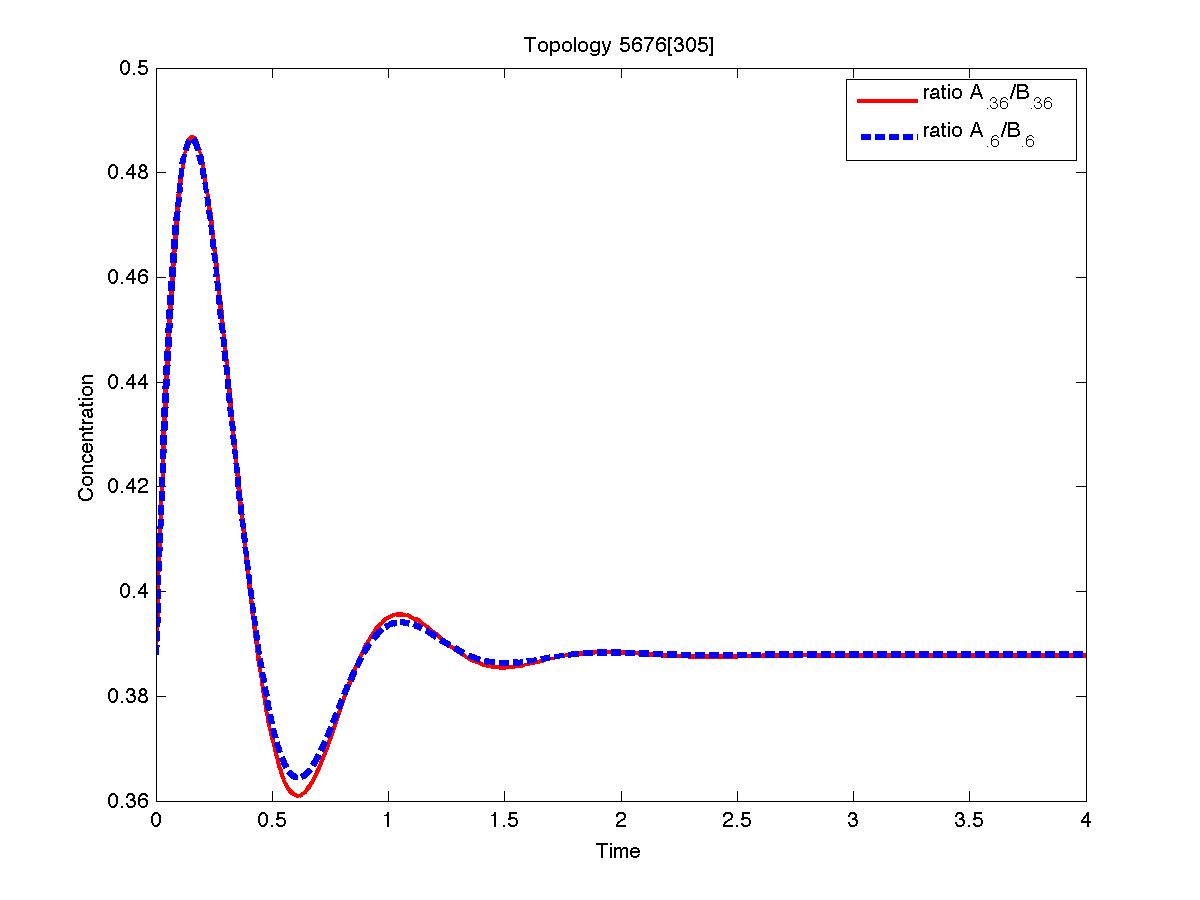}}\\   
\subfloat{\label{fig:f71}\includegraphics[width=0.55\textwidth]{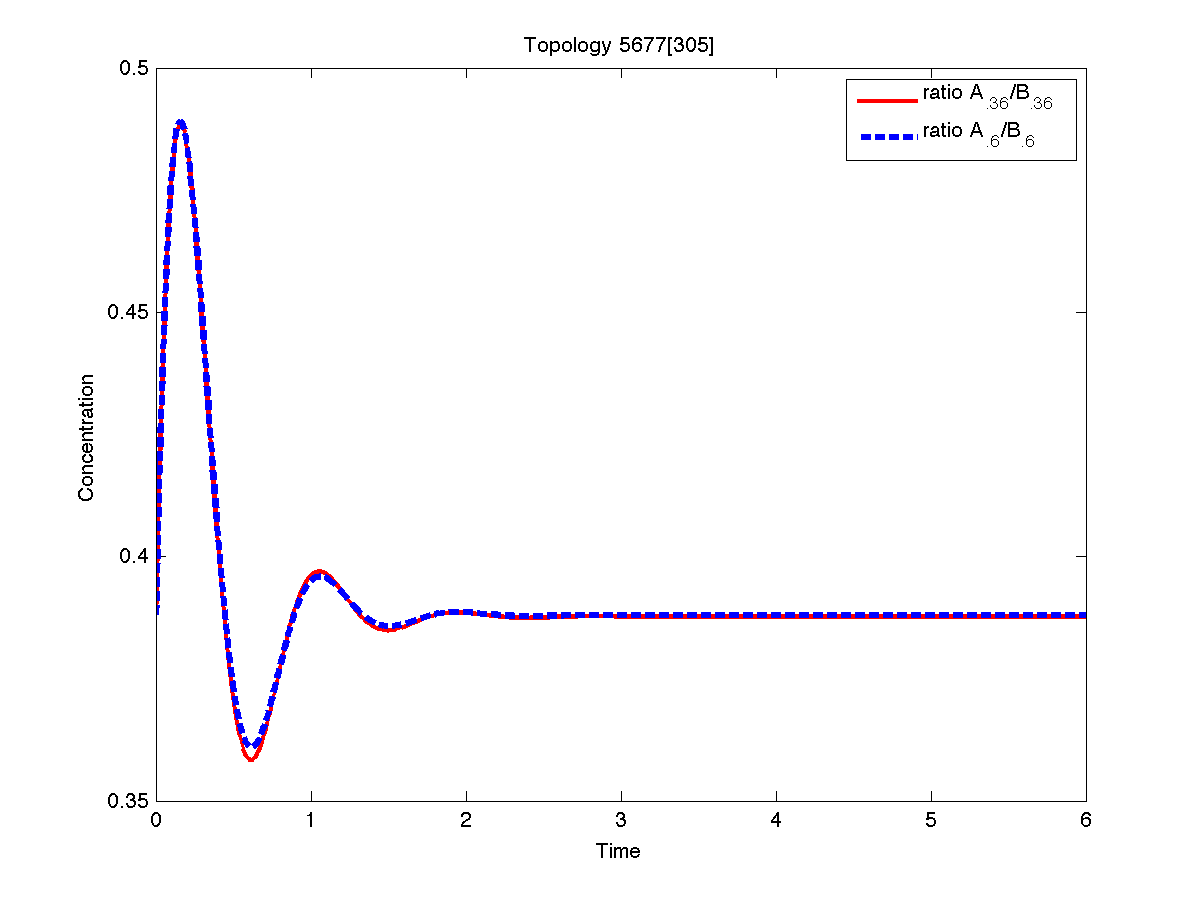}}                
\subfloat{\label{fig:f72}\includegraphics[width=0.55\textwidth]{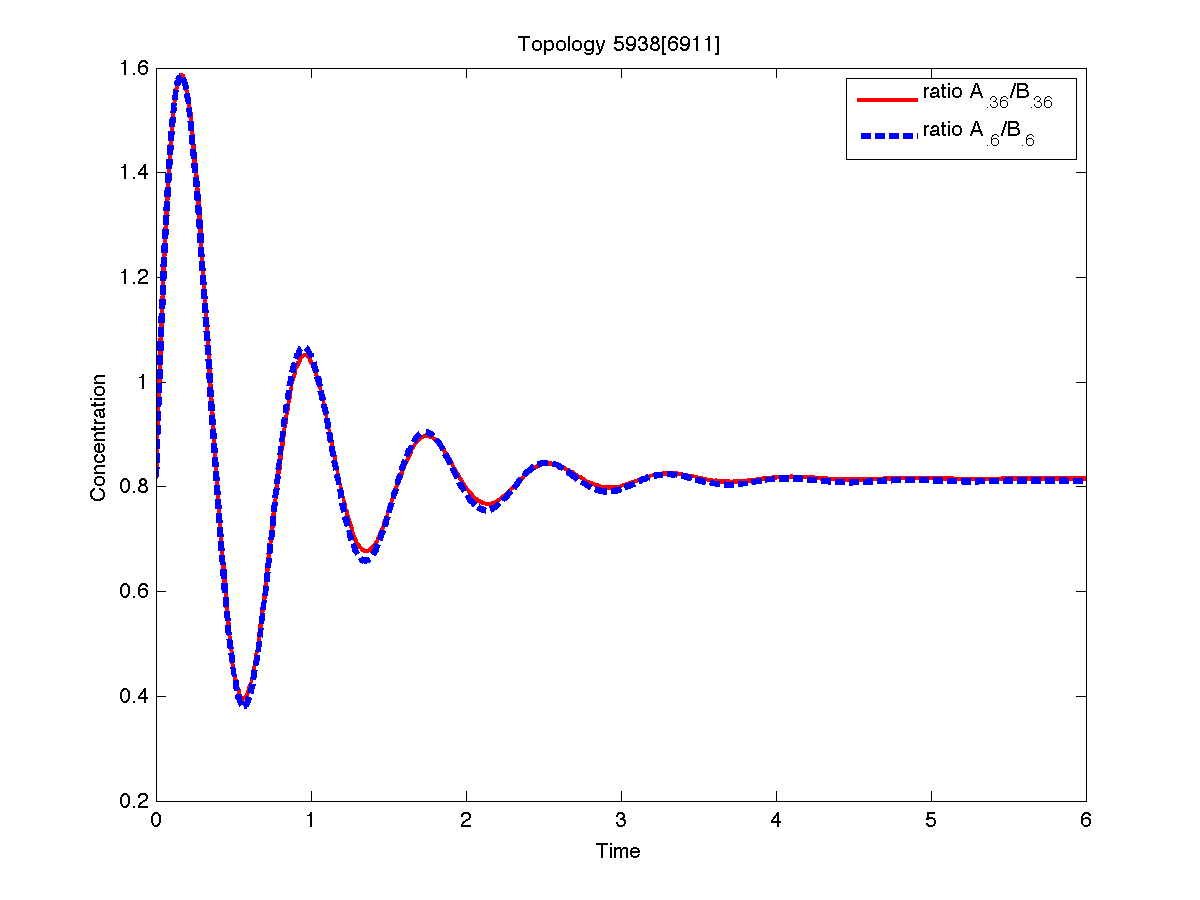}}        
\caption{$\xA(t)/ \xB(t)$ for circuits 7-12}
\end{figure}

\clearpage

\begin{figure}[ht]
  \centering
\subfloat{\label{fig:f81}\includegraphics[width=0.55\textwidth]{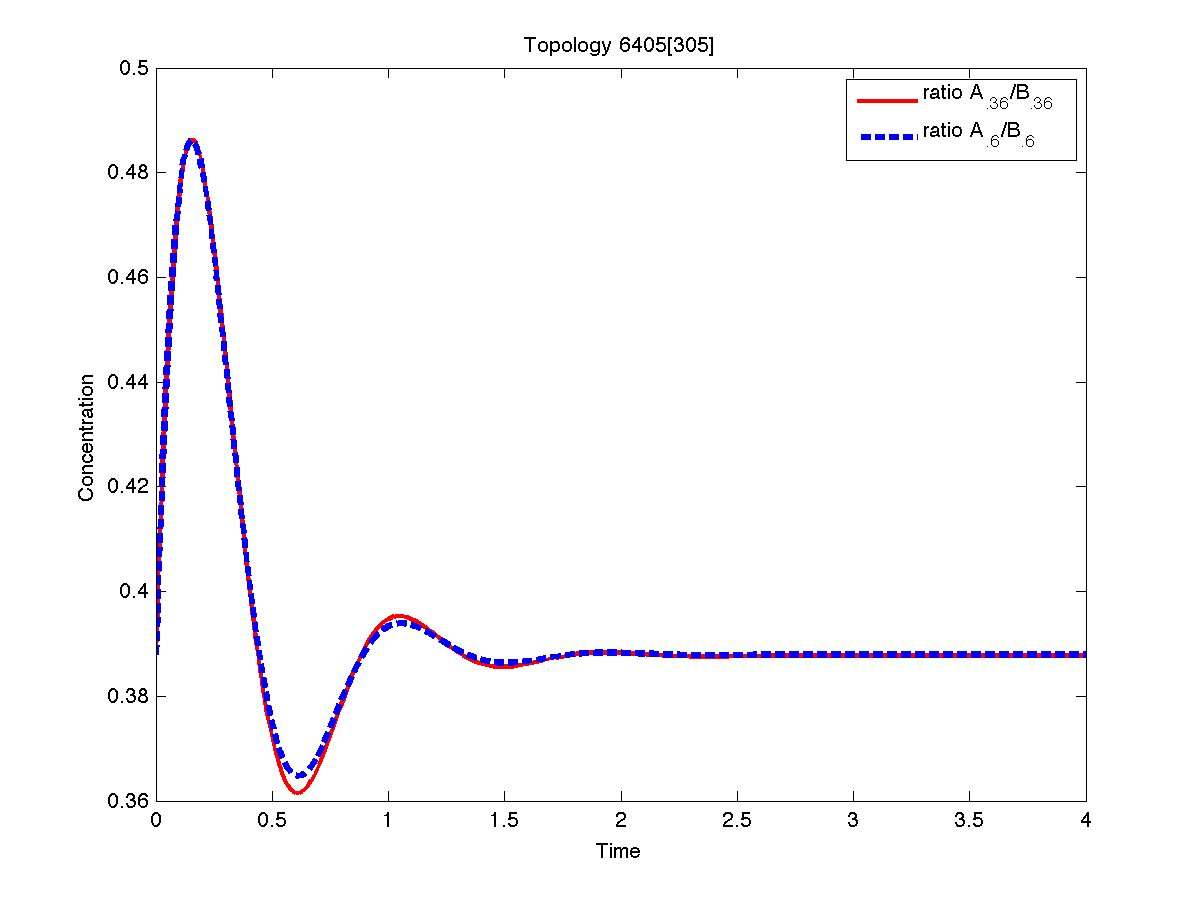}}                
\subfloat{\label{fig:f82}\includegraphics[width=0.55\textwidth]{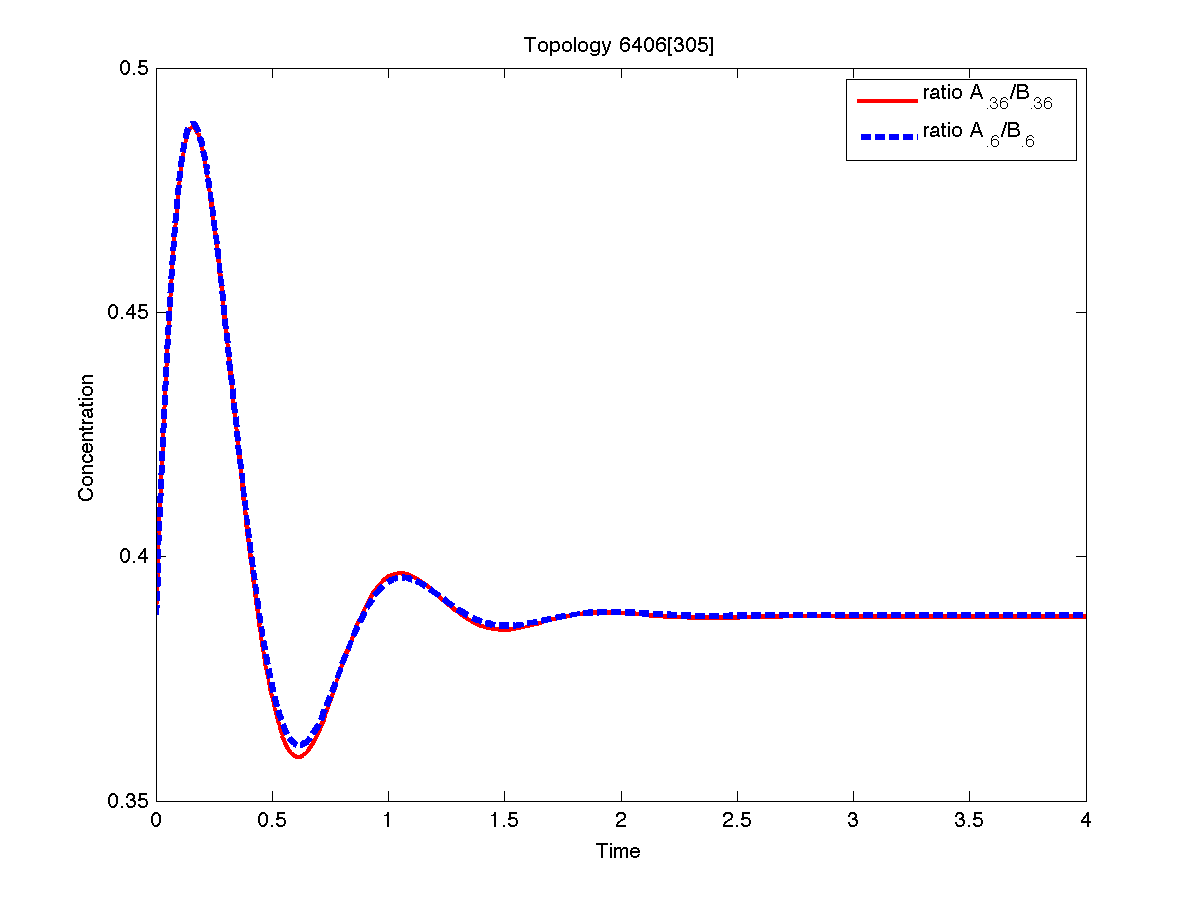}}  \\
\subfloat{\label{fig:f491}\includegraphics[width=0.55\textwidth]{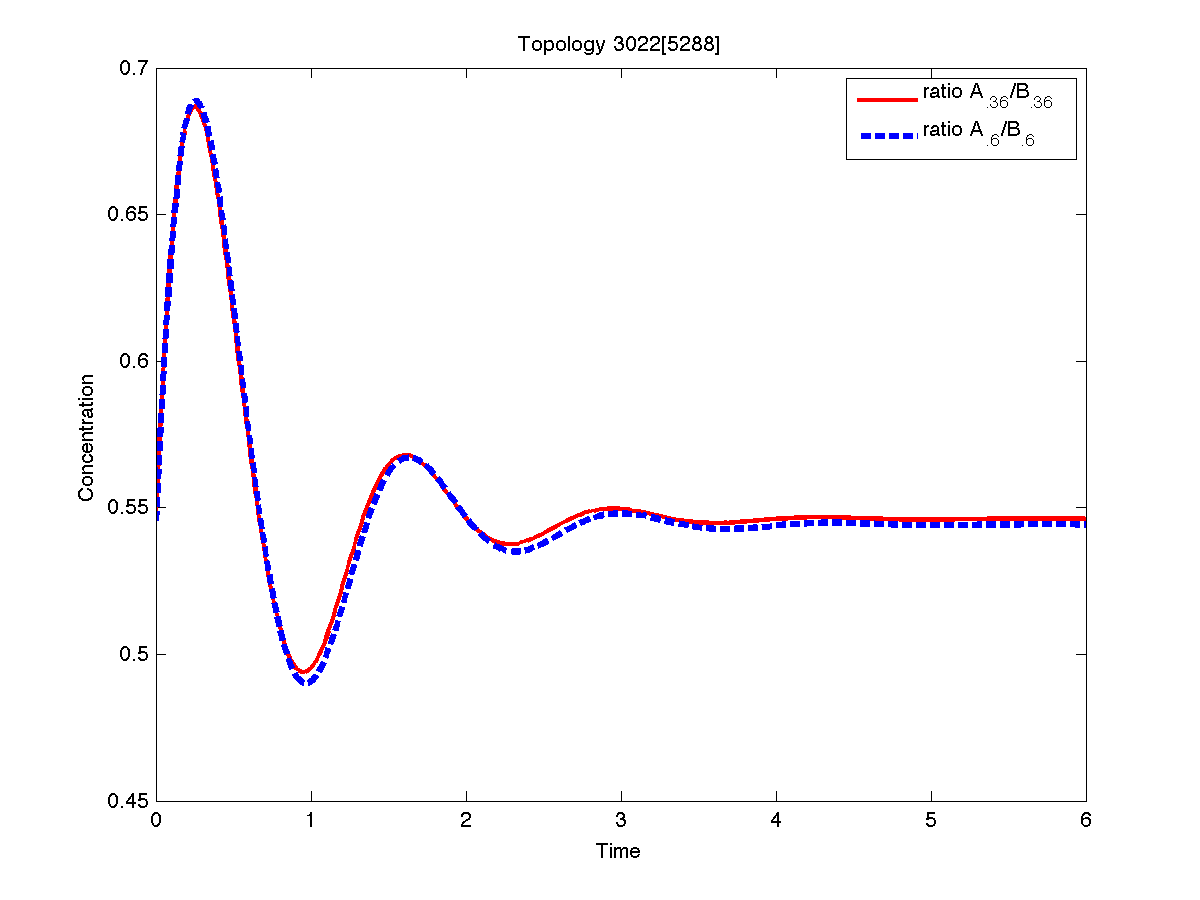}}   
\subfloat{\label{fig:f492}\includegraphics[width=0.55\textwidth]{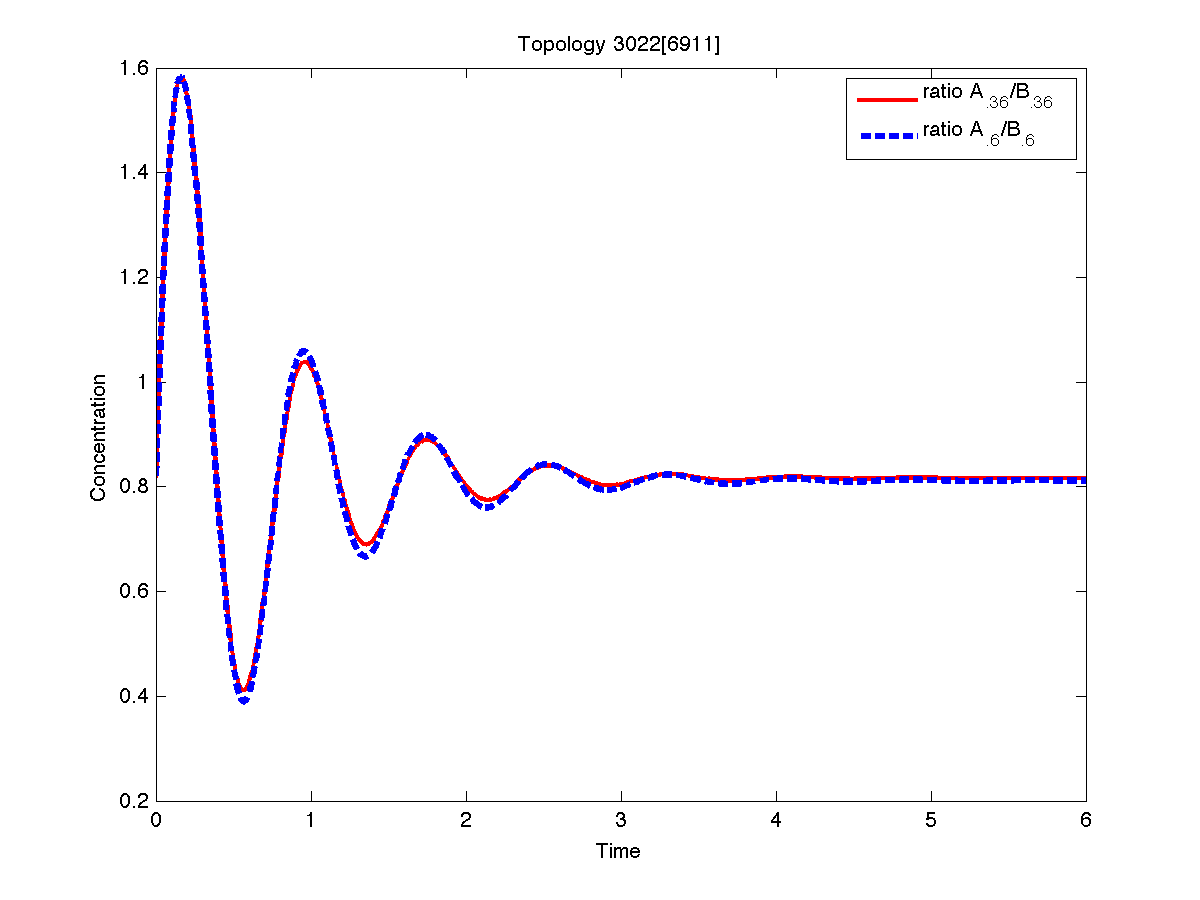}}\\    
\subfloat{\label{fig:f5101}\includegraphics[width=0.55\textwidth]{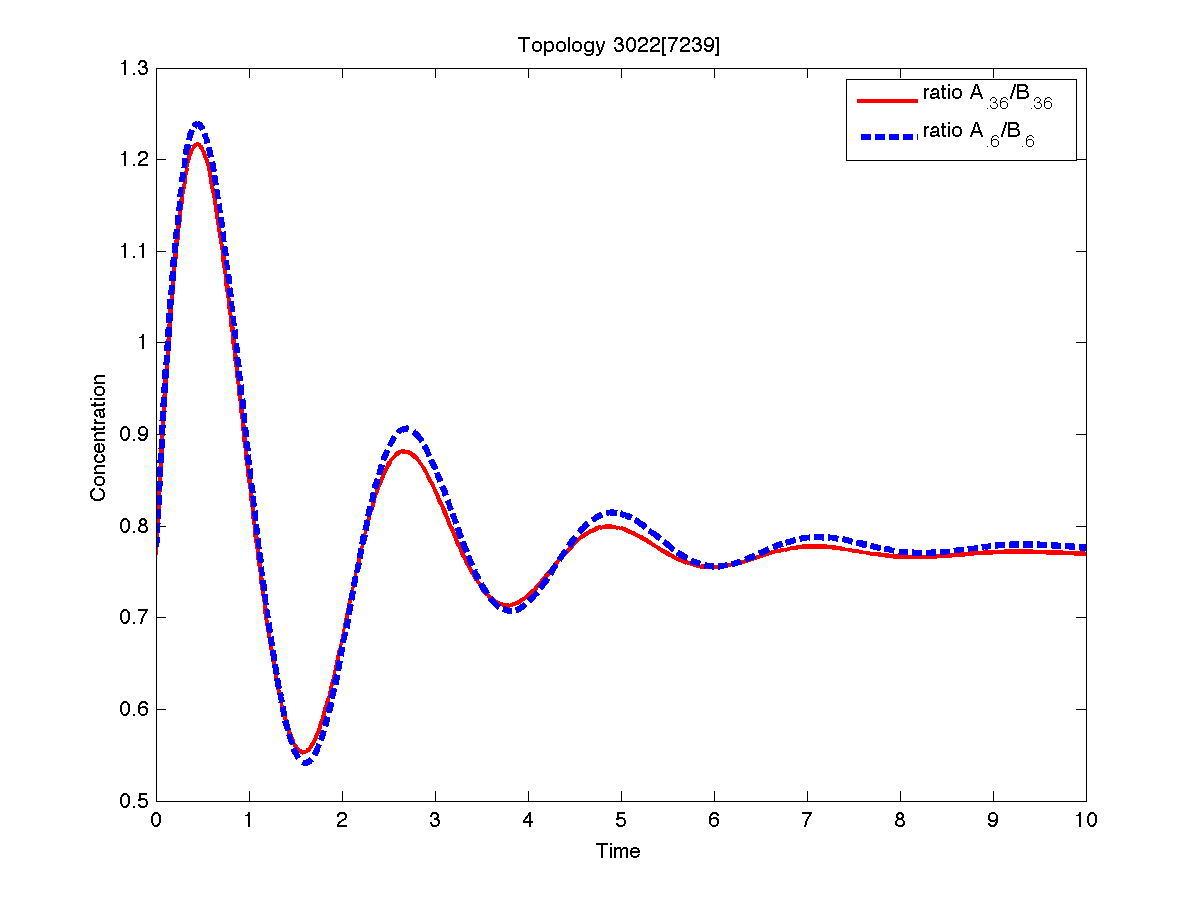}}                
\subfloat{\label{fig:f5102}\includegraphics[width=0.55\textwidth]{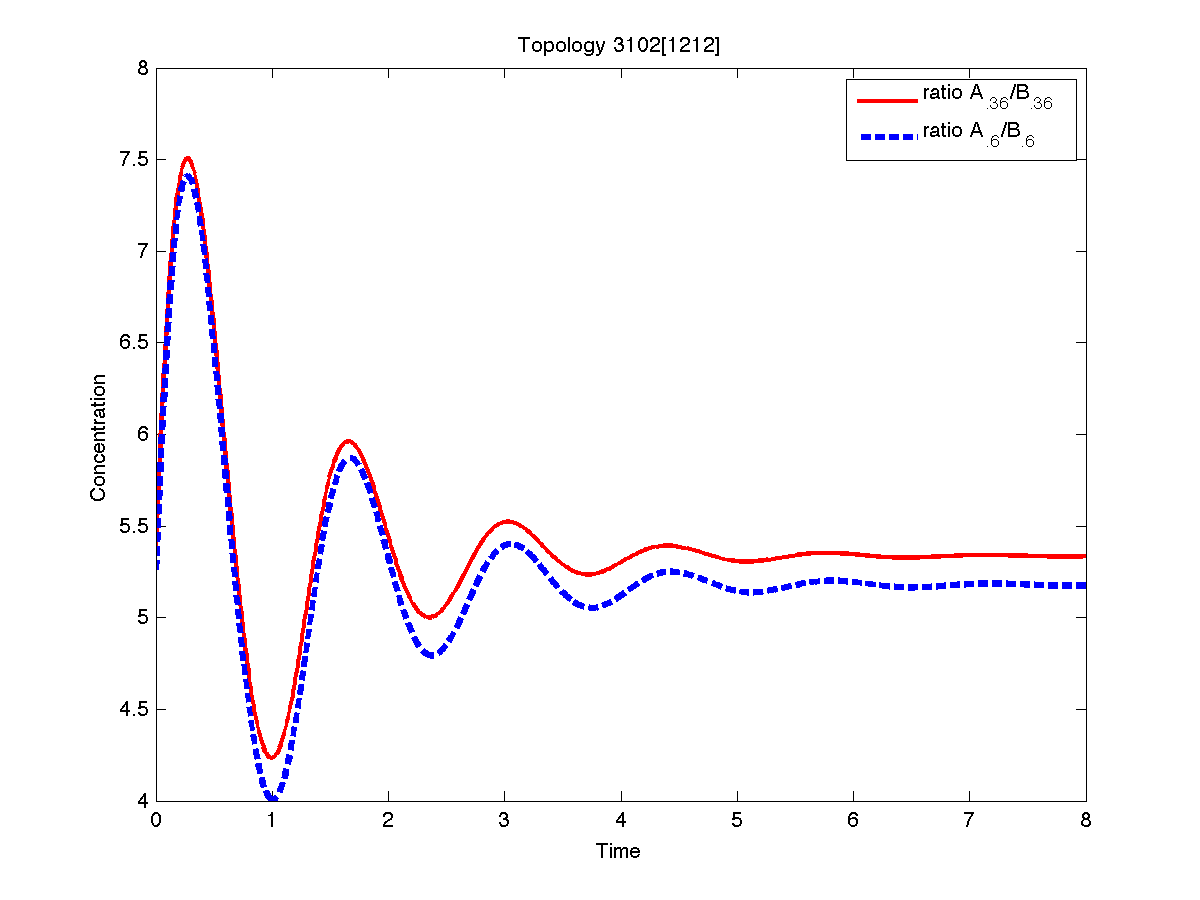}}            
\caption{$\xA(t)/ \xB(t)$ for circuits 13-18}
\end{figure}

\clearpage

\begin{figure}[hb]
  \centering
\subfloat{\label{fig:f5111}\includegraphics[width=0.55\textwidth]{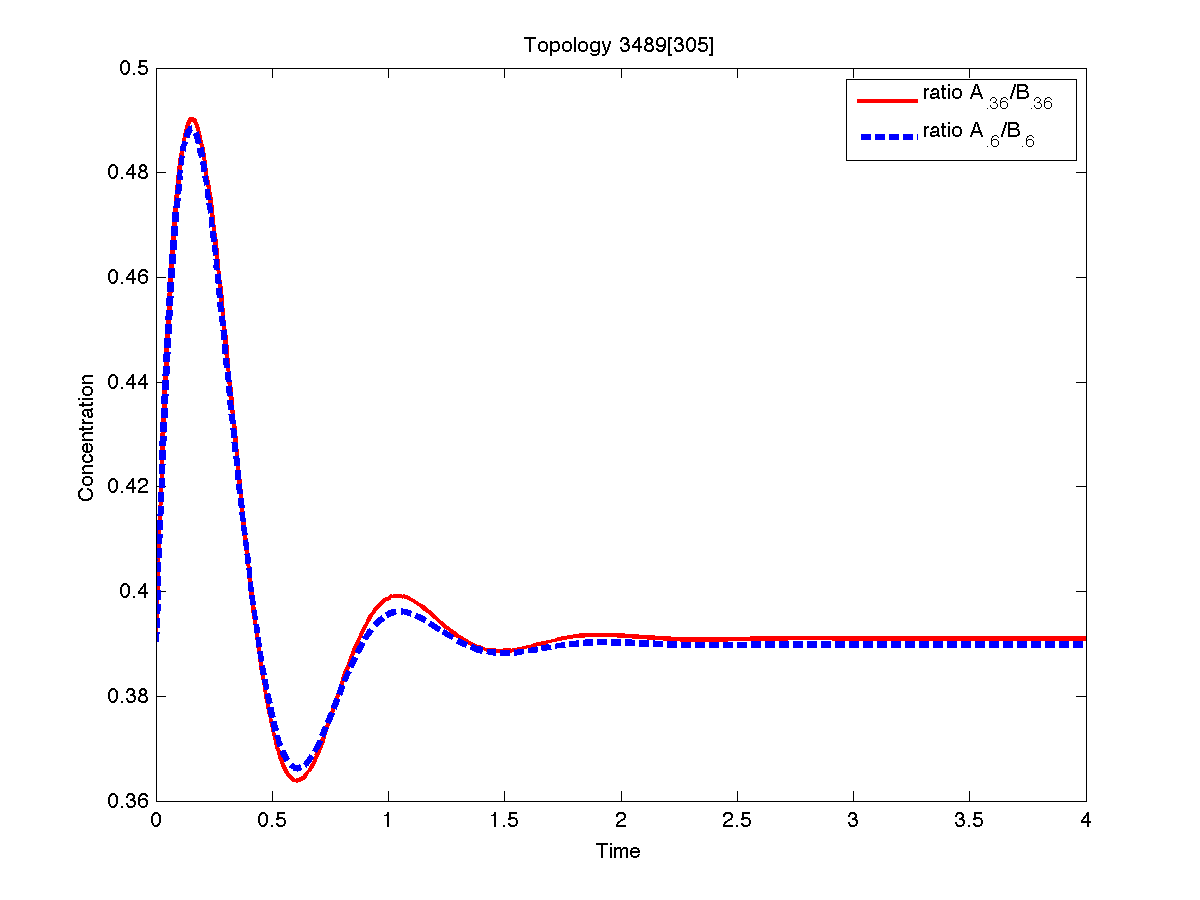}}                
\subfloat{\label{fig:f5112}\includegraphics[width=0.55\textwidth]{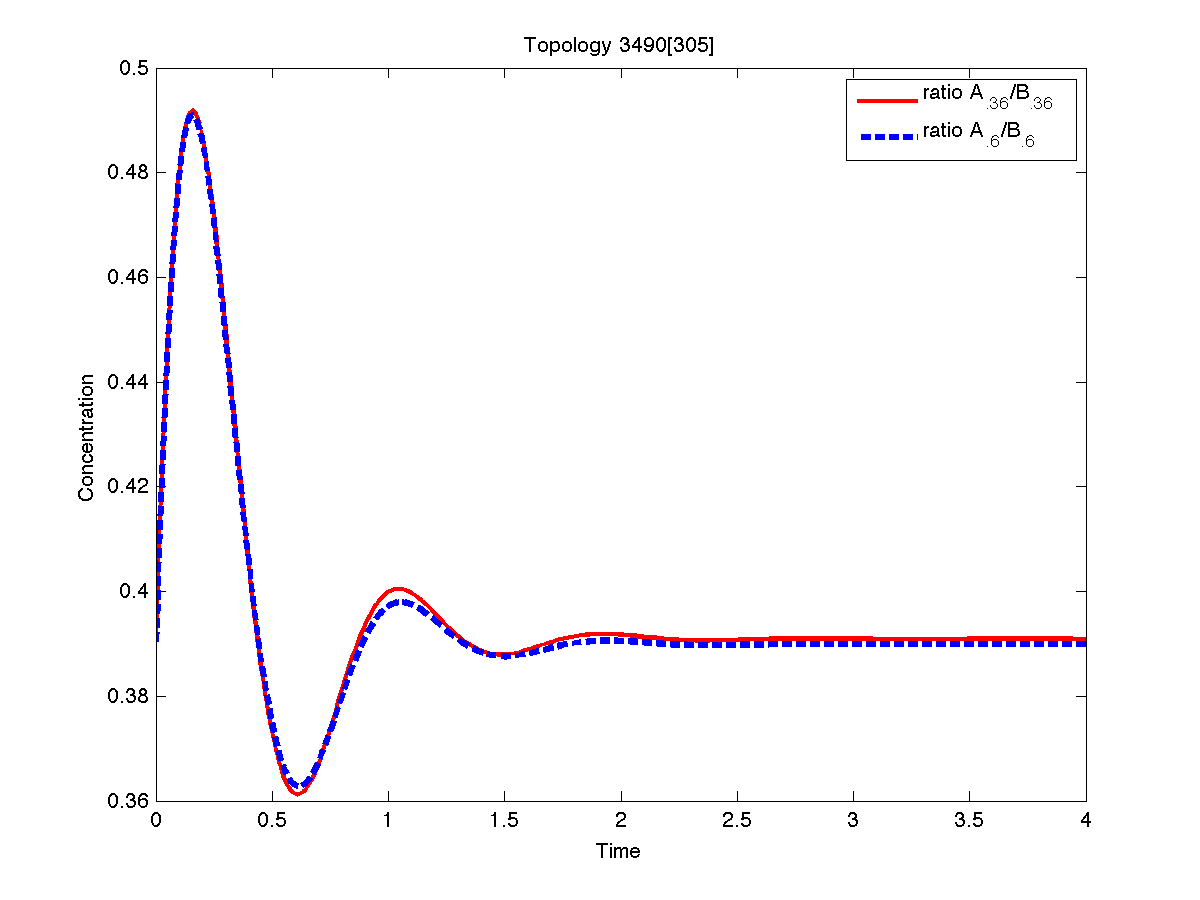}}\\        
\subfloat{\label{fig:6121}\includegraphics[width=0.55\textwidth]{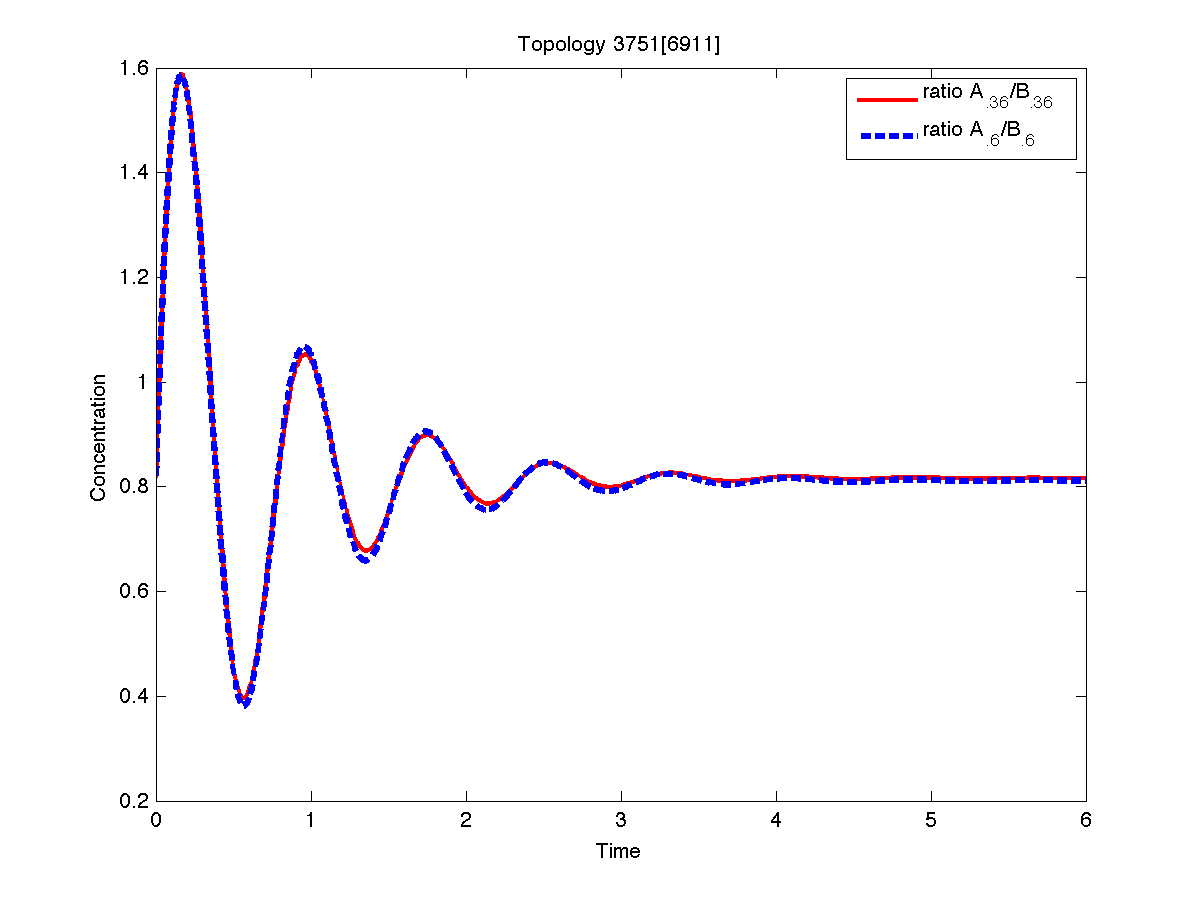}}                
\subfloat{\label{fig:f122}\includegraphics[width=0.55\textwidth]{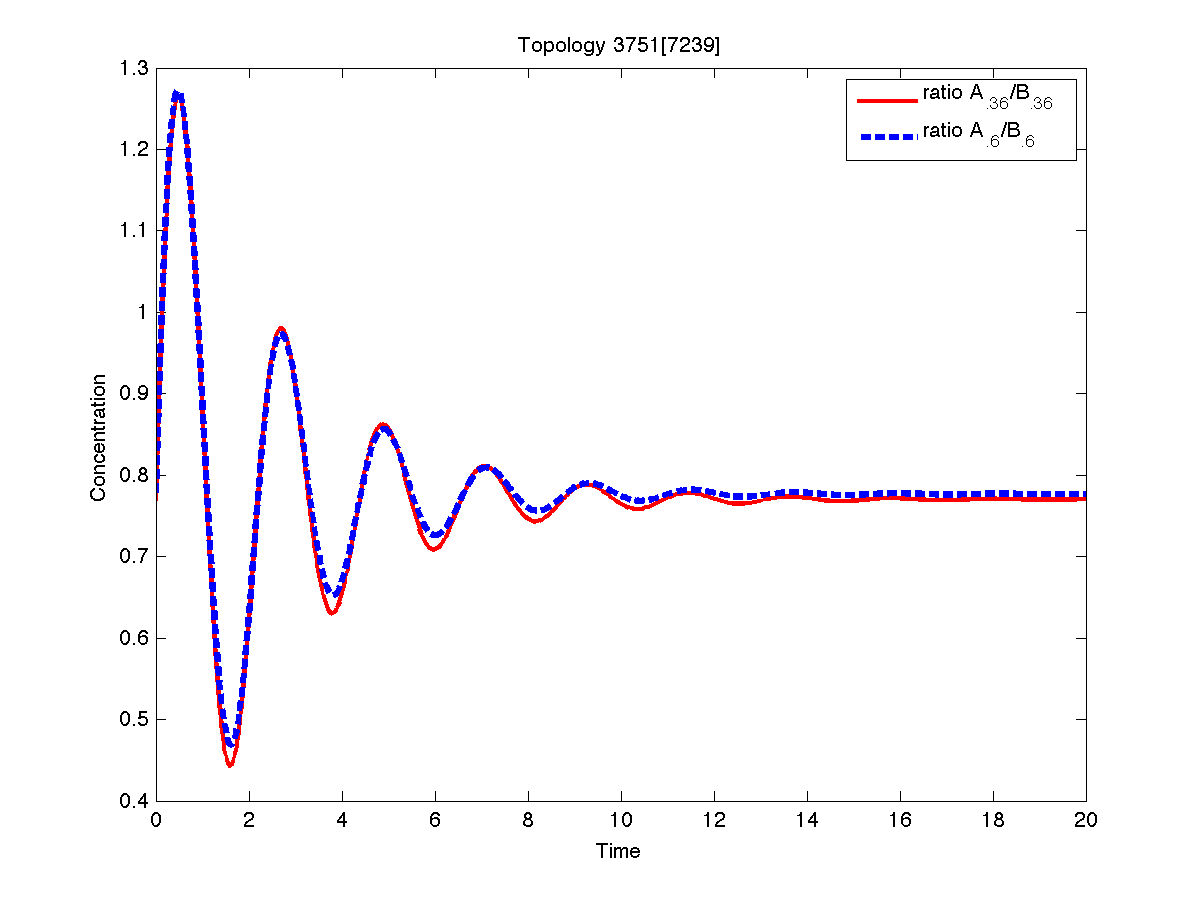}}\\
\subfloat{\label{fig:f6131}\includegraphics[width=0.55\textwidth]{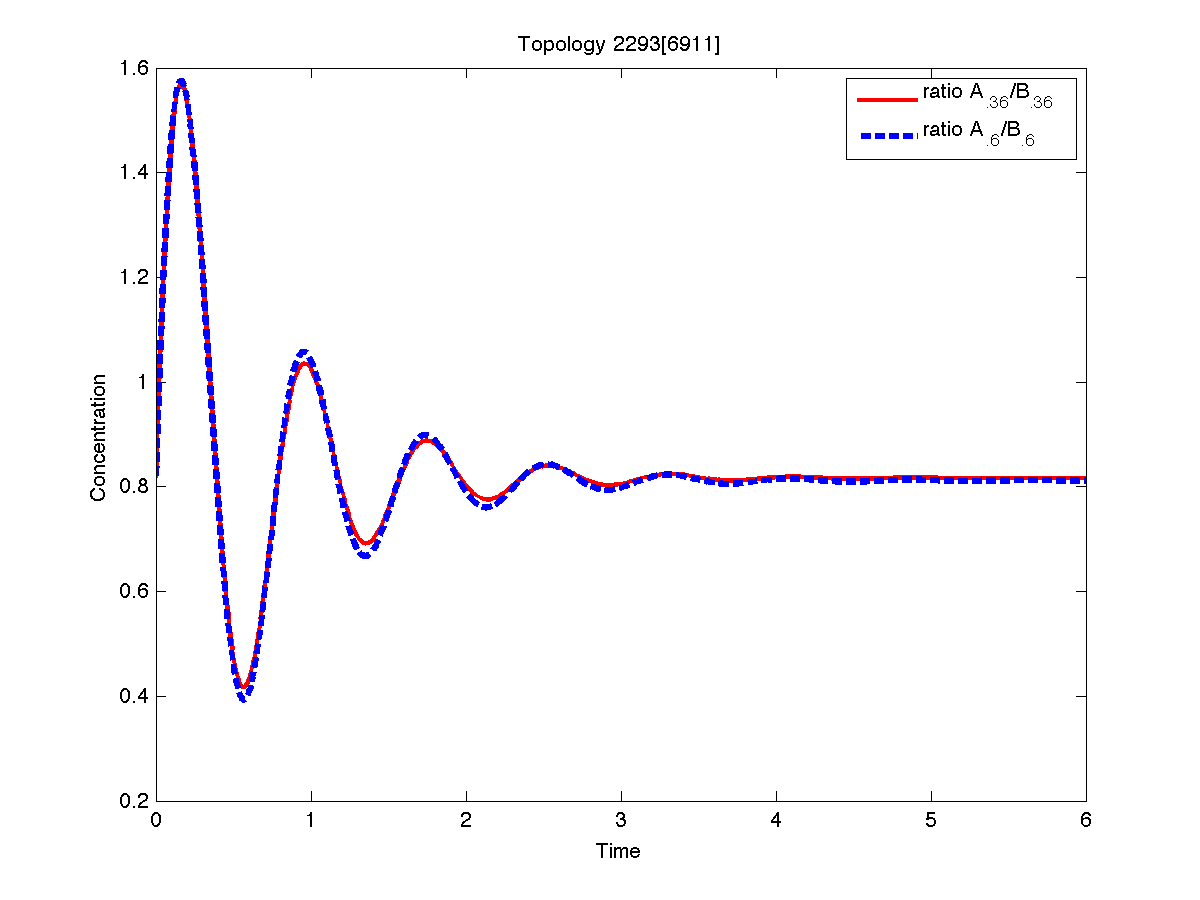}}                
  \subfloat{\label{fig:f7132}\includegraphics[width=0.55\textwidth]{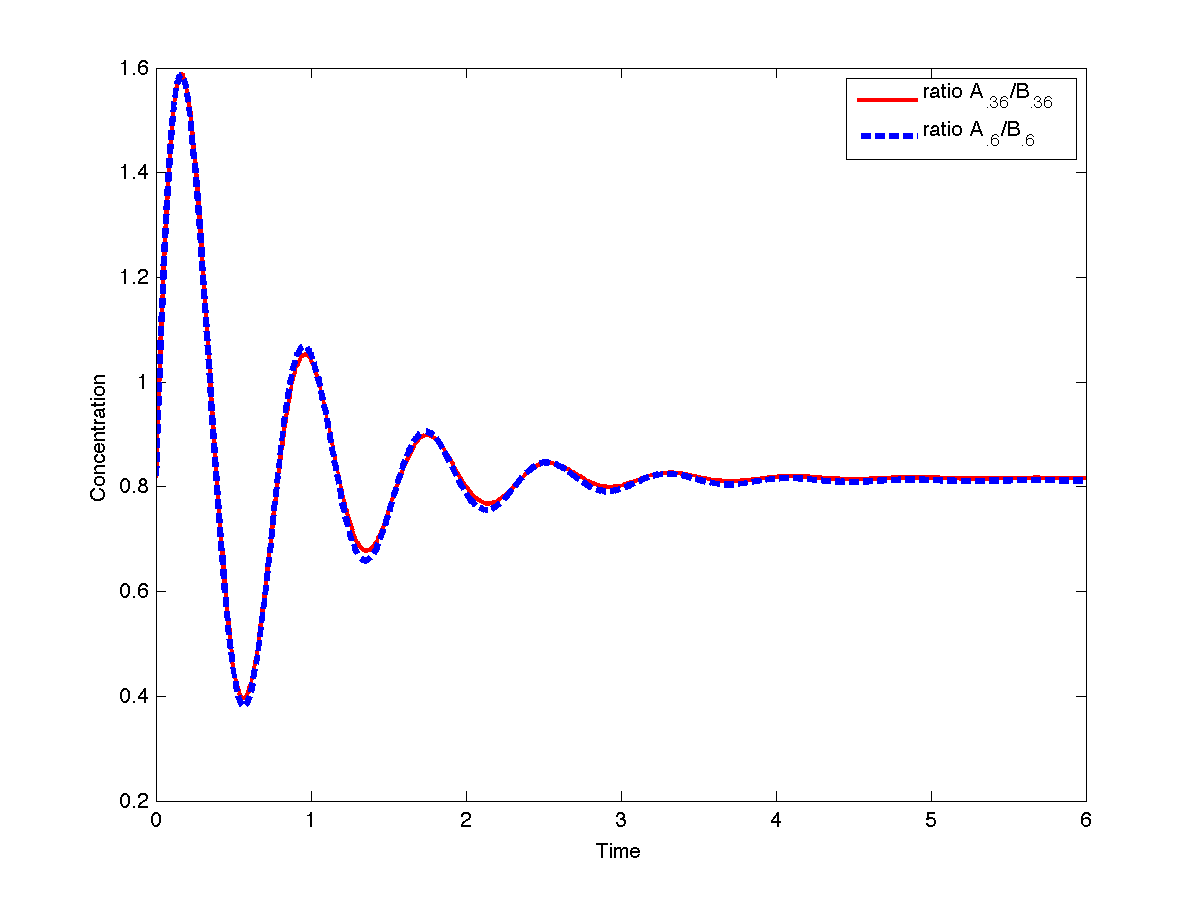}}                
\caption{$\xA(t)/ \xB(t)$ for circuits 19-24}
\end{figure}

\clearpage

\begin{figure}[ht]
  \centering
\subfloat{\label{fig:f7141}\includegraphics[width=0.55\textwidth]{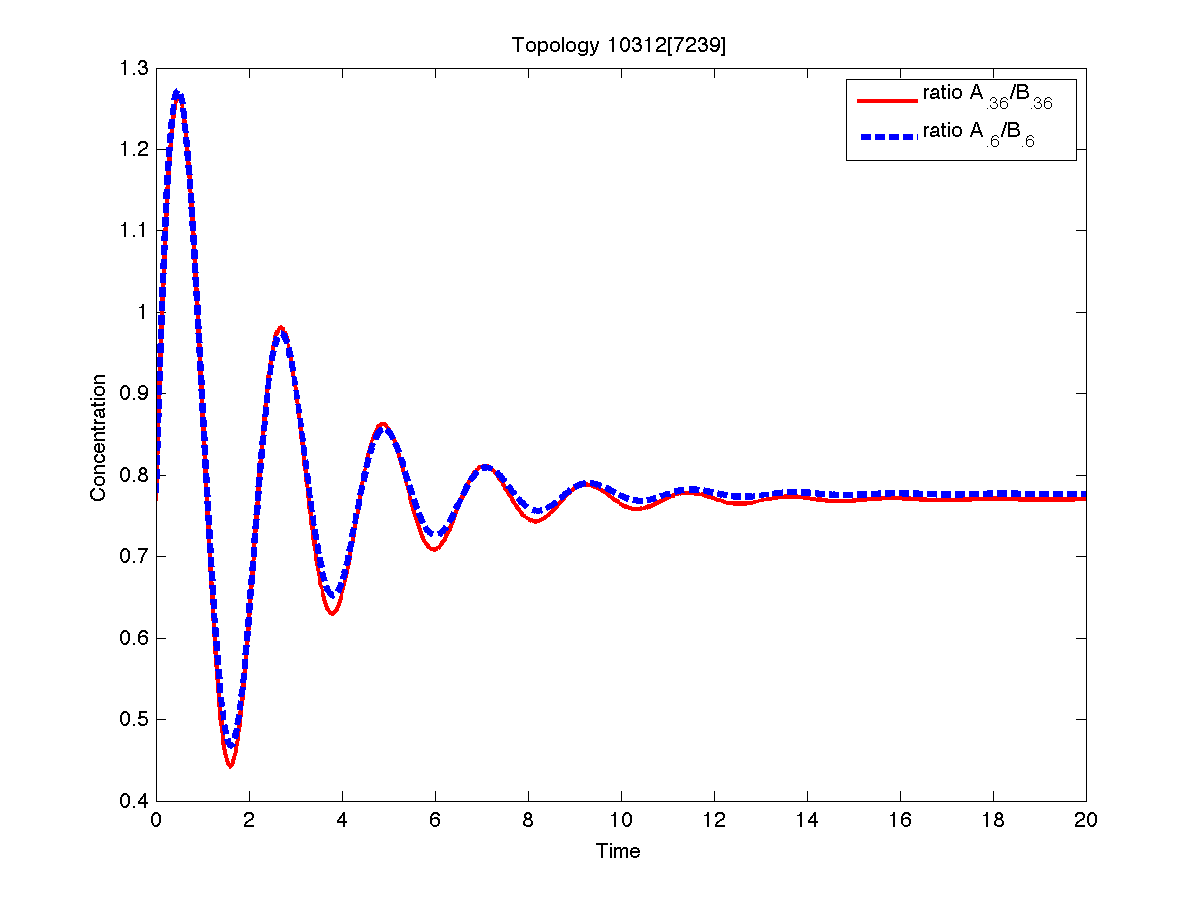}}                
\caption{$\xA(t)/ \xB(t)$ for circuit 25}
\end{figure}

%%%%%%%%%%%%%%%%%%%%%%%%%%%%%%%%%%%%%%%%%

\newpage
\clearpage
\section{Tables}

In this section the following three tables  for the 25 identified ASI  circuits are shown:

\begin{itemize}

\item  Table 1. Relative differences in linearization matrices corresponding
  to different linearizations, ${\cal A}_{0.3}={\cal A}(0.3)$, ${\cal
    A}_{0.4}={\cal A}(0.4)$, \ldots , ${\cal B}_{0.6}={\cal B}(0.6),$ rounded
  to 3 decimal places. The corresponding expressions are given by: 
\[
{\cal A}^{\mbox{err}}_{ij} = \sum_{u=0.3,0.4,0.5,0.6} 
\abs{\frac{({\cal A}_u)_{ij}-({\cal A}_{0.45})_{ij}}
          {({\cal A}_{0.45})_{ij}}}
\]
and  similarly for ${\cal B}^{\mbox{err}}$.
These relative differences are very small. The entries in the table are of the following form: ${\cal A}^{\mbox{err}}$ displayed as $[a_{11}\  a_{12} ; a_{21}\   a_{22}]$ and ${\cal B}^{\mbox{err}}$  displayed as $[b_{1}\  b_{2}]^T$.
\item Table 2. Relative error between original (nonlinear) system with an initial state $\xi =(\xA,\xB)$ corresponding to $u=0.3$, and applied input
$u=0.36$, and the approximation is $\xi + z(t)$, where $z$   solves the linear system with an initial condition of zero and a constant input of $0.06$.
Additionally,  we provide  relative errors  between the original (nonlinear) system with an initial state  corresponding to $u=0.5$, and applied input of
$u=0.6$, and the approximation given by $\xi +z(t)$, where $z$
solves the linear system with an initial condition of zero and a constant
input of $0.1$. The corresponding expressions are given by:
${\xA}_{max,u=0.36}^{err}=
\max_{t \geq 0}\abs{ \frac{ {\xA}_{0.36}^{\mbox{\tiny L}}(t) -
    {\xA}_{0.36}^{\mbox{\tiny N} }(t) } { {\xA}_{0.36}^{\mbox{\tiny N}} (t)}},$ \\
${\xA}_{max,u=0.6}^{err}=\max_{t \geq 0}
  \abs{ \frac{ {\xA}_{0.6}^{\mbox{\tiny L}}(t) 
  -{\xA}_{0.6}^{\mbox{\tiny N} }(t) } { {\xA}_{0.6}^{\mbox{\tiny N}} (t)}},$\\
where N denotes the nonlinear system, and L denotes the linear system.\\
We define similarly for  ${\xB}_{max,u=0.36}^{err}$ and ${\xB}_{max,u=0.6}^{err}.$
\item Table 3. Homogeneity property of the states $\xA$ and $\xB$. For a constant input $u$, it holds that $\sigma(pu)\approx p\sigma(u)$, where $\sigma(u)$ is a unique steady state $(\xA,\xB).$

\end{itemize}

\clearpage
\begin{table}[t]
\begin{center}
\begin{tabular}{||p{2.0cm}|p{4.5cm}|p{2.0cm}||}
\hline
\textbf{Circuit} & \textbf{${\cal A}^{\mbox{err}}$} & \textbf{${\cal B}^{\mbox{err}}$} \\ \hline
1&	$[0.069\  0.004; 0\  0.005] $ &  $[0.002\   0]^T $ \\ \hline
2 	&$[0.084\   0.006; 0.019\   0.015] $ &  $[0.004\   0]^T $ \\ \hline
3	&$[0.069\    0.004; 0\   0.005] $  & $[ 0.002\   0]^T$\\ \hline
4	&$[0.114\   0.007; 0.011\  0.003] $ &  $[ 0.002\ 0]^T$\\ \hline
5	&$[0.045\   0.003; 0.01\  0.033] $  & $[0 \ 0]^T$\\ \hline
6	       &$[0.075\   0.012; 0.021\  0.012] $  & $[0.015\ 0]^T$\\ \hline
7	       &$[0.057\   0.012; 0.021\   0.012] $  & $[0.012\ 0]^T$\\ \hline
8	       &$[0.055\  0.012; 0.019 \  0.009] $  & $[0.016\ 0]^T$\\ \hline
9        &$[0.069\  0.004; 0\   0.005] $   &$[0.002\ 0]^T$\\ \hline
10	   & $[0.037\   0.022; 0.009\  0.0707] $  & $[0.002\ 0]^T$ \\ \hline
11	 &$[0.037\   0.022; 0.007\   0.009] $   &$[0.002\ 0]^T$\\ \hline
12	 &$[0.025\   0.029; 0.007\  0.006] $   &$[0.012\ 0]^T$\\ \hline
13	 &$[0.037\   0.022; 0.009\  0.007] $   &$[0.002\ 0]^T$\\ \hline
14	 &$[0.036\   0.022; 0.007\  0.009] $   &$[0.002\ 0]^T$\\ \hline
15	 &$[0.07\ 0.004; 0\ 0.005] $   &$[0.002\ 0]^T$\\ \hline
16	  &$[0.07\  0.004; 0\   0.005] $   &$[0.002\ 0]^T$\\ \hline
17        &$[0.073\   0.012; 0.017\  0.009] $   &$[0.015\ 0]^T$\\ \hline
18	&$[ 0.051\   0.004; 0\   0.005] $   &$[0.002\ 0]^T$\\ \hline
19 &$[ 0.066\   0.013; 0.018\   0.009] $   &$[0.015\ 0]^T$\\ \hline
20   & $[0.048\   0.013; 0.018\   0.009] $  & $[ 0.016\ 0]^T$\\ \hline
21	& $[  0.051\  0.004; 0\   0.005] $   &$[ 0.002\ 0]^T$\\ \hline
22	  &$[ 0.233\   0; 0.011\  0.003] $   &$[ 0.002\ 0]^T$\\ \hline
23	  &$[ 0.069\   0.004; 0\   0.005] $   &$[ 0.002\ 0]^T$ \\ \hline
24	   &$[ 0.051\   0.004; 0\   0.005] $   &$[ 0.002\ 0]^T$\\ \hline
25	 &$[ 0.233\  0; 0.011\   0.003] $   &$[0.002\ 0]^T $\\ \hline
\end{tabular}
 \caption{Relative error in linearization matrices}
 \label{tab1}
\end{center}
\end{table}

\clearpage
\begin{table}[t]
\begin{center}
\begin{tabular}{||p{2.0cm}|p{2.0cm}|p{2.0cm}|p{2.0cm}|p{2.0cm}||}
\hline
\textbf{Circuit} & \textbf{${\xA}_{max,u=0.36}^{err}$}\   \  \  & \textbf{${\xB}_{max,u=0.36}^{err}$}   \  \  &  \textbf{${\xA}_{max,u=0.6}^{err}$}\   \  \  & \textbf{${\xB}_{max,u=0.6}^{err}$} 
\\ \hline
1	 &0.055 &0.011 &0.028 &0.005\\ \hline
2	 &0.008 &0.007 &0 &0.002\\ \hline
3	&0.055 &0.010 &0.028 &0.005 \\ \hline
4	&0.03 &0.007 &0.012  &0.004 \\ \hline
5	&0.031 &0.006 &0.003 &0 \\ \hline
6     &0.015 &0.016 &0.011 &0.005 \\ \hline
7      &0.023 &0.021 &0.005 &0.004 \\ \hline
8     &0.023 &0.021 &0.004  &0.004 \\ \hline
9	&0.055 &0.01 &0.028 &0.005 \\ \hline
10	&0.097 &0.020 &0.081 &0.016 \\ \hline
11	&0.010 &0.020 &0.084 &0.016 \\ \hline
12	&0.033 &0.021 &0.024 &0.010 \\ \hline
13	&0.097 &0.020 &0.081 &0.016\\ \hline
14	&0.010 &0.02 &0.084 &0.016 \\ \hline
15	&0.056 &0.010 &0.028 &0.005 \\ \hline
16	&0.056 &0.010 &0.028 &0.005\\ \hline
17     &0.027 &0.022 &0.004 &0.004 \\ \hline
18	&0.047 &0.010  &0.028 &0.006 \\ \hline
19       &0.027 &0.023 &0.005 &0.004 \\ \hline
20	         &0.023  &0.021 &0.005 &0.004 \\ \hline
21	&0.04   &0.009 &0.034   &0.004 \\ \hline
22	&0.116 &0.027  & 0.05 &0.013 \\ \hline
23	&0.055 & 0.010 &0.028 &0.005 \\ \hline
24	&0.045  &0.01  &0.027  &0.005 \\ \hline
25	&0.117  & 0.03 &0.05  &0.013 \\ \hline
\end{tabular}
 \caption{Relative error between nonlinear and linearized system}
 \label{tab2}
\end{center}
\end{table}

\begin{table}[t]
 \begin{center}
\begin{tabular}{||p{2.0cm}|p{2.5cm}|p{2.5cm}|p{2.5cm}|p{2.5cm}||}
\hline
\textbf{Circuit} & \textbf{$\sigma(u_{0.3})/0.3$}\   \  \  & \textbf{$\sigma(u_{0.4})/0.4$}   \  \  &  \textbf{$\sigma(u_{0.5})/0.5$}\   \  \  & \textbf{$\sigma(u_{0.6})/0.6$} 
\\ \hline
1     &$(0.195,  0.239)$   &$(0.193,  0.237)$   &$(0.192, 0.236)$  &$(0.19,  0.234)$ \\ \hline
2     &$(0.199, 0.364)$ &$(0.197,   0.359)$  &$(0.194,   0.356)$  &$(0.192, 0.353)$\\ \hline
3	&$(0.195, 0.239)$  &$(0.193, 0.237)$  &$(0.192, 0.236)$  &$(0.191, 0.234)$\\ \hline 
4       &$(0.132,  0.172)$  &$(0.131,   0.170)$  &$(0.131,  0.169)$  &$(0.13,  0.168)$ \\ \hline
5	&$(0.591,  0.11)$ &$(0.58,   0.109)$  &$( 0.57,  0.109)$ &$(0.561, 0.108)$\\ \hline
6	         &$(0.206,   0.526)$ &$(0.198, 0.507)$  &$(0.192, 0.493)$ &$( 0.188,  0.481)$  \\ \hline
7	         &$(0.208, 0.529)$  &$(0.2,   0.512)$  &$( 0.194,  0.498)$ &$(0.19,  0.486)$\\ \hline
8	         &$(0.206,  0.530)$ &$(0.199,  0.512)$   &$(0.193,  0.499)$ &$(0.189,  0.486)$ \\ \hline
9     &$(0.195, 0.239)$  &$(0.194, 0.237)$  &$(0.192,  0.236)$  &$(0.190, 0.234)$ \\ \hline
10	&$(0.078, 0.083)$ &$(0.075, 0.08)$ &$(0.073, 0.078)$ &$(0.071,  0.076)$ \\ \hline
11	&$(0.077, 0.083)$ &$(0.074, 0.08)$  &$(0.072,  0.078)$ &$(0.071,  0.076)$ \\ \hline
12	&$(0.153, 0.09)$  &$(0.145,  0.086)$   &$(0.139,  0.082)$  &$(0.135, 0.08)$\\ \hline
13	&$(0.078, 0.083)$  &$(0.075, 0.08)$  &$(0.073, 0.078)$ &$(0.071, 0.076)$\\ \hline
14	&$(0.077, 0.083)$  &$(0.074, 0.08)$   &$(0.072, 0.078)$ &$(0.071, 0.076)$ \\ \hline
15      &$(0.195, 0.239)$  &$(0.193,  0.237)$   &$(0.191,  0.235)$  &$(0.190, 0.234)$ \\ \hline
16	 &$(0.195, 0.239)$  &$(0.193,   0.237)$   &$(0.191, 0.236)$ &$(0.19, 0.234)$\\ \hline
17	         &$(0.204, 0.526)$   &$(0.197, 0.508)$  &$(0.191, 0.494)$  &$(0.186,  0.48)$ \\ \hline
18	&$(0.196, 0.24)$  &$(0.193,  0.238)$ &$(0.192,  0.236)$ &$(0.19,  0.235)$\\ \hline
19	         &$(0.205,  0.528)$  &$(0.197,  0.509)$  &$(0.192,   0.494)$ &$(0.187,  0.481)$ \\ \hline
20	         &$(0.206,  0.532)$ &$(0.199,  0.513)$ &$(0.193,  0.5)$ &$(0.189,  0.487)$\\ \hline
21	&$(0.196,  0.24)$  &$(0.194, 0.237)$  &$(0.192, 0.236)$ &$(0.191,   0.235)$\\ \hline
22	&$(0.136,  0.177)$  &$(0.134, 0.173)$  &$(0.133,  0.171)$ &$(0.132,  0.17)$ \\ \hline
23	&$(0.195, 0.239)$  &$(0.193, 0.237)$  &$(0.192,  0.236)$   &$(0.191, 0.234)$ \\ \hline
24	&$(0.196,  0.240)$   &$(0.194,  0.237)$  &$(0.192, 0.236)$  &$(0.190, 0.235)$\\ \hline
25	&$(0.136,  0.178)$  &$(0.134, 0.173)$  &$(0.133, 0.171)$  &$(0.132, 0.17)$ \\ \hline
\end{tabular}
 \caption{$\sigma(u)/u$ for constant inputs $u=0.3, 0.4, 0.5, 0.6$}
 \label{tab3}
\end{center}
\end{table}

\end{document}